\documentclass[phd, titlesmallcaps, examinerscopy, copyrightpage, foronline]{mqthesis}

\usepackage{rotating}  %used to rotate things like pictures or things you \put down
\usepackage{theorem}   %for mathematical theorems and lemmas
\usepackage{bm}  %bold fond inside equations, eg. $\bm{\alpha}$
\usepackage{amsmath,amsfonts,amssymb}
\usepackage{booktabs}  %Makes nice tables in books (sorta like RevTeX can)
\usepackage{pdfsync}   %used to sync pdf output inorder to flip from tex file to pdf and back
\usepackage{fix-cm} % Caracteres de cualquier tamaño!!!
\usepackage[utf8]{inputenc}
\usepackage[english]{babel}
\usepackage{bbm} 
\usepackage{color}% Paquete para usar letras de colores

\newcommand{\braket}[2]{\langle #1\vert #2\rangle}
\newcommand{\mean}[1]{\left\langle #1\right\rangle}
\newcommand{\abs}[1]{\vert #1\vert}
\newcommand{\norma}[1]{\parallel #1\parallel}
\newcommand{\Det}[1]{\left\vert #1\right\vert}
\newcommand{\TensorT} {\mathcal{T\!\!\!\!T}}
\renewcommand{\Vec}[1] {\textbf{#1}}
\newcommand{\derpar}[2]{\frac{\partial #1}{\partial #2}}
\newcommand{\Li}{{\rm Li}}
\newcommand{\tr}{\text{Tr}}
\newcommand{\re}[1]{\mathbb{R}\text{e}\left[#1\right]}
\newcommand{\im}[1]{\mathbb{I}\text{m}\left[#1\right]}

\ifpdf
    \pdfcatalog { /PageMode (/UseOutlines) /OpenAction (fitbh)  }
\fi

\begin{document}

\frontmatter

%%%%% Acknowledgements, titlepage, abstract, list of publications
\title{Casimir Effect in systems in and out of Equilibrium -- Efecto Casimir en sistemas en equilibrio y fuera del equilibrio}

\ifthenelse{\boolean{foronline}}{
  \author{\href{mailto:pablo.rodriguez@fis.ucm.es}{Pablo Rodr\'iguez L\'opez}}
  \department{Física Aplicada I (Termología)}
}{
  \author{Pablo Rodr\'iguez L\'opez}
  \department{F\'isica Aplicada I (Termolog\'ia)}
}

%%%%% Submition date if different from creation date
% \submitdate{March 2010}

% honours students will want to use the keyword honours instead of phd :)
% or override the \degreetext variable with
% something appropriate like:
% \renewcommand{\degreetext}%
% {in partial fulfilment of the Degree of Bachelor of Science with Honours}
% also easy to changes this to be for Masters degrees.

\addcontentsline{toc}{chapter}{Title}   % Asi aparece el Título en el propio indice
\titlepage

\chapter{Agradecimientos}

No hay palabras para agradecer suficientemente a Ricardo Brito, con el que he compartido los años que ha llevado producir esta tesis. Ricardo ha sido un mentor y un amigo, lleno de una paciencia y comprensión infinitas. Por supuesto, también quiero agradecer al departamento de Física Aplicada I (Termología) de la UCM (Universidad Complutense de Madrid), por la inmejorable acogida y por los esfuerzos hechos para que tuviera un espacio donde realizar esta tesis, no hay manera de sentirse más acogido.

Lo mismo se puede decir de Thorsten Emig, desde que llegué a Colonia, ha sido un placer trabajar y aprender a su lado. También quiero agradecer al Instituto de Física Teórica de la Universidad de Colonia, Alemania (Institut fuer Theoretische Physik, Koeln) su hospitalidad durante las distintas estancias que realicé allí.

Quiero agradecer a Rodrigo Soto, Adolfo Grushin y Alberto Cortijo el tiempo de\-di\-ca\-do, con ellos es un placer hablar de Física, siempre se aprende algo nuevo.

Agradezco también a mis compañeros del GISC por todo su apoyo, especialmente a Ricardo, Juan, Luis, Edgar y Jordan, con los que he compartido conferencias, cañas y charlas, formales e informales, habéis hecho que este camino valga la pena.

No existen palabras para expresar el agradecimento y cari\~no que siento hacia mis padres, Germán y Carmi\~na. Gracias a vosotros, a vuestro constante apoyo y confianza sin límite, todo esto ha sido posible. También quiero agradecer a mi hermana María y a su marido, Antoine, los sabios consejos que siempre me han sabido dar, y a mi sobrinita Inés, por la alegría que rebosa de la que nos hace a todos partícipes.

No puedo evitar acordarme aquí de mis abuelos, Vicente, por desgracias del destino nunca llegué a conocerte. Eduardo, Mercedes, Jexuxa, no pasa un día sin que me acuerde de vosotros, y ahora que ya no estáis, me doy cuenta de lo mucho que os hecho de menos. Espero que estéis donde quiera que estéis, os podáis sentir orgullosos de mí. Yo lo estoy de vosotros.

Por supuesto que me acuerdo del resto de mi familia: mis padrinos, mis primos, tíos, sobrinos \dots. La lista es afortunadamente interminable, muchas gracias a todos vosotros.

Y qué decir de mis amigos, Javier, Eduardo y Jacobo especialmente, de la gente de la Facultad y de Hypatia en particular. Y de toda la gente del Calasancio (esas clases de baile de Alfredo), vale la pena cada minuto que se pasa a vuestro lado.

Por último, mi más sincero agradecimiento a tí, Bea. El motivo último de esta tesis ha sido la búsqueda de la felicidad, y a medida que la he ido escribiendo he aprendido que esta felicidad no está completa si no es contigo. Gracias por acompañarme en este camino.

Mientras que estaba escribiendo esta tesis, mi abuela Jexuxa falleció tras una larga enfermedad. Esta tesis se la dedico especialmente a ella. Descanse en Paz.

Durante el desarrollo de esta tesis he recibido financiación del programa de becas de Formación de Personal Universitario (FPU), número de referencia de la beca: 2006-01310 del Ministerio de Educación y Ciencia (MEC). Dicho programa también financió una de las estacias en Colonia. También he recibido financión de los proyectos: Grupo Interdisciplinar de Sistemas Complejos (MODELICO-CM), Grupo Interdisciplinar de Sistemas Complejos. Modelización y Caracterización (Ref: GR58/08), Fluctuaciones en sistemas físicos complejos (Ref: CCG07/UCM/ESP-2870), Fluctuaciones en Sistemas Físicos de no equilibrio. Aplicación a Medios Granulares (Ref: PR34/07-15859), Modelización, Simulación y Análisis de Sistemas Complejos (MOSAICO) y Orden y Fluctuaciones en Sistemas Complejos (Ref: FIS2004/271).

\chapter{Acknowledgements}

I do not have the words to thank my supervisor Ricardo Brito with whom I have shared the long years it took to produce this thesis. Ricardo has been a mentor and a friend, always full of patience and comprehension. Of course, I also thank to the department of Física Aplicada I (Termología) (Applied Physics I (Thermology)) of the UCM (Universidad Complutense de Madrid), for the excellent hospitality and the efforts made to have a place to carry out this thesis, there is no way to feel most welcome.

The same can be said about Thorsten Emig, since I arrived in Cologne, it has been a pleasure to work and learn at your side. I also thank the Institute of Theoretical Physics at the University of Cologne, Germany (Institut fuer Theoretische Physik, Koeln) their hospitality during the fellowships that I made there.

I want to thank to Rodrigo Soto, Adolfo Grushin and Alberto Cortijo the pleasure to talk about physics with them, you always learn something new with them.

I also thank my colleagues of GISC for their support, especially Ricardo, Juan, Luis, Edgar and Jordan, with whom I shared conferences, drinks and talks, both formal and informal, you have made this journey worthwhile.

No words can express the acknowledgments and affection I feel toward my parents, Germán and Carmi\~na. Thanks to them, their continued support and confidence without limitation, all this has been possible. I also thank my sister María and her husband, Antoine, the wise advices that they have always been able to give me, and my niece Inés, she always makes us participants of her full joy.

I can not avoid thinking of my grandparents, as my grandfather Vicente, whom for misfortunes of fate I never got to meet him. Eduardo, Mercedes, Jexuxa, not a day goes by without remembering you, and now that you are not here with us, I realize how much I miss you. I hope that wherever you are, you can feel proud of me. I am proud of you.

Of course I remember the rest of my family, my godparents, my cousins, uncles, nephews, \dots. Fortunately the list is endless, thank you very much to all of you.

And what about my friends, Javier, Eduardo and Jacobo especially, the people of the Faculty and Hypatia in particular. And all the people of Calasancio (those dance classes with Alfredo), every minute spent at your side is worth.

Finally, my deepest thanks to you, Bea. The reason of this dissertation has been the pursuit of happiness, and as I was writing I learned that happiness is incomplete without you. Thanks for joining me in this way.

While I was writing this thesis, my grandmother Jexuxa died after a long illness. This thesis is dedicated specifically to her. Rest in Peace.

During the development of this thesis I was supported by a FPU grant (Becas de Formación de Personal Universitario), the reference number of the scholarship: 2006-01310 of the Ministry of Education and Science (MEC). This program also funded one of the fellowships in Cologne. I have also received financing from the projects: Grupo Interdisciplinar de Sistemas Complejos (MODELICO-CM), Grupo Interdisciplinar de Sistemas Complejos. Modelización y Caracterización (Ref: GR58/08), Fluctuaciones en sistemas físicos complejos (Ref: CCG07/UCM/ESP-2870), Fluctuaciones en Sistemas Físicos de no equilibrio. Aplicación a Medios Granulares (Ref: PR34/07-15859), Modelización, Simulación y Análisis de Sistemas Complejos (MOSAICO) y Orden y Fluctuaciones en Sistemas Complejos (Ref: FIS2004/271).

\tableofcontents
\addcontentsline{toc}{chapter}{Table of contents}   % Asi aparece el indice en el propio indice
%\listoffigures
%\listoftables

\mainmatter

%%%%% Introduction
\begin{savequote}[10cm] % this sets the width of the quote
\sffamily
``Once you start asking questions, innocence is gone.'' 
\qauthor{Mary Astor}
\end{savequote}

\chapter{Introduction}

\section{Introduction and objectives}

\subsubsection{Historical introduction}
The field of Casimir effect, or fluctuation induced forces, was started by two seminal papers published in 1948. In the first one~\cite{VdW-int.electrica}, Casimir and Polder calculated the energy between two polarizable neutral atoms in the retarder limit (considering a finite speed of light $c$). In the second one~\cite{Casimir_Placas_Paralelas}, Casimir studied the energy between two perfect metal plates facing each other. These calculations were the generalization of London energies~\cite{Energia_London} when a finite velocity of light $c$ is taken into account. In addition, these forces were predicted by Van der Waals in the context of the equation of state of gases which has his name~\cite{Van_der_Waals_Thesis}. Since them, this effect has been studied along the years with an exponential growth in the number of papers published.

For a historical review of this field of knowledge, see~\cite{4_Proceedings_QFEXT}\cite{Historia_Casimir_01}, and see~\cite{Book_Casimir_Physics_2011}\cite{RevModPhys.81.1827}\cite{BordagBook}\cite{RevModPhys.82.1887}\cite{Resource_Letter_Milton}\cite{KardarGolestanian} \cite{Johnson_Review_Numerical_Casimir}\cite{Review_Casimir} as a short list of recent reviews in the field.

%Since the first calculation of the retarded limit of energy between polarizable atoms by Casimir and Polder~\cite{VdW-int.electrica}, and the seminal work of Casimir of the energy between perfect metal plates~\cite{Casimir_Placas_Paralelas}, this effect has been studied long along the years.
%For a historical review of this field of knowledge, see~\cite{4_Proceedings_QFEXT}\cite{Historia_Casimir_01}, and see~\cite{Book_Casimir_Physics_2011}\cite{RevModPhys.81.1827}\cite{BordagBook}\cite{RevModPhys.82.1887}\cite{Resource_Letter_Milton} \cite{KardarGolestanian}\cite{Johnson_Review_Numerical_Casimir}\cite{Review_Casimir} as a short list of recent reviews in the field.
%These calculations were the generalization of London energies~\cite{Energia_London} when a finite velocity of light $c$ is taken into account. These forces were predicted by Van der Waals in the context of the equation of state of gases which has his name~\cite{Van_der_Waals_Thesis}. 

The description of this effect is as follows: Two neutral polarizable atoms in vacuum placed at a distance $d$ from each other, attract because the vacuum fluctuations induce polarizabilities in them, as a consequence, the induced polarizations interact and an attractive force between the neutral atoms appears. When effects due to finite speed of light $c$ are not important, the force is proportional to $\frac{\hbar\omega_{R}}{d^{7}}$, where $\hbar$ is the reduced Planck constant, and $\omega_{R}$ is the frequency of resonance of the atoms. This is the result obtained by London~\cite{Energia_London}. When the effect of finite $c$ becomes important, we are in the so called retarded limit, where the decay of the force with the distance is more pronounced (the force decays with $d^{-8}$ instead $d^{-7}$), and is proportional to $\hbar c\alpha_{1}\alpha_{2}$, where $\alpha_{i}$ is the electric polarizability of the $i^{\text{\underline{th}}}$ atom. As this force depends on the induced polarizations, it is expected a dependence with the temperature. In fact, when $\lambda_{T} = \frac{\hbar c}{k_{B}T}$ tends to zero (it is the so called classical limit, or high temperature regime), the force is proportional to $k_{B}T$ and reduces the decay to $d^{-7}$ again.

Casimir energy had a theoretical importance in the early times of Quantum Field Theory (QFT) as a consequence of the reality of the zero point energy~\cite{Book_Casimir_Physics_2011}.

Lifshitz extended the Casimir calculus of perfect metal plates at zero temperature to the more general case of dielectric plates at any finite temperature~\cite{Lifshitz}, which is one of the most famous and more used formulas of Casimir effect even nowadays. In contrast to the description of Casimir effect as a consequence of the zero point energy, Lifshitz proposed that this effect comes from fluctuating sources of the electromagnetic field inside the interacting objects.

There were another important contributions in the field, the result of Boyer~\cite{Boyer_Casimir_Spherical_Shell} of the Casimir energy of an spherical shell has also an historical importance, as the derivation of Balian and Duplantier~\cite{BalianDuplantier-I}\cite{BalianDuplantier-II}, just to cite a couple of them.

In Ref.~\cite{Renne1971125}, the multiscattering formalism was derived for the first time. It was later used by Kardar~\cite{PhysRevLett.67.3275}, and nowadays it leads to its modern form~\cite{Kardar-Geometrias-Arbitrarias}\cite{RE09}\cite{Casimir_forces_in_a_T-operator_approach}\cite{Reyn06}\cite{Wirzba}, which accounts with $N$ dielectrics of any geometry and nature at any given temperature $T$. The multiscattering formalism will be briefly introduced in Chap.~\ref{Cap_II-02}, because it will be widely used in the part II of this Thesis.

In the second part of this Thesis, we center our study in the multiscattering formalism of the EM Casimir effect. %, we study the appearances of nonmonotonicities in Casimir force between two objects because the presence of a third one, or because the force is correlated with a nonmonotonous Casimir entropy. We also derive the PSA from the multiscattering formalism, which lead us to obtain the N points potential formalism and to study the conditions of existence of Casimir repulsion and equilibrium points because of Casimir effect in Topological Insulators. Finally we apply this multiscattering formalism to a geometry not studied before in Casimir physics as is it the case of non parallel cylinders. This is the first case to our knowledge that two non--compact objects lead to a non--trace class operator.

\subsubsection{Numerical and Analytical Approximations to Casimir Effect}

Several approaches have been proposed to obtain reliable results of Casimir effect. The Pairwise Summation Approximation (PWS in the literature, PSA here)~\cite{Lifshitz}\cite{Golestanian}\cite{Derivacion_caso_ee_por_Milton}, which consists of the assumption of a pairwise behavior of Casimir energy. Then, the Casimir energy between two objects is obtained as the sum of the relative Casimir energy between their constituents. It can be demonstrated that PSA is a good approximation in the diluted limit~\cite{Rodriguez-Lopez_PSA}, but it gives spurious infinities when it is applied to metals. The Proximity Force Approximation (PFA)~\cite{Derjaguin_PFA}\cite{Semiclassical_and_PFA_Schaden}\cite{Reynaud_beyond_PFA}\cite{Blocki1977427}, or Derjaguin Approximation, consists of assuming that the surface of each arbitrary body consists of infinitesimal plates. Then the Casimir energy between the objects is the surface integral of these infinitesimal Casimir results. It is a valid approximation when objects are close together and their curvature radius are small. The great disadvantage of PFA is that it is an uncontrolled approximation. It is not clear at a quick glance the orientation of the infinitesimal plates, there are several criteria which lead to different results~\cite{Worldline_Approach_Thesis}: (i) Plates perpendicular to the union axis between the center of the two objects; (ii) Minimum distance between the points; or (iii) plates perpendicular to the union axis between the center of one object with the nearest point of the surface of the other.
Semiclassical approximations to Casimir effect lead to PFA results~\cite{Semiclassical_and_PFA_Schaden}, and Optical Paths approximation~\cite{Optical_Approach_I}\cite{Optical_Approach_II} lead also to results of the same nature.
From the stress--tensor formalism of Casimir effect, several numerical methods have been proposed~\cite{PhysRevLett.99.080401}, but the modern formulation of Multiscattering formalism has led to several new methods~\cite{PhysRevLett.103.040401}\cite{Worldline_Approach_Thesis}. For a modern review in numerical calculations of Casimir effect, see~\cite{Johnson_Review_Numerical_Casimir}.

\subsubsection{Experimental Results}

Early experiments carried out in the late 50's provided at best a qualitative support for an attractive force~\cite{Sparnaay}, but they were quantitative inconclusive, because of an experimental error of 100\%. The difficulty of the experiments has multiple causes. The first of all is that Casimir force is weak, and it has a microscopic nature (in general), because its proportionality to $\hbar c$. The second one is that the experimentalists have to take into account a plethora of corrections associated to the phenomenon, as the rugosity of materials, their dielectric properties, finite temperature effects, \dots and, finally, there is the difficulty to obtain the desired geometry of two microscopic parallel plates near to contact, but without being in touch.

In 1997, Lamoreaux~\cite{PhysRevLett.78.5} with a torsion pendulum, and in 1998, U. Mohideen and A. Roy ~\cite{PhysRevLett.81.4549} measured the Casimir force between a plate and a sphere with a cantilever. Several experiments followed to that one and started a new era of interest in Casimir effect experiments and theory p.~e.~\cite{Decca}.

Several more experiments have been performed in agreement with the theory: Casimir levitation~\cite{Capasso_Casimir_Repulsivo}, Casimir effect in critical systems~\cite{Bechinger}, \dots

\subsubsection{Ubiquity of Casimir Effect}

Even from the early times, Casimir effect has been considered a consequence of the quantum fluctuations of the EM field. Casimir himself claimed that, speaking with Böhr about this phenomena, Böhr claimed that this effect \textit{must have something to do with the zero--point energy}~\cite{4_Proceedings_QFEXT}.

In fact, this effect has always been related with fluctuating fields and with the Van der Waals force, proposed by Van der Waals himself to justify the existence of Phase transitions in classical Statistical Physics~\cite{Libro_Tejero}.

After that, it has been proposed the existence of Casimir force between objects immersed in fluctuating fields as: QED, critical systems~\cite{fisher1978wall}, systems in their thermodynamical equilibrium~\cite{Ajdari}, fluids~\cite{boersma:539}, granular media~\cite{BritoSotoGM}, out of equilibrium systems~\cite{Bartolo}\cite{BritoSotoPRE}\cite{Buenzli-Soto}, avalanches~\cite{PhysRevE.83.061301}, capilar Casimir effect~\cite{PhysRevLett.93.155302}\cite{Multiscattering_Casimir_Colloids_Thesis}, colloids~\cite{Bechinger}, \dots

All these systems share a property, there are fluctuating fields which induce an effective force between intrusions.

In the first part of this Thesis, we propose a dynamical formulation of Casimir effect from Langevin PDE's which models the fluctuations of the background field. This approach is not new in the sense that it has been used previously to model the Casimir effect in systems in thermodynamic equilibrium, as the case of liquid crystals~\cite{Ajdari}. Our formulation allows us to identify what these systems of different nature have in common, which is the existence of fluctuations, and thus it make explicit the relationship of the Casimir force with them. In fact, it allows to define the Casimir force as \textit{the response of a fluctuanting medium to the breakdown of the translation symmetry because of the presence of intrusions in that medium.}

%In a system in equilibrium, that means that we must have at least two objects so that Casimir force appears because of third Newton's law of motion, but this condition is not necessary fulfilled in the non--equilibrium case~\cite{Buenzli-Soto}.

\subsubsection{Related Physical Phenomena}
In this Thesis, our study is centered in the Casimir effect which appears between static intrusions in a surrounding field, which is placed in a steady state induced by a noise.

This is equivalent to assume quasi-static systems in Thermodynamics, but it does not have to be true when the intrusions move, because the assumption that we have placed in a steady state is broken. The simple non--free motion of a particle in a medium can break the status of thermodynamic equilibrium.

Therefore, we do not study the case when the intrusions have a non zero ve\-lo\-city~\cite{Intravaia_non_eq_and_dynamic_Casimir_effects}, when the steady state changes with time~\cite{Dean2}, or about other possible out of equilibrium situations~\cite{PhysRevLett.106.210404}.

Casimir effect is not the only physical phenomena associated to fluctuations.

In fact, there exists the Dynamical Casimir effect~\cite{Moore_Casimir_Dinamico}\cite{PhysRevLett.62.1742}, which consists of the generation of thermal photons of accelerated intrusions (recently measured in~\cite{Experimento_Casimir_Dinamico_Arxiv}), it is also proposed a vacuum friction of intrusions at constant velocity from each other~\cite{Intravaia_non_eq_and_dynamic_Casimir_effects}.

There also exists the Brownian motion of intrusions, also originated by fluctuations of the surrounding field, but independent of the Casimir effect. In fact, we could say that, if Brownian particles interact, it is due to the Casimir effect, because the fluctuations of the surrounded field can generate the erratic motion of the Brownian particle at the same time that the constriction of such fluctuations induce a Casimir force between the particles. It also appears induced conductivity because of the proximity of objects~\cite{PhysRevLett.106.210404}, and a plethora of other phenomena associated to fluctuations, as the problem of Dark energy, constrictions to modifications of Newton's Law of gravitation\dots

As we will see in the second part of this Thesis, Casimir effect is a generalization of Statistical Physics when we include the effect of boundaries in our system.

\section{Structure of the thesis dissertation}
The research presented here is structured in two blocks that conform the two parts of this Thesis:

\begin{itemize}

\item Part I: The dynamic formalism of Casimir effect is presented and studied here. This model complements the already known models in equilibrium systems, but this let us extend the study of Casimir effect to non--equilibrium systems. Then it allows us to define the Casimir force as \textit{the response of a fluctuanting medium to the breakdown of the translation symmetry because of the presence of intrusions in that medium.}

\begin{itemize}
\item In Chapter \ref{Dynamical approach to the Casimir effect}, we present the dynamical formalism of Casimir forces between intrusions originated from the fluctuations of the surrounding media, which is necessary to extend the study of Casimir effect to non--equilibrium systems. In fact, it is explained why, in general non equilibrium steady states, we need such dynamical formalism and we define the stochastic variable \textit{Stochastic Casimir force over a given $\alpha$ body}, which let us, by the use of stochastic calculus, define the mean value of this stochastic variable as the Casimir force itself. As an advantage, such formalism avoids the appearance of spurious divergences in the theory, because quantities independent of the distance between objects are automatically removed. As a particular example, we study the Casimir force between parallel plates immersed in a reaction--difussion media, and the case of parallel plates immersed in a nematic liquid crystal with an attenuation term because of the presence of an external magnetic field, in equilibrium and out of equilibrium cases. We also study the Casimir force between parallel plates immersed in a system whose fluctuations follow a non Hermitian evolution, something forbidden in equilibrium systems. The theory recovers the classical equilibrium Casimir force as a particular case, and admits generalizations to non--linear dynamics of the background (critical phenomena, Ginzburg--Landau, KPZ, \dots) and to more general noises (multiplicative noises, Levy noises, \dots).

\item In Chapter \ref{Chap: Stochastic Quantization and Casimir forces.}, by the use of the Stochastic formulation of QFT, also called Parisi--Wu formalism, we extend the dynamic formalism of Casimir effect to the important case of electromagnetic field subject to quantum fluctuations. As an example, we use this formalism to obtain the Casimir force between two perfect metal pistons of arbitrary given section at any temperature, and the limits of short and long distances. These results are consistent with results in certain limits of the literature. The use of the dynamical formalism also let us to obtain the variance of the Casimir force between pistons at a any given temperature. This result is more difficult to obtain from other formulations of the theory of Casimir effect. As a conclusion, this Chapter show us how Casimir effect appears in QED because of quantum fluctuations of EM field on an explicit form, and it is nothing else but a particular case of fluctuation--induced forces. In addition to that, we obtained the explicit equivalence between the Stress--Tensor formalism of Casimir effect, the partition function formalism, and the dynamic formalism of Casimir effect in the particular case of equilibrium presented in this Thesis.

\item In Chapter \ref{Chap: Extension of Langevin formalism to higher temporal derivatives and its application to Casimir effect.}, we present another generalization of the dynamical formalism. Already in Chapter \ref{Dynamical approach to the Casimir effect} it was shown how to generalize the dynamical formalism to non--linear fields and to other kind of more general noises, but the generalization presented here is in the temporal evolution of the fluctuations. We want to know what happens if a steady state is reached from a dynamic whose order of the temporal dynamic is greater than one and how would be modified the Casimir force in such non equilibrium steady state. As a result, we obtained the two point correlation function of the field in such steady states, and therefore we were able to obtain the Casimir force between parallel plates in those systems. It is important to note that this is not a useless result, when fluctuations are of an external origin, they can affect to fields whose temporal evolution was described by temporal derivatives of order greater than one, as it is the case of classical Electromagnetic field subject to random currents described by a white noise. Such case is qualitative different of the Parisi--Wu formalism studied in Chapter \ref{Chap: Stochastic Quantization and Casimir forces.}.
\end{itemize}

\item Part II: The multiscattering formalism of EM Casimir effect is studied here in order to obtain results with a physical importance.

\begin{itemize}
\item In Chapter \ref{Cap_II-02}, we give a short introduction to the multiscattering formalism, required to understand how this formalism is derived and how we use it in the following Chapters.

\item In Chapter \ref{chb2}, we study the superposition principle in Casimir effect and the appearance of nonmonotonicities in the Casimir force between two objects because of the presence of a third one in the system. Is it the Casimir force over an object equal to the sum of pairwise Casimir force of this object with the rest $N-1$ objects of the system?. The answer is not. Casimir force does not follow a superposition principle, neither to usual EM for example. In particular, with a modified multiscattering formalism (which takes into account the images of the objects), we obtained how the Casimir force between two neutral atoms and between two perfect metal spheres varies when a perfect metal plate is placed in the system. As a result, the Casimir force between compact objects, and between a compact object and the plate not only depends on the distance with the third object, but also it is nonmonotonous with such distance. Sometimes, force increases with the distance (but it do not reverse sign) because of the presence of a third object in the system.

\item In Chapter \ref{chb3}, we derive the Pairwise Summation Approximation (PSA here, PWS in the literature) for the EM field from the multiscattering formalism. PSA is the tree level of a Born expansion of the Casimir effect in the dielectric properties of the objects. This approach allows us to reobtain several results of the theory, for example the limits of high and low temperature of the Casimir energy between two objects. But we also obtain new results, as the energy of two objects at any given temperature and the magnetic contribution to PSA. We have also studied the superposition principle, which is obtained at the tree level of PSA, but broken by its first perturbation. From this formalism we do not only reobtain the $N$ point potential formalism of Power and Thirunamachandran, but we also are able to extend such formalism to magnetic couplings and finite temperature cases. Last, but not least, we study the Casimir effect between dielectric with magneto--electric coupling in general, and between Topological Insulators (TI) in special. These new state of matter materials leads to attractions, repulsions and even to the appearance of equilibrium distance points. By the use of PSA, we obtain a good approximation of the Casimir energy between objects with a general magnetoelectric coupling, in particular for TI, and we derive a criteria for the appearance of all the behaviors described here.

\item In Chapter \ref{Chap: Casimir Energy and Entropy in the Sphere--Sphere Geometry}, we study the entropy for a system of two metal spheres with three different models: perfect metal model, plasma model and Drude model. This work is motivated by the appearance of negative entropies because of Casimir effect between two Drude plates and between a sphere and a plate, both perfect metal. As a result, we obtained that an interval of negative entropies appears between perfect metal spheres and for some parameters of the plasma model, for a range of distances and temperatures. In the case of perfect metal spheres, there exists a minimum critical distance from which this anomalous behavior disappears. For Drude spheres and for plasma model with large enough penetration length this effect simply does not appear. A thermodynamical study of this system is performed and, as a consequence, we claim that the only consequence of this range of negative entropies is a nonmonoticity of the Casimir force with the temperature.

\item In Chapter \ref{Chap: Casimir energy between non parallel cylinders}, we obtain the Casimir energy for a system not studied before, it is the case of non--parallel cylinders. To do so, we obtain the translation matrix of non parallel and translated cylindrical coordinates basis, for the scalar and EM fields. In this system, the energy does not longer scale with the length $L$ of the cylinders, but with the cosecant of the angle $\theta$ between their axis. For perfect metal cylinders in the high $T$ limit, we find a case when the theorem of existence of a trace--class operator generator of Casimir energy is not applicable and is not fulfilled~\cite{Kenneth_and_Klich}. Anyway, we are able to avoid this anomalous result evaluating the force instead of the energy in this case.

%\item \OJO{In Chapter 10, we obtain the Casimir energy between an sphere and a cylinder with the multiscattering formalism. This is a system with an increasing experimental interest for which the multiscattering formalism was not already developed. In addition to that, this is the last piece to complete the theory of Casimir effect between plates, cylinders and spheres, because we have all the tools to transform from one coordinate system to any other.}

\end{itemize}
\end{itemize}

The main results of this Thesis are summarized in the conclusions Section.

%%%%% Other chapters in here

% Parte del Modelo de autofunciones
\chapter*{}
\markboth{}{} 
\addcontentsline{toc}{chapter}{Part I: Casimir effect in systems subject to noise}

\begin{center}
\vspace{-400pt}
\fontsize{300}{360}\selectfont {\sc\makebox[200pt][c]{I}}
\end{center}
\vspace{60pt}
\begin{center}
\fontsize{30}{36}\selectfont{\sc\expandafter{Casimir effect in\\systems subject to noise}}
\end{center}
% Aproximacion dinamica al calculo de fuerzas de Casimir con ejemplos, en y fuera del equilibrio
\begin{savequote}[10cm] % this sets the width of the quote
\sffamily
``The same equations have the same solutions.'' 
\qauthor{Richard P. Feynman}
\end{savequote}

\chapter{Dynamical approach to the Casimir effect}\label{Dynamical approach to the Casimir effect}
\graphicspath{{01-Casimir_Langevin/ch1/Figuras/}}

In this Chapter we present a mathematical model for the evaluation of fluctuation induced forces which appears in objects immersed in fluctuating media. 
We discuss the interest and the generality of these systems. They are ubiquitous in nature, as well as previous work in the literature. This will lead to understand why we call to the forces acting between bodies mediated by a fluctuating medium as Casimir forces. We will see that in the case where we have an environment in thermodynamic equilibrium, we could use all the tools of Statistical Physics and Thermodynamics to study this phenomenon, so partition functions, free energies and thermodynamic formalism will be used, whereas if we consider a medium in a steady state but outside the thermodynamic equilibrium, thermodynamics potentials are meaningless. The partition function, characteristic of the problem, may be useless or even unsolvable, so we only have the option of a dynamic system formalism.

There are many systems in nature which are subjected to fluctuations, of thermal or quantum origin.
For such systems, under certain physical conditions, Casimir forces, created by the confinement of fluctuations, exist
and have been calculated (see, e.g., \cite{KardarGolestanian}).
The usual way to obtain the Casimir forces uses equilibrium techniques and is therefore valid only for
systems in thermodynamic equilibrium. This means that the fluctuations must satisfy a fluctuation--dissipation
theorem that guarantees the existence of an equilibrium state, as discussed below.
Casimir forces for these systems are calculated in the spirit of the original work of H.\,G.~Casimir for the
electromagnetic case~\cite{Casimir_Placas_Paralelas}. The method takes as a starting point the Hamiltonian of the
system, from which the partition function $Z=\int \exp(-\beta H)$ is calculated,
either directly or using functional integration~\cite{Kardar-Geometrias-Arbitrarias}. In the calculation of the
partition function one must take into account the boundary conditions, that is, the macroscopic bodies
which are immersed in the system. The partition function of the system will have
different values for
different configurations, e.g., different separations of the objects.
Once the partition function has been obtained, its logarithm provides the free energy $F$.
The final step required to obtain the Casimir force is the calculation of the pressure
as the difference in the free energy when the configurations of the macroscopic bodies change
(for example, changing their position, distance or sizes). For instance, in the usual Casimir case of
forces between two flat parallel plates at separation $L$, the force per unit area is given by
$F_{C}/A=-\partial F/\partial L$.

The second approach also takes as a starting point the Hamiltonian of the system. However, in this approach
the Casimir force is derived not from the free energy but from
the stress tensor $\mathbb{T}$, which is integrated over the surface of the
macroscopic bodies and then averaged over the thermal Boltzmann distribution of the
associated Hamiltonian $\exp(-\beta H)$.
The approach based on the stress tensor has been taken by several authors~\cite{Cardy}\cite{Bartolo}\cite{Krech}\cite{Dean1}.
In fact, both approaches are equivalent and valid for equilibrium systems only, as will be demonstrated in Sect.~\ref{Appendix C: Equivalencia formalismos}.
The reason is that both are based on properties which are only valid in
equilibrium situations. The former uses the thermodynamic relation for the pressure as the derivative of the
free energy with respect to the volume, and the latter uses the Boltzmann distribution
function, which is only valid for systems in equilibrium.

On the other hand, other authors have developed a dynamical approach~\cite{Ajdari}\cite{Bartolo}\cite{CasimirGranular} \cite{Dean2}.
Here the starting point is an evolution equation for the considered field(s), supplemented
with a noise source term, so that the evolution of the field takes the form of a
Langevin equation. Once this equation is solved, the field is inserted into the expression
for the pressure and the average over the noise is taken. As we will see in the next Section,
if the noise is of internal origin,
say thermal or quantum, this description reduces to the equilibrium one, because of the fluctuation--dissipation theorem.
However, the noise does not necessarily have to be internal but can have an external origin~\cite{Sagues}, for instance,
a system in a fluctuating temperature gradient~\cite{Gollub}, subjected to
external energy injection such as vibration~\cite{Cattuto} or electrically driven convection~\cite{Brandt}, light incident on
a photosensitive medium~\cite{photosensitive}, or spatially and/or temporally correlated noise, as considered in Ref.~\cite{Bartolo}. Recently, Ref.~\cite{NajafiGolestanian} generalized this method to a nonequilibrium
temperature gradient. In none of these cases can the equilibrium approach
be applied. Also, the internal dynamics cannot satisfy the condition of detailed balance, and therefore the internal noise is not described by the  fluctuation--dissipation relation. In both cases, it is only possible to calculate Casimir forces via the dynamical approach.
In all these nonequilibrium cases a common feature shared with the equilibrium Casimir force is that the origin is the limitation of the fluctuation spectrum at large wavelengths. They are, therefore, conceptually different from other fluctuation-induced phenomena such as ratchets or Brownian motors that act at small length scales.

The plan of this Chapter is as follows.
We start in Sect.~\ref{sec.2} by presenting the Langevin equation subjected to a general noise. We stress
the differences between the cases when the Langevin equation derives from an energy
functional or not, and discuss the implications of the fluctuation--dissipation theorem.
Section~\ref{Casimir forces from the average stress tensor} derives the Casimir force from the stress tensor, while Sect.~\ref{Computation of Casimir forces} calculates the actual Casimir force by substituting the solution of the Langevin equation into the stress tensor.

The subsequent Sections~\ref{Reaccion-Difusion}, \ref{Cristales-Liquidos}, and \ref{2campos_no_hermitico} are devoted to the application of the formalism to different physical systems and different nonequilibrium conditions, that is, different ways of violating the fluctuation--dissipation theorem.
In particular, Sect.~\ref{Reaccion-Difusion} studies a reaction--diffusion system with three types of noise: (1)~a noise uncorrelated in
space and time, (2)~a noise exponentially correlated in time, and (3)~a spatially homogeneous noise, only fluctuating in time. Section~\ref{Cristales-Liquidos} is devoted to the study of a liquid crystal, with an equilibrium noise, satisfying the fluctuation--dissipation theorem and therefore in an equilibrium situation. We continue with a temporally correlated noise and finish the Section with a maximally correlated noise. Finally, to illustrate the power of the method, we apply it to a two-field system where the evolution equation is non-Hermitian in Sect.~\ref{2campos_no_hermitico}.
Usual approaches, based on equilibrium properties, have no applicability in this case. We finish with some conclusions.

The contents of this Chapter is based on the work published in~\cite{PhysRevE.83.031102}.
\section{Equilibrium and nonequilibrium fluctuations}\label{sec.2}
The most widely used tool to study the dynamics of fluctuations is the Langevin equations and its related Fokker--Planck
equation. There is a wide literature on this subject, in particular using Langevin equations; see, for example, Refs.~\cite{deGroot}\cite{Gardiner}\cite{Risken}\cite{vanKampen}\cite{Zwanzig}.

Let us consider a linear stochastic differential equation for the field $\phi(\textbf{r},t)$,
\begin{equation}\label{Langevin}
\partial_{t}\phi = - \mathcal{M}\phi + \xi(\textbf{r},t),
\end{equation}
which is a generalization of the Langevin equation to spatially extended systems.
In this equation, $\mathcal{M}$ is an operator (usually differential)
that can be Hermitian or non-Hermitian.
The operator does not depend on the field $\phi$, so the Langevin
equation \eqref{Langevin} is linear. To simplify notation, we have assumed Langevin equations without memory, but the generalization to memory kernels is direct.

For the important case of critical fluctuations, which leads to the so called Critical Casimir effect, the correct associated Langevin equation is a Ginzburg-Landau equation with a white additive noise term, then the hypothesis of linearity of $\mathcal{M}$ operator fails in this case. It has consequences in the study of critical Casimir effect with this formalism, but the linearity of $\mathcal{M}$ is enough general to restrict us to this hypothesis, as will be shown.

The term $\xi(\textbf{r},t)$ is a Gaussian noise that represents the random
or stochastic force acting over the field $\phi$,
and therefore it is the source of fluctuations for $\phi$. It is customary
to assume that the noise is Gaussian, and its averages are
\begin{eqnarray}
\mean{\xi(\textbf{r},t)} & = & 0, \label{noises} \\
\mean{\xi(\textbf{r},t)\xi(\textbf{r}',t')} & = & \mathcal{Q}\delta(\textbf{r} - \textbf{r}')\delta(t - t') = h(\textbf{r} - \textbf{r}')\delta(t - t'), \nonumber
\end{eqnarray} 
where $\mathcal{Q}$ is a Hermitian operator that can contain differential and integral terms. Differential terms characterize noise of conserved quantities, and integral terms noises with spatial correlations. The application of this operator to the Dirac delta function produces the spatial correlation distribution $h$.
The noise here is uncorrelated in time, although temporal correlations will also be considered below.

Equation~\eqref{Langevin} admits a solution for an initial condition $\phi^{0}(\textbf{r})$ as
\begin{equation}\label{solutionLangevin}
\phi(\textbf{r},t) = e^{- \mathcal{M}t}\phi^{0}(\textbf{r}) + e^{ - \mathcal{M}t}\int_{0}^{t} d\tau \,
e^{\mathcal{M}\tau}\xi(\textbf{r},\tau).
\end{equation}
In the limit $t\to\infty$, $\phi(\textbf{r},t)$ reaches a stationary state iff
$e^{-\mathcal{M}t}\phi^{0}\to 0$ . This implies that the eigenvalues of $\mathcal{M}$
must have positive real parts.

From the Langevin equation, one can construct a functional
Fokker--Planck equation for the probability distribution $P$
of the field $\phi$. The technique is standard (see, e.g., \cite{Gardiner}), and its solution (which is not normalizable) is a
Gaussian of the form
\begin{equation}\label{solutionFP}
P[\phi] = \frac{1}{\sqrt{\Det{\mathcal{K}}}} e^{- \frac{1}{2}\int d\textbf{r}\phi\mathcal{K}\phi},
\end{equation}
where the Hermitian operator $\mathcal{K}$ is the solution of the functional Lyapunov equation~\cite{Zwanzig}
\begin{equation}\label{matrixg}
\mathcal{M}\mathcal{K}^{-1} + \mathcal{K}^{-1}\mathcal{M}^{+} = \mathcal{Q},
\end{equation}
where $\mathcal{M}^{+}$ denotes the adjoint of $\mathcal{M}$.
The probability distribution $P[\phi]$ depends both on the
matrix $\mathcal{M}$ and also on the intensity of the fluctuations $\mathcal{Q}$ via Eq.~\eqref{matrixg}.

Unfortunately, although it is known that the Lyapunov equation \eqref{matrixg} is solvable for finite dimensional systems~\cite{larin2008solutions}, to the infinite dimensional case is unknown by us whether a general solution of equation \eqref{matrixg} exists, or even if can be solved analytically. While this question were without an answer, we will not be able to solve the equation \eqref{matrixg} analytically, so that we are not capable to obtain an analytical closed form for the $\mathcal{K}$ operator as a function of $\mathcal{M}$ and $\mathcal{Q}$.

The partition function, or the normalization constant of the probability distribution $P[\phi]$, is defined as the integral over the whole space of definition of our random variable, which in the case of the field $\phi$ can be written as the functional integral
\begin{equation}\label{funcion_particion_phi_no_eq}
\mathcal{Z} = \int\mathcal{D}\phi(\textbf{r},t) P[\phi(\textbf{r},t)] = \int\mathcal{D}\phi(\textbf{r},t)\frac{1}{\sqrt{\Det{\mathcal{K}}}}e^{- \frac{1}{2}\int d\textbf{r}\phi\mathcal{K}\phi},
\end{equation}
For all the above, we have a situation in which, for steady states in general, we know there exists a well defined partition function, but despite knowing that this is solvable as a Gaussian functional integral, it is not useful to us, because in general we do not know the functional $\mathcal{K}$.

What happens now if the system is at equilibrium? For this case there exists an energy functional $F$
(that can be either the entropy~\cite{deGroot}, a Lyapunov functional~\cite{Ojalvo}, the Hamiltonian,
 or a free energy~\cite{HH},$\cdots$)
which is an integral over space of a local functional $\mathcal{F}$,
which depends on the field $\phi$ and its gradients.
The evolution equation for $\phi$ can be obtained by generalization to the continuum of the Thermodynamics of
irreversible process (see, e.g., Chapter~VII of~\cite{deGroot}). This theory relates the time evolution
of the fields with its conjugated variables $\Phi$, or the so-called thermodynamic forces, as
\begin{equation}\label{Onsager}
\partial_{t}\phi = - \mathcal{L}\Phi + \xi(\textbf{r},t).
\end{equation}
Here $\mathcal{L}$ is the (symmetric) Onsager operator, which is a generalization to continuum of the
Onsager matrix of transport coefficients. It is also called the dissipation matrix.
The second law of Thermodynamics requires that the real parts of the Onsager operator eigenvalues are positive in order to
guarantee local increase of entropy.
The fields $\Phi$ appearing in \eqref{Onsager} are the conjugated variables
of $\phi$ and can be derived from the functional $F$ as
\begin{equation}\label{OnsagerF}
\Phi = \frac{\delta F}{\delta \phi}.
\end{equation}
If the fluctuations are small and the system is far from a phase transition, we can assume that
$\delta F/\delta \phi$ is linear in the field $\phi$. This implies that $F[\phi]$ is bilinear in
the field $\phi$:
\begin{equation}\label{Funcional_Energia_Lineal}
F[\phi] = \int d\textbf{r}\, \mathcal{F}(\phi,\nabla \phi, \ldots) \equiv  \frac{1}{2}\int d\textbf{r}\,\phi\mathcal{G}\phi,
\end{equation}
and the equation above defines the operator $\mathcal{G}$.
Although the definition of $\mathcal{F}$ may not be unique, $\mathcal{G}$ is unique.
For instance the two local energy functions $\mathcal{F}_{1} = \abs{\nabla\phi}^{2}/2$ and
$\mathcal{F}_{2} = -\phi\Delta\phi/2$ yield the same $\mathcal{G}$ operator: $\mathcal{G}=\Delta$.
In addition to that, $\mathcal{G}$ must be positive definitive in order to guarantee the existence
of a minimum of the free energy. It can also be chosen to be Hermitian, because
the antisymmetric part (if there is any) does not contribute to the free energy.
In this way, the corresponding Langevin equation far away of a Phase Transition is linear in $\phi$. Therefore, we can write that
\begin{equation}\label{defG}
\frac{\delta F}{\delta \phi} = \mathcal{G}\phi.
\end{equation}

Combination of Eqs.~\eqref{Onsager} and \eqref{defG} allows us to write the evolution equation for $\phi$ as
\begin{equation}\label{OnsagerFinal}
\partial_{t}\phi = - \mathcal{L}\mathcal{G}\phi + \xi(\textbf{r},t).
\end{equation} 

If the system is at equilibrium, the well-known \textit{Fluctuation--Dissipation Theorem}~\cite{Kubo}
imposes that the intensity of the noise, given by $\mathcal{Q}$, must be related to the Onsager operator $\mathcal{L}$ by
\begin{equation}\label{FDtheorem.Q}
\mathcal{Q} = k_{B}T(\mathcal{L}+\mathcal{L}^{+}).
\end{equation}

Equation~\eqref{OnsagerFinal} is formally equal to Eq.~\eqref{Langevin}, with the operator $\mathcal{M}$
 given by $\mathcal{M}=\mathcal{L}\mathcal{G}$.
As both $\mathcal{G}$ and $(\mathcal{L}+\mathcal{L}^{+})$ are Hermitian and definitive positive, it can be shown that the eigenvalues of
$\mathcal{M}$ have positive real parts, even though $\mathcal{M}$ can be
non-Hermitian or undefined (as in the case of the linear hydrodynamic equations)
(see~\cite{deGroot}, Chapter~V)~\cite{HornJohnson}.
In both equilibrium and nonequilibrium dynamics we will assume that the real part of the spectrum of $\mathcal{M}$
is strictly positive, i.e., there are no neutral modes as happens when there is
continuous symmetry breaking~\cite{Forster} or critical phenomena~\cite{Critical1}\cite{Critical2}.

In equilibrium, the fluctuation--dissipation relation has drastic consequences for the solution of the Langevin equation
associated to Eq.~\eqref{OnsagerFinal}. The equation \eqref{matrixg} is now written
\begin{equation}\label{matrixgEQ}
\mathcal{LGK}^{-1} + \mathcal{K}^{-1}\mathcal{GL}^{+} =k_BT  (\mathcal{L} + \mathcal{L}^{+}),
\end{equation}
which admits the solution $\mathcal{K}= \beta\mathcal{G}$. Once substituted
into Eq.~\eqref{solutionFP}, the probability
distribution is given by the exponential of the functional $F$ multiplied by $\beta = (k_{B}T)^{-1}$.
More precisely,
\begin{equation}\label{Peq}
P[\phi] = \frac{1}{\mathcal{Z}} e^{ - \beta F[\phi]},
\end{equation}
where $\mathcal{Z}$ is the partition function or the normalization constant of $P[\phi]$.
Given this probability distribution $P[\phi]$, we can now calculate the average of any dynamical variable $A(\phi)$ as
\begin{equation}\label{Faverage}
\mean{A} = \int\mathcal{D}\phi A(\phi)P[\phi].
\end{equation}
In particular, the average of the functional $F$ can be calculated as
\begin{equation}\label{Faverage2}
\mean{F} = - \frac{\partial\ln(\mathcal{Z})}{\partial\beta}.
\end{equation}
Equations~\eqref{Peq} and \eqref{Faverage2} are only valid for equilibrium systems,
for which an energy functional exists and the fluctuation--dissipation theorem is valid.
However, if the system is out of equilibrium, the probability distribution is
not the exponential of $F$ and therefore its average is not given in terms of
the partition function $\mathcal{Z} $.

\section{Casimir forces from the average stress tensor}\label{Casimir forces from the average stress tensor}
How is this discussion related to the calculation of Casimir forces?
The Casimir force is normally calculated for equilibrium situations, that is,
when the noise is of thermal origin and the fluctuation--dissipation theorem is
satisfied.
One way to calculate the Casimir force is by the evaluation of the stress
tensor $\mathbb{T}_{ij}$.
From the functional $F$, the stress tensor is calculated as~\cite{LandauLifshitz}
\begin{align}\label{stress}
\mathbb{T}_{ij}&= \mathbbm{1}_{ij}\mathcal{F} - \nabla_{i}\phi\frac{\partial\mathcal{F}}{\partial\nabla_{j}\phi} - 
2\nabla_{ik}\phi\frac{\partial\mathcal{F}}{\partial\nabla_{kj}\phi} + \dots\nonumber \\
\mathbb{T} & \equiv  \TensorT[\phi,\phi,\textbf{r}],
\end{align} 
which allows the definition of the symmetric bilinear stress tensor operator $\TensorT$.
For isotropic systems, the local stress is simply given by the diagonal components of the stress
tensor, or by one-third of its trace. Usual forms of $\TensorT$ are $\lambda\phi(\Vec{r})^2$ times the identity matrix or a tensorial product of gradient as in liquid crystals, but it can be also nonlocal, as in~\cite{Tarazona}.

Because of the intrinsically fluctuating nature of the fields, the stress tensor has
to be averaged over the random fields $\xi( {\bf r},t)$ or the probability distribution (\ref{solutionFP}).
Once we have the averaged stress tensor, the Casimir force over a body $\alpha$ of surface $S_{\alpha}$ is obtained as the sum over all the points of the surface of the local pressure over each point because the external fluctuanting medium, it is
\begin{equation}\label{FCasimir1}
\Vec{F}_{C} = - \oint_{S_{\alpha}} \mean{\mathbb{T}(\textbf{r})}\cdot\,\hat{\textbf{n}}\, dS.
\end{equation}
In components it is written as
\begin{equation}\label{FCasimir}
\Vec{F}_{C}^{i} = - \oint_{S_{\alpha}} \mean{\mathbb{T}^{ij}(\textbf{r})}\cdot\,\hat{\textbf{n}}_{j}\, dS,
\end{equation}
where the integral extends over the surface of the embedded bodies and the vector $\hat{\textbf{n}}$ is a unit vector normal to the surface, pointing inward the body. 

As in the original Casimir calculation, the geometry that will be considered throughout this paper
consists of two parallel, infinite plates, perpendicular to the $x$-axis, separated by distance $L_{x}$ (Fig.~\ref{fig.plates}).
In this geometry, the Casimir force per unit area on the plates is then the difference between the normal stress on the interior
and exterior side, where the latter is obtained by taking the limit $L'_{x}\to\infty$. 
The force per unit area on the left plate is
\begin{equation}\label{FCA}
F_{C}/A = \left [\mean{\mathbb{T}_{xx}(x=0;L_{x})} - \lim_{L'_{x}\to\infty} \mean{\mathbb{T}_{xx}(x=0;L'_{x})} \right].
\end{equation}
The interpretation is that, if $F_{C}/A$ is negative, the plates repel each other, while if it is positive, an attraction between the plates appears. Note that this is not the conventional interpretation of signs.
\begin{figure}[htb]
\includegraphics[width=\columnwidth]{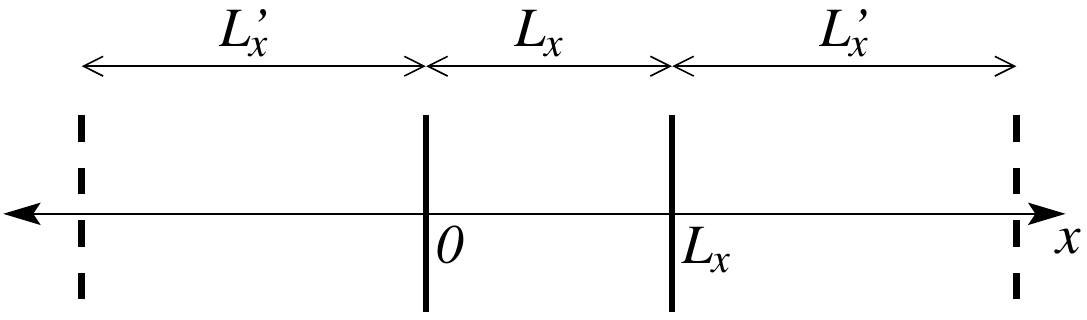}
\caption{Parallel-plate geometry used to compute the Casimir force. The system is confined between plates located at $x=0$ and $x=L_{x}$. Additional plates are located at distances $L_{x}'$ from these plates, and finally the limit $L_{x}'\to\infty$ is taken to mimic an infinite system.}
\label{fig.plates}
\end{figure}
 
Let us discuss Eq.~\eqref{FCasimir} for equilibrium and nonequilibrium situations.
In the former case, i.e., a system in equilibrium, the average of the stress tensor
can be taken in two ways: as an average over the probability distribution
given by Eq.~(\ref{Peq}), or as an average over the fluctuating
term $\xi(\textbf{r},t)$. Equilibrium Thermodynamics guarantees that both averages are the same.
In contrast, in a system out of equilibrium, we are left with one option,
the average over the noise $\xi(\textbf{r},t)$, because $\mathcal{K}$ cannot be obtained in general.
As mentioned, the system can be out of equilibrium if the fluctuation--dissipation relation is not satisfied.
In this case, there still exists a functional $F$ (from which the Langevin equation is constructed), and the
stress tensor can be defined via Eq.~\eqref{stress}. Then, the average in \eqref{FCasimir} has to be taken over the noise.
Finally, a more complex situation is when the Langevin equation is in its most
general form, i.e., Eq.~\eqref{Langevin}, without $\mathcal{M}$ deriving from an Onsager matrix
and a functional $F$. In this case, the stress tensor cannot be constructed from Eq.~\eqref{stress},
and one must appeal to other considerations in order to construct a stress tensor. One can use a microscopic
analysis of momentum transfer, kinetic theory, or invoke, for instance, the existence of a hydrostatic
pressure from which the Casimir force can be derived. We will assume, therefore, that it will always be possible
to build the stress tensor operator $\TensorT$ and, then, compute the Casimir force.

\section{Computation of Casimir forces}\label{Computation of Casimir forces}
In this Section we will develop a formalism, valid for both equilibrium and nonequilibrium systems, that allows us to compute the average stress tensor and therefore the Casimir force. We will assume that the dynamics close to the stationary state is described by the dynamical equation \eqref{Langevin}, where the noise term is assumed to be Gaussian with vanishing mean. We assume that the noise has temporal and spatial correlations, but no cross--correlations,
\begin{equation}\label{correlated.noise}
\mean{\xi(\textbf{r}, t)\xi^{*}(\textbf{r}',t')} = h(\textbf{r}-\textbf{r}') c(t-t'). 
\end{equation}
Note that we have assumed a dynamical model whose deterministic part
is local in time (no memory) but that the noise can have some memory.
This possibility is not allowed by the fluctuation--dissipation theorem, and
therefore the system is automatically put out of equilibrium.
A necessary condition to recover equilibrium is a local correlation in time,
although this condition is not sufficient, as shown in Sect.~\ref{sec.2}.

To solve \eqref{Langevin} we construct the left and right eigenvalue problems of $\mathcal{M}$
with the appropriate boundary conditions over the immersed bodies. Although we will consider the
case of two parallel plates, the formalism developed in this Section is completely general.
The left and right eigenvalue problems read
\begin{eqnarray} \label{ev}
\mathcal{M}    f_{n}\left(\textbf{r}\right) &=& \mu_{n}    f_{n}\left(\textbf{r}\right),\\
\mathcal{M}^{+}g_{n}\left(\textbf{r}\right) &=& \mu_{n}^{*}g_{n}\left(\textbf{r}\right),
\end{eqnarray}
with the boundary conditions provided by $\mathcal{M}$
(which are the same as those of $\mathcal{L}$ if the dynamics derives from a free energy functional).
The left and right eigenfunctions are orthogonal under the scalar
product, i.e., $\mean{ g_{n}\vert f_{m}} = \int{d\textbf{r} g^{*}_{n}(\textbf{r}) f_{m}(\textbf{r}) }$;
that is, under appropriate normalization, $\mean{ g_{n}\vert f_{m}} = \delta_{nm}$.
We can project the field and the noise over the left eigenvalues
\begin{equation}\label{functions-factoritation}
\phi(\textbf{r},t) = \sum_{n}\phi_{n}(t)f_{n}(\textbf{r}),
\quad
\xi(\textbf{r},t)  = \sum_{n}\xi_{n}(t) f_{n}(\textbf{r}),
\end{equation}
where $\phi_{n}(t)=\mean{g_{n}\vert \phi(t)}$ and $\xi_{n}(t)=\mean{g_{n}\vert \xi(t)}$.
By inserting these expressions \eqref{functions-factoritation} into the evolution equation \eqref{Langevin} we get the evolution equation of each mode $\phi_{n}(t)$ as
\begin{equation}\label{Dinamica_Langevin_modo_n}
\partial_{t}\phi_{n}(t) = -\mu_{n}\phi_{n}(t) + \xi_{n}(t).
\end{equation}

In order to solve this equation, we obtain its correspondent Green function
\begin{equation}
\left(\partial_{t} + \mu_{n}\right)G(t,t') = \delta(t - t'),
\end{equation}
which is
\begin{equation}
G(t,t') = e^{-(t - t')\mu_{n}}\Theta(t - t'),
\end{equation}
then the analytical solution of \eqref{Dinamica_Langevin_modo_n} is
\begin{equation}
\phi_{n}(t) = \int_{-\infty}^{\infty}dt'G(t,t')\xi_{n}(t').
\end{equation}
\begin{equation}\label{temporal-mode-evolution}
\phi_{n}(t) = \int_{-\infty}^{t}d\tau e^{-(t - \tau)\mu_{n}}\xi_{n}(\tau) = e^{-t\mu_{n}}\phi_{n}(0) + \int_{0}^{t}d\tau e^{-(t - \tau)\mu_{n}}\xi_{n}(\tau).
\end{equation}
The first term $e^{-\mu_{n}t}\phi_{n}(0)$ is a transient term that vanishes for times
longer than $t\gg\frac{1}{\text{Re}(\mu_{n})}$, so that the average of each mode over the noise $\xi$ is zero in this limit.

To compute the average stress tensor at each point, we need to compute $\mean{\TensorT[\phi,\phi,\Vec{r}]}$.
Expanding on the eigenvalue basis we get that, in the steady state,
\begin{eqnarray}
\lim_{t\to\infty}\mean{\mathbb{T}(\textbf{r},t)} & = & \lim_{t\to\infty}\sum_{m,n}\mean{\phi_{n}(t) \phi^{*}_{m}(t)} \TensorT_{nm}(\textbf{r}),
\end{eqnarray}
where $\TensorT_{nm}(\Vec{r})=\TensorT[f_{n},f_{m}^{*},\Vec{r}]$.

The cross-average of the mode amplitudes is obtained from \eqref{temporal-mode-evolution} and in the stationary regime
[$t\gg 1/\text{Re}(\mu_n),1/\text{Re}(\mu_m)$] can be written as
\begin{equation}\label{two-field-modes-average}
\lim_{t\to\infty}\mean{\phi_{n}(t)\phi_{m}^{*}(t)} = \lim_{t\to\infty}e^{ - (\mu_{n} + \mu_{m}^*)t}\int_{-\infty}^{t} d\tau_{1}\int_{-\infty}^{t}d\tau_{2}e^{\mu_{n}\tau_{1} + \mu_{m}^{*}\tau_{2}}
\mean{\xi_{n}(\tau_{1})\xi_{m}^{*}(\tau_{2})}.
\end{equation}
Therefore, we need to calculate the correlation of the $n$ and $m$ components of the noise
\begin{equation}\label{correlation.noise.modes}
\mean{\xi_{n}(\tau_{1})\xi_{m}^{*}(\tau_{2})} = \int\!\! d\textbf{r}_{1}\! \int\!\! d\textbf{r}_{2} g_{n}^{*}(\textbf{r}_{1})g_{m}(\textbf{r}_{2})\mean{\xi(\textbf{r}_{1},\tau_{1})\xi^{*}(\textbf{r}_{2},\tau_{2})}. 
\end{equation}
Substituting Eq.~\eqref{correlated.noise} into Eq.~\eqref{correlation.noise.modes} and the result into Eq.~\eqref{two-field-modes-average}, we obtain (see Appendix \ref{Apendice-Resolucion.promedio.modos.phi_n.phi_m}),
\begin{equation} \label{average-square-field}
 \lim_{t\to\infty}\mean{\phi_{n}(t)\phi_{m}^{*}(t)} = h_{nm}\frac{\widetilde{c}(\mu_{n}) + \widetilde{c}(\mu^{*}_{m})}{\mu_{n} + \mu^{*}_{m}},
\end{equation}
where
\begin{equation}\label{h_nm}
h_{nm} = \int d\textbf{r}_1\int d\textbf{r}_{2}g^*_{n}(\textbf{r}_{1})h(\textbf{r}_{1} - \textbf{r}_{2})g_{m}(\textbf{r}_{2})=\mean{g_n\vert {\cal Q} g_m}
\end{equation}
and $\widetilde{c}$ is the Laplace transform of $c$.
Finally, the local average of the stress tensor in the stationary regime, where transients have been eliminated and the value of the stress tensor is independent of time, is given by
\begin{equation}
\mean{\mathbb{T}(\textbf{r})} = \sum_{nm}\frac{\widetilde{c}(\mu_{n})+\widetilde{c}(\mu^{*}_{m})}{\mu_{n} + \mu^{*}_{m}}h_{nm} \TensorT_{nm}(\textbf{r}).
\end{equation}
This expression is generally divergent when summed over all eigenfunctions. This divergence comes from the highest eigenvalues (corresponding to small wavelengths) and is due to consider the mesoscopic dynamics given by Eq.~\eqref{Langevin}, valid for all wavelengths.
However, it is only valid above a certain minimal distance (the atomic or molecular length, for example).
There are some techniques to avoid this divergence. For instance, a short-wavelength cutoff could be introduced as in Ref.~\cite{BritoSotoPRE}, but here we will use regularization techniques similar to the Riemann zeta function used in the electrodynamic case~\cite{Casimir_Placas_Paralelas}.

Using the previous expression, the conditions under which Casimir forces exist in an equilibrium system can be deduced. As mentioned above, if the dynamics is local in time, the fluctuation--dissipation theorem implies that the noise terms must not have memory either, then $c(t - t') = \delta(t - t')$ and therefore $\widetilde{c}(\mu)=1/2$. Also, the equilibrium relation between noise autocorrelation and Onsager matrices given in Eq.~\eqref{FDtheorem.Q} implies that $h_{nm} =k_{B}T(\mu_{n} + \mu_{m}^{*})\mean{g_{n}\vert\mathcal{G}^{-1}g_{m}}$, and therefore the equilibrium average stress tensor simplifies to
\begin{equation}\label{Caso termico}
\mean{\mathbb{T}_{\rm eq}(\textbf{r})} = k_{B}T\sum_{n,m} \mean{g_{n}\vert\mathcal{G}^{-1}g_{m}} \TensorT_{nm}(\textbf{r}).
\end{equation}
If the free energy functional $F$ depends only on $\phi$ but not on its derivatives, the stress tensor operator $\TensorT_{nm}(\textbf{r})$ turns out to be isotropic and is given by $\TensorT_{nm}(\textbf{r}) = f_{n}\mathcal{G}f_{m}^{*}\mathbb{I}_{3\times3}$ \eqref{stress}. Then, thanks to the completeness of the basis, the stress tensor can be further simplified to $\mean{\mathbb{T}_{\rm eq}(\textbf{r})} =k_{B}T\delta(\textbf{r})$. This expression, once properly regularized, gives a stress that is independent of system size; that is, the stress is not renormalized by the fluctuations in a size-dependent way and therefore no Casimir force can be developed. On the contrary, if the stress tensor is not isotropic, as in the case of liquid crystals, the result is not trivial and Casimir forces can develop, as shown in~\cite{Casimir-liquidcrystals}.

All these equations provide expressions for the average fields and fluctuations, expressed in terms of the eigenvalues and eigenvectors of the problem, which encode the information of the evolution equation together with the boundary conditions.

To summarize, in this Section we have proven that the Casimir force over a body is given by
\begin{equation} \label{Force}
\Vec{F}_{C} = - \sum_{nm}\frac{\widetilde{c}(\mu_{n}) + \widetilde{c}(\mu^{*}_{m})}{\mu_{n} + \mu^{*}_{m}}h_{nm}
\oint_{S}\TensorT_{nm}(\textbf{r})\cdot \hat{{\bf n}}\, dS,
\end{equation}
which is the main result of this Thesis. It shows how to derive the Casimir force from the
dynamical equations for the field $\phi$ subjected to any kind of noise. It is obtained by diagonalizing the
evolution operator of the field, and projecting the noise correlation and the stress tensor over the set of eigenfunctions.
This approach provides the Casimir force for both equilibrium and nonequilibrium systems.

Equation~\eqref{Force} shows the well-known nonadditive character of the Casimir force: neither the eigenvalues nor the eigenfunctions for different boundary conditions are easily related. They cannot be written as a sum of the eigenvalues and eigenfunctions of each different problem. This behavior is studied in Chapters \ref{chb2} and \ref{chb3} of this Thesis for the electromagnetic case.

The rest of the Chapter deals with applications of Eq.~\eqref{Force} to different physical systems, in both equilibrium and nonequilibrium situations.

\section{Reaction--diffusion systems}\label{Reaccion-Difusion}
To show how this formalism works, we calculate the Casimir pressure between
two plane, infinite plates separated by distance $L_{x}$ immersed in a medium described by a quadratic free energy
\begin{equation}
F[\phi] = \int d\textbf{r}\,\frac{f_{0}}{2}\phi^{2}(\textbf{r}).
\end{equation}
The multiplicative constant $f_{0}$ can be absorbed into $\phi$, and we will eliminate it in what follows.
The dynamics is described by two transport processes: relaxation and diffusion;
that is, the Onsager operator is $\mathcal{L} = \lambda - D\nabla^{2}$, where $\lambda$ and $D$ are the
transport coefficients (and consequently, positive) associated with the two irreversible processes of relaxation and diffusion, respectively.
The resulting equation is
\begin{equation}\label{eq.readdif}
\derpar{\phi}{t} = -\lambda \phi + D\nabla^{2} \phi  +\xi(\textbf{r},t). 
\end{equation}
Fluctuation--dissipation is satisfied if the noise is delta correlated in time and the space correlation function is
\begin{equation}
h(\textbf{r}) = 2k_{B}T(\lambda - D\nabla^{2})\delta(\textbf{r}).
\end{equation}
As the energy functional for this system is $\phi^{2}/2$, without terms with spatial derivatives, the stress tensor is identical to the local energy functional.
Also, the dynamic operator is Hermitian, implying that eigenvalues are real and that there is no need to distinguish between left and right eigenfunctions.
In order to obtain Casimir forces the appropriate boundary conditions are of Neumann type, as Dirichlet boundary condition would imply trivial vanishing forces.

We need to solve the eigenfunction problem for the spatial part of the dynamics given by Eq.~\eqref{ev} in order to calculate the average of the fields that will lead to the Casimir pressure over the plates. So, we have to solve the eigenfunction problem given by Eq.~(\ref{ev}) with $\mathcal{M}= \lambda - D\nabla^{2}$ obeying Neumann boundary conditions (no-flux boundary conditions), $\partial_{x}\phi(0,y,z) = \partial_{x}\phi(L_{x},y,z) = 0$.

The normalized eigenfunctions are characterized by three indices $n_{x}$, $n_{y}$, and $n_{z}$, denoted as a whole by $n$, and their form is
\begin{eqnarray}\label{autofunciones-Neumann}
& f_{n}(\textbf{r}) =  \sqrt{\frac{1}{V}}e^{i\textbf{k}_{\parallel}\cdot\textbf{r}_{\parallel}} &
\hspace{0.5cm}\text{if}\hspace{0.2cm}n_{x}=0\nonumber\\
& f_{n}(\textbf{r}) =  \sqrt{\frac{2}{V}}\cos\left(k_{x}x \right) e^{i\textbf{k}_{\parallel}\cdot\textbf{r}_{\parallel}} &
\hspace{0.5cm}\text{if}\hspace{0.2cm}n_{x}\geq 1.
\end{eqnarray}
Here, $\textbf{r}_{\parallel} = y\hat{\textbf{y}} + z\hat{\textbf{z}}$ and $\Vec{k}_{\parallel} = k_{y}\hat{\textbf{y}} + k_{z}\hat{\textbf{z}}$.
The eigenvalues are
\begin{equation}\label{eigenvalues}
\mu_{n} = D\left(k_{x}^{2} + k_{y}^{2} + k_{z}^{2} + k_{0}^{2}\right),
\end{equation}
where $k_{x} = \frac{\pi}{L_{x}}n_{x}$,
$k_{y} = \frac{2\pi}{L_{y}}n_{y}$, and
$k_{z} = \frac{2\pi}{L_{z}}n_{z}$,
with $n_{x}=0,1,2,\dots$ and $(n_{y},n_{z})\in\mathbb{Z}^{2}$. The quantity $k_{0}^{-1}  =  \sqrt{D/\lambda}$ is the characteristic correlation length of the system.

The average stress is then
\begin{equation}\label{TensorTReacDif}
\mean{\mathbb{T}_{xx}(\mathbf{r})} = \frac{1}{2D}\sum_{nm}\frac{h_{nm}
\left[\widetilde{c}(\mu_{n}) + \widetilde{c}(\mu_{m})\right]}{\Vec{k}_{n}^{2} + \Vec{k}_{m}^{2}  +2 k_{0}^{2}}f_{n}(\textbf{r})f^{*}_{m}(\textbf{r}). 
\end{equation}
This expression needs to be regularized, otherwise it is divergent.
The divergence, as explained in~\cite{BritoSotoPRE} \cite{BritoSotoGM}, is due to the application of the mesoscopic model \eqref{eq.readdif} up to very large wavevectors. Conceptually, the stress could be regularized by considering generalized hydrodynamic models valid for high wavevectors leading to finite stresses, but as the Casimir forces have their origin in the limitation of the fluctuation at small wavevectors, this is not necessary and other procedures are available. There are various regularization methods that allow the isolation of the divergent term that is independent of the plate separation and therefore cancels out in the computation of the Casimir force. The regularization method used in this manuscript is based on the Elizalde function detailed in the Appendix \ref{Appendix: Elizalde zeta function}.

To obtain quantitative predictions, we consider specific cases for the noise correlations.

%%%%%%%%%%%%%%%%%%%%%%%%%%%%%%%%%%%%%%%%%%%
\subsection{Uncorrelated noise in time and space}
We first consider the case of a noise with vanishing correlation time and length, and intensity $\Gamma$, i.e.,
\begin{equation}\label{doublewhitenoise}
\mean{\xi(\textbf{r}, t) \xi(\textbf{r}',t')} = \Gamma \delta(\textbf{r}-\textbf{r}')\delta(t-t'). 
\end{equation}
This noise correlation, without the $-\nabla^{2}\delta(\textbf{r})$ term,
automatically puts the system out of equilibrium \eqref{FDtheorem.Q}, as $\mathcal{Q}\neq \mathcal{L} + \mathcal{L}^{+}$. The addition of such a term would have led to a stress that was independent of plate separation, not producing a Casimir force~\cite{CasimirGranular}. Therefore, we consider the effect of the nonequilibrium noise \eqref{doublewhitenoise} on Casimir forces. In this case, $h_{nm}(\widetilde{c}(\mu_{n})+\widetilde{c}(\mu_{m}))=\Gamma\delta_{nm}$ and the double sum in \eqref{TensorTReacDif} is reduced.
On the surface of a plate and applying the limit $L_{y}, L_{z}\rightarrow\infty$, the stress is given by
\begin{align}
\mean{\mathbb{T}_{xx}(0)} &= \frac{\Gamma}{16\pi^{2}L_{x}D}\int_{-\infty}^{\infty}\!dk_{y}\int_{-\infty}^{\infty}\!dk_{z}\sum_{n_{x}\in\mathbb{Z}}\frac{1}{\left(\frac{\pi n_{x}}{L_{x}}\right)^{2} + k_{y}^{2} + k_{z}^{2} + k_{0}^{2}} \nonumber\\
& = \frac{\Gamma}{8\pi L_{x}D} \int_{0}^{\infty}dkk\sum_{n_{x}\in\mathbb{Z}}\frac{1}{\left(\frac{\pi n_{x}}{L_{x}}\right)^{2} + k^{2} + k_{0}^{2}}, \label{T.reacdif.whitenoise}
\end{align}
where polar coordinates in the $y$- and $z$-components have been used. Note that the original sum over $n_x$ in \eqref{TensorTReacDif} runs only over $\mathbb{N}$, but the form of the normalizations of the eigenfunctions \eqref{autofunciones-Neumann} allows extension of the sum over $\mathbb{Z}$ with a prefactor of $1/2$.

\begin{figure}%[H]
\begin{center}
\includegraphics[width=\columnwidth]{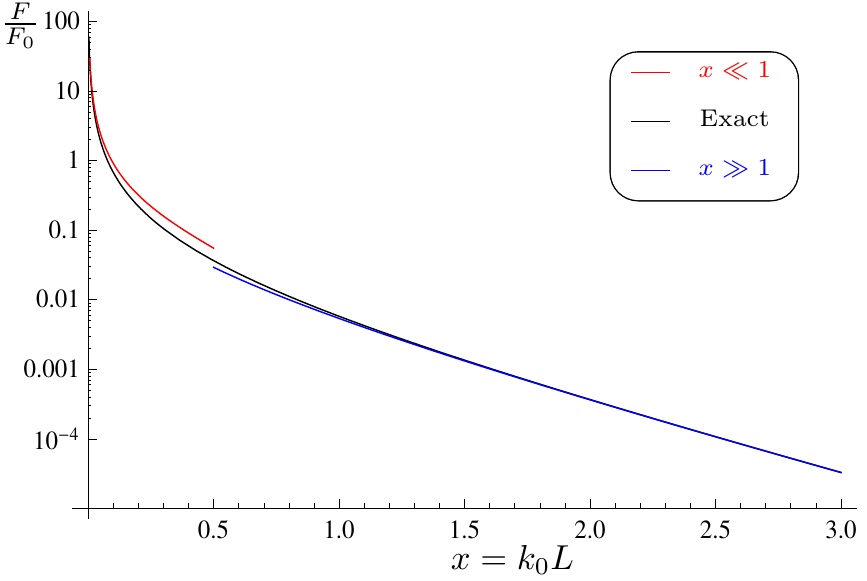}
\caption{\label{Grafica_Reaccion_Difusion_Blanco}Casimir force in units of $F_{0} = \frac{\Gamma k_{0}}{8\pi D}$ in dimensionless units of distance $x = k_{0}L$ over a plate immersed in a reaction--diffusion media subject to white noise of intensity $\Gamma$ in the presence of another plate at a distance $L$. The exact result is the black curve. The short (Eq.~\eqref{fza.cliq.white}) and long (Eq.~\eqref{fza.cliq.white_k0L_grande}) distance limits are the red and blue curve respectively.}
\label{Energia_2_esferas_metalicas}
\end{center}
\end{figure}

This expression is divergent, so it must be regularized. In
order to do so, we use the Chowla--Selberg expression shown in Eq.~\eqref{Chowla-Selberg 1d}.
The parameters are $s = 1$ , $\alpha = \pi/L_{x}$, and $\omega^{2} = k^{2} + k_{0}^{2}$.
The first term in the sum of \eqref{Chowla-Selberg 1d} equals $L_{x}/\sqrt{k^{2} + k_{0}^{2}}$,
which combined with the prefactor in Eq.~\eqref{T.reacdif.whitenoise} yields a term which is independent of $L_{x}$,
and therefore its contribution to the stress tensor cancels in virtue of Eq.~\eqref{FCA}. It must be remarked that the size-independent term is actually divergent if the continuous model is assumed to be valid for any wavevector.
Then, we are left with the infinite sum of modified Bessel functions $K_{1/2}$. This sum can be performed analytically,
with the result
\begin{equation}\label{reacc}
F_{C}/A  = \frac{\Gamma}{4\pi D}\int_{0}^{\infty}dk\frac{k}{\sqrt{k^{2} + k_{0}^{2}}}\frac{1}{e^{2\sqrt{k^{2} + k_{0}^{2}}L_{x}} - 1} = - \frac{\Gamma k_{0}}{8\pi D}\frac{\ln(1-e^{-2k_{0}L_{x}})}{k_{0}L_{x}}.
\end{equation}
Let us note that, because the divergence was eliminated, we could have interchanged the integral with the summation of the modified Bessel functions to obtain the same result.
At distances long compared with the correlation length, that is $L_{x}\gg k_{0}^{-1}$, the force decays as
\begin{equation}\label{fza.rd.white_k0L_grande}
F_{C}/A = \frac{\Gamma}{8\pi D L_{x}} e^{-2 k_{0}L_{x}}.
\end{equation}
In the opposite limit, when the plates are at a distance much smaller than the correlation length, or $L_{x}\ll k_{0}^{-1}$, the force is
\begin{equation}\label{fza.rd.white_k0L_small}
F_{C}/A = - \frac{\Gamma}{8\pi DL_{x}}\log\left(k_{0}L_{x}\right).
\end{equation}
Equation~\eqref{reacc}, as well as Eq.~\eqref{fza.rd.white_k0L_small}, shows that the Casimir force diverges if the correlation length tends to infinity, i.e., if $k_{0}\to 0$.
This result was obtained in~\cite{BritoSotoPRE} using a regularizing kernel technique. The force, as well as short and long distance limits are plotted in Fig.~\ref{Grafica_Reaccion_Difusion_Blanco}.

Eq.~\eqref{reacc} shows that Casimir force diverges if correlation length tends to infinity ($k_{0}\to 0$). This is a general property of Casimir forces in Fluctuation--Reaction systems, and its origin is the fact that, when $k_{0}\to 0$, does not exist any counter-term which could dampens the field fluctuations in \eqref{eq.readdif}. The total mass, defined as $M(t) = \int d\textbf{r}\phi(\textbf{r},t)$ perform an unbounded random walk.
This problem disappears if we add higher order terms to the dynamics, as a $\phi^{4}$ and/or a $\abs{\nabla\phi}^{4}$ term in the energy functional, as done in Ginzburg Landau theory of Critical Phenomena \cite{HH}. In this case, it also would be needed Renormalization Group techniques \cite{KrechBook}.

%%%%%%%%%%%%%%%%%%%%%%%%%%%%%%%%%%%%%%%%%%%
\subsection{Temporally correlated noise}\label{temporalcorrelation.reacdif}
We next consider the case of a noise that is delta correlated in space but has exponential correlation in time
\begin{equation}\label{noise.tempcorr}
\mean{\xi\left(\textbf{r},t\right)\xi\left(\textbf{r}',t'\right)} = \Gamma\delta\left(\textbf{r} - \textbf{r}'\right)\left(1+\frac{a}{2}\right)e^{-a\abs{t - t'}}, 
\end{equation}
where the factor $(1+\frac{a}{2})$ allows both the white noise limit ($a\to\infty$) and the quenched noise limit ($a\to 0$) to be taken.
Again, the delta correlation in space leads to a term $\delta_{nm}$ that eliminates one summation in the stress at the plate, which is then given by
\begin{equation}
\mean{\mathbb{T}_{xx}(0)} = \frac{\left(1+\frac{a}{2}\right)\Gamma}{2V}\sum_{n}\frac{1}{\mu_{n} + a}\frac{1}{\mu_{n}}.
\end{equation}
If $a> 0$, we can factorize the quotient as
\begin{equation}
\mean{\mathbb{T}_{xx}(0)} = \frac{\left(1+\frac{a}{2}\right)\Gamma}{2aV}\sum_{n}\left[\frac{1}{\mu_{n}} - \frac{1}{\mu_{n} + a}\right],
\end{equation}
with $\mu_{n} = k_{x}^{2} + k_{y}^{2} + k_{z}^{2} + k_{0}^{2}$ as before. We note that this stress
is the difference between the Casimir stress of two systems with a white temporal noise \eqref{T.reacdif.whitenoise} of
intensity $(1+a)\Gamma/a$, the first one with $k_{0}^{2} = \frac{\lambda}{D}$ and
the second one with $k_{1}^{2} = \frac{\lambda}{D} + \frac{a}{D}$. Then, the stress on the plate is
\begin{equation}\label{aaa}
\mean{\mathbb{T}_{xx}(0)}  = -\frac{\Gamma\left(1+\frac{a}{2}\right)}{4a\pi DL_{x}}\ln\left(\frac{1 - e^{-2k_{0}L_{x}}}{1 - e^{-2k_{1}L_{x}}}\right).
\end{equation}
The Casimir force per unit surface on the plate is given just by this expression, because the stress
on the unbounded side vanishes [as shown by taking the limit $L_{x}\to\infty$ in Eq.~\eqref{aaa}].
Finally, we can reobtain the white noise limit if $a\rightarrow\infty$.

The case $a\to 0$ corresponds to the quenched limit, where static sources of noise are randomly distributed in space. The average normal stress at the wall is
\begin{equation}
\mean{\mathbb{T}_{xx}(0)} = \frac{\Gamma}{2V}\sum_{n}\frac{1}{\mu_{n}^{2}}.
\end{equation}
Taking the limit $L_{y},L_{z}\to\infty$ and using polar coordinates,
\begin{equation}
\mean{\mathbb{T}_{xx}(0)} = \frac{\Gamma}{4\pi L_{x}D^{2}}\int_{0}^{\infty}dkk \sum_{n_{x}\in\mathbb{Z}}\left(\left(\frac{\pi n_{x}}{L_{x}}\right)^2 + k^{2} + k_{0}^{2} \right)^{-2}.
\end{equation}
Although this expression is finite and does not require a regularization procedure, the size-independent contribution can be eliminated using the same regularization procedure as before, using Eq.~\eqref{Chowla-Selberg 1d} with $s = 2$. The result is
\begin{equation}
F_{C}/A = \frac{\Gamma}{4\pi D^{2}}\int_{0}^{\infty} dk\frac{k}{\omega^{3}}\frac{e^{2\omega L_{x}}\left(2\omega L_{x}  + 1\right) - 1}
{\left(e^{2\omega L_{x}} - 1\right)^{2}},
\end{equation}
where $\omega = \sqrt{k^{2} + k_{0}^{2}}$. After carrying out the integral, we obtain
\begin{equation}
F_{C}/A = \frac{\Gamma}{4\pi D^{2}k_{0}}\frac{1}{e^{2k_{0}L_{x}} - 1}.
\end{equation}
We remark that this system is not dynamically fluctuating, because the noise is quenched and the transients have been eliminated. Nevertheless, it creates a Casimir force whose origin is the same as previously considered in the sense that the presence of the second plate limits the spectrum of possible fluctuations, and therefore the renormalized stresses on the two sides of the plate are different.

%%%%%%%%%%%%%%%%%%%%%%%%%%%%%%%%%%%%%%%%%%%
\subsection{Maximally spatially correlated noise}
As a final case, we consider the situation in which the medium is perturbed externally by a spatially homogeneous noise, with vanishing correlation time. This could be the case when a rapidly fluctuating external field is applied to the medium.
\begin{equation} \label{maxspatialcorrnoise}
\mean{\xi\left(\textbf{r},t\right)\xi\left(\textbf{r}',t'\right)} = \Gamma \delta\left(t - t'\right).
\end{equation}
Applying the same computation procedure as in the other cases, the average local stress on each side of the plates is
\begin{equation}
\mean{\mathbb{T}_{xx}(0)}   = \frac{\Gamma}{2D}\frac{1}{2k_{0}^{2}} = \frac{\Gamma}{4\lambda},
\end{equation}
which is independent of the plate separation. Therefore, the Casimir force vanishes in this case.

%%%%%%%%%%%%%%%%%%%%%%%%%%%%%%%%%%%%%%%%%%%
%%%%%%%%%%%%%%%%%%%%%%%%%%%%%%%%%%%%%%%%%%%
\section{Liquid crystals}\label{Cristales-Liquidos}
The existence of Casimir forces in liquid crystals has been known for some time now~\cite{Casimir-liquidcrystals}.
In this Section we apply the formalism presented in Sect.~\ref{sec.2} to a nematic crystal, obtaining
the known Casimir force for an equilibrium situation, and expressions for the force for some
nonequilibrium conditions.
The free energy functional of a nematic liquid crystal~\cite{deGennes} can be written in terms of a planar field $\phi$ as
\begin{equation}\label{F.liqcry}
F = \int d\textbf{r}\left[\frac{\kappa_{1}}{2} \phi^{2} + \frac{\kappa_{2}}{2}(\nabla\phi)^{2}\right],
\end{equation}
where we have assumed that the director vector is written in terms of the field $\phi$ as $\hat{n} = (\sin\phi,0,\cos\phi)$, together with the one-constant approximation (proportional to $\kappa_{2}$).
The first term in Eq.~\eqref{F.liqcry} comes from a magnetic field directed along the $z$-axis, whose intensity is absorbed into $\kappa_{1}$.

The simplest dynamical model is obtained with a single relaxational transport coefficient, with the Onsager operator $\mathcal{L}=\lambda$, leading to
\begin{equation}\label{eq.liqcry}
\derpar{\phi}{t} = -\lambda\kappa_{1}\phi + \lambda\kappa_{2}\nabla^{2}\phi + \xi,
\end{equation}
which is identical in form to \eqref{eq.readdif}, but with three main differences: the form of the fluctuation--dissipation relation to be in equilibrium, the stress tensor, and the possible boundary conditions that produce Casimir forces.
Fluctuation--dissipation is realized, according to Sect.~\ref{sec.2}, if the noise satisfies
\begin{equation}
\mean{\xi(\textbf{r},t)\xi(\textbf{r}',t')} = 2k_{B}T\lambda\delta(\textbf{r} - \textbf{r}')\delta(t - t'),
\end{equation}
i.e., is purely nonconservative.

Due to the presence of the gradient terms in the free energy
functional \eqref{F.liqcry}, the stress tensor is not isotropic, and
therefore even in equilibrium Casimir forces can appear.
Using Eq.~\eqref{stress} the $xx$ component of the stress tensor is
\begin{equation}
\mathbb{T}_{xx}= \frac{\kappa_{1}}{2}\phi^{2} + \frac{\kappa_{2}}{2}\left(\derpar{\phi}{y}\right)^{2} + \frac{\kappa_{2}}{2}\left(\derpar{\phi}{z}\right)^{2} - \frac{\kappa_{2}}{2}\left(\derpar{\phi}{x}\right)^{2}.
\end{equation}
It is then possible to develop Casimir forces by imposing either Dirichlet or Neumann boundary conditions.
Dirichlet boundary conditions are equivalent to the strong anchoring conditions, i.e., $\phi=0$ over the surfaces, and will be the case studied here.
 Casimir forces with Neumann boundary conditions can be easily extracted from the Dirichlet ones.

In this case of Dirichlet boundary conditions, the eigenfunctions of the
operator $\mathcal{M}= \lambda\left[\kappa_{1} -\kappa_{2}\nabla^{2}\right]$ are given by
$ f_{n}(\textbf{r}) = \sqrt{\frac{2}{V}}\sin\left(k_{x}x \right) e^{i\textbf{k}_{\parallel}\cdot\textbf{r}_{\parallel}}$
with eigenvalues
\begin{equation}\label{eigenvalues2}
\mu_{n} = \lambda\kappa_{2}( k_{x}^{2} + k_{y}^{2} + k_{z}^{2} + k_{0}^{2}),
\end{equation}
where $ k_{x} = \frac{\pi}{L_{x}}n_{x}$, $k_{y}=\frac{2\pi}{L_{y}}n_{y}$, $k_{z}=\frac{2\pi}{L_{z}}n_{z}$, and $k_{0} = \sqrt{\frac{\kappa_{1}}{\kappa_{2}}}$, with indices $n_{x}\in\mathbb{N}$ and $(n_{y},n_{z})\in\mathbb{Z}^{2}$.

Because of the boundary conditions, the $xx$ component of the stress tensor at the plates is simply given by
\begin{equation}
\mathbb{T}_{xx}(0)= - \frac{\kappa_{2}}{2}\left(\derpar{\phi}{x}\right)^2.
\end{equation}

As in the case of the reaction--diffusion system, we will consider different types of noise correlations that, as will be shown below, produce Casimir forces of different character.

%%%%%%%%%%%%%%%%%%%%%%%%%%%%%%%%%%%%%%%%%%%
\subsection{Uncorrelated noise in time and space}
We consider an uncorrelated noise as described by \eqref{doublewhitenoise}. This case can be considered as in equilibrium with a temperature given by $\Gamma = 2k_{B}T\lambda$. Again the double sum in Eq.~\eqref{Force} can be reduced, and the stress tensor on the surface of a plate is given by
\begin{equation}\label{stress_tensor_cristal_liquido_ruido_blanco1}
\mean{\mathbb{T}_{xx}(0)} = - \frac{\Gamma}{2V\lambda} \sum_{n}\frac{k_{x}^{2}}{k_{x}^{2} + k_{y}^{2} + k_{z}^{2} + k_{0}^{2}}.
\end{equation}
Applying the limit $L_{y}, L_{z}\rightarrow\infty$, we obtain
\begin{equation}\label{stress_tensor_cristal_liquido_ruido_blanco2}
\mean{\mathbb{T}_{xx}(0)} = \frac{-\Gamma}{16 \pi^{2}L_{x}\lambda} \int_{-\infty}^{\infty}\!dk_{y}\int_{-\infty}^{\infty}\!dk_{z} \sum_{n_{x}\in\mathbb{Z}}\frac{\left(\frac{\pi n_{x}}{L_{x}}\right)^{2}}{\left(\frac{\pi n_{x}}{L_{x}}\right)^{2} + k_{y}^{2} + k_{z}^{2} + k_{0}^{2}}.
\end{equation}

\begin{figure}%[H]
\begin{center}
\includegraphics[width=\columnwidth]{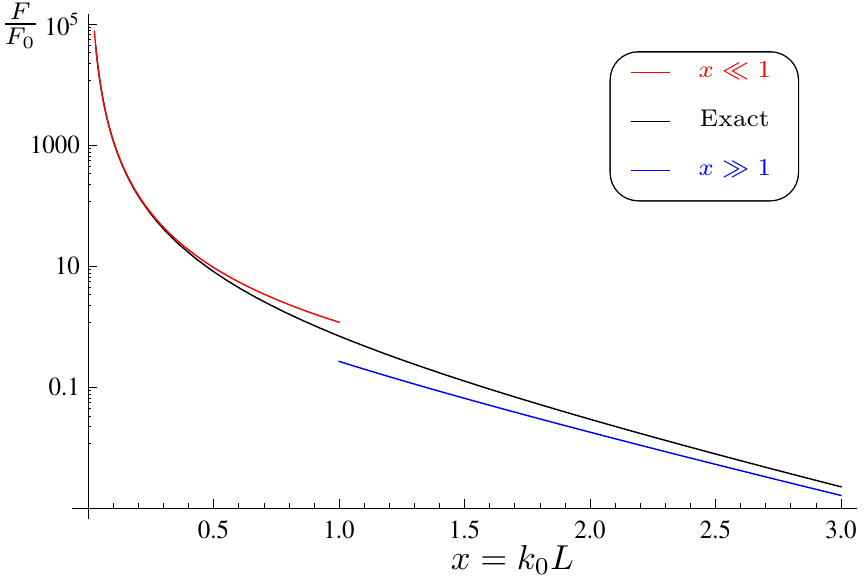}
\caption{\label{Grafica_Nematico_Blanco_Adimensional}Casimir force in units of $F_{0} = \frac{\Gamma k_{0}^{3}}{16\pi\lambda}$ in dimensionless units of distance $x = k_{0}L$ over a plate immersed in a liquid crystal subject to white noise of intensity $\Gamma$ in the presence of another plate at a distance $L$. The exact result is the black curve. The short (Eq.~\eqref{fza.cliq.white}) and long (Eq.~\eqref{fza.cliq.white_k0L_grande}) distance limits are the red and blue curve respectively.}
%\textbf{•}\label{Energia_2_esferas_metalicas}
\end{center}
\end{figure}

Using polar coordinates and regularizing the resulting expression using Eqs.~\eqref{Relacion de Y con Z 2} and \eqref{Chowla-Selberg 1d} with $s = 1$,
\begin{eqnarray}\label{Fuerza_Nematico_Blanco}
F_{C}/A & = & \frac{\Gamma}{4\pi\lambda}\int_{0}^{\infty}dkk\frac{\sqrt{k^{2} + k_{0}^{2}}}{e^{2\sqrt{k^{2} + k_{0}^{2}}L_{x}} - 1}\nonumber \\
        & = & \frac{\Gamma}{16\pi\lambda L_{x}^{3}}\left[\Li_{3}(e^{-2k_{0}L_{x}}) + 2k_{0}L_{x}\Li_{2}(e^{-2k_{0}L_{x}}) + 2k_{0}^{2}L_{x}^{2}\Li_{1}(e^{-2k_{0}L_{x}})\right],
\end{eqnarray}
where $\Li_{s}(z) = \sum_{n=1}^{\infty}\frac{z^{n}}{n^{s}}$ is the polylogarithm function. This force, as well as the short and long distance limits are plotted in Fig~\ref{Grafica_Nematico_Blanco_Adimensional}. At distances long compared with the correlation length, that is $L_{x}\gg k_{0}^{-1}$, the force decays as
\begin{equation}\label{fza.cliq.white_k0L_grande}
F_{C}/A = \frac{\Gamma k_{0}^{2}}{8\pi\lambda L_{x}} e^{-2 k_{0}L_{x}}.
\end{equation}
In the opposite limit, when the plates are at a distance much smaller than the correlation length, or $L_{x}\ll k_{0}^{-1}$, the force is
\begin{equation}\label{fza.cliq.white}
F_{C}/A = \frac{\Gamma}{16\pi\lambda}\frac{\zeta(3)}{ L_{x}^{3}}.
\end{equation}
This result has already been obtained in the context of liquid crystals in~\cite{Ajdari} if we use the
fluctuation--dissipation theorem. It is also the high-temperature limit of the electromagnetic Casimir force between two perfect metal plates~\cite{Review_Casimir}.

%%%%%%%%%%%%%%%%%%%%%%%%%%%%%%%%%%%%%%%%%%%
\subsection{Temporally correlated noise}
We consider the temporally correlated noise described in Eq.~\eqref{noise.tempcorr}. By
using the eigenfunctions of the Dirichlet problem, the stress tensor over a plate takes the value
\begin{equation}
\mean{\mathbb{T}_{xx}(0)} = - \frac{\kappa_{2}\Gamma\left(1 + \frac{a}{2}\right)}{V}\sum_{n}
\frac{1}{\mu_{n} + a}\frac{k_{x}^{2}}{\mu_{n}}.
\end{equation}
For any $a\neq 0$, the same factorization method as used in Sect.~\ref{temporalcorrelation.reacdif} can be used, leading to a Casimir force per unit surface equal to
\begin{equation}
F_{C}/A = \frac{\left(1 + \frac{a}{2}\right)\Gamma}{8\pi a\lambda L_{x}^{3}}\left[\begin{array}{c}
\phantom{-} \Li_{3}(e^{-2k_{0}L_{x}}) + 2k_{0}L_{x}\Li_{2}(e^{-2k_{0}L_{x}}) + 2k_{0}^{2}L_{x}^2\Li_{1}(e^{-2k_{0}L_{x}})\\
- \Li_{3}(e^{-2k_{1}L_{x}}) - 2k_{1}L_{x}\Li_{2}(e^{-2k_{1}L_{x}})  - 2k_{1}^{2}L_{x}^2\Li_{1}(e^{-2k_{1}L_{x}})
\end{array}\right],
\end{equation}
where $k_{0}^{2} = \frac{\kappa_{1}}{\kappa_{2}}$ and $k_{1}^{2} = \frac{\kappa_{1}}{\kappa_{2}} + \frac{a}{\lambda\kappa_{2}}$.
In the limit of infinite correlation length we have $k_{0}\rightarrow 0$ and $k_{1}\rightarrow\sqrt{\frac{a}{\lambda\kappa_{2}}}$, from which we obtain
\begin{equation}
F_{C}/A = \frac{\left(1 + \frac{a}{2}\right)\Gamma}{8\pi a\lambda L_{x}^{3}}
\left[\zeta(3) - \Li_{3}(e^{-2 k_{1}L_{x}}) - 2k_{1}L_{x}\Li_{2}(e^{-2k_{1}L_{x}}) - 2k_{1}^{2}L_{x}^{2}\Li_{1}(e^{-2k_{1}L_{x}})\right].
\end{equation}
The presented result should be compared with~\cite{Bartolo}, where the same system was studied, but a different answer was given \cite{Fournierprivate}. At long distances it decays as in the case of white noise \eqref{fza.cliq.white} with a prefactor $(1 + \frac{a}{2})/a$.

For $a\to 0$, we obtain the quenched limit of the stress tensor at the plates
\begin{equation}
\mean{\mathbb{T}_{xx}(0)}  = \frac{-\Gamma}{4\pi\lambda^{2}\kappa_{2}L_{x}}
\int_{0}^{\infty}dkk \sum_{n_{x}\in\mathbb{Z}}\frac{\left(\frac{n_{x}\pi}{L_{x}}\right)^{2}}{\left(\left(\frac{n_{x}\pi}{L_{x}}\right)^2 + k^{2} + k_{0}^{2} \right)^{2}}.
\end{equation}
This expression is regularized using Eqs.~\eqref{Relacion de Y con Z 2} and \eqref{Chowla-Selberg 1d} with $s = 2$, resulting in
\begin{equation}
F_{C}/A = \frac{\Gamma}{4\pi\lambda^{2}\kappa_{2}}\int_{0}^{\infty} dk\frac{k}{\omega}\frac{\left(1 - e^{2\omega L_{x}} + 2\omega L_{x}
e^{2\omega L_{x}}\right)}{\left(e^{2\omega L_{x}} - 1\right)^{2}},
\end{equation}
where $\omega = \sqrt{k^{2} + k_{0}^{2}}$. This integral can be carried out to obtain the Casimir force as
\begin{equation}
F_{C}/A = \frac{\Gamma}{4\pi\lambda^{2}\kappa_{2}L_{x}}\frac{k_{0}L_{x}}{e^{2k_{0}L_{x}} - 1}.
\end{equation}
In the limit of infinite correlation length we have
\begin{equation}
F_{C}/A = \frac{\Gamma}{8\pi\lambda^{2}\kappa_{2}L_{x}},
\end{equation}
and in the limit of small correlation length ($k_{0}L_{x}\gg 1$) we obtain
\begin{equation}
F_{C}/A = \frac{\Gamma k_{0}}{4\pi\lambda^{2}\kappa_{2}}e^{-2k_{0}L_{x}}.
\end{equation}

%%%%%%%%%%%%%%%%%%%%%%%%%%%%%%%%%
\subsection{Maximally spatially correlated noise}
\begin{figure}[h]
\begin{center}
\includegraphics[width=\columnwidth]{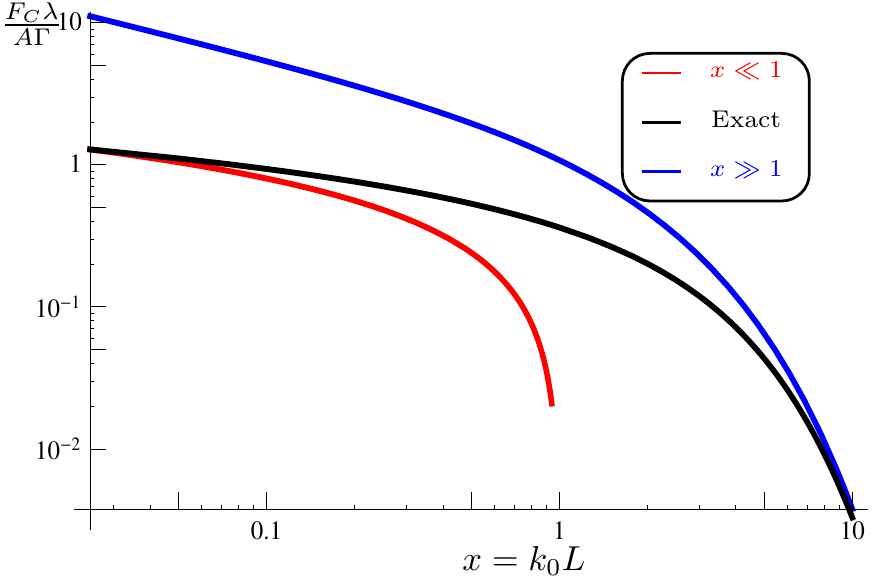}
\caption{\label{Serie_Ruido_Espacial}Casimir force in a liquid crystal subject to a spatially correlated noise in dimensionless units of distance $x = k_{0}L$ over a plate immersed in a liquid crystal subject to a spatially correlated noise of intensity $\Gamma$ in the presence of another plate at a distance $L$. The exact result is the black curve (Eq.~\eqref{Eq:Serie_Ruido_Espacial}). The short (Eq.~\eqref{Eq:Serie_Ruido_Espacial_Cerca}) and long (Eq.~\eqref{Eq:Serie_Ruido_Espacial_Lejos}) distance limits are the red and blue curve respectively.}
\end{center}
\end{figure}

As a final case we consider a noise that is rapidly fluctuating in time but that is homogeneous in space, described by the correlation \eqref{maxspatialcorrnoise}. In this case, $h_{nm}$ is not diagonal but is given by
\begin{equation}
h_{nm} = 2V\Gamma \frac{[1 - (-1)^{n_{x}}][1 - (-1)^{m_{x}}]}{\pi^{2}n_{x}m_{x}}\delta_{n_{y}0}\delta_{n_{z}0}\delta_{m_{y}0}\delta_{m_{z}0}.
\end{equation}
The stress on the plates is then given by
\begin{equation}
\mean{ \mathbb{T}_{xx}(0)} = - \frac{2\Gamma}{\lambda\pi^{2}}\sum_{n_{x},m_{x}=1}^{\infty}\frac{[1 - (-1)^{n_{x}}][1 - (-1)^{m_{x}}]}{n_{x}^{2} + m_{x}^{2} + 2\left(\frac{k_{0} L_{x}}{\pi}\right)^{2}}.
\end{equation}
As in the double summation above only odd values of $n$ and $m$ are summed, it can be
expressed in terms of the
Elizalde zeta function over odd numbers, defined as
\begin{equation}
Z_{I}(\alpha ,\beta ,\omega ,s) = \sum_{n,m\in\mathbb{Z}}\frac{1}{\left(\alpha^{2}(2n + 1)^{2} + \beta^{2}(2m + 1)^{2} + \omega^{2}\right)^{s}},
\end{equation}
which can be written in terms  of four Elizalde zeta functions.
Using the asymptotic expansion of the Elizalde zeta functions given in Eq.~\eqref{Eliz} with $p=2$ and $s=1$, the Casimir force can be expressed as an infinite sum of Bessel functions $K_{0}(x) $ with different values of $x$. The divergent terms, given by the first term in Eq.~\eqref{Eliz}, are independent of $L_{x}$, so the final expression is finite and given by
\begin{equation}\label{Eq:Serie_Ruido_Espacial}
F_{C}/A = - \frac{\Gamma}{\lambda}\sum_{n\in\mathbb{Z}}\sum_{m\in\mathbb{Z}}^{\hspace{-0.6cm}(n,m)\neq(0,0)}\left[ \begin{array}{l}
 \phantom{+}K_{0}\left(2\sqrt{2}k_{0}L_{x}\sqrt{n^{2} + m^{2}}\right)\\
 - \frac{1}{2}K_{0}\left(2\sqrt{2}k_{0}L_{x}\sqrt{\frac{n^{2}}{4} + m^{2}}\right)\\
 - \frac{1}{2}K_{0}\left(2\sqrt{2}k_{0}L_{x}\sqrt{n^{2} + \frac{m^{2}}{4}}\right)\\
 + \frac{1}{4} K_{0}\left(2\sqrt{2}k_{0}L_{x}\sqrt{\frac{n^{2}}{4} + \frac{m^{2}}{4}}\right)
\end{array} \right].
\end{equation}
Two limiting cases can be considered to clarify this result. First, in the limit of long distances $k_{0}L_{x}\gg 1$, the Casimir force is given by
\begin{equation}\label{Eq:Serie_Ruido_Espacial_Lejos}
F_{C}/A = \frac{\Gamma}{2\lambda}\sqrt{\frac{\sqrt{2}\pi}{k_{0}L}}e^{-\sqrt{2}k_{0}L},
\end{equation}
whereas in the opposite limit of long correlation length $k_0L\ll 1$, the result is
\begin{equation}\label{Eq:Serie_Ruido_Espacial_Cerca}
F_{C}/A = - \alpha\frac{\Gamma}{\lambda}\log\left(k_{0}L\right),
\end{equation}
with $\alpha = \frac{1}{16}\left( 7 + 12\sqrt{5} - 20\sqrt{2}\right)\approx 0.346784$, obtained as a Taylor expansion of the associated two dimensional Abel--Plana formula of Eq.~\eqref{Eq:Serie_Ruido_Espacial}. In the limit of infinite correlation length this result diverges. Eq~\eqref{Eq:Serie_Ruido_Espacial}, and the limits of long a short correlation length are represented in Fig.~\ref{Serie_Ruido_Espacial}.

%%%%%%%%%%%%%%%%%%%%%%%%%%%%%%%%%%%%%%%%%%%
%%%%%%%%%%%%%%%%%%%%%%%%%%%%%%%%%%%%%%%%%%%
\section{Two non hermitian fields system}\label{2campos_no_hermitico}
In the two systems we have considered so far (reaction--diffusion \ref{Reaccion-Difusion} and liquid crystals \ref{Cristales-Liquidos}), the dynamics is described by a Hermitian operator and therefore the potential of the method described herein is not fully evident. In this Section, we build a more complex system, described by a model with two fields (which could be temperature and concentration, for example) coupled in a non-symmetric way. For simplicity and to be concrete we will consider that the fields $\psi_{1}$ and $\psi_{2}$ are scalar,  subject to Neumann boundary conditions, and with eigenfunctions described by the Fourier modes \eqref{autofunciones-Neumann}. In Fourier space, the dynamic equation is
\begin{equation}
\derpar{}{t}  \begin{pmatrix} \psi_{1}\\ \psi_{2} \end{pmatrix} =
- \begin{pmatrix} \alpha_{\textbf{k}} & 0 \\ \beta_{\textbf{k}} & \gamma_{\textbf{k}}\end{pmatrix}
\begin{pmatrix} \psi_{1}\\ \psi_{2} \end{pmatrix}
+ \begin{pmatrix} \xi_{1}\\ \xi_{2} \end{pmatrix}.
\end{equation}
The noises are assumed to be white with different correlation intensities (allowing one of them to be set equal to zero later) and no cross-correlation
\begin{align}
\mean{\xi_{1}(\textbf{r}, t) \xi_{1}(\textbf{r}',t')} &= \Gamma_{1} \delta(\textbf{r}-\textbf{r}')\delta(t-t'),\nonumber \\
\mean{\xi_{2}(\textbf{r}, t) \xi_{2}(\textbf{r}',t')} &= \Gamma_{2} \delta(\textbf{r}-\textbf{r}')\delta(t-t'),\nonumber \\
\mean{\xi_{1}(\textbf{r}, t) \xi_{2}(\textbf{r}',t')} &= 0.
\end{align}
Finally, the stress tensor is assumed to be isotropic, depending only on the fields as
\begin{equation}
\mathbb{T}_{xx} = \kappa_{1}\psi_{1}^{2} + \kappa_{2}\psi_{2}^{2}.
\end{equation}
As the dynamic matrix is non-Hermitian, the left and right eigenmodes are different, being given by
\begin{align}
f_{1,\Vec{k}}(\textbf{r}) &= \sqrt{\frac{2}{V}}\cos\left(k_{x}x \right) e^{i\textbf{k}_{\parallel}\cdot\textbf{r}_{\parallel}} \begin{pmatrix} 1 \\ \frac{\beta_{\textbf{k}}}{\alpha_{\textbf{k}}-\gamma_{\textbf{k}}} \end{pmatrix},\nonumber \\
f_{2,\Vec{k}}(\textbf{r}) &= \sqrt{\frac{2}{V}}\cos\left(k_{x}x \right) e^{i\textbf{k}_{\parallel}\cdot\textbf{r}_{\parallel}} \begin{pmatrix} 0 \\ 1 \end{pmatrix},\nonumber \\
g_{1,\Vec{k}}(\textbf{r}) &= \sqrt{\frac{2}{V}}\cos\left(k_{x}x \right) e^{i\textbf{k}_{\parallel}\cdot\textbf{r}_{\parallel}} \begin{pmatrix} 1 \\ 0 \end{pmatrix},\nonumber \\
g_{2,\Vec{k}}(\textbf{r}) &= \sqrt{\frac{2}{V}}\cos\left(k_{x}x \right) e^{i\textbf{k}_{\parallel}\cdot\textbf{r}_{\parallel}} \begin{pmatrix} -\frac{\beta_{\textbf{k}}}{\alpha_{\textbf{k}}-\gamma_{\textbf{k}}} \\ 1, \end{pmatrix}
\end{align}
with eigenvalues
\begin{equation}
\mu_{1,\Vec{k}} = \alpha_{\textbf{k}},\hspace{2.0cm}
\mu_{2,\Vec{k}} = \gamma_{\textbf{k}}.
\end{equation}
Using these eigenmodes, the different elements needed $h_{nm}$ \eqref{h_nm} to compute the Casimir pressure are
\begin{align}
h_{i\Vec{k},j\Vec{q}} &= \delta_{\Vec{k}\Vec{q}}
\begin{pmatrix}
\Gamma_{1} & - \Gamma_{1}\frac{\beta_{\textbf{k}}}{\alpha_{\textbf{k}} - \gamma_{\textbf{k}}} \\
- \Gamma_{1}\frac{\beta_{\textbf{k}}}{\alpha_{\textbf{k}} - \gamma_{\textbf{k}}} & \Gamma_{2} - \Gamma_{1}^{2}\frac{\beta_{\textbf{k}}^{2}}{(\alpha_{\textbf{k}}-\gamma_{\textbf{k}})^{2}}
\end{pmatrix}, \nonumber \\
T_{i\Vec{k},j\Vec{q}} &= \delta_{\Vec{k}\Vec{q}}
\begin{pmatrix}
\kappa_{1} + \kappa_{2}\frac{\beta_{\textbf{k}}^2}{(\alpha_{\textbf{k}}-\gamma_{\textbf{k}})^{2}} & \kappa_{2}\frac{\beta_{\textbf{k}}}{\alpha_{\textbf{k}}-\gamma_{\textbf{k}}}\\
 \kappa_{2}\frac{\beta_{\textbf{k}}}{\alpha_{\textbf{k}} - \gamma_{\textbf{k}}} & \kappa_{2}
\end{pmatrix} .
\end{align}
After simple algebra, the stress tensor on the plates is obtained as
\begin{equation}
\mean{\mathbb{T}_{xx}(0)} = \frac{2}{V} \sum_{\Vec{k}} \left[\begin{array}{l}
\frac{\Gamma_{1}(\kappa_{1}+\kappa_{2}\beta_{\textbf{k}}^2/(\alpha_{\textbf{k}}-\gamma_{\textbf{k}})^2}{2\alpha_{\textbf{k}}}\frac{2\Gamma_{1} \kappa_{2} \beta_{\textbf{k}}^2/(\alpha_{\textbf{k}}-\gamma_{\textbf{k}})^2}{\alpha_{\textbf{k}}+\gamma_{\textbf{k}}}\\
 - \frac{(\Gamma_{1} \beta_{\textbf{k}}^2/(\alpha_{\textbf{k}}-\gamma_{\textbf{k}})^2 +\Gamma_{2}) \kappa_{2}}{2\gamma_{\textbf{k}}}\end{array}
\right],
\end{equation}
which for specific models (that is, specific values of $\alpha_{\textbf{k}}$, $\beta_{\textbf{k}}$, and $\gamma_{\textbf{k}}$) could be computed and regularized to obtain the Casimir force on the plates.

To show the kind of results that can be obtained we consider the simple reaction--diffusion two-field model $\alpha_{\textbf{k}}=\lambda_1+Dk^{2}$, $\beta_{\textbf{k}}=\lambda_{12}$, and $\gamma_{\textbf{k}}=\lambda_2+Dk^{2}$, with noise intensities $\Gamma_{1} = \Gamma$ and $\Gamma_{2}=0$, representing the system
\begin{align}
\derpar{\psi_{1}}{t} &= -\lambda_{1}\psi_{1} + D\nabla^{2}\psi_{1} + \xi, \nonumber \\
\derpar{\psi_{2}}{t} &= -\lambda_{2}\psi_{2} + D\nabla^{2}\psi_{2} - \lambda_{12}\psi_{1} .
\end{align}
Caution should be taken to avoid the case $\lambda_{1} = \lambda_{2}$, for which the dynamic matrix is not diagonalizable and a Jordan block appears. The method developed in this article is not directly applicable, but the generalization is simple. Also, neither $\lambda_{1}$ or $\lambda_{2}$ can vanish, because there is no damping term to make the nonconservative noise vanish and the fields would perform an unbounded random walk.
Furthermore, we assume that the stress only depends on $\psi_{2}$, i.e., $\mathbb{T}_{xx}=\kappa\psi_{2}^{2}$. Therefore, any eventual Casimir force is produced by the fluctuations of the second field which are produced by the coupling with the first field. The stress on the plates is finally
\begin{equation}
\mean{\mathbb{T}_{xx}(0)} = \frac{\Gamma\kappa\lambda_{12}^{2}}{2V} \sum_{\Vec{k}} \left[(\lambda_{1} + Dk^{2})(\lambda_{2} + Dk^{2})(\lambda_{1} + \lambda_{2} + 2Dk^{2})\right]^{-1}. \nonumber
\end{equation}
As $\lambda_{1}\neq\lambda_{2}$, we can apply partial fraction decomposition to obtain
\begin{equation}
\mean{\mathbb{T}_{xx}(0)} = \frac{\Gamma\kappa\lambda_{12}^{2}}{2D V(\lambda_{1} - \lambda_{2})^{2}} \sum_{\Vec{k}}\left[\frac{1}{k^{2} + k_{1}^{2}} + \frac{1}{k^{2} + k_{2}^{2}} - \frac{2}{k^{2} + k_{3}^{2}}\right], \nonumber
\end{equation}
where $k_{1}^{2} = \lambda_{1}/D$, $k_{2}^{2} = \lambda_{2}/D$, and $k_{3}^{2} = (\lambda_{1} + \lambda_{2})/2D$. Then, we can perform each infinite sum as in the case of scalar white noise to obtain the Casimir force as
\begin{equation}
\mean{\mathbb{T}_{xx}(0)} = \frac{- \Gamma\kappa\lambda_{12}^{2}}{4\pi DL_{x}(\lambda_{1} - \lambda_{2})^{2}}\ln\left(\frac{\left(1 - e^{-2k_{1}L_{x}}\right)\left(1 - e^{-2k_{2}L_{x}}\right)}{\left(1 - e^{-2k_{3}L_{x}}\right)^{2}}\right). \nonumber
\end{equation}
It is interesting to note that, if $\lambda_{1} = 0$ and/or $\lambda_{2} = 0$, this Casimir force diverges.

\section{Discussion and conclusions}
In this Chapter, we have developed a formalism to study Casimir forces in classical systems out of equilibrium based on the stochastic dynamical equations of the system under study. The equilibrium case is recovered as a particular limit where the fluctuation--dissipation theorem is valid.

In particular, we study the interaction which appears between intrusions in a medium subject to any kind of noise. The study is restricted to additive noise and non quantum systems. The method only relies on the stochastic evolution equation of the field in the medium, and information about the interaction between the medium and the intrusions, as given by the boundary conditions of the fields at the surface of the bodies. No assumptions are made regarding any characteristic of the noise, which could be internal of external, thermal or induced, white or colored, additive or multiplicative, and even non-Gaussian.

This formalism reduces to the classical thermal Casimir effect when the medium is subjected to an additive Gaussian white noise with autocorrelation amplitude $\mathcal{Q} = k_{B}T(\mathcal{L}+\mathcal{L}^{+})$ and its dynamics is described by a Hermitian operator, as shown in \eqref{Caso termico}.

Eq.~\eqref{Force} is the main result of the Chapter. It can be used to obtain the Casimir force (or fluctuation induced forces) for the steady state of any system, in Equilibrium or not equilibrium situations, described by a noise term which can be correlated or not in space, colored or white in time and for any geometry in principle. Furthermore, as shown in Sect.~\ref{2campos_no_hermitico}, it is possible study systems whose dynamics would be given by non Hermitian evolution equations, or even by dynamical operators explicitly time dependent~\cite{Dean1}~\cite{Dean2}. In principle, we can study time dependent Casimir forces with this formalism. Eq.~\eqref{Force} can be also used as a numerical computation tool of Casimir forces in realist situations:
Eigenvalues problem can be numerically solved for any geometry and, after introduce the results of this problem in Eq.~\eqref{Force} and perform the numerical sum, we obtain a numerical result for the Casimir force. All this procedure must be done carefully in order to avoid spurious numerical divergences. In fact, an implementation of the problem can be easily done with a Finite Element Method implementation software. First of all, we would solve numerically the eigenvalue problem of the geometry over study with the boundary condition of each body and, to carry out the (divergent) series over eigenvalues, we add a regularization kernel to the series. The introduction of the regularizing kernel transform the considered divergent series into an asymptotic series. Therefore we have to carry out this sum with the specific techniques of the asymptotic series to obtain the correct result of the series~\cite{Sergio}.
We also used the formalism presented here to probe that Casimir forces nulls in equilibrium isotropic systems, and it is generally non zero for any other situation.

We have obtained an exact formula for the Casimir force felt by a body \eqref{Force}, which shows how to derive the Casimir force for any geometrical configuration and noise. Equation \eqref{Force} can be used to obtain non-equilibrium induced self forces over asymmetric bodies, as shown in \cite{Buenzli-Soto}. It can also provide a numerical tool useful to evaluate Casimir forces for complicated geometries. 

Along this Chapter, we have used the  formula to obtain the force between parallel plates in different media (in a reaction--diffusion model \ref{Reaccion-Difusion} and in liquid crystals \ref{Cristales-Liquidos}) under the influence of different kinds of Gaussian noises, i.e., white noise to recover the thermal case already studied in the literature, and noises with nonzero spatial or temporal correlation lengths, where different forces appears.

Finally we have shown an example of the evaluation of Casimir forces in a system with non-Hermitian evolution dynamics, which was an intractable problem until the development of the formalism presented herein.

%%%%%%%%%%%%%%%%%%%%%%%%%%%%%%%%%%%%%%%%%%%
%%%%%%%%%%%%%%%%%%%%%%%%%%%%%%%%%%%%%%%%%%%
\section{Appendix}
\subsection{Appendix A: Elizalde zeta function}\label{Appendix: Elizalde zeta function}
% Ponemos el contador de las ecuaciones a cero:
%\setcounter{equation}{0}
%\renewcommand{\theequation}{2A.\arabic{equation}}

The computation of the Casimir forces makes use of the asymptotic expansion of the Elizalde zeta function, which is defined as~\cite{Elizalde}
\begin{equation}\label{Definicion_funcion_zeta_Elizalde}
Z_{p}(s,A,\omega) = \sum_{\Vec{n}\in\mathbb{Z}^{p}}\frac{1}{\left(\Vec{n}^{T}\cdot A\cdot\Vec{n} + \omega^{2}\right)^{s}},
\end{equation}
we are mainly interested in the case of diagonal matrix $A$. This function admits an asymptotic expansion valid for all complex $s$~\cite{springerlink:10.1007/s002200050472}
\begin{eqnarray}\label{Eliz}
Z_{p}(s,A,\omega) & = & \frac{\pi^{\frac{p}{2}}\Gamma(s - \frac{p}{2})}{\Gamma(s)\sqrt{\Det{A}}}\omega^{p - 2s}\\
& + & \frac{2\pi^{s}\omega^{\frac{p}{2} - s}}{\Gamma(s)\sqrt{\Det{A}}} {\sum_{\Vec{n}\in\mathbb{Z}^{p}}}'\left(\Vec{m}^{T}A^{-1}\Vec{m}\right)^{\frac{2s - p}{4}}
 K_{s - \frac{p}{2}}\left(2\pi\omega\sqrt{\Vec{m}^{T}A^{-1}\Vec{m}}\right).\nonumber
\end{eqnarray}
where $K_{\nu}(z)$ is the inhomogeneous Bessel function of the second kind or Macdonald function. When $p = 2$, this is called the Chowla--Selberg formula. The demonstration of this formula is as follows. Using the known integral
\begin{equation}
\int_{0}^{\infty}dt\,t^{s-1}e^{-at} = \frac{\Gamma(s)}{a^{s}},
\end{equation}
Elizalde zeta function \eqref{Definicion_funcion_zeta_Elizalde} can be written as
\begin{equation}\label{Definicion_funcion_zeta_Elizalde3}
Z_{p}(s,a_{i},\omega) = \frac{1}{\Gamma(s)}\int_{0}^{\infty}dt\,t^{s-1}\sum_{\Vec{n}\in\mathbb{Z}^{p}}e^{-t\omega^{2}}e^{-t\left(\Vec{n}^{T}\cdot A\cdot\Vec{n}\right)},
\end{equation}
We also need the Poisson theorem
\begin{equation}\label{Formula_Poisson}
\sum_{\Vec{n}\in\mathbb{Z}^{p}}\delta\left(\Vec{x} - \Vec{n}\right) = \sum_{\Vec{m}\in\mathbb{Z}^{p}}e^{2\pi i\Vec{m}\cdot\Vec{x}}.
\end{equation}
If we multiply on both sides of the equality by $f(\Vec{x}) = e^{-t\left(\Vec{n}^{T}\cdot A\cdot\Vec{n}\right)}$, and perform an integration over $\Vec{x}\in\mathbb{R}^{p}$, we obtain
\begin{equation}\label{Formula_Poisson2}
\sum_{\Vec{n}\in\mathbb{Z}^{p}}e^{-t\left(\Vec{n}^{T}\cdot A\cdot\Vec{n}\right)} = \frac{\pi^{\frac{p}{2}}t^{-\frac{p}{2}}}{\sqrt{\Det{A}}}\sum_{\Vec{m}\in\mathbb{Z}^{p}}e^{-\frac{\pi^{2}}{t}\Vec{m}^{T}\cdot A^{-1}\cdot\Vec{m}}.
\end{equation}
After the substitution of \eqref{Formula_Poisson2} into \eqref{Definicion_funcion_zeta_Elizalde3}, and separating the term $\Vec{m} = \Vec{0}$, we obtain 
\begin{equation}\label{Definicion_funcion_zeta_Elizalde4}
Z_{p}(s,a_{i},\omega) = \frac{1}{\Gamma(s)}\int_{0}^{\infty}dt\,t^{s-1}e^{-t\omega^{2}}\frac{\pi^{\frac{p}{2}}t^{-\frac{p}{2}}}{\sqrt{\Det{A}}}
\left[ 1 +
{\sum_{\Vec{m}\in\mathbb{Z}^{p}}}'e^{-\frac{\pi^{2}}{t}\Vec{m}^{T}\cdot A^{-1}\cdot\Vec{m}}\right].
\end{equation}
Finally, after carrying out the $t$-integral with 
\begin{equation}
\int_{0}^{\infty}dt\,t^{\nu - 1}e^{- \frac{\alpha}{t}- \beta t} = 2\left(\frac{\alpha}{\beta}\right)^{\frac{\nu}{2}}K_{\nu}(2\sqrt{\alpha\beta}),
\end{equation}
and simplifying the result, we obtain Eq.~\eqref{Eliz}.
We are mainly interested in the case $p = 1$ with $\alpha > 0$, for which
\begin{align}\label{Chowla-Selberg 1d}
Z_{1}(s,\alpha,\omega) = & \sum_{n\in\mathbb{Z}}\frac{1}{\left(\alpha^{2}n^{2} + \omega^{2}\right)^{s}}\nonumber\\
                        = & \frac{\sqrt{\pi}\Gamma(s - \frac{1}{2})}{\Gamma(s)\alpha}\omega^{1 - 2s}
 + \frac{4\pi^{s}}{\Gamma(s)\alpha} \sum_{n = 1}^{\infty}\left(\frac{n}{\alpha\omega}\right)^{s - \frac{1}{2}}K_{s - \frac{1}{2}}\left(2\pi\omega\frac{n}{\alpha}\right).
\end{align}
When studying Casimir forces between plates with Dirichlet boundary conditions, the following series needs to be computed
\begin{equation}
Y_{1}(s,\alpha,\omega) = \sum_{n\in\mathbb{Z}}\frac{\alpha^{2}n^{2}}{\left(\alpha^{2}n^{2} + \omega^{2}\right)^{s}}.
\end{equation}
It is straightforward  to obtain
\begin{equation}\label{Relacion de Y con Z}
\sum_{n\in\mathbb{Z}}\frac{\alpha^{2}n^{2}}{\left(\alpha^{2}n^{2} + \omega^{2}\right)^{s}} = \sum_{n\in\mathbb{Z}}\frac{1}{\left(\alpha^{2}n^{2} + \omega^{2}\right)^{s-1}} - \omega^{2}\sum_{n\in\mathbb{Z}}\frac{1}{\left(\alpha^{2}n^{2} + \omega^{2}\right)^{s}},
\end{equation}
and therefore
\begin{equation}\label{Relacion de Y con Z 2}
Y_{1}(s,\alpha,\omega) = Z_{1}(s - 1,\alpha,\omega) - \omega^{2}Z_{1}(s,\alpha,\omega).
\end{equation}

%%%%%%%%%%%%%%%%%%%%%%%%%%%%%%%%%%%%%%%%%%%
\subsection{Appendix B: Derivation of Eq.~\eqref{average-square-field}}\label{Apendice-Resolucion.promedio.modos.phi_n.phi_m}
% Ponemos el contador de las ecuaciones a cero:
%\setcounter{equation}{0}
%\renewcommand{\theequation}{2B.\arabic{equation}}

In this Appendix we will obtain an expression for the cross-average of the mode amplitudes $\phi_{n}(t)$ in the steady state given in Eq.~\eqref{average-square-field} and derived from Eq.~\eqref{two-field-modes-average}
\begin{equation}\label{average-square-field2}
\lim_{t\to\infty}\mean{\phi_{n}(t)\phi_{m}^{*}(t)} = \lim_{t\to\infty}e^{ - (\mu_{n} + \mu_{m}^*)t}\int_{-\infty}^{t} d\tau_{1}\int_{-\infty}^{t}d\tau_{2}e^{\mu_{n}\tau_{1} + \mu_{m}^{*}\tau_{2}}
\mean{\xi_{n}(\tau_{1})\xi_{m}^{*}(\tau_{2})}.
\end{equation}
The problem is reduced to the evaluation of the cross-average of the noise amplitudes $\xi_{n}(t)$ and $\xi_{m}(t)$
\begin{equation}\label{correlation.noise.modes2}
\mean{\xi_{n}(\tau_{1})\xi_{m}^{*}(\tau_{2})} = \int\!\! d\textbf{r}_{1}\! \int\!\! d\textbf{r}_{2} g_{n}^{*}(\textbf{r}_{1})g_{m}(\textbf{r}_{2})\mean{\xi(\textbf{r}_{1},\tau_{1})\xi^{*}(\textbf{r}_{2},\tau_{2})}.
\end{equation}
Spatial and temporal noise correlations factorizes. We can use that to solve \eqref{correlation.noise.modes2} obtaining
\begin{equation}\label{correlation.noise.modes3}
\mean{\xi_{n}(\tau_{1})\xi_{m}^{*}(\tau_{2})} = c(\tau_{1} - \tau_{2})\int d\textbf{r}_{1} \int d\textbf{r}_{2} g_{n}^{*}(\textbf{r}_{1}) h(\Vec{r}_{1} - \Vec{r}_{2})g_{m}(\textbf{r}_{2}) = c(\tau_{1} - \tau_{2})h_{nm},
\end{equation}
where $h_{nm}$ is defined. Applying this result in \eqref{average-square-field2}, the cross-average of the mode amplitudes in the steady state is
\begin{align}
\lim_{t\to\infty}\mean{\phi_{n}(t)\phi_{m}^{*}(t)} & = \lim_{t\to\infty}\int_{-\infty}^{t} d\tau_{1}\int_{-\infty}^{t}d\tau_{2}
e^{-\mu_{n}(t-\tau_{1}) - \mu_{m}^*(t-\tau_{2})} c(\tau_{1} - \tau_{2})h_{nm}\nonumber\\
& = \phantom{\lim_{t\to\infty}}\int_{0}^{\infty} dz_{1}\int_{0}^{\infty}dz_{2}
e^{-\mu_{n}z_{1} - \mu_{m}^{*}z_{2}} c(z_{1} - z_{2})h_{nm},\label{correlation.noise.modes4}
\end{align}
where the change of variables $z_{1} = t - \tau_{1}$ and $z_{2} = t - \tau_{2}$ has been applied. After performing a second change of variables $u = z_{1} - z_{2}$ and $v = (z_{1} + z_{2})/2$, and using the parity $c(z) = c(-z)$, we obtain
\begin{align*}
\lim_{t\to\infty}\mean{\phi_{n}(t)\phi_{m}^{*}(t)} & = h_{nm}\int_{-\infty}^{\infty} du \int_{|u|/2}^{\infty}dv
e^{-\mu_{n}(v + u/2) - \mu_{m}^{*}(v - u/2)}c(u) \\
& = h_{nm}\int_{-\infty}^{\infty} du\, c(u) e^{-(\mu_{n} - \mu_{m}^{*})u/2}\int_{|u|/2}^{\infty}dv e^{-(\mu_{n}+\mu_{m}^{*})v} \\
& = h_{nm}\int_{-\infty}^{\infty} du\, c(u) \frac{e^{-(\mu_{n} - \mu_{m}^{*})u/2} e^{-(\mu_{n} + \mu_{m}^{*})|u|/2}}{\mu_{n} + \mu_{m}^{*}}\\
& = \frac{h_{nm}}{\mu_{n} + \mu_{m}^{*}}\left[\int_{0}^{\infty} du\, c(u) e^{-\mu_{n}u} + \int_{0}^{\infty} du\, c(u) e^{-\mu_{m}^{*}u} \right].
\end{align*}
If we define $\tilde{c}(p)$ as the Laplace transformation of $c(t)$, we obtain Eq.~\eqref{average-square-field}
\begin{equation}
\lim_{t\to\infty}\mean{\phi_{n}(t)\phi_{m}^{*}(t)} = h_{nm}\frac{\widetilde{c}(\mu_{n})+\widetilde{c}(\mu_{m}^{*})}{\mu_{n} + \mu_{m}^{*}}.
\end{equation}

% Formalismo de Parisi-Wu para el calculo de fuerzas de Casimir, caso de pistones y simetría cilíndrica y Fluctuaciones
\begin{savequote}[11cm] % this sets the width of the quote
\sffamily
``I think I can safely say that nobody understands quantum mechanics.'' 
\qauthor{Richard P. Feynman}% - The Character of Physical Law (1965) Ch. 6}
\end{savequote}

\chapter{Stochastic quantization and Casimir forces.}\label{Chap: Stochastic Quantization and Casimir forces.}
\graphicspath{{01-Casimir_Langevin/ch2/Figuras/}}

In this Chapter we show how the stochastic quantization method
developed by Parisi and Wu can be used to obtain electromagnetic (EM) Casimir
forces starting from the dynamical approach presented in Chapter \ref{Dynamical approach to the Casimir effect} of this Thesis.
Both quantum and thermal fluctuations are taken into account by
a Langevin equation for the field. The method allows the Casimir
force to be obtained directly, derived from the stress tensor instead of
the free energy.
It only requires the spectral decomposition of the Laplacian operator
in the given geometry. 
As an application, we compute the Casimir force on the plates of a
finite piston of arbitrary cross section. Fluctuations of the force are
also directly obtained, and it is shown that, in the piston case,
the variance of the force is twice the force squared.

Fluctuation-induced (Casimir) forces~\cite{Casimir_Placas_Paralelas}\cite{Review_Casimir} are currently attracting renewed attention, probably because of the
accessibility of small systems at the nano- and micro-scale, as it is
in these small systems where Casimir forces are revealed.
However, techniques to calculate forces for complicated geometries,
beyond the usual ones with high degrees of symmetry, are scarce.
Recently, a powerful technique to calculate EM
Casimir forces~\cite{Kardar-Geometrias-Arbitrarias}\cite{RE09} has been proposed. It is based
on a multiscattering technique and has been successfully applied to
many configurations, such as plates, cylinders, spheres, wedges, etc.~\cite{Casimir_Wedges}

Casimir forces have their origin in fluctuations of EM fields.
With that idea in mind, a method
that takes as its starting point the Langevin equation describing
the evolution of systems subjected to fluctuations has recently been proposed~\cite{PhysRevE.83.031102}.
This method was applied
to thermal fluctuations, and one of its advantage is that it is applicable to systems in
or out of equilibrium. However, its application to EM Casimir forces of quantum origin
has not yet been developed.

The goal of this Chapter is to apply the Langevin equation method to calculate forces
of quantum origin. The method is based on the stochastic quantization method developed
by Parisi and Wu~\cite{Parisi-Wu_original}\cite{Parisi-Wu_review}. They construct a Langevin equation for a given field
subjected to thermal fluctuations. However, this Chapter will show
how quantum fluctuations arise naturally within that method.

The plan of this Chapter is as follows.
We start in Sect.~\ref{Parisi-Wu formalism} by presenting the Stochastic quantization procedure or Parisi--Wu formalism, which let us define the quantum fluctuations of the EM field.
In Sect.~\ref{Casimir forces from the average stress tensor_PW} we dynamic formalism of Casimir effect is applied to the EM field subject to quantum--thermal fluctuations. As a result, a formula of the Casimir force over a given object is obtained.
In Sect.~\ref{Perfect metal pistons of arbitrary cross section}, the developed formalism is applied to the faces of a perfect metal metal piston of arbitrary cross section. As a result, the Casimir force over each plate is obtained for all temperatures. The short and large distance limits are studied in the zero and high temperature limits.
In Sect.~\ref{Fluctuation of Casimir forces}, the variance of the Casimir force is defined and obtained for the system under study.
We finish with some conclusions and three appendix.
In Appendix A (Sect.~\ref{Appendix A: Derivation of averaged stress tensor over the piston surface}), the intermediate calculations to obtain a formula for the Casimir force over a face of the arbitrary shape piston are presented.
In Appendix B (Sect.~\ref{Appendix B: Derivation of the fluctuations of the force.}), the formula of the variance of the Casimir force is obtained for linear surrounding field in equilibrium.
Finally, in Appendix C (Sect.~\ref{Appendix C: Equivalencia formalismos}), we present the equivalence between three formalism of Casimir effect: (1) the Partition function formalism; (2) the Stress--Tensor formalism and; (3) the Dynamical formalism presented in this Thesis.

The contents of this Chapter is based on the work published in~\cite{Rodriguez-Lopez_2}.
\section{Parisi-Wu formalism}\label{Parisi-Wu formalism}
The Casimir force is calculated in~\cite{PhysRevE.83.031102} via the stress tensor, $\mathbb{T}$, which must be averaged over the probability distribution
of the fields~$\phi$. Such a probability distribution $\rho = P[\phi]$ is given by:% \cite{TFT_Le_Bellac}:
\begin{equation}\label{Pphi}
\rho = \frac{1}{Z}e^{- S[\phi]/\hbar},
\end{equation}
where $S[\phi]$ is the action of the scalar or EM field with zero mass,
Wick-rotated in the time variable ($t = i\tau$)~\cite{Review_Casimir}\cite{Masujima}\cite{Thermal-Field-Theory}, i.e.,
\begin{equation}\label{PW2}
S[\phi]= -\frac{1}{2} \int_{0}^{\beta\hbar} d\tau \int d\textbf{r} \,
\phi^{*} \left(\frac{1}{c^{2}}\frac{\partial^{2}}{\partial \tau^{2}}+\nabla^{2}\right)\phi,
\end{equation}
where $\beta=1/k_BT$ and $Z$ is the partition function
$ Z= \int\mathcal{D}\phi\mathcal{D}\phi^{*}\, e^{-\beta S[\phi]}.$
For the bosonic case, the field~$\phi$ must obey periodic boundary conditions in time, that is,
$\phi(\tau + \beta\hbar,\textbf{r}) = \phi(\tau,\textbf{r})$.
The stress tensor is normally a bilinear form in the field $\mathbb{T}=\TensorT[\phi^{*},\phi,\textbf{r}]$, the expression that defines the stress tensor operator $\TensorT$.

Herein we use the formalism of Parisi and Wu~\cite{Parisi-Wu_original} to evaluate
the average of the stress tensor in an alternative way.
The idea of Parisi and Wu consists in a formulation of quantum mechanics or quantum field theory in terms of a stochastic process.
More precisely, a Langevin equation for the field~$\phi$ is written
as an evolution equation in an auxiliary time, which we will call pseudo-time, $s$.
In this description, the field depends on the new pseudo-time variable:
$\phi(\tau,\textbf{r})\to \phi(\tau,\textbf{r};s)$.
The Langevin equation takes the form~\cite{Lim2006269}
\begin{align}\label{eq.Parisi-Wu}
\partial_{s}\phi(\tau ,\textbf{r};s) & = - \frac{\delta S[\phi]}{\delta\phi} + \eta(\tau ,\textbf{r};s)\nonumber \\
& = \left(\frac{1}{c^{2}}\frac{\partial^{2}}{\partial\tau^{2}} + \nabla^{2}\right) \phi + \eta(\tau ,\textbf{r};s).
\end{align}
The term $\eta(\tau ,\textbf{r};s)$ is the source of fluctuations, given by
a~Gaussian white noise satisfying the fluctuation--dissipation relation~\cite{Kubo}
\begin{eqnarray}\label{eq.noise}
\mean{\eta (\tau ,\textbf{r};s)} & =& 0, \\
\mean{\eta (\tau ,\textbf{r};s)\eta (\tau' ,\textbf{r}';s')}& =& 2k_{B}T \delta(\tau - \tau')\delta(\textbf{r} - \textbf{r}')\delta(s-s'). \nonumber
\end{eqnarray}

The associated Fokker--Planck equation (also called advanced Kolmogorov Equation) in pseudo-time $s$ for the probability distribution $\rho$ is
\begin{equation}
\partial_{s}\rho - \frac{\delta}{\delta\phi}\left[\frac{\delta S[\phi]}{\delta\phi} + \frac{1}{\beta}\frac{\delta}{\delta\phi}\right]\rho = 0,
\end{equation}
where we have used the properties of the Gaussian noise given in Eq.~\eqref{eq.noise}. The steady state in pseudo-time is reached in the limit $s\to\infty$, where $\partial_{s}\rho = 0$, then $\rho$ obeys in the steady state
\begin{equation}\label{Steady_State_Equation_For_Parisi_Wu_Formalism}
\frac{\delta}{\delta\phi}\left[\frac{\delta S[\phi]}{\delta\phi} + \frac{1}{\beta}\frac{\delta}{\delta\phi}\right]\rho = 0.
\end{equation}
Eq.~\eqref{Steady_State_Equation_For_Parisi_Wu_Formalism} is a linear variational equation in $\rho$, which admits the solution
\begin{equation}\label{Steady_State_Equation_For_Parisi_Wu_Formalism}
\rho = \frac{1}{Z}e^{-\beta S[\phi]}.
\end{equation}
Then the solution of the Langevin equation in the stationary limit $s\to \infty$ reproduces the probability distribution
given by Eq.~\eqref{Pphi}~\cite{Masujima}.

Having an expression for the stress tensor and a Langevin equation for the field, we can follow the procedure
recently developed in~\cite{PhysRevE.83.031102} to obtain the Casimir force. The field~$\phi$ is written as
(and a similar decomposition
for the noise~$\eta$ with coefficients $\eta_{nm}$):
\begin{equation}\label{eq.phi}
\phi (\tau ,\textbf{r};s)=\sum_{n,m} \phi_{nm}(s) g_{m}(\tau) f_{n}(\textbf{r}), 
\end{equation}
where $f_{n}$ and $g_{m}$ are the eigenfunctions:
\begin{equation}\label{eq.eigenfunctions}
\nabla^{2} f_{n}(\textbf{r}) = - \lambda_{n}^{2} f_{n}(\textbf{r})
\hspace{2cm}
\frac{1}{c^{2}}\frac{\partial^{2}}{\partial \tau^{2}}g_{m}(\tau) = - \omega^{2}_{m} g_{m}(\tau),
\end{equation}
which are orthogonal under the $\text{L}^{2}$ scalar product in space
or time. The above expression indicates that $f_{n}(\textbf{r}) $ and $\lambda_{n}^{2}$
encode the spatial configuration of the system, that is, the position of the bodies and their boundary conditions.
In a similar fashion, $g_{m}(\tau)$ and $\omega^{2}_{m}$ contain the (Wick-rotated) time dependence.
As we are considering a bosonic field that obeys periodic boundary conditions in~$\tau$,
the eigenvalues are $\omega_{m} = 2\pi m/\beta\hbar c$, $m\in\mathbb{Z}$, and the normalized eigenfunctions are
$g_{m}(\tau)=\exp(i\omega_m\tau)/\sqrt{\beta\hbar}$. Then, the coefficients $\phi_{nm}$ satisfy the
differential equation
\begin{equation}
\frac{d\phi_{nm}(s)}{ds}=-  \left[\lambda_{n}^{2} + \omega_{m}^{2}\right]\phi_{nm}(s)+ \eta_{nm}(s),
\end{equation}
which can be integrated to give
\begin{equation}\label{phi_en_funcion_de_eta}
\phi_{nm}(s)=\int_{-\infty}^s d{\sigma} e^{(\lambda_{n}^{2} + \omega_{m}^{2})(\sigma - s)} \eta_{nm}(\sigma),
\end{equation}
where the noise coefficients satisfy, from Eq.~\eqref{eq.noise}
\begin{equation}\label{correlacion_modos_ruido}
\mean{\eta_{nm}(s)\eta^{*}_{n'm'}(s')} = 2k_{B}T\delta(s-s')\delta_{nn'}\delta_{mm'}.
\end{equation}

\section{Casimir forces from the average stress tensor}\label{Casimir forces from the average stress tensor_PW}
The formalism developed in~\cite{PhysRevE.83.031102} allows the average stress tensor to be calculated by substituting expression
\eqref{eq.phi} into the stress tensor and taking the average
over the fluctuations, $\eta(\tau,\textbf{r};s)$,
in the limit $s\to\infty$.
\begin{equation}
\mean{\mathbb{T}(\textbf{r})} = \lim_{s\to\infty}\mean{\TensorT[\phi(s),\phi(s),\textbf{r}]} = \lim_{s\to\infty}\sum_{n_{1},m_{1}}\sum_{n_{2},m_{2}}\mean{\phi_{n_{1},m_{1}}(s)\phi_{n_{2},m_{2}}^{*}(s)}\TensorT[f_{n_{1}},f_{n_{2}}^{*},\textbf{r}].
\end{equation}
The two functions $\phi_{nm}$ are replaced by their values in terms of the 
fluctuations, given by Eq.~\eqref{phi_en_funcion_de_eta}, resulting into a double integral over two pseudo-times.
Then, the average $\mean{\mathbb{T}(\textbf{r})}$ is carried out using Eq.~\eqref{correlacion_modos_ruido}, that eliminates
one integral, with the final result:
\begin{equation}\label{eq.T}
\mean{\mathbb{T}(\textbf{r})} = \frac{1}{\beta}\sum_{nm}\frac{\TensorT_{nn}}{\lambda_{n}^{2} + \omega_{m}^{2}},
\end{equation}
where $\TensorT_{nm}(\textbf{r})=\TensorT[f_n,f_m^*,\textbf{r}]$.

Furthermore, the sum over the temporal eigenvalues, $\omega_m$, can be carried out to obtain the result
\begin{equation}\label{eq.T2}
\mean{\mathbb{T}(\textbf{r})} = \frac{\hbar c}{2}\sum_{n}\frac{\TensorT_{nn}}{\lambda_{n}}\left[1 + \frac{2}{e^{\beta \hbar c \lambda_{n}} - 1}\right],
\end{equation}
which can be related with the quantum fluctuation--dissipation theorem for the EM field~\cite{QFDT_Weber}.
Finally, to obtain the Casimir force over a~certain body, the stress tensor must be integrated over the
surface~$\Omega$ that defines the object
\begin{equation}\label{eq.FCint}
\textbf{F}_{C} = \oint_{\Omega}\mean{\mathbb{T}(\textbf{r})}\cdot d\textbf{S}.
\end{equation}

Equation~\eqref{eq.T2} is the main result of this Chapter. It
gives an expression for the quantum Casimir force including the
effects of a finite, non-vanishing temperature, in terms of the eigenvalues and eigenfunctions of the
Laplacian operator. This expression has been easily obtained using the
stochastic quantization approach to a quantum field, together with the formalism of the Langevin equation
calculation of fluctuation-induced forces~\cite{PhysRevE.83.031102}.

Let us note that, as it occurs in most of the calculations of the Casimir Force, 
the expression for $\mean{\mathbb{T}(\textbf{r})}$ in Eq.~\eqref{eq.T2} is generally 
divergent at every point of space $\textbf{r}$ if the sum over eigenvalues runs to infinity. 
However, the expression for the force, that is obtained integrating over the surface of 
the body, Eq.~\eqref{eq.FCint}, is finite in the sense of divergent series, as numerical calculations show. It means that the integration over the body regularizes the divergences of the averaged stress tensor. If we assume that 
such regularization is carried out mode by mode, we can interchange the integration over the surface 
and the summation over eigenvalues, to obtain,
\begin{equation}\label{eq.T3}
\textbf{F}_{C} = \frac{\hbar c}{2}\sum_{n}\frac{1}{\lambda_{n}}\left[1 + \frac{2}{e^{\beta \hbar c \lambda_{n}} - 1}\right]
\oint_{\Omega}\TensorT_{nn}(\textbf{r})\cdot d\textbf{S}.
\end{equation}
which is a finite result. Therefore, the interchange of the integral and summation
regularizes the Casimir force, avoiding the use of ultraviolet cutoffs.

As we show herein, this provides a new method to calculate Casimir forces for a~given geometry by diagonalizing the
Laplace operator. So, this approach is suitable for numerical calculations of Casimir forces
in complicated, realistic geometries. Moreover, this method provides the force directly, not
as a difference of the free energy with respect to a reference state, which is
somehow difficult to establish. Also, it can be used as a starting point for a
perturbative theory for, e.g., non-flat geometries, rough surfaces or similar problems.

The expression for the Casimir force, Eq.~\eqref{eq.T2}, allows one to evaluate the quantum limit, that is, by setting the
temperature equal to zero, where the second summand inside the brackets in Eq.~\eqref{eq.T2} vanishes, i.e.,
\begin{equation} \label{eq.FCT0}
\lim_{T\to 0}\mean{\mathbb{T}(\textbf{r})} = \frac{\hbar c}{2}\sum_{n}\frac{\TensorT_{nn}}{\lambda_{n}}.
\end{equation}
In the opposite, classical limit, when $\hbar \to 0$, a Taylor expansion of the square bracket in Eq.~\eqref{eq.T2}
gives
\begin{equation}\label{eq.FCh0}
\lim_{\hbar \to 0}\mean{\mathbb{T}(\textbf{r})} = k_{B}T\sum_{n}\frac{\TensorT_{nn}} {\lambda_{n}^{2}}.
\end{equation}
This expression for the Casimir force has been used for classical systems~\cite{PhysRevE.83.031102},
such as liquid crystals~\cite{Ajdari} or reaction--diffusion systems~\cite{BritoSotoPRE}.
The two limits, quantum \eqref{eq.FCT0} and thermal \eqref{eq.FCh0}, show
that the driving force of the fluctuations has different origin. In the first case, the presence of the factor~$\hbar$
indicates the quantum nature of the fluctuations, whereas in the second case, the factor $k_{B}T$
reveals its thermal origin.

\section{Perfect metal pistons of arbitrary cross section}\label{Perfect metal pistons of arbitrary cross section}
\begin{figure}[htb]
\includegraphics[width=0.9\columnwidth,angle=0]{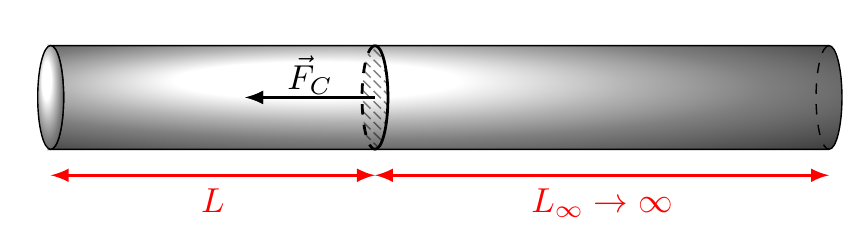}
\caption{\label{Figura_Piston_Circular}Geometry of the perfect metal piston system, for the particular case of circular cross section. The the cavity and the piston are perfect metals. The piston is at a finite distance $L$ of one of the faces of the cavity, and at a distance $L_{\infty}$ of the other face. This distance tends to infinity in the calculations of the Chapter, but studies of finite $L_{\infty}$ are straightforward.}
\label{fig.plates}
\end{figure}

Let us consider a piston of arbitrary cross section, area $A$ and perimeter $P$ made of a perfectly conducting metal surface~\cite{Hertzberg}.
Two flat conducting plates of the same cross section of the piston are placed at a~distance~$L$ apart along the $x$ direction.
The plates are perpendicular to the surface of the cylinder.
We have represented the geometry under study for the particular case of a circular piston in Fig.~\ref{Figura_Piston_Circular}
We calculate the Casimir force for these plates by applying the formalism developed in this Chapter.
First, we solve the eigenvalue problem for the EM field and apply Eqs.~\eqref{eq.T} and \eqref{eq.FCint} to obtain the force.
For the EM field the normal component of the stress tensor  reads
\begin{equation}
\mathbb{T}_{xx}=E_{x}^{2} + B_{x}^{2} - \frac{1}{2}\textbf{E}^{2}- \frac{1}{2}\textbf{B}^{2},
\end{equation}
which has to be averaged over the noise and then integrated over the surface of the plates, as shown in Eq.~\eqref{eq.FCint} and Fig.~\ref{Figura_Piston_Circular}. These technical calculations are left to the Appendix A \ref{Appendix A: Derivation of averaged stress tensor over the piston surface} of this Chapter.
For perfectly conducting plates, the boundary conditions are:
$\textbf{E}\times\textbf{n} = \textbf{0}$ and $\textbf{B}\cdot \textbf{n}=0$, where $\textbf{n}$ is the normal
vector at the surface.
In this geometry, the EM field can be decomposed into transverse electric ($\text{TE}$)
and transverse magnetic ($\text{TM}$) modes, which are discussed independently~\cite{Jackson}.

In Coulomb gauge ($\nabla\cdot\textbf{A} = 0$), the electric $\textbf{E}$ and magnetic $\textbf{B}$ fields are derived from the potential vector as $\textbf{E} = - \partial_{t}\textbf{A}$ and $\textbf{B} = \nabla\times\textbf{A}$.

For the $\text{TM}$ modes, the magnetic field is transversal to the $x$ direction, and
the vector potential $\textbf{A}$ for $\text{TM}$ modes can be written (in Coulomb gauge) as
\begin{align}\label{Modos_TM}
\textbf{A}_{\text{TM}} & = ( -C \nabla_{\perp}^{2} D, \partial_{x} C\, \partial_{y} D, \partial_{x} C\, \partial_{z}D) e^{-i\omega t},\nonumber\\
\textbf{E}_{\text{TM}} & = i\omega( -C \nabla_{\perp}^{2} D, \partial_{x} C\, \partial_{y} D, \partial_{x} C\, \partial_{z}D) e^{-i\omega t},\nonumber\\
%\textbf{B}_{\text{TM}} & = ( 0,  -C\partial_{z}\nabla_{\perp}^{2}D - \partial_{x}^{2}C\partial_{z}D,  \partial_{x}^{2}C\partial_{y}D + C\partial_{y}\nabla_{\perp}^{2}D ) e^{-i\omega t}.\\
\textbf{B}_{\text{TM}} & = - \omega^{2}( 0, - C\partial_{z}D, C\partial_{y}D ) e^{-i\omega t}.
\end{align}
Here $\omega^{2} = k_{x}^{2} + \lambda_{n}^{2}$, where the fields $C(x)$ and $D(\textbf{r}_\perp)$ satisfy
\begin{eqnarray}\label{eq.Cg}
\partial_{x}^{2}C(x)=&-k_{x}^{2}C(x), &{\text{\ (Neumann BC on\ }} x=0,L{\text{)}} \nonumber  \\
\nabla_{\perp}^{2}D_{n} (\textbf{r}_\perp)=& - \lambda_{n}^{2}D_{n}(\textbf{r}_\perp),  &{\text{\ (Dirichlet BC on  }} {\cal S} {\text{)}},
\end{eqnarray}
where ${\cal S}$ is the surface of the cylinder and $\textbf{r}_\perp=(y,z)$.
For the $\text{TE}$ set, the electric field is transversal to $x$, so the vector potential is
\begin{align}\label{Modos_TE}
\textbf{A}_{\text{TE}} & = i\omega( 0, -S\,\partial_{z}N, S\, \partial_{y}N  ) e^{-i\omega t},\nonumber\\
\textbf{E}_{\text{TE}} & = \omega^{2}( 0, -S\,\partial_{z}N, S\, \partial_{y}N  ) e^{-i\omega t},\nonumber\\
\textbf{B}_{\text{TE}} & = i\omega( S\nabla_{\perp}^{2}N, - \partial_{x}S\, \partial_{y}N, - \partial_{x}S\, \partial_{z}N ) e^{-i\omega t},
\end{align}
where the functions $S(x)$ and $N(\textbf{r}_\perp)$ satisfy Eqs.~\eqref{eq.Cg}
with the opposite boundary conditions: Dirichlet for $S(x)$ and Neumann for $N(\textbf{r}_\perp)$.
However, in this case, we must exclude the constant eigenfunction, with eigenvalue $\lambda_n^2$ equal to zero, as
it gives that $\textbf{A}_{\text{TE}}=\textbf{0}$ and then $\textbf{E}_{\text{TE}} = \textbf{B}_{\text{TE}} = \textbf{0}$.

Substitution of the $\text{TE}$ modes into the expression for the stress tensor, and integration over one side of the
plates gives, after a long but straightforward calculation left to the Appendix A \ref{Appendix A: Derivation of averaged stress tensor over the piston surface}, left to an expression of the Casimir force over one side of the piston as
\begin{equation}\label{12}
\int_{\parbox{0.2cm}{\tiny{\text{1 side}}}}\mean{\mathbb{T}_{xx}^{\text{TE}}}dS_{x} =
- \frac{1}{\beta L}\sum_{m\in\mathbb{Z}}\sum_{n_{x}=1}^{\infty}\sum_{n}
\frac{k_{x}^{2}}{\omega_{m}^{2} + k_{x}^{2} + \lambda_{n}^{2}},
\end{equation}
where $k_{x}^{2}= (n_{x}\pi/L)^2$, and $\omega_{m}$ are the Matsubara frequencies defined after Eq.~\eqref{eq.eigenfunctions}.
This expression is one of the main results of this Chapter, because we use it to derive the Casimir force from noise averages instead zeta regularizations or optical paths.
For the $\text{TM}$ modes, one obtains exactly the same expression, but $\lambda_{n}$ are the eigenvalues of the two-dimensional (2D) Laplacian
with Neumann boundary conditions. We will denote the complete set of eigenvalues of
the Laplacian with Neumann (excluding the zero eigenvalue)
and Dirichlet boundary conditions by the index~$p$.
The expression above is the equivalent of Eq.~\eqref{eq.T} when the spectrum can be split into a~longitudinal and transversal part, that is, $\lambda_{n}^{2} = k_{x}^{2} + \lambda_{p}^{2}$.
These series are divergent, but the net Casimir force,
which is the difference between the force on the two sides of the plate, is finite.

The sum over the variable $n_{x}$ in Eq.~\eqref{12} can be carried out with the help of the
Chowla--Selberg summation formula~\cite{Elizalde}. This formula extracts the divergent, $L$-independent
part of the summation, which cancels when the integral in Eq.~\eqref{12} is
performed for both sides of the piston, resulting in
\begin{equation}\label{FT}
F_{C} = -k_{B}T\sum_{p}\sum_{m\in \mathbb{Z}}\sum_{n=1}^{\infty}
\sqrt{m^{2}\Lambda^{2} + \lambda_{p}^{2}} \, e^{-2 L n \sqrt{m^{2}\Lambda^{2} + \lambda_{p}^{2}} }.
\end{equation}
Here, $\Lambda = 2\pi k_{B}T/\hbar c $ is the inverse thermal wavelength. If we carry out the sum over $n$, the force results on the more compact form
\begin{equation}
F_{C} = -k_{B}T\sum_{p}\sum_{m\in \mathbb{Z}}
\frac{\sqrt{m^{2}\Lambda^{2} + \lambda_{p}^{2}}}{e^{2 L\sqrt{m^{2}\Lambda^{2} + \lambda_{p}^{2}}} - 1}.
\end{equation}
In Ref.~\cite{Marachevsky} it was obtained a formula for the free energy of the configuration considered here, that, after differentiation with respect to the distance between the plates, leads to the force above.
We can proceed to evaluate the Casimir force at $T=0$, that is, the purely EM case
without thermal corrections.
This case is obtained by noting that, when $T\to 0$ (equivalent to the limit $\Lambda\to 0$ in Eq.~\eqref{FT}), the sum over~$m$ can be replaced by an integral by the use of Abel--Plana formula \cite{Libro-Residuos}. Computing the integral for finite~$\Lambda$ and taking the limit $\Lambda\to 0$, the result is
\begin{equation}\label{eq.FCBessel}
F_{C}= - \frac{\hbar c}{2\pi}\sum_{p,n}\lambda_{p}^{2}\left[ K_{0}(2n L \lambda_{p}) + K_{2}(2n L\lambda_{p})\right],
 \end{equation}
where $K_{\alpha}(x)$ is the modified Bessel function of order~$\alpha$.
This expression gives the finite or regularized Casimir force between two plates
at distance~$L$. The precise geometry of the plates enters into the double
set of eigenvalues of the Laplacian (with Neumann and Dirichlet boundary conditions) $\lambda_{p}$.

\subsection{Short distance limit}
For short distances, however, the summation above can be calculated
without explicitly knowing the eigenvalues of the Laplacian. Such eigenvalues
must scale with the inverse of the typical size of the piston. So, for
distances~$L$ much smaller than the section of the piston, we can
replace the sum over the eigenvalues, $\lambda_{p}$, by an integral, using the
asymptotic expression for the density of states of the Laplacian in two dimensions for each set of Dirichlet or Neumann BC~\cite{Hertzberg}\cite{SPECTRA_FINITE_SYSTEMS}:
\begin{equation}
\rho(k) = \frac{A}{2\pi}k\theta(k) + \eta\frac{P}{4\pi}\theta(k) + \chi\delta(k),
\end{equation}
where $\theta(k)$ is the step function, $A$ is the area of the piston, $P$ its perimeter, $\eta = +1$ for Neumann BC and $\eta = -1$ for Dirichlet BC and $\chi$ depends on the curvature of the perimeter and on the number and angle of vertices as
\begin{equation}
\chi = \frac{1}{24}\sum_{i}\left[\frac{\pi}{\alpha_{i}} - \frac{\alpha_{i}}{\pi}\right] + \frac{1}{12}\sum_{j}\int_{\gamma_{j}}\kappa(\gamma_{j})d\gamma_{j},
\end{equation}
where $\alpha_{i}$ is the angle of each piston's vertex and $\kappa(\gamma_{j})$ is the curvature of each smooth section of the perimeter.
The resulting integrals can be performed to obtain the Casimir force contribution of each mode as
\begin{align}
F_{T=0}^{\text{TE}} & = - \hbar c \left[\frac{\pi^{2}}{480L^{4}}A + \frac{\zeta(3)}{32\pi L^{3}}P + \frac{\pi}{24L^{2}}\left(\chi - 1\right)\right],\nonumber\\
F_{T=0}^{\text{TM}} & = - \hbar c \left[\frac{\pi^{2}}{480L^{4}}A - \frac{\zeta(3)}{32\pi L^{3}}P + \frac{\pi}{24L^{2}}\chi\right],
\end{align}
where the contribution of the zero eigenvalue for the Neumann BC problem must be explicitly dropped because, as discussed above, it does not contribute to the Casimir force. The global Casimir force is the sum of the two polarizations contributions~\cite{Hertzberg}
\begin{equation}\label{eq.Near}
%F_{T=0} = - \frac{\hbar c\pi^{2}}{240L^{4}}A,
F_{T=0} = F_{T=0}^{\text{TE}} + F_{T=0}^{\text{TM}} =  - \hbar c \left[\frac{\pi^{2}}{240L^{4}}A + \frac{\pi}{24L^{2}}\left(2\chi - 1\right)\right],
\end{equation}
which reduces to the well-known result of the EM Casimir force for infinite  parallel plates~\cite{Casimir_Placas_Paralelas}.

\subsection{Large distance limit}
In the opposite limit, when~$L$ is much larger than the typical size of
the plate, the argument of the Bessel functions in Eq.~\eqref{eq.FCBessel} is much larger than one.
Because of the exponential behavior of the Bessel functions,
only the smallest eigenvalue~$\lambda_{1}$ contributes to the sum, with the result
\begin{equation}\label{eq.Far}
F_{T=0} = - \frac{\hbar c}{2\sqrt{\pi L}}g_{1}\lambda_{1}^{3/2}e^{-2L\lambda_{1}},
\end{equation}
result directly proportional to the degeneration $g_{1}$ of $\lambda_{1}$.
Here, a counterintuitive result is obtained.
One would expect that the thin piston would tend to the known one-dimensional (1D) Casimir force,
but instead an exponential decay of the force is found. The known 1D result would be obtained if the zero eigenvalue
were considered. However, this eigenvalue is excluded, as it 
leads to a vanishing eigenfunction. 
Therefore, the 1D Casimir force cannot be obtained as the limit from three dimensions (3D) to 1D.

At intermediate distances, that is, $L$ comparable to the size of the plates, one must
solve the eigenvalue problem. To illustrate the intermediate behavior and the transition from
Eq.~\eqref{eq.Near} to \eqref{eq.Far}, we study the case of a circular cylinder of radius~$R$.
In this case, the eigenvalues are the zeros of the Bessel function $J_{\nu}(r)$ and its derivative, with degeneration $g_{n} = 2$ except in their lower zero ($\lambda_{D} = 2.40482$ for the Dirichlet problem, and $\lambda_{N} = 0$ for the Neumann case)
The dependence with the radius of the cylinder is 
$\lambda_{p}^{2}(R) = \lambda_{p}^{2}(R=1)/R^{2}$, so the Casimir force as expressed in Eq.~\eqref{eq.FCBessel}
is a function of $L/R$ when the force is multiplied by $R^{2}$.
Figure \ref{fig.plates} shows  the results  for the Casimir force, Eq.~\eqref{eq.FCBessel},
 when $N=1$, $N=10$, $N=100$, $N=1,\!000$ and $N=10,\!000$ eigenvalues (taking  into account their
degeneration).
We have also plotted the two limiting results: (i) the 3D Casimir force \eqref{eq.Near},
valid for $L\ll R$ with an algebraic behavior: $F_{C} R^{2}\propto (L/R)^{-4}$ (solid  line), (ii) 
the far distance limit, given by Eq.~\eqref{eq.Far} with the smallest eigenvalue
$\lambda_{1}^2(R)\simeq\left(\frac{1.84118}{R}\right)^{2} = 3.38996/R^{2}$ with degeneracy $g_{1} = 2$,
that is valid for $L\gg R$ (dotted   line).
The transition between both regimes is observed at  $L\simeq R$.
As expected, when few eigenvalues are summed, for instance $N=10$, 
the resulting force is only valid in the limit of long distances. As the number of
eigenvalues increase, the numerical result approaches the 3D Casimir force. 
For $N=1,000$ eigenvalues, we have excellent results for $L/R>0.05$.
We remark, however, that the full curve can be obtained by only considering $N\approx 100$ 
eigenvalues for large distances and matching this numerical result with the  
asymptotic expression \eqref{eq.Near} for short distances, 
with a crossover distance $L\approx0.3 R$.

\begin{figure}[htb]
\includegraphics[width=0.9\columnwidth,angle=0]{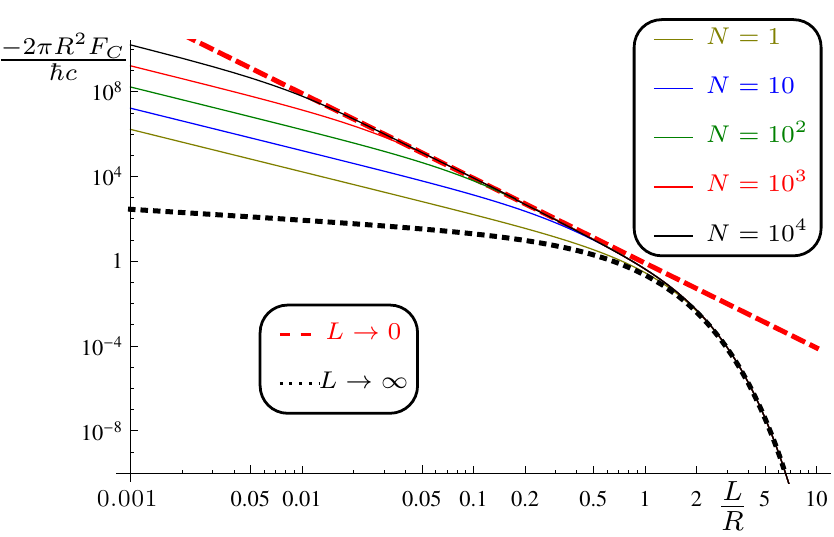}
\caption{\label{Fuerza_Casimir_Piston_Circular}Adimensional Casimir force on the plates of a~cylindrical piston of circular cross section of radius~$R$ as a function of the distance~$L$ between the plates.
Different curves are obtained by summing $N$
eigenvalues of the Laplacian, using Eq.~\eqref{eq.FCBessel}), where $N$ is
indicated in the legend. Dashed line is the short distance limit and dotted line is the far distance result Eq.~\eqref{eq.Far}.
For shorter distances we would need even more eigenvalues to converge to the analytical known limit given in Eq.~\eqref{eq.Near}.}
\label{fig.plates}
\end{figure}

\section{Thermal Casimir force}
In a similar fashion, we can calculate the thermal Casimir force when $\hbar \to 0$,
or $\Lambda\to\infty$. Then, in Eq.~\eqref{FT}, only the term with $m=0$
is different from zero. The sum over $n$ can be done, with the result
\begin{equation}\label{eq.ant}
F_{\hbar=0} = - k_{B}T\sum_{p}\frac{\lambda_{p}}{e^{2L\lambda_{p}} - 1}.
\end{equation}
For short ($L\!\ll\! R$) and long ($L\!\gg\! R$) distances, Eq.~\eqref{eq.ant} reads
\begin{align}
F_{\hbar=0} &= - k_{B}T\left[\frac{\zeta(3)}{4\pi L^{3}}A + \frac{1}{2L}\left(2\chi - 1\right)\right],\nonumber\\
F_{\hbar=0} &= - k_{B}T g_{1}\lambda_{1}e^{-2L\lambda_{1}},
\end{align}
where $g_{1}$ is the degeneration of the smallest non zero eigenvalue $\lambda_{1}$.
Here the Area--correction term for short distances forces should be read carefully, because the sum over $n$ and the integration over $k$ do not commute and gives different results. If sum over $n$ is done first, we obtain the result shown above, but if the integration over $k$ is done first, the term proportional to $\left(2\chi - 1\right)$ is zero.

\section{Fluctuation of Casimir forces}\label{Fluctuation of Casimir forces}
The Casimir force has its origin in fluctuations, so it is a
fluctuating quantity itself. The formalism developed in this Chapter allows the calculation of fluctuations of the force,
defined as
\begin{equation}\label{Fluc}
\sigma_{F}^{2} = \oint_{\Omega}\oint_{\Omega}\mean{[\mathbb{T}(\textbf{r}_{1})\cdot d\textbf{S}_{1}][\mathbb{T}(\textbf{r}_{2})\cdot d\textbf{S}_{2}]} - F_{C}^{2}.
\end{equation}
Calculation of $\sigma_{F}^{2}$ requires the correlation of a product of four noises,
$\eta$, which, because of the Gaussian nature of $\eta$, factorizes into three products
of pairs of noises, as given by Eq.~\eqref{eq.noise}.

For the piston geometry considered in this Chapter, the fluctuations of the force can be obtained by evaluating
Eq.~\eqref{Fluc}, technical details are left to Appendix B \ref{Appendix B: Derivation of the fluctuations of the force.} of this Chapter. In this case, and because of the geometry of the problem, each
of the summands that appear in the four-point correlations of the noise
gives~$F_{C}^{2}$. Therefore, the fluctuations of the force, for any temperature and cross section, are simply
\begin{equation}\label{Varianza_Fuerza_Casimir}
\sigma_{F}^{2} = 2 F_{C}^{2}.
\end{equation}
Similar fluctuations have been obtained for a~purely thermal force~\cite{Fournier} and
measured in~\cite{Cattuto} by means of numerical simulation in a hydrodynamical system.
The fact that the fluctuation of the force is as large as the force itself is
a signature of fluctuation-induced forces.

\section{Equivalence between dynamical approach to Ca\-si\-mir effect, Stress Tensor formalism and Partition function formalism}\label{Appendix C: Equivalencia formalismos}
In this Section, we demonstrate the equivalence of the dynamical formalism of Casimir effect in the particular case of equilibrium systems with the Stress--Tensor formalism~\cite{Rodriguez:2007a} of Casimir effect and with the partition function formalism, from which the multiscattering formalism is derived~\cite{Kardar-Geometrias-Arbitrarias}.

As seen in Chaper \ref{Dynamical approach to the Casimir effect} of this Thesis, for equilibrium systems, the fluctuations of the surrounding field follows the next Langevin PDE given in Eq.~\eqref{OnsagerFinal}:
\begin{equation}
\partial_{t}\phi = - \mathcal{L}\mathcal{G}\phi + \xi(\textbf{r},t),
\end{equation}
where $\mathcal{L}$ is the Onsager operator associated with transport coefficients, $\mathcal{G}$ is the density functional of the energy functional of our problem given in Eq.~\eqref{Funcional_Energia_Lineal}, it is
\begin{equation}\label{Funcional_Energiav2}
F[\phi] = \frac{1}{2}\int dv\phi\mathcal{G}\phi = \int dv \mathbb{F},
\end{equation}
and $\xi(\textbf{r},t)$ is a zero mean Gaussian equilibrium noise, with autocorrelation (Eqs.~\eqref{noises} and \eqref{FDtheorem.Q})
\begin{equation}
\mean{\xi(\textbf{r},t)\xi(\textbf{r}',t')} = k_{B}T\left(\mathcal{L} + \mathcal{L}^{+}\right)\delta(\textbf{r} - \textbf{r}')\delta(t - t').
\end{equation}
In this model, $\phi$ is constrained because of the boundary conditions of each $\alpha$ intrusion, which in the more general linear case are of Robin type:
\begin{equation}
\left(D_{\alpha}\phi(\textbf{s}) + N_{\alpha}\partial_{n}\phi(\textbf{s})\right)\vert_{\textbf{s}\in S_{\alpha}} = 0,
\end{equation}
where $S_{\alpha}$ is the surface of each $\alpha$ intrusion, and $D_{\alpha}$ and $N_{\alpha}$ are constants which defines the kind of boundary condition of the surface $S_{\alpha}$ ($D_{\alpha} = 0$ and $N_{\alpha}\neq 0$ for Neumann Boundary Conditions, $N_{\alpha} = 0$ and $D_{\alpha}\neq 0$ for Dirichlet Boundary Conditions and $N_{\alpha} \neq 0$ and $D_{\alpha}\neq 0$ for general Robin Boundary Conditions). Then we can define the spatial eigenproblem
\begin{align}
& \mathcal{L}\mathcal{G}f_{n}(\textbf{r}) = \mu_{n}f_{n}(\textbf{r}),
& \left(D_{\alpha}f_{n}(\textbf{s}) + N_{\alpha}\partial_{n}f_{n}(\textbf{s})\right)\vert_{\textbf{s}\in S_{\alpha}} = 0,
\end{align}
where $\mathcal{L}\mathcal{G}$ is a hermitian operator because our system is in equilibrium, then the right eigenvalue is $g_{n}(\textbf{r}) = f_{n}^{*}(\textbf{r})$, and $\braket{f^{*}_{n}}{f_{m}} = \int_{\Omega}d\textbf{r} f^{*}_{n}(\textbf{r})f_{m}(\textbf{r}) = \delta_{nm}$.

It can be demonstrated that the partition function of our problem is (Eqs.~\eqref{funcion_particion_phi_no_eq} and \eqref{Peq})
\begin{equation}
\mathcal{Z} = \int\mathcal{D}\phi e^{-\beta F[\phi]}.
\end{equation}
As $F[\phi]$ is a quadratic functional over $\phi$, $\mathcal{Z}$ is a Gaussian functional integral, then the functional integral can be carried out to give~\cite{RE09}
\begin{equation}
\mathcal{Z} = \frac{1}{\sqrt{\Det{\mathcal{G}}}}.
\end{equation}
As we have a system in equilibrium at a constant temperature, we can define the Helmholtz free energy of the system in terms of the partition function of the system as
\begin{equation}
-\beta\mathcal{F} = \log\left(\mathcal{Z}\right) = - \frac{1}{2}\log\Det{\mathcal{G}}.
\end{equation}
Using this determinant in term of the eigenvalues of the linear operator $\mathcal{G}$, and the properties of the logarithm, the Helmholtz free energy is~\cite{elizalde1995ten}\cite{elizalde1994zeta}
\begin{equation}\label{Energia_Libre_Helmholtz_Determinante_Espectral}
\mathcal{F} = \frac{k_{B}T}{2}\sum_{n = 1}^{\infty}\log\left(\lambda_{n}\right),
\end{equation}
where the functional determinant is well defined here because the spectral zeta regularization principle is used here. The quantities $\lambda_{n}$ are defined in the eigenproblem
\begin{align}
& \mathcal{G}f_{n}(\textbf{r}) = \lambda_{n}f_{n}(\textbf{r}),
& \left(D_{\alpha}f_{n}(\textbf{s}) + N_{\alpha}\partial_{n}f_{n}(\textbf{s})\right)\vert_{\textbf{s}\in S_{\alpha}} = 0,
\end{align}
where we have assumed that operators $\mathcal{LG}$ and $\mathcal{G}$ share eigenfunctions (that is the case in the examples considered in this Thesis), and the relationship between $\lambda_{n}$ and $\mu_{n}$ is assumed to be $\mu_{n} = L_{n}\lambda_{n}$, where $\mathcal{L}f_{n} = L_{n}f_{n}$. These assumptions will be understood at the end of the Section.

From Eq.~\eqref{Energia_Libre_Helmholtz_Determinante_Espectral}, we define the Casimir force as
\begin{equation}\label{Fuerza_Casimir_desde_Determinante_Espectral}
\textbf{F}_{C} = - \partial_{L}\mathcal{F} = \frac{k_{B}T}{2}\sum_{n = 1}^{\infty}\frac{\partial_{L}\lambda_{n}}{\lambda_{n}}.
\end{equation}
The problem is to obtain an expression for the derivative of an eigenvalue in terms of an internal parameter. To do so, we need an integral representation of the eigenvalue $\lambda_{n}$, that can be easily obtained from the eigenvalue problem. If we left--multiply the eigenvalue problem by $f_{m}^{*}(\textbf{r})$ and carry out an integration over all the space, the eigenvalue can be written as
\begin{equation}\label{Expresion_Integral_autovalor}
\lambda_{n} = \int_{\Omega}d\textbf{r}f_{n}^{*}(\textbf{r})\mathcal{G}f_{n}(\textbf{r}) = 2\int_{\Omega}d\textbf{r}\mathbb{F}[f_{n}].
\end{equation}
Following~\cite{Ajdari}, let us consider the variation of $\lambda_{n}$ under a displacement $\delta\vec{\ell}$. If we have into account that a displacement $\delta\vec{\ell}$ of a surface implies not only a variation of the field in the bulk, but also a variation of the bulk itself,
\begin{equation}\label{Variacion_autovalor_1}
\delta\lambda_{n} = 2\int_{\Omega}d\textbf{r}\left[\frac{\partial\mathbb{F}}{\partial f_{n}}\delta f_{n} + \frac{\partial\mathbb{F}}{\partial\nabla f_{n}}\delta\nabla f_{n}\right] + 2\oint_{\partial\Omega}ds\,\hat{\textbf{n}}\cdot\delta\vec{\ell}\,\mathbb{F}.
\end{equation}
We integrate by parts the second summand of the volume integral, obtaining
\begin{equation}\label{Variacion_autovalor_2}
\delta\lambda_{n} = 2\int_{\Omega}d\textbf{r}\left[\frac{\partial\mathbb{F}}{\partial f_{n}} - \nabla\left(\frac{\partial\mathbb{F}}{\partial\nabla f_{n}}\right)\right]\delta f_{n} + 2\oint_{\partial\Omega}ds\,\hat{\textbf{n}}\cdot\left[\frac{\partial\mathbb{F}}{\partial\nabla f_{n}}\delta f_{n} + \delta\vec{\ell}\,\mathbb{F}\right].
\end{equation}
The integrand of the volume integral vanishes, because the equilibrium state is a minimum of the free energy, and the integrand is the Euler--Lagrange equations of the field derived from the energy functional of our problem given in Eq.~\eqref{Funcional_Energiav2}. In addition to that, we can approximate at first order in $\delta\vec{\ell}$ that
\begin{equation}\label{Variacion_autovalor_3}
\delta f_{n}(L) = f_{n}(L + \delta\vec{\ell}) - f_{n}(L) = \nabla f_{n}(L)\cdot\delta\vec{\ell} + \mathcal{O}\left(\delta\vec{\ell}^{2}\right),
\end{equation}
then
\begin{equation}\label{Variacion_autovalor_4}
\delta\lambda_{n} = 2\oint_{\partial\Omega}ds\,\hat{\textbf{n}}_{\mu}\cdot\left[\delta^{\mu\nu}\mathbb{F} - \nabla^{\mu}f_{n}\frac{\partial\mathbb{F}}{\partial\nabla^{\nu}f_{n}}\right]\cdot\delta\vec{\ell}_{\nu}.
\end{equation}
We observe that the tensor operator under brackets is the stress tensor as defined in Eq.~\eqref{stress}, then
\begin{equation}\label{Variacion_autovalor_5}
\frac{\delta\lambda_{n}}{\delta\vec{\ell}_{\nu}} = 2\oint_{\partial\Omega}ds\,\hat{\textbf{n}}_{\mu}\cdot\TensorT^{\mu\nu}[f_{n}, f_{n}^{*}, \textbf{s}].
\end{equation}
Using this result in Eq.~\eqref{Fuerza_Casimir_desde_Determinante_Espectral}, we arrive to our final result
\begin{equation}\label{Fuerza_Casimir_desde_Determinante_Espectral_2}
\textbf{F}_{C}^{\nu} = - \partial_{L}\mathcal{F} = \sum_{n = 1}^{\infty}\frac{k_{B}T}{\lambda_{n}}\oint_{\partial\Omega}ds\,\hat{\textbf{n}}_{\mu}\cdot\TensorT^{\mu\nu}[f_{n}, f_{n}^{*}, \textbf{s}],
\end{equation}
which is the Eq.~\eqref{Force} for equilibrium systems.

Note that, when $\mathcal{L}\neq \mathbbm{1}$, the Casimir force is in this case, following Eq.~\eqref{Force}:
\begin{equation}
\Vec{F}_{C} = - \sum_{n = 1}^{\infty}\frac{k_{B}T\left(L_{n} + L_{n}^{*}\right)}{\mu_{n} + \mu_{n}^{*}}
\oint_{\partial\Omega}ds\,\hat{\textbf{n}}_{\mu}\cdot\TensorT^{\mu\nu}[f_{n}, f_{n}^{*}, \textbf{s}].
\end{equation}
Using the imposed assumption $\mu_{n} = L_{n}\lambda_{n}$, the Casimir force is in this case, following Eq.~\eqref{Force}:
\begin{equation}
\Vec{F}_{C} = - \sum_{n = 1}^{\infty}\frac{k_{B}T}{\lambda_{n}}
\oint_{\partial\Omega}ds\,\hat{\textbf{n}}_{\mu}\cdot\TensorT^{\mu\nu}[f_{n}, f_{n}^{*}, \textbf{s}],
\end{equation}
which is the same force as the one given in in Eq.~\eqref{Fuerza_Casimir_desde_Determinante_Espectral_2}, but with the opposite criterium of sign because of the criteria of signs given in Eq~\eqref{FCA}. Note that if the operator $\mathcal{G}$ is isotropic in space then its eigenvalues cannot have any information about the geometry of the system ($\lambda_{n}\neq f(L)$), and therefore the Casimir force vanishes in this special case, as explained in Sect.~\ref{Computation of Casimir forces} of this Thesis.

If we use the spectral representation of the Green function of the problem
\begin{equation}\label{Fuerza_Casimir_desde_Determinante_Espectral_3}
\mathcal{G}(\textbf{r},\textbf{r}') = \sum_{n}\frac{f_{n}^{*}(\textbf{r})f_{n}(\textbf{r}')}{\lambda_{n}},
\end{equation}
we recover the Stress--Tensor formalism of equilibrium Casimir effect with the Fluctuation--Dissipation Theorem already applied~\cite{Rodriguez:2007a}. Then, in this Section we have demonstrated the equivalence between the Stress--Tensor formalism, the partition function formalism (from which the multiscattering formalism is derived) and the dynamical approach to equilibrium Casimir effect.

\section{Conclusions}
We have shown in this Chapter that stochastic quantization
together with the Langevin formalism provides a new approach to calculate Casimir
forces in the quantum electrodynamics (QED) case, including thermal effects. The starting point is the calculation
of the Casimir force via the stress tensor, which is a function of the
fluctuating fields. Parisi and Wu derived a Langevin equation to describe
the dynamics of such fields, which can be integrated to give an expression
for the force. The method presented herein
is quite simple, and avoids some technical complication of other approaches.
Moreover, it calculates the force directly, instead of the free energy.
Another advantage is that it provides a numerical method to calculate
the Casimir force in complicated geometries, such as those of interest for microelectromechanical systems (MEMS)
devices. The method only requires spectral decomposition of the Laplacian
operator in the given geometry, and summation of the eigenvalues and the value of the eigenfunction
along the boundary of the object, as shown in Eqs.~\eqref{eq.T2} and \eqref{eq.FCint}.
Quantum ($T\to 0$)  and classical ($\hbar\to 0 $) limits are recovered by Eqs.~\eqref{eq.FCT0} and \eqref{eq.FCh0} respectively. 

In addition, the equivalence of the dynamical formalism in equilibrium cases with the partition function formalism and with the stress-tensor formalism have been obtained and, thanks to the use of the dynamical formalism applied here, we have been able to define a variance of the Casimir force and to give a general expression of it. This formula have let us to obtain the variance of the arbitrary section perfect metal piston at any temperature.

\section{Appendices}
\subsection{Appendix A: Derivation of averaged stress tensor over the piston surface}\label{Appendix A: Derivation of averaged stress tensor over the piston surface}
In this appendix we will obtain the integral over one side of the piston of the average over the noise of the stress energy tensor given in Eq.~\eqref{12} as a sum over $n_{x}$ and $\lambda_{n}$. Because of the system geometry, it is evident that $y$ and $z$ components are zero, then the Casimir force \eqref{eq.FCint} over one side of the piston is defined as
\begin{equation}
\textbf{F}_{C,x} = \int_{\mathcal{P}}\mean{\mathbb{T}_{xx}}ds = \int_{\mathcal{P}}ds\mean{E_{x}^{2} + B_{x}^{2} - \frac{1}{2}\textbf{E}^{2} - \frac{1}{2}\textbf{B}^{2}},
\end{equation}
where $\mathcal{P}$ is the surface of one side of the piston. The total Casimir force is the sum of the contributions of both sides of the piston. Then we need the two point correlation function of $\textbf{A}$, which is obtained from Eq.~\eqref{eq.T} as
\begin{equation}
\mean{A_{i}(\textbf{r})A_{j}(\textbf{r}')} = \frac{1}{\beta}\sum_{nm}\frac{1}{\omega_{m}^{2} + \lambda_{n}^{2}}\frac{a_{i,n}(\textbf{r})a_{j,n}^{*}(\textbf{r}')}{\norma{\textbf{a}_{n}}^{2}},
\end{equation}
where $a_{i,n}(\textbf{r})$ is the $i$ component of the eigenvector $\Delta\textbf{a}_{n}(\textbf{r}) = - \lambda_{n}^{2}\textbf{a}_{n}(\textbf{r})$. Then, for the electric and magnetic fields we have
\begin{equation}
\mean{E_{i}(\textbf{r})E_{j}(\textbf{r}')} = \frac{1}{\beta}\sum_{nm}\frac{1}{\omega_{m}^{2} + \lambda_{n}^{2}}\frac{e_{i,n}(\textbf{r})e_{j,n}^{*}(\textbf{r}')}{\norma{\textbf{a}_{n}}^{2}},
\end{equation}
\begin{equation}
\mean{B_{i}(\textbf{r})B_{j}(\textbf{r}')} = \frac{1}{\beta}\sum_{nm}\frac{1}{\omega_{m}^{2} + \lambda_{n}^{2}}\frac{b_{i,n}(\textbf{r})b_{j,n}^{*}(\textbf{r}')}{\norma{\textbf{a}_{n}}^{2}},
\end{equation}
where $\textbf{e}_{n}(\textbf{r}) = -i\omega\textbf{a}_{n}(\textbf{r})$ and $\textbf{b}_{n}(\textbf{r}) = \nabla\times\textbf{a}_{n}(\textbf{r})$. These expressions are the Quantum Fluctuation--Dissipation theorem for the EM field~\cite{QFDT_Weber}.

From \eqref{eq.Cg} and its equivalent result for $\text{TM}$ modes, it is immediate to obtain that the functions $C(x)$ and $S(x)$ of $\text{TM}$ and $\text{TE}$ modes respectively, are $S(x) = \sin(k_{x}x)$ and $C(x) = \cos(k_{x}x)$, with $k_{x}^{2}= (n_x\pi/L)^2$. Then, for $Z(x) = S(x) = C(x)$, we obtain
\begin{equation}
\int_{0}^{L}dx Z^{2}(x) = \frac{L}{2}.
\end{equation}
In addition to that, if $g_{n}$ is solution of the eigenproblem $\nabla_{\perp}^{2}g_{n} = - \lambda_{\perp,n}^{2}g_{n}$ over the surface of one side of the piston $\mathcal{P}$, then a product with $g_{n}^{*}$ and an integration over $\mathcal{P}$ gives
\begin{equation}\label{Formula_Integral_Autovalores}
\lambda_{\perp,n}^{2} = \frac{1}{\norma{g_{n}}^{2}}\int_{\mathcal{P}}ds \left(\nabla_{\perp}g_{n}\right)^{2},
\end{equation}
where the norm $\norma{g_{n}}$ is defined as
\begin{equation}
\norma{g_{n}}^{2} = \int_{\mathcal{P}}ds \abs{g_{n}}^{2}.
\end{equation}
We separate the study in polarizations because the geometry of the system does not mix them.

\subsubsection{TE modes}
The norm of $\textbf{A}_{\text{TE}}$ is defined as
\begin{equation}\label{Definicion_norma_A}
\norma{\textbf{A}_{\text{TE}}}^{2} = \int_{0}^{L}dx\int_{\mathcal{P}}ds \abs{\textbf{A}_{\text{TE}}}^{2}.
\end{equation}
Then, having into account Eq.~\eqref{Modos_TE}, it is easy to obtain
\begin{equation}\label{norma_A_TE}
\norma{\textbf{A}_{\text{TE}}}^{2} = \frac{L}{2}\lambda_{\perp}^{2}\omega^{2}.
\end{equation}
We obtain $\int_{\mathcal{P}}\mean{E_{\text{TE},x}}^{2}ds = 0$ because we are working with $\text{TE}$ modes, and \\ $\int_{\mathcal{P}}\mean{B_{\text{TE},x}}^{2}ds = \int_{\mathcal{P}}\mean{\textbf{E}_{\text{TE}}}^{2}ds = 0$ because these integrals are proportional to $\sin(k_{x}x)$ over the plate. The contribution of the magnetic field is
\begin{equation}\label{Contribucion_B_TE}
\int_{\mathcal{P}}ds\mean{\textbf{B}_{\text{TE}}}^{2} = \frac{1}{\beta}\sum_{nm}\frac{1}{\omega_{m}^{2} + \lambda_{n}^{2}}\frac{\int_{\mathcal{P}}ds \textbf{b}_{n}(s)\textbf{b}_{n}^{*}(s)}{\norma{\textbf{a}_{n}}^{2}}.
\end{equation}
We study this case in detail as an example of the calculations involved in this Appendix. The rest of integrals are obtained on a similar way. Using Eq.~\eqref{Modos_TE}, we obtain
\begin{equation}
\int_{\mathcal{P}}ds \textbf{b}_{n}(s)\textbf{b}_{n}^{*}(s) = \omega^{2}\int_{\mathcal{P}}ds\left[S^{2}\left(\nabla_{\perp}^{2}N\right)^{2} + \left(\partial_{x}S\right)^{2}\left(\partial_{y}N\right)^{2} + \left(\partial_{x}S\right)^{2}\left(\partial_{z}N\right)^{2}\right],
\end{equation}
where $S(x) = \sin(k_{x}x)$, then over the plate we have $S(x) = 0$ and \\ $\partial_{x}S(x) = \lim_{x\to 0}k_{x}\cos(k_{x}x) = k_{x}$. Then
\begin{equation}
\int_{\mathcal{P}}ds \textbf{b}_{n}(s)\textbf{b}_{n}^{*}(s) = \omega^{2}k_{x}^{2}\int_{\mathcal{P}}ds\left[\left(\partial_{y}N\right)^{2} + \left(\partial_{z}N\right)^{2}\right] = \omega^{2}k_{x}^{2}\int_{\mathcal{P}}ds\left(\nabla_{\perp}N\right)^{2}.
\end{equation}
Using Eq.~\eqref{Formula_Integral_Autovalores}, the integral is carried out to
\begin{equation}\label{1_side_B_TE}
\int_{\mathcal{P}}ds \textbf{b}_{n}(s)\textbf{b}_{n}^{*}(s) = \omega^{2}k_{x}^{2}\lambda_{\perp}^{2}.
\end{equation}
Finally, using Eqs.~\eqref{norma_A_TE} and \eqref{1_side_B_TE} in Eq.~\eqref{Contribucion_B_TE}, we obtain the contribution to the Casimir force of the $\text{TE}$ polarization as
\begin{equation}\label{Expresion_Fuerza_TE_Pistones}
\textbf{F}_{C,x}^{\text{TE}} =  - \frac{1}{2}\int_{\mathcal{P}}ds\mean{\textbf{B}_{\text{TE}}}^{2} = - \frac{1}{\beta L}\sum_{nm}\frac{k_{x}^{2}}{\omega_{m}^{2} + \lambda_{n}^{2}}.
\end{equation}

\subsubsection{TM modes}
The norm of $\textbf{A}_{\text{TM}}$ is, using the definition of Eq.~\eqref{Definicion_norma_A} and Eq.~\eqref{Modos_TM}:
\begin{equation}
\norma{\textbf{A}_{\text{TM}}}^{2} = \frac{L}{2}\lambda_{\perp}^{2}\omega^{2}.
\end{equation}
Obviously, $\int_{\mathcal{P}}\mean{B_{\text{TM},x}}^{2}ds = 0$ because we are working with $\text{TM}$ modes. The rest of integrals over one side of the piston are obtained as
\begin{equation}
\int_{\mathcal{P}}\mean{E_{\text{TM},x}}^{2}ds = \frac{2}{\beta L}\sum_{nm}\frac{\lambda_{\perp}^{2}}{\omega_{m}^{2} + \lambda_{n}^{2}},
\end{equation}
\begin{equation}
\int_{\mathcal{P}}\mean{\textbf{E}_{\text{TM}}}^{2}ds = \frac{2}{\beta L}\sum_{nm}\frac{\lambda_{\perp}^{2}}{\omega_{m}^{2} + \lambda_{n}^{2}},
\end{equation}
\begin{equation}
\int_{\mathcal{P}}\mean{\textbf{B}_{\text{TM}}}^{2}ds = \frac{2}{\beta L}\sum_{nm}\frac{\omega^{2}}{\omega_{m}^{2} + \lambda_{n}^{2}},
\end{equation}
where we have used that $C(x) = \cos(k_{x}x)$, then over the plate $C(0) = 1$ and $\partial_{x}C(0) = \lim_{x\to 0}-k_{x}\sin(k_{x}x) = 0$. Then, the contribution to the Casimir force of the $\text{TM}$ polarization, having into account all non zero contributions is
\begin{equation}\label{Expresion_Fuerza_TM_Pistones}
\textbf{F}_{C,x}^{\text{TM}} = - \frac{1}{\beta L}\sum_{nm}\frac{k_{x}^{2}}{\omega_{m}^{2} + \lambda_{n}^{2}}.
\end{equation}
Eqs.~\eqref{Expresion_Fuerza_TE_Pistones} and \eqref{Expresion_Fuerza_TM_Pistones} are used to obtain the Casimir force over a piston shown in Eq.~\eqref{FT}.

\subsection{Appendix B: Derivation of the fluctuations of the force.}\label{Appendix B: Derivation of the fluctuations of the force.}
In this appendix we are going to obtain a formula for the variance of the Casimir force from Eq.~\eqref{Fluc} and derive the result given in Eq.~\eqref{Varianza_Fuerza_Casimir} for pistons.

In the model presented here, the dynamics of the fluctuations of the medium is given by the SDE (Eqs.~\eqref{Langevin} and~\eqref{eq.Parisi-Wu})
\begin{equation}\label{SDE_fluctuaciones}
\partial_{s}\phi = \left(\frac{1}{c^{2}}\frac{\partial^{2}}{\partial\tau^{2}} - \mathcal{M}\right)\phi + \eta(\tau,\textbf{r},s),
\end{equation}
where $\eta(\tau,\textbf{r},s)$ is a zero mean Gaussian noise with autocorrelation
\begin{equation}\label{autocorrelacion_ruido_en_varianza}
\mean{\eta(\tau,\textbf{r},s)\eta(\tau',\textbf{r}',s')} = \Gamma\delta\left(\tau - \tau'\right)\delta\left(\textbf{r} - \textbf{r}'\right)\delta\left(s - s'\right).
\end{equation}
Therefore, Eq.~\eqref{SDE_fluctuaciones} generalizes the SDEs given in Eq.~\eqref{eq.Parisi-Wu} when $\mathcal{M} = - \Delta$ and in Eq.~\eqref{Langevin} when we consider the pseudo-time as the real time $s = t$ and we drop the dependence on $\tau$.

We have assumed that each object is subject to an stochastic force $\textbf{F}$, which is a stochastic variable defined as
\begin{equation}
\textbf{F}^{i} = \oint_{s} \mathbb{T}^{ij}\cdot\textbf{n}_{j}ds,
\end{equation}
and whose mean is the so defined Casimir force (Eq.~\eqref{FCasimir})
\begin{equation}\label{Definicion_media}
\textbf{F}_{C}^{i} = \mean{\textbf{F}^{i}} = \oint_{s} \mean{\mathbb{T}^{ij}}\cdot\textbf{n}_{j}ds.
\end{equation}
The average is performed over the stress tensor because we are assuming that the surface of the bodies does not fluctuate.
Then the variance of the Casimir force is defined as the variance of the stochastic force $\textbf{F}$ as
\begin{equation}\label{Definicion_varianza}
\sigma_{F^{i}}^{2} = \mean{\oint_{s} \mathbb{T}^{ij}\cdot\textbf{n}_{j}ds\oint_{r} \mathbb{T}^{ik}\cdot\textbf{n}_{k}dr} - \mean{\oint_{s} \mathbb{T}^{ij}\cdot\textbf{n}_{j}ds}^{2}.
\end{equation}
In the case of scalar of EM fields, $\mathbb{T}^{ij}$ is bilinear in the field, then the integrand of the Casimir force is proportional to the two point correlation function. We also need the four points correlation function in order to obtain the variance. In this derivation, we will assume that the spatial operator $\mathcal{M}$ is an hermitian operator, then its eigenvalues $\mu_{n}$ are real and left and right eigenfunctions are equal, then we have the spatial and temporal eigenproblems
\begin{equation}\label{eigenproblem_varianza}
\mathcal{M}f_{n}(\textbf{r}) = \mu_{n}f_{n}(\textbf{r}),
\hspace{2.0cm}
\frac{1}{c^{2}}\frac{\partial^{2}}{\partial\tau^{2}}g_{m}(\tau) = - \omega_{m}g_{m}(\tau),
\end{equation}
which are a generalization of Eq.~\eqref{eq.eigenfunctions} as explained above. This eigenvalue expansion let us factorize the general solution of Eq.~\eqref{SDE_fluctuaciones} in the same way as in Eq.~\eqref{eq.phi} as
\begin{equation}\label{eq.phiv2}
\phi (\tau ,\textbf{r};s)=\sum_{n,m} \phi_{nm}(s) g_{m}(\tau) f_{n}(\textbf{r}), 
\end{equation}
the same factorization must be applied to the noise. %Then each mode $\phi_{n}(t)$ can be obtained as\begin{equation}\label{Def_Modo_n_v1}\phi_{n}(t) = \int d\textbf{r}f_{n}^{*}(\textbf{r})\phi(\textbf{r},t).\end{equation}
We can apply this factorization to Eq.~\eqref{SDE_fluctuaciones} to obtain
\begin{equation}
\partial_{s}\phi_{nm} = - \left[ \omega_{m}^{2} + \mu_{n}\right]\phi_{nm} + \eta_{nm}(s),
\end{equation}
whose general solution is
\begin{equation}\label{Def_Modo_n_v2}
\phi_{nm}(s) = \int_{-\infty}^{s}d\sigma e^{\left(\omega_{m}^{2} + \mu_{n}\right)(\sigma - s)}\eta_{nm}(\sigma).
\end{equation}
The four points correlation function in the steady state is defined as
\begin{align}\label{4_points_correlation_function_definition}
\mean{\phi(\textbf{r}_{1},\tau_{1})\phi(\textbf{r}_{2},\tau_{2})\phi(\textbf{r}_{3},\tau_{3})\phi(\textbf{r}_{4},\tau_{4})} & = 
\lim_{s\to\infty}\sum_{n_{1},m_{1}}\sum_{n_{2},m_{2}}\sum_{n_{3},m_{3}}\sum_{n_{4},m_{4}}\mean{\prod_{k=1}^{4}\phi_{n_{k},m_{k}}(s)}\times\nonumber\\
& \times \prod_{k=1}^{4}f_{n_{k}}(\textbf{r}_{k})g_{m_{k}}(\tau_{k}),
\end{align}
where the average of the product of four modes over the noise is obtained by the use of Eq.~\eqref{Def_Modo_n_v2} as
\begin{align}\label{varianza_larga_1}
\lim_{s\to\infty}\mean{\prod_{k=1}^{4}\phi_{n_{k},m_{k}}(s)} & = \lim_{s\to\infty} \int_{-\infty}^{s}d\sigma_{1}\int_{-\infty}^{s}d\sigma_{2}\int_{-\infty}^{s}d\sigma_{3}\int_{-\infty}^{s}d\sigma_{4}\times\nonumber\\
& \times e^{\left(\omega_{m_{1}}^{2} + \mu_{n_{1}}\right)(\sigma_{1} - s)}e^{\left(\omega_{m_{2}}^{2} + \mu_{n_{2}}\right)(\sigma_{2} - s)}e^{\left(\omega_{m_{3}}^{2} + \mu_{n_{3}}\right)(\sigma_{3} - s)}e^{\left(\omega_{m_{4}}^{2} + \mu_{n_{4}}\right)(\sigma_{4} - s)}\times\nonumber\\
& \times\mean{\eta_{n_{1}m_{1}}(\sigma_{1})\eta_{n_{2}m_{2}}(\sigma_{2})\eta_{n_{3}m_{3}}(\sigma_{3})\eta_{n_{4}m_{4}}(\sigma_{4})}.
\end{align}
Eq.~\eqref{varianza_larga_1} depends on the average over the noise of the product of four factorization terms of the noise itself. The four point correlation of the modes of the noise is obtained by the use of the definition given in Eq.~\eqref{Def_Modo_n_v2} as
\begin{align}\label{varianza_larga_2}
& \mean{\eta_{n_{1}m_{1}}(\sigma_{1})\eta_{n_{2}m_{2}}(\sigma_{2})\eta_{n_{3}m_{3}}(\sigma_{3})\eta_{n_{4}m_{4}}(\sigma_{4})} = \nonumber\\ 
& \int d\textbf{r}_{1}\int d\textbf{r}_{2}\int d\textbf{r}_{3}\int d\textbf{r}_{4}
\int_{0}^{\hbar\beta} d\tau_{1}\int_{0}^{\hbar\beta} d\tau_{2}\int_{0}^{\hbar\beta} d\tau_{3}\int_{0}^{\hbar\beta} d\tau_{4}\times\nonumber\\
& \times f_{n_{1}}(\textbf{r}_{1})f_{n_{2}}(\textbf{r}_{2})f_{n_{3}}(\textbf{r}_{3})f_{n_{4}}(\textbf{r}_{4})
g_{m_{1}}(\tau_{1})g_{m_{2}}(\tau_{2})g_{m_{3}}(\tau_{3})g_{m_{4}}(\tau_{4})
\times\nonumber\\
& \times \mean{\eta(\textbf{r}_{1},\tau_{1},\sigma_{1})\eta(\textbf{r}_{2},\tau_{2},\sigma_{2})\eta(\textbf{r}_{3},\tau_{3},\sigma_{3})\eta(\textbf{r}_{4},\tau_{4},\sigma_{4})}.
\end{align}
Having into account that $\xi(\textbf{r},t)$ is a Gaussian noise, the four points correlation function of the noise is given as
\begin{equation}\label{varianza_larga_3}
\mean{\eta(1)\eta(2)\eta(3)\eta(4)}
\hspace{-2pt}=\hspace{-2pt} \mean{\eta(1)\eta(2)}\hspace{-2pt}\mean{\eta(3)\eta(4)} 
\hspace{-2pt}+\hspace{-2pt} \mean{\eta(1)\eta(3)}\hspace{-2pt}\mean{\eta(2)\eta(4)}
 \hspace{-2pt}+\hspace{-2pt} \mean{\eta(1)\eta(4)}\hspace{-2pt}\mean{\eta(2)\eta(3)}\hspace{-2pt},
\end{equation}
where we have used the compact notation $\eta(i) = \eta(\textbf{r}_{i},\tau_{i},\sigma_{i})$.
Then, we use the autocorrelation noise given in Eq.~\eqref{autocorrelacion_ruido_en_varianza} to obtain
\begin{equation}\label{varianza_larga_4}
\mean{\eta(1)\eta(2)\eta(3)\eta(4)}  = \Gamma^{2}\tilde{\delta}(1-2)\tilde{\delta}(3-4) + \Gamma^{2}\tilde{\delta}(1-3)\tilde{\delta}(2-4) + \Gamma^{2}\tilde{\delta}(1-4)\tilde{\delta}(2-3),
\end{equation}
where we have used the compact notation $\tilde{\delta}(i - j) = \delta(\textbf{r}_{i} - \textbf{r}_{j})\delta(\tau_{i} - \tau_{j})\delta(\sigma_{i} - \sigma_{j})$.
With the result given by Eq.~\eqref{varianza_larga_4}, we can carry out the integration over the space of Eq.~\eqref{varianza_larga_2} by the use of the spatial Dirac delta to obtain
\begin{align}\label{varianza_larga_5}
& \mean{\eta_{n_{1}m_{1}}(\sigma_{1})\eta_{n_{2}m_{2}}(\sigma_{2})\eta_{n_{3}m_{3}}(\sigma_{3})\eta_{n_{4}m_{4}}(\sigma_{4})} 
 = \nonumber\\
& \Gamma^{2}\delta_{n_{1},n_{2}}\delta_{m_{1},m_{2}}\delta_{n_{3},n_{4}}\delta_{m_{3},m_{4}}\delta(\sigma_{1} - \sigma_{2})\delta(\sigma_{3} - \sigma_{4}) \nonumber\\
& + \Gamma^{2}\delta_{n_{1},n_{3}}\delta_{m_{1},m_{3}}\delta_{n_{2},n_{4}}\delta_{m_{2},m_{4}}\delta(\sigma_{1} - \sigma_{3})\delta(\sigma_{2} - \sigma_{4}) \nonumber\\
& + \Gamma^{2}\delta_{n_{1},n_{4}}\delta_{m_{1},m_{4}}\delta_{n_{2},n_{3}}\delta_{m_{2},m_{3}}\delta(\sigma_{1} - \sigma_{4})\delta(\sigma_{2} - \sigma_{3}).
\end{align}
With the last result given in Eq.~\eqref{varianza_larga_5}, we can carry out the integrals over the pseudo--time $s$ of Eq.~\eqref{varianza_larga_1} obtaining the last intermediate step of the calculation as
\begin{align}\label{varianza_larga_6}
& \lim_{s\to\infty}\mean{\prod_{k=1}^{4}\phi_{n_{k},m_{k}}(s)} = \nonumber\\
& \Gamma^{2}\frac{\delta_{n_{1},n_{2}}\delta_{m_{1},m_{2}}\delta_{n_{3},n_{4}}\delta_{m_{3},m_{4}}}{\left[\left(\omega_{m_{1}}^{2} + \mu_{n_{1}}\right) + \left(\omega_{m_{2}}^{2} + \mu_{n_{2}}\right)\right]\left[\left(\omega_{m_{3}}^{2} + \mu_{n_{3}} \right) + \left(\omega_{m_{4}}^{2} + \mu_{n_{4}}\right)\right] }\nonumber\\
& + \Gamma^{2}\frac{\delta_{n_{1},n_{3}}\delta_{m_{1},m_{3}}\delta_{n_{2},n_{4}}\delta_{m_{2},m_{4}}}{\left[\left(\omega_{m_{1}}^{2} + \mu_{n_{1}}\right) + \left(\omega_{m_{3}}^{2} + \mu_{n_{3}}\right)\right]\left[\left(\omega_{m_{2}}^{2} + \mu_{n_{2}} \right) + \left(\omega_{m_{4}}^{2} + \mu_{n_{4}}\right)\right] }\nonumber\\
& + \Gamma^{2}\frac{\delta_{n_{1},n_{4}}\delta_{m_{1},m_{4}}\delta_{n_{2},n_{3}}\delta_{m_{2},m_{3}}}{\left[\left(\omega_{m_{1}}^{2} + \mu_{n_{1}}\right) + \left(\omega_{m_{4}}^{2} + \mu_{n_{4}}\right)\right]\left[\left(\omega_{m_{2}}^{2} + \mu_{n_{2}} \right) + \left(\omega_{m_{3}}^{2} + \mu_{n_{3}}\right)\right] }.
\end{align}
Then the four points correlation function defined in Eq.~\eqref{4_points_correlation_function_definition} results in
\begin{align}\label{4_points_correlation_function_result}
& \mean{\phi(\textbf{r}_{1},\tau_{1})\phi(\textbf{r}_{2},\tau_{2})\phi(\textbf{r}_{3},\tau_{3})\phi(\textbf{r}_{4},\tau_{4})} = \nonumber\\
& \frac{\Gamma^{2}}{4}\sum_{n_{1}m_{1}}\sum_{n_{3}m_{3}}\frac{f_{n_{1}}(\textbf{r}_{1})f_{n_{1}}(\textbf{r}_{2})}{\left(\omega_{m_{1}}^{2} + \mu_{n_{1}}\right)}\frac{f_{n_{3}}(\textbf{r}_{3})f_{n_{3}}(\textbf{r}_{4})}{\left(\omega_{m_{3}}^{2} + \mu_{n_{3}} \right)}g_{m_{1}}(\tau_{1})g_{m_{1}}(\tau_{2})g_{m_{3}}(\tau_{2})g_{m_{3}}(\tau_{4})\nonumber\\
& + \frac{\Gamma^{2}}{4}\sum_{n_{1}m_{1}}\sum_{n_{2}m_{2}}\frac{f_{n_{1}}(\textbf{r}_{1})f_{n_{1}}(\textbf{r}_{3})}{\left(\omega_{m_{1}}^{2} + \mu_{n_{1}}\right)}\frac{f_{n_{2}}(\textbf{r}_{2})f_{n_{2}}(\textbf{r}_{4})}{\left(\omega_{m_{2}}^{2} + \mu_{n_{2}}\right)}g_{m_{1}}(\tau_{1})g_{m_{1}}(\tau_{3})g_{m_{2}}(\tau_{2})g_{m_{2}}(\tau_{4})\nonumber\\
& + \frac{\Gamma^{2}}{4}\sum_{n_{1}m_{1}}\sum_{n_{2}m_{2}}\frac{f_{n_{1}}(\textbf{r}_{1})f_{n_{1}}(\textbf{r}_{4})}{\left(\omega_{m_{1}}^{2} + \mu_{n_{1}}\right)}\frac{f_{n_{2}}(\textbf{r}_{2})f_{n_{2}}(\textbf{r}_{3})}{\left(\omega_{m_{2}}^{2} + \mu_{n_{2}}\right)}g_{m_{1}}(\tau_{1})g_{m_{1}}(\tau_{4})g_{m_{2}}(\tau_{2})g_{m_{2}}(\tau_{3}).
\end{align}
Here, we assume that $\tau_{1} = \tau_{2} = \tau_{3} = \tau_{4} = \tau$, then the dependence on $g_{m_{k}}(\tau)$ is simplified to
\begin{align}\label{4_points_correlation_function_result_12}
\mean{\phi(\textbf{r}_{1},\tau)\phi(\textbf{r}_{2},\tau)\phi(\textbf{r}_{3},\tau)\phi(\textbf{r}_{4},\tau)} &
 = \frac{\Gamma^{2}}{4}\sum_{n_{1}m_{1}}\sum_{n_{3}m_{3}}\frac{f_{n_{1}}(\textbf{r}_{1})f_{n_{1}}(\textbf{r}_{2})}{\left(\omega_{m_{1}}^{2} + \mu_{n_{1}}\right)}\frac{f_{n_{3}}(\textbf{r}_{3})f_{n_{3}}(\textbf{r}_{4})}{\left(\omega_{m_{3}}^{2} + \mu_{n_{3}} \right)}\nonumber\\
& + \frac{\Gamma^{2}}{4}\sum_{n_{1}m_{1}}\sum_{n_{2}m_{2}}\frac{f_{n_{1}}(\textbf{r}_{1})f_{n_{1}}(\textbf{r}_{3})}{\left(\omega_{m_{1}}^{2} + \mu_{n_{1}}\right)}\frac{f_{n_{2}}(\textbf{r}_{2})f_{n_{2}}(\textbf{r}_{4})}{\left(\omega_{m_{2}}^{2} + \mu_{n_{2}}\right)}\nonumber\\
& + \frac{\Gamma^{2}}{4}\sum_{n_{1}m_{1}}\sum_{n_{2}m_{2}}\frac{f_{n_{1}}(\textbf{r}_{1})f_{n_{1}}(\textbf{r}_{4})}{\left(\omega_{m_{1}}^{2} + \mu_{n_{1}}\right)}\frac{f_{n_{2}}(\textbf{r}_{2})f_{n_{2}}(\textbf{r}_{3})}{\left(\omega_{m_{2}}^{2} + \mu_{n_{2}}\right)}.
\end{align}
We are interested in the special case with $\textbf{r}_{1} = \textbf{r}_{2} = \textbf{r}$ and $\textbf{r}_{3} = \textbf{r}_{4} = \textbf{r}'$, then
\begin{align}\label{4_points_correlation_function_result_2}
\mean{\phi(\textbf{r},\tau)\phi(\textbf{r},\tau)\phi(\textbf{r}',\tau)\phi(\textbf{r}',\tau)} &
 = \frac{\Gamma^{2}}{4}\sum_{n_{1}m_{1}}\sum_{n_{2}m_{2}}\frac{f_{n_{1}}(\textbf{r})f_{n_{1}}(\textbf{r})}{\left(\omega_{m_{1}}^{2} + \mu_{n_{1}}\right)}\frac{f_{n_{2}}(\textbf{r}')f_{n_{2}}(\textbf{r}')}{\left(\omega_{m_{2}}^{2} + \mu_{n_{2}}\right)}\nonumber\\
& + \frac{\Gamma^{2}}{2}\sum_{n_{1}m_{1}}\sum_{n_{2}m_{2}}\frac{f_{n_{1}}(\textbf{r})f_{n_{1}}(\textbf{r}')}{\left(\omega_{m_{1}}^{2} + \mu_{n_{1}}\right)}\frac{f_{n_{2}}(\textbf{r}')f_{n_{2}}(\textbf{r})}{\left(\omega_{m_{2}}^{2} + \mu_{n_{2}}\right)}.
\end{align}
Finally we obtain from Eqs.~\eqref{4_points_correlation_function_result_2} and \eqref{Definicion_varianza} the variance as
%\begin{align}\label{4_points_correlation_function_result_2}\sigma_{F}^{2} = \frac{\Gamma^{2}}{2}\sum_{n}\sum_{m}\frac{f_{n}(\textbf{r})f_{n}(\textbf{s})}{\mu_{n}}\frac{f_{m}(\textbf{r})f_{m}(\textbf{s})}{\mu_{m}}.\end{align}
\begin{align}\label{Definicion_varianza_en_modos}
& \sigma_{F}^{2} = \\
& \frac{\Gamma^{2}}{2}\sum_{n_{1}m_{1}}\sum_{n_{2}m_{2}}\frac{1}{\left(\omega_{m_{1}}^{2} + \mu_{n_{1}}\right)}\frac{1}{\left(\omega_{m_{2}}^{2} + \mu_{n_{2}}\right)}\oint_{\textbf{r}}\TensorT[f_{n}(\textbf{r}),f_{m}(\textbf{r})]\cdot d\textbf{r}\oint_{\textbf{r}'}\TensorT[f_{m}(\textbf{r}'),f_{n}(\textbf{r}')]\cdot d\textbf{r}',\nonumber
\end{align}
where the second term in Eq.~\eqref{Definicion_varianza} has been simplified with the first term of Eq.~\eqref{4_points_correlation_function_result_2}. It is an integral formula of the the variance of the Casimir force over a given body, one of the main results of this Thesis. It is possible to carry out the sums over $m_{1}$ and $m_{2}$ using $\omega_{m} = \frac{2\pi}{\hbar c\beta}m$ with $m\in\mathbb{Z}$ and $\Gamma = 2k_{B}T$, obtaining
\begin{align}
\sum_{m\in\mathbb{Z}}\frac{2k_{B}T}{\omega_{m}^{2} + \mu} = \frac{\hbar c}{\sqrt{\mu}}\left[1 + \frac{2}{e^{\frac{\hbar c\sqrt{\mu}}{k_{B}T}} - 1}\right].
\end{align}
Therefore, the variance can be written as
\begin{align}\label{Definicion_varianza_en_modos_2}
\sigma_{F}^{2} = & \frac{1}{2}\sum_{n}\sum_{m}
\frac{\hbar c}{\sqrt{\mu_{n}}}\left[1 + \frac{2}{e^{\frac{\hbar c\sqrt{\mu_{n}}}{k_{B}T}} - 1}\right]
\frac{\hbar c}{\sqrt{\mu_{m}}}\left[1 + \frac{2}{e^{\frac{\hbar c\sqrt{\mu_{m}}}{k_{B}T}} - 1}\right]\times\nonumber\\
& \times\oint_{\textbf{r}}\TensorT[f_{n}(\textbf{r}),f_{m}(\textbf{r})]\cdot d\textbf{r}\oint_{\textbf{r}'}\TensorT[f_{m}(\textbf{r}'),f_{n}(\textbf{r}')]\cdot d\textbf{r}',
\end{align}
and the low (quantum) and high (classic) temperature limits of this formula are straightforward to obtain as
\begin{align}\label{Definicion_varianza_en_modos_quantum}
\lim_{T\to 0}\sigma_{F}^{2} = \frac{(\hbar c)^{2}}{2}\sum_{n}\sum_{m}
\frac{1}{\sqrt{\mu_{n}\mu_{m}}}\oint_{\textbf{r}}\TensorT[f_{n}(\textbf{r}),f_{m}(\textbf{r})]\cdot d\textbf{r}\oint_{\textbf{r}'}\TensorT[f_{m}(\textbf{r}'),f_{n}(\textbf{r}')]\cdot d\textbf{r}'.
\end{align}
\begin{align}\label{Definicion_varianza_en_modos_classic}
\lim_{\hbar\to 0}\sigma_{F}^{2} = 2(k_{B}T)^{2}\sum_{n}\sum_{m}
\frac{1}{\mu_{n}\mu_{m}}\oint_{\textbf{r}}\TensorT[f_{n}(\textbf{r}),f_{m}(\textbf{r})]\cdot d\textbf{r}\oint_{\textbf{r}'}\TensorT[f_{m}(\textbf{r}'),f_{n}(\textbf{r}')]\cdot d\textbf{r}'.
\end{align}
In the special case $\mathcal{M} = - \Delta$, we have $\mu_{n} = \lambda_{n}^{2}$. As a particular case, we obtain the variance of the Casimir force of a perfect metal piston of arbitrary section. In this particular case, $f_{n}(\textbf{r}) = f_{n}(\textbf{r}')$ and they are constant over each side of the piston. In addition to that, $\TensorT[f_{n}(\textbf{r}),f_{m}(\textbf{r})]$ is a bilinear form for the EM field, then we can swap $n$ and $m$ indices because they are dummy indices to obtain
\begin{align}\label{Varianza_piston_EM_1}
\sigma_{F}^{2} = & 2\sum_{n}\sum_{m}
\frac{\hbar c}{2\lambda_{n}}\left[1 + \frac{2}{e^{\frac{\hbar c\lambda_{n}}{k_{B}T}} - 1}\right]
\frac{\hbar c}{2\lambda_{m}}\left[1 + \frac{2}{e^{\frac{\hbar c\lambda_{m}}{k_{B}T}} - 1}\right]\times\nonumber\\
& \times\oint_{\textbf{r}}\TensorT[f_{n}(\textbf{r}),f_{n}(\textbf{r})]\cdot d\textbf{r}\oint_{\textbf{r}'}\TensorT[f_{m}(\textbf{r}'),f_{m}(\textbf{r}')]\cdot d\textbf{r}',
\end{align}
This result admits the simplification
\begin{align}\label{Varianza_piston_EM_2}
\sigma_{F}^{2} = & 2\left(\sum_{n}
\frac{\hbar c}{2\lambda_{n}}\left[1 + \frac{2}{e^{\frac{\hbar c\lambda_{n}}{k_{B}T}} - 1}\right]
\oint_{\textbf{r}}\TensorT[f_{n}(\textbf{r}),f_{n}(\textbf{r})]\cdot d\textbf{r}\right)^{2}.
\end{align}
Using the definition of Casimir force given in Eq.~\eqref{Definicion_media}, we obtain that, for any temperature and for pistons of arbitrary section, the variance of the Casimir force is twice the Casimir force itself
\begin{equation}
\sigma_{F_{x}}^{2} = 2 \textbf{F}_{C,x}^{2},
\end{equation}
which is the result presented in Eq.~\eqref{Varianza_Fuerza_Casimir}, valid for all perfect metal pistons with arbitrary section and for any given temperature.
The same result for the variance is obtained for classical systems by the application of the same procedure as shown above to Eq.~\eqref{Definicion_varianza_en_modos_classic} for any given spatial (linear) operator $\mathcal{M}$ whose eigenvalues are $\mu_{n}$.

As in the case of evaluation of Casimir forces, we do not need any cutoff to obtain a finite result for the variance of the Casimir force.

The interchange of the integral and summation regularizes the variance Casimir force on the same fashion as the Casimir force was regularizated too, avoiding also in this case the use of ultraviolet cutoffs. Other regularizations, that in some cases may lead to non--universal forces or fluctuations are, for instance, the subtraction of the vacuum stress tensor \cite{Ford} or by averaging the stress tensor over a finite area or a finite time~\cite{Barton2}\cite{Barton3}.

Therefore, we have found an universal form for the fluctuations, as opposite to other authors \cite{Fournier}\cite{Barton}. 
The difference has its origin in that we do not compute the stress fluctuations or the fluctuations of the force on each side of a plate.
Rather, we first compute the total fluctuating force on the body (which is finite) and then we compute its variance.

% Extensión a sistemas cuyas fluctuaciones están descritas por derivadas temporales mayores que uno
%\begin{savequote}[10cm] % this sets the width of the quote
%\sffamily
%``Physics is to math what sex is to masturbation.'' 
%\qauthor{Richard P. Feynman}
%\end{savequote}
\begin{savequote}[12cm] % this sets the width of the quote
\sffamily
``In mathematics you don't understand things. You just get used to them.'' 
\qauthor{John Von Neumann}
\end{savequote}

\chapter{Extension of Langevin formalism to higher temporal derivatives and its application to Casimir effect.}\label{Chap: Extension of Langevin formalism to higher temporal derivatives and its application to Casimir effect.}
\graphicspath{{01-Casimir_Langevin/ch3/Figuras/}}

In this Chapter we present another application of our model of evaluation of fluctuation induced forces to a generalization of Langevin dynamics.

Generalizations of Stochastical Differential Equations is an active field in last years. For example, the Langevin Difference Equation has an evident interest in numerical simulations. The replacement of Gaussian noises by L\'evi noises is a fundamental tool in the study of turbulence and generalizations of Fluctuation--Dissipation theorem. Also the study of multiplicative noises have interest in noise--induced phase transitions.

The generalization presented here has a different nature. We will obtain the two point correlation function of a field whose temporal dynamic is described by a general differential functional. i.e. by a linear functional on temporal derivatives of an integer order greater or equal to one (that is the case considered before). The case of fractional derivatives is left to a future work~\cite{Lim2006269}.

Usual Langevin equations are a fundamental tool in the study of equilibrium and non equilibrium Thermodynamics, but there exist cases when Langevin equations cannot describe the relaxation dynamics of our systems. The case of the electromagnetic field subject to stochastic sources is a simple first example, note that it is a different problem from the case considered in Chapter~\ref{Chap: Stochastic Quantization and Casimir forces.}, where a first order temporal derivative describes the evolution of the field in a virtual time, while in this case, a second order temporal derivative (characteristic of the wave equation) describes the evolution of the filed, and its sources are descibed by a white noise.

We will begin with a description of the dynamics in Sect.~\ref{Langevin equation with higher temporal derivatives}, followed by the derivation of the Green function of the field  in Sect.~\ref{Section_Derivacion_Funcion_Green}. In Sect.~\ref{Correlation function in the steady state}, we will obtain the two point correlation function in the steady state and compare it with the Green function of the spatial part of the dynamics. In Sect.~\ref{Casimir force between infinite parallel plates}, as an example of the generality of the proposed model of evaluation of Casimir forces, we will obtain the Casimir force between two parallel plates immersed in these fluctuating media.
We finish this Chapter with a Discussion in Sect.~\ref{chapI3:Discussion}.

The contents of this Chapter have not been published anywhere.
\section{Langevin equation with higher temporal derivatives}\label{Langevin equation with higher temporal derivatives}
Let us consider the generalization of the linear stochastic differential equation \eqref{Langevin} for the field $\phi(\textbf{r},t)$,
\begin{equation}\label{Generalized_Langevin_function}
F(\partial_{t})\phi(\textbf{r},t) = - \mathcal{M}\phi(\textbf{r},t) + \xi(\textbf{r},t),
\end{equation}
where $\mathcal{M}$ is a spatial linear operator and $F(\partial_{t})$ is an entire function in $\partial_{t}$. The term $\xi(\textbf{r},t)$ is a Gaussian noise. Here we assume that $\xi(\textbf{r},t)$ is a zero mean Gaussian white noise characterized by the autocorrelation 
\begin{equation}\label{Autocorrelacion_ruido_blanco}
\mean{\xi(\textbf{r},t)\xi^{*}(\textbf{r}',t')} = \Gamma\delta(\textbf{r} - \textbf{r}')\delta(t - t').
\end{equation}
Generalization to colored Gaussian noises can be performed following the same reasoning that in Chapter \ref{Dynamical approach to the Casimir effect} adapted to the formalism presented here.

When $F(x) = x$, we recover the usual Langevin dynamics given in Eq.~\eqref{Langevin}, and the two point correlation function of the field in the steady state is given by
\begin{equation}
\lim_{t\to\infty}\mean{\phi(\textbf{r},t)\phi^{*}(\textbf{r}',t)} = \frac{\Gamma}{2} \sum_{\vec{k}}\frac{f_{n}(\textbf{r})g_{n}^{*}(\textbf{r}')}{\mu_{n}},
\end{equation}
where $f_{n}(\textbf{r})$ are the solutions of the spatial left and right eigenproblem
\begin{align}\label{Spatial_eigenproblem}
\mathcal{M}f_{n}(\textbf{r}) &= \mu_{n}f_{n}(\textbf{r}),\\
\mathcal{M}^{+}g_{n}(\textbf{r}) &= \mu_{n}^{*}g_{n}(\textbf{r}).
\end{align}

\section{Green function of the Generalized Langevin Equation}\label{Section_Derivacion_Funcion_Green}
As an intermediate step of the derivation of the two point correlation function of the field $\phi$ in the steady state, we need the Green function of Eq.~\eqref{Generalized_Langevin_function} without noise. This Green function obeys
\begin{equation}\label{Green_function_of_Generalized_Langevin_function}
\left[F(\partial_{t}) + \mathcal{M}\right]G\left(\textbf{r},\textbf{r}',t,t'\right) = \delta\left(\textbf{r} - \textbf{r}'\right)\delta(t - t').
\end{equation}
After a symmetric Fourier transformation on time $t$ ($t'$ is not transformed) and factorizing the spatial part of the dynamics by the use of the expansion of eigenfunctions given by Eq.~\eqref{Spatial_eigenproblem}, we obtain that the Fourier transform on time $t$ of the expansion terms of the Green function obeys
\begin{equation}
\left[F(-i\omega) + \mu_{n}\right]\hat{G}_{n}(\textbf{r},\textbf{r}',\omega, t') = \frac{e^{i\omega t'}}{\sqrt{2\pi}}f_{n}(\textbf{r})g^{*}_{n}(\textbf{r}'),
\end{equation}
where the expansion of spatial Dirac delta on eigenfunctions has been used. Then the complete Green function can be written as the sum of fractions
\begin{equation}\label{Funcion_Green_no_desarrollada}
\hat{G}(\textbf{r},\textbf{r}',\omega, t') = \frac{e^{i\omega t'}}{\sqrt{2\pi}}\sum_{n\in\mathbb{Z}}\frac{f_{n}(\textbf{r})g_{n}^{*}(\textbf{r}')}{F(-i\omega) + \mu_{n}}.
\end{equation}
As a crucial step, we use the Mittag--Leffler theorem \cite{Libro-Residuos}. So that these fractions can be decomposed as a sum of simple fractions in $\omega$, obtaining
\begin{equation}
\hat{G}(\textbf{r},\textbf{r}',\omega, t') = \frac{e^{i\omega t'}}{\sqrt{2\pi}}\sum_{n\in\mathbb{Z}}\left[\sum_{m=1}^{M}\frac{R_{m}^{n}}{-i\omega + \omega_{m}^{n}}\right]f_{n}(\textbf{r})g_{n}^{*}(\textbf{r}').
\end{equation}
Here:
\begin{itemize}
\item $\left\lbrace \omega_{m}^{n}\right\rbrace_{m}^{n\in\mathbb{Z}}$ is the countable family of solutions of the dispersion relation of the problem $F(-i\omega) = \mu_{n}$ for each spatial mode, i.e., they are poles of $\hat{G}$ in the variable $\omega$.
\item $\left\lbrace R_{m}^{n}\right\rbrace_{m}^{n\in\mathbb{Z}}$ are the correspondent residue for each pole $\omega_{m}^{n}$ of $\hat{G}$. It is important to observe that, if $F(-i\omega)$ has units of $s^{-n}$, then the Green function has units $s^{n}$. As a consequence, these residue must have units of $s^{n-1}$ because of dimensional coherence.
\end{itemize}

Then we have obtained the Green function in the temporal frequency space. We have to invert the Fourier transform in order to obtain the Green function in the correct representation in time as
\begin{equation}\label{Resultado_Green_function_of_Generalized_Langevin_function}
G(\textbf{r},\textbf{r}', t, t') = \sum_{n\in\mathbb{Z}}{\sum_{m=1}^{M}}'\left[R_{m}^{n}e^{-\omega_{m}^{n}(t - t')}\theta(\text{sign}\left(\re{\omega_{m}^{n}}\right)(t - t'))\right]f_{n}(\textbf{r})g_{n}^{*}(\textbf{r}'),
\end{equation}
where the tilde means that modes with real part equal to zero contribute with a $1/2$ weight. This is the main result of the Section. We observe that these Green functions do not explode at infinite time because modes are splitted in three families depending of the sign of their real part:
\begin{itemize}
\item If $\re{\omega_{m}^{n}} > 0$, then this mode is relevant for all $t > t'$.
\item If $\re{\omega_{m}^{n}} < 0$, then this mode is relevant for all $t < t'$.
\item If $\re{\omega_{m}^{n}} = 0$, then the inverse Fourier transform gives an undamped term $\frac{1}{2}\theta(t - t')$.
\end{itemize}
When $F(x)$ is an entire function on $x$ with a zero of order $N>1$, the Mittag--Leffler theorem \cite{Libro-Residuos} says that the fraction that we obtained in Eq.~\eqref{Funcion_Green_no_desarrollada} can be expanded to
\begin{equation}
\frac{e^{i\omega t'}}{\sqrt{2\pi}}\frac{1}{(-i\omega + \omega_{N})^{N}} = \frac{e^{i\omega t'}}{\sqrt{2\pi}}\sum_{n=1}^{N}\frac{R_{N,n}}{(-i\omega + \omega_{N})^{n}},
\end{equation}
whose inverse Fourier transform results in
\begin{equation}
\mathcal{F}^{-1}\left[\frac{e^{i\omega\, t'}}{(-i\omega + \omega_{N})^{N}}\right](t) = e^{-\omega_{N}(t - t')}\theta(\re{\omega_{N}})(t - t'))\sum_{n=1}^{N}R_{N,n}\frac{(t - t')^{n - 1}}{(n - 1)!}.
\end{equation}

In the important particular case of the usual Langevin equation with first order temporal derivative, the known Green function can be derived as a special case of Eq.~\eqref{Resultado_Green_function_of_Generalized_Langevin_function} as
\begin{equation}
G(\textbf{r},\textbf{r}', t, t') = \sum_{n\in\mathbb{Z}}e^{-\omega_{n}(t - t')}f_{n}(\textbf{r})g_{n}^{*}(\textbf{r}').
\end{equation}

\section{Correlation function in the steady state}\label{Correlation function in the steady state}
Once the Green function of the generalized Lanvegin equation has been derived, we are able to obtain the correlation function of the field in the steady state. This is the key quantity to obtain, because the average of any measurable quantity in the steady of the field $\phi$ can be written in terms of this correlation function.

We can write the general solution of the generalized Lanvegin equation \eqref{Generalized_Langevin_function} in terms of the Green function of the field \eqref{Resultado_Green_function_of_Generalized_Langevin_function} as
\begin{equation}
\phi(\textbf{r},t) = \int_{-\infty}^{\infty}dt'\int d\textbf{r}' G(\textbf{r},\textbf{r}',t,t')\xi(\textbf{r}',t').
\end{equation}
The correlation function in the steady state is defined as
\begin{equation}\label{Definicion_funcion_correlacion_en_el_steady_state}
\mean{\phi(\textbf{r})\phi^{*}(\textbf{r}')} = \lim_{t\to\infty}
\lim_{\begin{minipage}{0.5cm}${\scriptstyle{t'\to t}}$\\[-2mm]${\scriptstyle{t' > t}}$\end{minipage}}
\mean{\phi(\textbf{r},t)\phi^{*}(\textbf{r}',t')}.
\end{equation}
This average must be obtained from
\begin{align}
& \mean{\phi(\textbf{r},t)\phi^{*}(\textbf{r}',t')} = \\
& \mean{\int d\textbf{r}''\int_{-\infty}^{\infty} dt''\, G(\textbf{r},\textbf{r}'',t,t'')\xi(\textbf{r}'',t'')\int d\textbf{r}''' \int_{-\infty}^{\infty} dt'''\, \xi^{*}(\textbf{r}''',t''')G^{*}(\textbf{r}',\textbf{r}''',t',t''')}.\nonumber
\end{align}
By the use of Eq.~\eqref{Autocorrelacion_ruido_blanco}, this integral can be reduced to
\begin{equation}
\mean{\phi(\textbf{r},t)\phi^{*}(\textbf{r}',t')} = \Gamma\int d\textbf{r}''\int dt''\, G(\textbf{r},\textbf{r}'',t,t'')G^{*}(\textbf{r}',\textbf{r}'',t',t'').
\end{equation}
Using Eq.~\eqref{Resultado_Green_function_of_Generalized_Langevin_function} and carrying out the integrations in time and space (where the orthogonality of $f_{n}(\textbf{r})$ has been used), we obtain that the two point correlation function is
\begin{align}
& \mean{\phi(\textbf{r},t)\phi^{*}(\textbf{r}',t')} = \Gamma\sum_{n\in\mathbb{Z}}{\sum_{m=1}^{M}}'{\sum_{m'=1}^{M}}'R_{m}^{n}R^{n *}_{m'}f_{n}(\textbf{r})f_{n}^{*}(\textbf{r}')\times\\
& \times\int dt'\,\left[e^{-\omega_{m}^{n}(t - t')}e^{-\omega_{m'}^{n *}(t - t')}\theta(\text{sign}\left(\re{\omega_{m}^{n}}\right)(t - t'))\theta(\text{sign}\left(\re{\omega_{m'}^{n}}\right)(t - t'))\right].\nonumber
\end{align}
Finally, by the use of the definition of the correlation function in the steady state given in Eq.~\eqref{Definicion_funcion_correlacion_en_el_steady_state}, we obtain after a straightforward calculus
\begin{equation}\label{Resultado_Correlation_function_of_Generalized_Langevin_function_steady_state}
\mean{\phi(\textbf{r})\phi^{*}(\textbf{r}')} = \Gamma\sum_{n\in\mathbb{Z}}{\sum_{\omega_{m'}^{n} \geq 0}^{M}}'{\sum_{\omega_{m}^{n} \geq 0}^{M}}'\frac{R_{m}^{n}R^{n *}_{m'}}{\omega_{m}^{n} + \omega_{m'}^{n *}}f^{*}_{n}(\textbf{r})f_{n}(\textbf{r}'),
\end{equation}
where the tilde means that modes with real part equal to zero contribute with a $1/2$ weight. Modes with $\re{\omega_{m}^{n}} < 0$ do not contribute to the final result because we have assumed causality when we imposed the condition $t' > t$.
When $\im{\omega_{m}^{n}} \neq 0$, having into account that the sum is symmetrical in $\{m,m'\}$, the result is reduced to
\begin{align}\label{Resultado_Correlation_function_of_Generalized_Langevin_function_steady_statev2}
& \mean{\phi(\textbf{r})\phi^{*}(\textbf{r}')} = \\
& \Gamma\sum_{n\in\mathbb{Z}}{\sum_{\omega_{m'}^{n} \geq 0}^{M}}'{\sum_{\omega_{m}^{n} \geq 0}^{M}}'\frac{\re{R_{m}^{n}R^{n *}_{m'}}({\omega_{m}^{n}}' + {\omega_{m'}^{n}}') + \im{R_{m}^{n}R^{n *}_{m'}}({\omega_{m}^{n}}'' - {\omega_{m'}^{n}}'')}{({\omega_{m}^{n}}' + {\omega_{m'}^{n}}')^{2} + ({\omega_{m}^{n}}'' - {\omega_{m'}^{n}}'')^{2}}
f^{*}_{n}(\textbf{r})f_{n}(\textbf{r}'),\nonumber
\end{align}
where we have splitted the modes in real and imaginary parts, $\omega_{m}^{n} = {\omega_{m}^{n}}' + i{\omega_{m}^{n}}''$. Equation \eqref{Resultado_Correlation_function_of_Generalized_Langevin_function_steady_statev2} is the main result of this Chapter. When modes with $\re{\omega_{m}^{n}} = 0$ appear, we have to take into account that they contribute with a $1/2$ weight, and terms with $\re{\omega_{m}^{n}} = \re{\omega_{m'}^{n}} = 0$ do not contribute at all, because $\lim_{x\to 0}\frac{x}{x^{2} + y^{2}} = 0$.

In the important particular case of the usual Langevin equation with first order temporal derivative, the correlation function in the steady state can be derived as a special case of Eq.~\eqref{Resultado_Correlation_function_of_Generalized_Langevin_function_steady_state} as
\begin{equation}
\mean{\phi(\textbf{r})\phi^{*}(\textbf{r}')} = \Gamma\sum_{n\in\mathbb{Z}}\frac{f_{n}(\textbf{r})f_{n}^{*}(\textbf{r}')}{\mu_{n} + \mu_{n}^{*}}.
\end{equation}
Another interesting case is the correlation function of a field subject to the wave equation in the presence of white noise, in this case we have
\begin{equation}\label{Generalized_Langevin_function}
\frac{\partial_{t}^{2}}{c^{2}}\phi(\textbf{r},t) - \Delta\phi(\textbf{r},t) = \xi(\textbf{r},t).
\end{equation}
In the temporal frequency domain, the inverse of the wave operator admits the factorization
\begin{equation}
\frac{- c^{2}}{\omega^{2} - c^{2}k_{n}^{2}} = \frac{c^{2}}{2c\abs{\textbf{k}_{n}}}\left[\frac{1}{\omega + c\abs{\textbf{k}_{n}}} - \frac{1}{\omega - c\abs{\textbf{k}_{n}}}\right],
\end{equation}
then $\omega_{m}^{n} = c\abs{\textbf{k}_{n}}$ are the positive modes of the dispersion relation, and $R_{m}^{n} = - \frac{c}{2\abs{\textbf{k}_{n}}}$ is its correspondent residue, then the correlation function in the steady state is given by
\begin{equation}
\mean{\phi(\textbf{r})\phi^{*}(\textbf{r}')} = \frac{\Gamma\, c}{8}\sum_{n\in\mathbb{Z}}\frac{f_{n}(\textbf{r})f_{n}^{*}(\textbf{r}')}{\abs{\textbf{k}_{n}}^{3}},
\end{equation}
where $k_{n} > 0$ is imposed because causality. Similar results can be obtained for even greater orders of the temporal derivative. It is easy to verify that, if $F(x) = \left(\frac{x}{c_{p}}\right)^{p}$ and we include a mass term $m^{2}\phi$ in the left part of Eq.\eqref{Generalized_Langevin_function}, then the correlation function in the steady state has the form
\begin{equation}\label{2_points_Correlation_function_derivada_temporal_orden_p}
\mean{\phi(\textbf{r})\phi^{*}(\textbf{r}')} = \frac{\Gamma\, c_{p}}{C(p)}\sum_{n\in\mathbb{Z}}\frac{f_{n}(\textbf{r})f_{n}^{*}(\textbf{r}')}{\left(\textbf{k}_{n}^{2} + m^{2}\right)^{2 - \frac{1}{p}}},
\end{equation}
where $C(p)$ is a non monotonous increasing function in $p$, too complicated to be shown here. For $p=\{1,\cdots,6\}$ it is easy to obtain $C(p) = \{2, 8, 18, 32, 50, 36\}$.

\section{Casimir force between infinite parallel plates}\label{Casimir force between infinite parallel plates}
As an application of the developed formalism, we are going to calculate the Casimir force between infinite parallel plates, located at $x = 0$ and $x = L$, immersed in a medium $\phi(\textbf{r},t)$, whose fluctuations in the steady state are described by the generalized Langevin equation
\begin{equation}
\left(\frac{\partial_{t}\phi}{c_{p}}\right)^{p} = \kappa_{2}\Delta\phi - \kappa_{1}\phi + \xi(\textbf{r},t),
\end{equation}
where $\xi(\textbf{r},t)$ is a white Gaussian noise of intensity $\Gamma$ and $p$ is a positive integer. We assume that our plates vanish the fluctuations of $\phi$ over their surfaces, then $\phi$ is subject to Dirichlet boundary conditions over the surface of each plate. In addition to that, we assume that the free energy is the one of a nematic liquid crystals in presence of an external magnetic field given in Eq.~\eqref{F.liqcry}.
Then the stress tensor and the used eigenfunctions are the ones given in Sect. \ref{Cristales-Liquidos}, but the average of the stress tensor over the surface of a plate is given in terms of the two point correlation function in the steady state of this stochastical process, given in Eq.~\eqref{2_points_Correlation_function_derivada_temporal_orden_p}.

The stress tensor on the surface of a plate is given by (see Eqs.~\eqref{stress_tensor_cristal_liquido_ruido_blanco1} and \eqref{stress_tensor_cristal_liquido_ruido_blanco2})
\begin{equation}
\mean{\mathbb{T}_{xx}(0)} = - \frac{\Gamma}{V C(p)} \sum_{n}\frac{k_{x}^{2}}{\left(k_{x}^{2} + k_{y}^{2} + k_{z}^{2} + k_{0}^{2}\right)^{2-\frac{1}{p}}}.
\end{equation}
Applying the limit $L_{y}, L_{z}\rightarrow\infty$, we obtain
\begin{equation}
\mean{\mathbb{T}_{xx}(0)} = \frac{-\Gamma}{8 \pi^{2}C(p)L_{x}} \int_{-\infty}^{\infty}\!dk_{y}\int_{-\infty}^{\infty}\!dk_{z} \sum_{n_{x}\in\mathbb{Z}}\frac{\left(\frac{\pi n_{x}}{L_{x}}\right)^{2}}{\left(\left(\frac{\pi n_{x}}{L_{x}}\right)^{2} + k_{y}^{2} + k_{z}^{2} + k_{0}^{2}\right)^{2-\frac{1}{p}}}.
\end{equation}
Using polar coordinates and regularizing the resulting expression using Eqs.~\eqref{Relacion de Y con Z 2} and \eqref{Chowla-Selberg 1d} with $s = 2-\frac{1}{p}$, we obtain an infinite series on modified Bessel functions, which cannot be carried up to a closed analytical result in general. In any case, it is possible to obtain a closed expression for the limits of large and small adimensional variable $x = k_{0}L$. At long distances compared with the correlation length, that is $L\gg k_{0}^{-1}$, the force decays as
\begin{equation}
F_{C}/A = \frac{\Gamma k_{0} e^{-2Lk_{0}}}{4\pi C(p)\Gamma\left(2 - \frac{1}{p}\right)}\left(\frac{k_{0}}{L}\right)^{1/p}.
\end{equation}
In the opposite limit, when the plates are at a distance much smaller than the correlation length, or $L\ll k_{0}^{-1}$, the force is
\begin{equation}
F_{C}/A = \frac{\Gamma \sin\left(\frac{\pi}{p}\right)\Gamma\left(\frac{2}{p}\right)}{2^{1+2/p}\pi^{2}(p-1)C(p)}\frac{\zeta\left(1 + \frac{2}{p}\right)}{L^{1 + \frac{2}{p}}}.
\end{equation}
If $p=1$, we recover the results given \eqref{fza.cliq.white_k0L_grande} and \eqref{fza.cliq.white} respectively. For the interesting case $p=2$, we obtain, at long distances compared with the correlation length $L\gg k_{0}^{-1}$, the force
\begin{equation}
F_{C}/A = \frac{\Gamma k_{0} e^{-2Lk_{0}}}{16\pi}\sqrt{\frac{k_{0}}{L}},
\end{equation}
and in the opposite limit $L\ll k_{0}^{-1}$, the force is
\begin{equation}
F_{C}/A = \frac{\Gamma}{192 L^{2}}.
\end{equation}

\section{Discussion}\label{chapI3:Discussion}
In this Chapter we have derived the Casimir force between parallel plated immersed in a medium whose fluctuations are described by a Langevin equation with a complex temporal evolution differential term. The result obtained here is an example of the kind of calculus for Casimir forces that we can perform within the dynamic formalism of Casimir forces developed in this Thesis.

Obviously, the Langevin equation showed here describes out of equilibrium systems, because the fluctuations in equilibrium require a first order temporal derivative to describe the temporal evolution of such fluctuations.

We have also found that the two point correlation function in the steady state is an essentially different object of the Green function of such field. These two functions are related both each other only for Hermitian systems whose temporal evolution is described by a first order temporal derivative.

The shape of the obtained Green functions is interesting. When an eigenvalue could produce a divergence (in the infinite future or past), the Green function cancels its contribution in the problematic section of time either future or past. Therefore, spatial operators $\mathcal{M}$ with negative eigenvalues can be treated in this formalism.

It is important to remark that in this Chapter we do not claim that some kind of steady state is reached for the systems studied here. We have studied what would be the Casimir force between plates if such steady state exists and it is described by a generalized Langevin equation in temporal derivatives.

% Parte de Multiscattering
\chapter*{}
\markboth{}{} 
\addcontentsline{toc}{chapter}{Part II: Multiscattering formalism of Casimir effect}

\begin{center}
\vspace{-400pt}
\fontsize{300}{360}\selectfont {\sc\makebox[200pt][c]{II}}
\end{center}
\vspace{60pt}
\begin{center}
\fontsize{30}{36}\selectfont{\sc\expandafter{Multiscattering formalism\\of Casimir effect}}
\end{center}

% Introduccion al metodo
\begin{savequote}[10cm] % this sets the width of the quote
\sffamily
``Nature's great book is written in mathematical language.'' 
\qauthor{Galileo Galilei}
\end{savequote}

\chapter{Introduction to the multiscattering formalism of Casimir effect}\label{Cap_II-02}
\graphicspath{{02-Casimir_Multiscattering/chb1_Introduccion_Multiscattering/}}

In this Chapter we present a short introduction to the multiscattering formalism of Casimir effect. This formalism was first introduced by Bulgac et. al. in \cite{Functional_Determinant_Method_2}. It was generalized to the electromagnetic (EM) Casimir effect by Emig et. al. in \cite{Kardar-Geometrias-Arbitrarias} and \cite{RE09}.

This formalism is general, and let us to obtain the Casimir energy of a system of $N$ generalized dielectrics of arbitrary shape for any given temperature. In particular, Casimir energy between perfect metal parallel plates~\cite{Casimir_Placas_Paralelas} and the more general Lifshitz formula~\cite{Review_Casimir} are derived from this formalism. London energy~\cite{Energia_London}, and retarded limit of Van der Waals energy given by Casimir and Polder in~\cite{VdW-int.electrica}, and the more general case of Feinberg and Sucher potential~\cite{VdW_int.magnetica} are covered too. The Casimir energy of a system inside a container can also be obtained~\cite{Cuerpo_dentro_de_otro}.

The contents of this Chapter is not an original work of the author of this Thesis. It is a theoretical introduction of the multiscattering formalism of Casimir effect, needed to perform the (original) results obtained in following Chapters of this Thesis. This introduction is based on~\cite{Kardar-Geometrias-Arbitrarias}\cite{RE09}\cite{Emig:2008ee}.
\section{Partition function description of the problem}
The system under study is described as follows. There is a thermal bath at a given temperature $T$ (the case $T = 0$ is included), within which there are $N$ stationary dielectrics. The EM action is
\begin{equation}\label{Accion_EM}
S\left[A_{\mu}\right] = \frac{1}{2}\int dx^{\lambda} A^{*}_{\mu}(x^{\lambda})\left[\frac{1}{c^{2}}\partial_{t}^{2} - \Delta\right]A^{\mu}(x^{\lambda}),
\end{equation}
and the boundary conditions of the EM fields are the continuity across the surface of the dielectric of the transversal components of $\textbf{E}$ and $\textbf{H}$ fields
\begin{equation}\label{Condiciones_Dielectrico}
\textbf{E}_{in}\times\hat{\textbf{n}} = \textbf{E}_{out}\times\hat{\textbf{n}},
\hspace{2.0cm}
\textbf{H}_{in}\times\hat{\textbf{n}} = \textbf{H}_{out}\times\hat{\textbf{n}}.
\end{equation}
Having into account that these boundary conditions are linear in electric and magnetic fields, and that these fields have a linear relationship with the potential vector $A^{\mu}$, then the partition function of the constrained $A^{\mu}$ field can be written symbolically as
\begin{equation}
\mathcal{Z} = \int \mathcal{D}A^{\mu}(x^{\lambda}) \prod_{\alpha=1}^{N}\delta\left[C_{\alpha}A^{\mu}(x^{\lambda})\right]
e^{\frac{i}{2\hbar}\int dx^{\lambda} A^{*}_{\mu}(x^{\lambda})\left[\frac{1}{c^{2}}\partial_{t}^{2} - \Delta\right]A^{\mu}(x^{\lambda})}.
\end{equation}
The functional Dirac delta for each dielectric $\alpha$ can be written, by means a functional Fourier transformation, as a functional integral over the induction currents $j_{\mu}^{\alpha}(x^{\lambda})$ inside the volume and surface of each dielectric $\alpha$. The origin of $j_{\mu}^{\alpha}(x^{\lambda})$ are the imposed boundary conditions over the surface of each dielectric.
\begin{equation}
\delta\left[C_{\alpha}A^{\mu}(x^{\lambda})\right] = \int \mathcal{D}j_{\mu}^{\alpha}(x^{\lambda})
e^{\frac{i}{\hbar}\int dx^{\lambda}j_{\mu}^{\alpha *}(x^{\lambda})A^{\mu}(x^{\lambda})}.
\end{equation}
Then the partition function of our problem is
\begin{equation}
\mathcal{Z} = \int \mathcal{D}A^{\mu}(x^{\lambda}) \prod_{\alpha=1}^{N}\int \mathcal{D}j_{\mu}^{\alpha}(x^{\lambda})
e^{\frac{i}{2\hbar}\int dx^{\lambda} A^{*}_{\mu}(x^{\lambda})\left[\frac{1}{c^{2}}\partial_{t}^{2} - \Delta\right]A^{\mu}(x^{\lambda}) + \frac{i}{\hbar}\sum_{\alpha=1}^{N}\int dx^{\lambda}j_{\mu}^{\alpha *}(x^{\lambda})A^{\mu}(x^{\lambda})}.
\end{equation}

%\section{Wick Rotation}
In order to carry out the functional integrals that appear in Quantum Field Theory (QFT), we must apply a Wick rotation of them, which also let us impose the temperature of the thermal bath explicitly. This rotation also connect the QFT with Statistical Physics by means of the Matsubara formalism~\cite{Thermal-Field-Theory}. Then we apply the Wick rotation by a change of variable to an adimensional imaginary time $x^{0} = ict$. Then the EM action \eqref{Accion_EM} Wick rotated is transformed to
\begin{equation}
\frac{i}{\hbar}S\left[A_{\mu}\right]_{x^{0} = ict} = - \frac{1}{2\hbar c}\int_{0}^{\beta\hbar c} dx^{0}\int d\textbf{x} A^{*}_{\mu}\left[\partial_{0}^{2} + \Delta\right]A^{\mu} = - \frac{1}{2\hbar c}S_{E}.
\end{equation}
Note that the same Wick transformation was applied in Chap.~\ref{Chap: Stochastic Quantization and Casimir forces.} of this Thesis, where the Parisi--Wu formalism was used to derive the Casimir force in the EM field.
Because of Matsubara formalism, bosonic fields (as EM or scalar field) are periodic in the interval $x^{0}\in\left[0, \beta\hbar c\right]$, then we can Fourier expand over this interval to obtain 
\begin{equation}\label{Factorizacion_tiempo_imaginario}
A^{\mu}(x^{0},\textbf{x}) = \sum_{n\in\mathbb{Z}}A^{\mu}_{n}(\textbf{x})e^{i\kappa_{n}x^{0}}.
\end{equation}
Periodicity condition over $x^{0}$ imposes $\kappa_{n} = \frac{2\pi}{\beta\hbar c}n$, with $n\in\mathbb{Z}$. In addition to that, if $n\neq 0$, then $A^{\mu}_{-n}(\textbf{x}) = A^{\mu *}_{n}(\textbf{x})$, then we can factorize the functional differential to
\begin{equation}
\mathcal{D}A^{\mu}(x^{0},\textbf{x}) = \mathcal{D}A^{\mu}_{0}(\textbf{x})\prod_{n=1}^{\infty}\mathcal{D}A^{\mu}_{n}(\textbf{x})\mathcal{D}A^{\mu *}_{n}(\textbf{x}),
\end{equation}
while the euclidean action is expanded to the infinity series
\begin{equation}
S_{E} = \beta\hbar c\sum_{n=0}^{\infty}\left(\int d\textbf{x}A^{\mu *}_{n}(\textbf{x})\left[\kappa_{n}^{2} + \Delta \right]A_{\mu,n}(\textbf{x}) + \sum_{\alpha=1}^{N}\int dx^{\lambda}j_{\mu, n}^{\alpha *}(\textbf{x})A^{\mu}_{n}(\textbf{x}) + c.c.\right),
\end{equation}
where $c.c.$ means the complex conjugate of the written expression. Then the partition function is factorized in Matsubara frequencies
\begin{equation}\label{Factorizacion_Z}
\mathcal{Z} = \prod_{n=0}^{\infty}\mathcal{Z}_{n},
\end{equation}
where each partition function has the general form
\begin{align}\label{Modo_Z_neq_0}
\mathcal{Z}_{n\neq 0} & = \int \mathcal{D}A_{n}^{\mu}(\textbf{x})\mathcal{D}A_{n}^{\mu *}(\textbf{x}) \prod_{\alpha=1}^{N}\int \mathcal{D}j_{n,\mu}^{\alpha}(\textbf{x})\mathcal{D}j_{n,\mu}^{\alpha *}(\textbf{x})\times\nonumber\\
& \times e^{- \frac{\beta}{2}\left[\int dx^{\lambda} A_{n,\mu}^{*}(\textbf{x})\left(\kappa_{n}^{2} + \Delta\right)A_{n}^{\mu}(\textbf{x}) + \sum_{\alpha=1}^{N}\int d\textbf{x}j_{n,\mu}^{\alpha *}(\textbf{x})A_{n}^{\mu}(\textbf{x}) + c.c.\right]}.
\end{align}
Here some caution is needed with the zero Matsubara frequency, because in this case we have just a real field without complex conjugate contribution. This will result in a $1/2$ weight in its contribution to the Casimir effect.
We can carry out these Gaussian integrals in $A_{n}^{\mu}(\textbf{x})$ to obtain
\begin{align}\label{Gaussian_Integral_Complex_Integrands}
& \int \mathcal{D}A_{n}^{\mu}(\textbf{x})\mathcal{D}A_{n}^{\mu *}(\textbf{x})
e^{- \frac{\beta}{2}\int d\textbf{x} A_{n,\mu}^{*}(\textbf{x})\left(\kappa_{n}^{2} + \Delta\right)A_{n}^{\mu}(\textbf{x}) - j_{n,\mu}^{*}(\textbf{x})A_{n}^{\mu}(\textbf{x}) + c.c.} = \nonumber\\
& \frac{1}{\Det{\kappa_{n}^{2} + \Delta}}e^{-\frac{1}{2\beta}\int d\textbf{x}\int d\textbf{y}j_{n,\mu}^{*}(\textbf{x})\mathcal{G}_{0}(\textbf{x},\textbf{y},\kappa_{n})j_{n}^{\mu}(\textbf{y}) + c.c.},
\end{align}
where $\mathcal{G}_{0}(\textbf{x},\textbf{y},\kappa_{n})$  is the dyadic Green function, which takes into account the continuity condition of the induction currents. It can be written for imaginary frequency as
\begin{equation}
\mathcal{G}_{0}(\textbf{x},\textbf{x}',\kappa) = G_{0}(\textbf{x},\textbf{x}',\kappa) -  \frac{1}{\kappa^{2}}\nabla\otimes\nabla 'G_{0}(\textbf{x},\textbf{x}',\kappa),
\end{equation}
where $G_{0}(\textbf{x},\textbf{x}',\kappa) = \frac{1}{4\pi}\frac{e^{-\kappa\abs{\textbf{x} - \textbf{x}'}}}{\abs{\textbf{x} - \textbf{x}'}}$ is the Wick rotated Green function for the scalar field. Then Eq.~\eqref{Modo_Z_neq_0} is reduced to
\begin{equation}\label{Modo_Z_neq_0_2}
\mathcal{Z}_{n\neq 0} = \prod_{\alpha=1}^{N}\int \mathcal{D}j_{n,\mu}^{\alpha}(\textbf{x}_{\alpha})\mathcal{D}j_{n,\mu}^{\alpha *}(\textbf{x}_{\alpha})
e^{-\frac{1}{2\beta}\sum_{\alpha=1}^{N}\sum_{\beta=1}^{N}\int_{\alpha} d\textbf{x}_{\alpha}\int_{\beta} d\textbf{x}_{\beta}\left[j_{n,\mu}(\textbf{x}_{\alpha})\mathcal{G}_{0}(\textbf{x}_{\alpha},\textbf{x}_{\beta},\kappa_{n})j_{n}^{\mu}(\textbf{x}_{\beta}) + c.c.\right]}.
\end{equation}

\section{Free energy of the system}
To solve Eq.~\eqref{Modo_Z_neq_0_2}, we apply a multipolar expansion of the dyadic Green function in a given base with label $\ell$ as
\begin{equation}\label{Expansion_Multiplor_Funcion_Geen_Dyadica}
\mathcal{G}_{0}(\textbf{x},\textbf{x}',\kappa_{n}) = \sum_{\ell}C_{\ell}\left[ \textbf{M}_{\ell}^{out}(\textbf{x},\kappa_{n})\otimes\textbf{M}_{\ell}^{* in}(\textbf{x}',\kappa_{n}) - \textbf{N}_{\ell}^{out}(\textbf{x},\kappa_{n})\otimes\textbf{N}_{\ell}^{* in}(\textbf{x}',\kappa_{n}) \right],
\end{equation}
where we are assuming that $\textbf{x} > \textbf{x}'$. The quantities $\textbf{M}_{\ell}(\textbf{x},\kappa_{n}) = \nabla\times\left[\phi_{\ell}(\textbf{x},\kappa_{n})\textbf{x}\right]$ and $\textbf{N}_{\ell}(\textbf{x},\kappa_{n}) = \frac{1}{\kappa}\nabla\times\textbf{M}_{\ell}(\textbf{x},\kappa_{n})$ are the vector multipoles at imaginary frequency, which correspond to the two polarizations of the EM field. $\text{TM}$ polarization is given by $\textbf{M}_{\ell}(\textbf{x},\kappa_{n})$, while $\text{TE}$ polarization is given by $\textbf{N}_{\ell}(\textbf{x},\kappa_{n})$. $\phi_{\ell}(\textbf{x},\kappa_{n})$ is the scalar multipole in the same coordinate system. We can find this Dyadic Green functions and their multipolar expansions for different coordinate systems in \cite{RE09}\cite{Casimir_Wedges}\cite{Morse_FeshbachI}\cite{Morse_FeshbachII}.

And we apply a change of variable from induction currents to external multipoles moments. These external multipoles moments for an object $\alpha$ are defined in terms of the induction currents as
\begin{equation}\label{Momentos_Multipolares_Externos}
Q_{M, \ell}^{\alpha}(\kappa_{n}) = \int d\textbf{x}_{\alpha} j_{\alpha}(\textbf{x}_{\alpha})\textbf{M}_{\ell}^{* in}(\textbf{x}_{\alpha},\kappa_{n}),
\hspace{1.0cm}
Q_{E, \ell}^{\alpha}(\kappa_{n}) = \int d\textbf{x}_{\alpha} j_{\alpha}(\textbf{x}_{\alpha})\textbf{N}_{\ell}^{* in}(\textbf{x}_{\alpha},\kappa_{n}).
\end{equation}
We will also need the internal multipole moments, defined as
\begin{equation}\label{Momentos_Multipolares_Internos}
\phi_{M, \ell}^{\alpha}(\kappa_{n}) = \int d\textbf{x}_{\alpha} j_{\alpha}(\textbf{x}_{\alpha})\textbf{M}_{\ell}^{* out}(\textbf{x}_{\alpha},\kappa_{n}),
\hspace{1.0cm}
\phi_{E, \ell}^{\alpha}(\kappa_{n}) = \int d\textbf{x}_{\alpha} j_{\alpha}(\textbf{x}_{\alpha})\textbf{N}_{\ell}^{* out}(\textbf{x}_{\alpha},\kappa_{n}).
\end{equation}
Then, following the formalism developed in \cite{Kardar-Geometrias-Arbitrarias}\cite{RE09}\cite{Emig_Casimir_caso_escalar}\cite{Emig:2008ee}, the action of the problem given in Eq.~\eqref{Modo_Z_neq_0_2} as
\begin{equation}
\frac{-1}{2\beta}\sum_{\alpha=1}^{N}\sum_{\beta=1}^{N}S_{\alpha\beta} = 
\frac{-1}{2\beta}\sum_{\alpha=1}^{N}\sum_{\beta=1}^{N}\int_{\alpha} d\textbf{x}_{\alpha}\int_{\beta} d\textbf{x}_{\beta}
\left[j_{\alpha,\mu}^{*}(\textbf{x}_{\alpha})\mathcal{G}_{0}(\textbf{x}_{\alpha},\textbf{x}_{\beta},\kappa_{n})j_{\beta}^{\mu}(\textbf{x}_{\beta}) + c.c.\right],
\end{equation}
is transformed in terms of external multipole moments as
\begin{equation}\label{Acciones_Total}
\frac{-1}{2\beta}\sum_{\alpha=1}^{N}\sum_{\beta=1}^{N}S_{\alpha\beta} = 
\frac{-1}{\beta}\sum_{\alpha=1}^{N}\sum_{\beta=1}^{N}
Q_{\alpha}^{*}\mathbb{M}^{\alpha\beta}Q_{\beta},
\end{equation}
with $\mathbb{M}$ matrix defined as
\begin{equation}\label{Definicion_Matriz_M}
\mathbb{M}^{\alpha\beta} = (\mathbb{T}^{\alpha})^{-1}\delta^{\alpha\beta} +  \left(1 - \delta^{\alpha\beta}\right)\mathbb{U}^{\alpha\beta}.
\end{equation}
The $\mathbb{M}^{\alpha\beta}$ matrix contains all the information of our system. Diagonal terms contain the inverse of scattering $\mathbb{T}$ matrix of each object, which take into account the geometry and dielectric properties of the considered object. Non diagonal terms contain translation $\mathbb{U}$ matrices, which take into account the relative positions and orientations between objects. If these objects are immersed inside a container, the formalism changes a bit \cite{Casimir_cuerpo_dentro_de_esfera}\cite{RE09}\cite{Wittmann}.

After applying the change of variable to external multipole to the partition function, Eq.~\eqref{Modo_Z_neq_0_2} is transformed to
\begin{equation}\label{Modo_Z_neq_0_2b}
\mathcal{Z}_{n\neq 0} = \prod_{P=E,M}\prod_{\alpha=1}^{N}\prod_{\ell=1}^{\infty}\int \mathcal{D}Q_{P,\ell}^{\alpha *}(\kappa_{n})\mathcal{D}Q_{P,\ell}^{\alpha}(\kappa_{n})
e^{\frac{-1}{\beta}Q_{\alpha}^{*}\mathbb{M}^{\alpha\beta}(\kappa_{n})Q_{\beta}}.
\end{equation}
This is a Gaussian integral, which can be carried out to give
\begin{equation}\label{Modo_Z_neq_0_3}
\mathcal{Z}_{n\neq 0} = \frac{1}{\Det{\mathbb{M}^{\alpha\beta}(\kappa_{n})}}.
\end{equation}
For the zero Matsubara frequency, we have
\begin{equation}\label{Modo_Z_eq_0_3}
\mathcal{Z}_{n = 0} = \frac{1}{\sqrt{\Det{\mathbb{M}^{\alpha\beta}(0)}}},
\end{equation}
because $A^{\mu}_{n=0}(\textbf{x})$ is real.

Helmholtz free energy is defined in terms of the partition function as
\begin{equation}\label{Energia_Libre_Helmholtz_1}
e^{-\beta\mathcal{F}} = \mathcal{Z} = \prod_{n=0}^{\infty}\mathcal{Z}_{n}
\hspace{1.0cm}
\Rightarrow
\hspace{1.0cm}
\mathcal{F} = - \frac{1}{\beta}\sum_{n=0}^{\infty}\log\left(\mathcal{Z}_{n}\right),
\end{equation}
where we have factorized $\mathcal{Z}$ in Matsubara frequencies given in Eq.~\eqref{Factorizacion_Z}. Using Eqs.~\eqref{Modo_Z_neq_0_3} and \eqref{Modo_Z_eq_0_3}, we obtain
\begin{equation}\label{Energia_Libre_Helmholtz_2}
\mathcal{F}_{T} = \frac{1}{\beta}{\sum_{n=0}^{\infty}}'\log\Det{\mathbb{M}^{\alpha\beta}(\kappa_{n})},
\end{equation}
where the prime indicates that the zero Matsubara frequency has a $1/2$ weight. Casimir energy is the excess of free energy because of the relative distance between objects. As a consequence, we regularize Eq.~\eqref{Energia_Libre_Helmholtz_2} eliminating the contribution to the free energy which does not depend on the distance. If we subtract to Eq.~\eqref{Energia_Libre_Helmholtz_2} the energy when the objects are infinitely far away form each other, the Casimir energy is defined as
\begin{equation}\label{Energia_Libre_Helmholtz_3}
\mathcal{F}_{T} = \frac{1}{\beta}{\sum_{n=0}^{\infty}}'\log\Det{\mathbb{M}^{\alpha\beta}(\kappa_{n})} - \frac{1}{\beta}{\sum_{n=0}^{\infty}}'\log\Det{\mathbb{M}_{\infty}^{\alpha\beta}(\kappa_{n})},
\end{equation}
with the definition $\mathbb{M}^{\alpha\beta}_{\infty}(\kappa_{n}) = \lim_{\abs{X_{\alpha\beta}}\to\infty}\mathbb{M}^{\alpha\beta}(\kappa_{n})$. This is a diagonal matrix whose determinant is equal to $\Det{\mathbb{M}^{\alpha\beta}_{\infty}(\kappa_{n})} = \prod_{\alpha=1}^{N}\Det{\tilde{\mathbb{T}}^{\alpha}}^{-1}$, because non-diagonal components are zero ($\lim_{\abs{X_{\alpha\beta}}\to\infty}\tilde{\mathbb{U}}^{\alpha\beta}(\kappa_{n}) = 0$). It is easy to see that Helmholtz free energy can be written as
\begin{equation}\label{Energia_Libre_Helmholtz_3}
\mathcal{F}_{T} = \frac{1}{\beta}{\sum_{n=0}^{\infty}}'\log\Det{\mathbbm{1} - \mathbb{N}(\kappa_{n})},
\end{equation}
where $\mathbb{N}^{\alpha\beta}(\kappa_{n})$ is a characteristic matrix of the system that goes to zero when $\kappa_{n}\to\infty$.
The $\left(\mathbbm{1} - \mathbb{N}^{\alpha\beta}\right)$ matrix can be written as
\begin{equation}\label{Definicion_Matriz_M}
\left(\mathbbm{1} - \mathbb{N}^{\alpha\beta}\right) = \delta^{\alpha\beta} +  \left(1 - \delta^{\alpha\beta}\right)\mathbb{T}^{\alpha}\mathbb{U}^{\alpha\beta}.
\end{equation}
By the use of Abel--Plana formula~\cite{Libro-Residuos}, this Casimir energy can be written as
\begin{equation}\label{Energia_Libre_Helmholtz_4}
\mathcal{F}_{T} = \frac{\hbar c}{2\pi}\int_{0}^{\infty} d\kappa\log\Det{\mathbbm{1} - \mathbb{N}(\kappa)} - \frac{\hbar c}{2\pi}\int_{0}^{\infty}\frac{d\kappa}{e^{\beta\hbar c\kappa} - 1}\frac{1}{2i}\log\left(\frac{\Det{\mathbbm{1} - \mathbb{N}(i\kappa)}}{\Det{\mathbbm{1} - \mathbb{N}(-i\kappa)}}\right).
\end{equation}
In the zero temperature limit ($T\to 0$), we obtain the celebrated formula for the quantum Casimir energy
\begin{equation}\label{Energia_Libre_Helmholtz_5}
\mathcal{F}_{0} = \frac{\hbar c}{2\pi}\int_{0}^{\infty} d\kappa\log\Det{\mathbbm{1} - \mathbb{N}(\kappa)},
\end{equation}
while in the classical limit ($\hbar\to 0$), equivalent to the high temperature limit ($T\to\infty$), the Casimir energy is given by the zeroth Matsubara frequency contribution
\begin{equation}\label{Energia_Libre_Helmholtz_6}
\mathcal{F}_{cl} = \frac{k_{B}T}{2}\log\Det{\mathbbm{1} - \mathbb{N}(0)}.
\end{equation}
In the following chapters, we will use the formulas showed above and variations of them to study different properties of EM Casimir, as the appareance of negative entropy because the Casimir effect between matel spheres, defferent geometries as the case of non--parallel cylinders or even different limits, as the Pairwise Summation Approximation.

% Nonmonotonic forces between spheres in presence of a plate
\begin{savequote}[5cm] % this sets the width of the quote
\sffamily
``Three's a crowd.'' 
\qauthor{Anonymous}
\end{savequote}

\newcommand{\cZ}{\mathcal Z}
\newcommand{\cD}{\mathcal D}
\newcommand{\cC}{\mathcal C}
\newcommand{\cG}{\mathcal G}
\newcommand{\cE}{\mathcal E}
\newcommand{\br}{\mathbf r}
\newcommand{\bk}{\mathbf k}
\newcommand{\bn}{\mathbf n}
\newcommand{\bp}{\mathbf p}
\newcommand{\bz}{\mathbf z}
\newcommand{\bA}{\mathbf A}
\newcommand{\bE}{\mathbf E}
\newcommand{\bB}{\mathbf B}
\newcommand{\bD}{\mathbf D}
\newcommand{\bJ}{\mathbf J}
\newcommand{\bH}{\mathbf H}
\newcommand{\bM}{\mathbf M}
\newcommand{\bN}{\mathbf N}
\newcommand{\bY}{\mathbf Y}
\newcommand{\bQ}{\mathbf Q}
\newcommand{\bP}{\mathbf P}
\newcommand{\bphi}{\mbox{\boldmath$\phi$}}
\newcommand{\bpsi}{\mbox{\boldmath$\psi$}}
\newcommand{\hr}{\hat{\mathbf r}}
\newcommand{\hz}{\hat{\mathbf z}}

\newcommand{\ra}{\rightarrow}
\newcommand{\pd}{\partial}
\newcommand{\til}[1]{\tilde{#1}}

\definecolor{BrickRed}{cmyk}{0,0.89,0.94,0.28}%%%PANTONE 1805
\definecolor{MidnightBlue}{cmyk}{0.98,0.13,0,0.43}%%%PANTONE 302
\definecolor{DarkGreen}{rgb}{0,0.7,0.1}

\chapter{Three-body Casimir effects and non-monotonic forces}\label{chb2}
\graphicspath{{02-Casimir_Multiscattering/chb2_Plano_2_esferas/Figuras/}} 

In this Chapter, we study the not pairwise additive behavior of Casimir interactions.
  This non pairwise additive property leads
  to collective effects that we study for a pair of objects near a
  conducting wall. We employ a scattering approach~\cite{Emig:2008ee} to compute the
  interaction in terms of fluctuating multipoles. The wall can lead to
  a non-monotonic force between the objects. For two atoms with
  anisotropic electric and magnetic dipole polarizabilities we
  demonstrate that this non-monotonic effect results from a
  competition between two and three body interactions. By including
  higher order multipoles we obtain the force between two macroscopic
  metallic spheres for a wide range of sphere separations and
  distances to the wall.

A hallmark property of dispersion forces is their non-additivity which
clearly distinguishes them from electromagnetic forces between charged
particles \cite{Parsegian:2005eu}.
Work on the interactions between multiple objects is limited mostly to atoms 
or small particles which are described well in dipole approximation \cite{Power:1982cs}.
This approximation cannot be used for macroscopic objects at separations
that are comparable to their size since higher order multipole
fluctuations have to be included \cite{Feinberg:1974xw}\cite{Kardar-Geometrias-Arbitrarias}.
In such situations, also other common ``additive'' methods such as 
 proximity or two-body-interaction approximations fail. 
Three-body effects for macroscopic bodies have been studied in quasi two-dimensional (2D)
 geometries that are composed of parallel perfect metal cylinders of quadratic
\cite{Rodriguez:2007a} or circular \cite{Rahi:2008bv}\cite{Rahi:2008kb}
cross section and parallel sidewalls. 
For this setup non-monotonic forces have been found and interpreted as resulting
from a competition between electric and magnetic polarizations which are decoupled
for quasi 2D geometries of perfect metal structures.
In this Chapter we investigate collective 3-body effects between compact objects,
including anisotropic polarizabilities, and a wall in three dimensions
using the multiscattering approach \cite{Kardar-Geometrias-Arbitrarias}\cite{Emig:2008ee}\cite{RE09}.
This allows us to observe the influence of polarization coupling and anisotropy on non-monotonic effects.

\begin{figure}[htb]
\begin{center}
\includegraphics[width=0.5\linewidth]{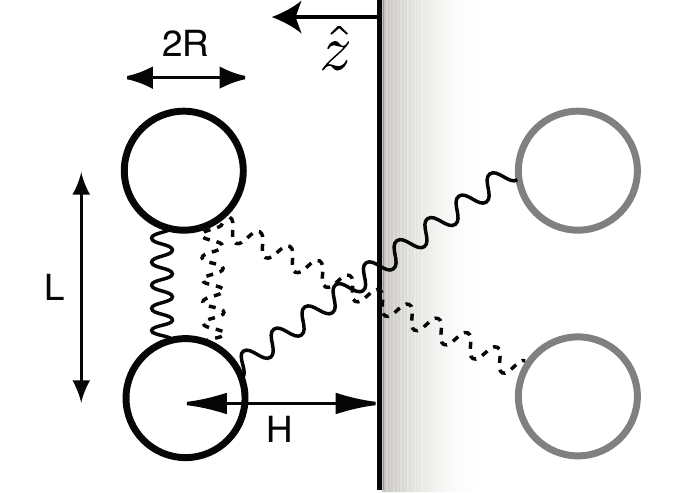}
\end{center}
  \caption{\label{fig:1}Geometry of the two-sphere/atom and sidewall
    system. Shown are also the mirror images (grey) and two- and
    three-body contributions (solid and dashed curly lines, respectively).}
\end{figure}

We consider the retarded Casimir interaction between a pair of atoms
with anisotropic electric and magnetic polarizabilities near a conducting wall, see Fig.~\ref{fig:1}.
We identify a competition between 2- and 3-body effects and prove that this leads to a
non-monotonic dependence of the force between the atoms on the wall
separation $H$ for each of the four possible polarizations of
fluctuations (electric/magnetic and parallel/perpendicular to the
wall) separately.
For isotropic polarizabilities we find that only the force component due to electric fluctuations is non-monotonic in $H$.
%This findings suggest the possibility to engineer the monotonicity properties of the force by suitable tailoring of the polarization tensors via the shape and material composition of macroscopic particles.

For atoms, magnetic effects are almost always rather small in the retarded limit. 
Contrary to this, for conducting macroscopic objects contributions from electric and magnetic multipole fluctuations are comparable.
To study the effect of higher-order multipoles, we consider also two perfect metal spheres near a wall, see Fig.~\ref{fig:1}.
Based on consistent analytical results for large separations and numerical computations at smaller distances we
find a non-monotonic dependence of the force between the spheres on $H$.
Unlike for atoms, this effect occurs at sufficiently large sphere separations only.

We follow this plan for the Chapter:
In Sect.~\ref{General Approach_chapII-2} we present the formalism used to describe the system of 2 compact objects (asymmetric atoms or perfect metal spheres) in presence of a perfect metal plate. It is a variation of the multiscattering formalism~\cite{Kardar-Geometrias-Arbitrarias}\cite{RE09} presented in \cite{Emig:2008ee}, where the multiscattering formalism and image method are used together.
In Sect.~\ref{Two atoms_chapII-2} the appearance of nonmnotonic forces between anisotropic atoms because the presence of a perfect metal plate is studied.
In Sect.~\ref{Two macroscopic metallic spheres_chapII-2} we study the appearance of nonmonotonic forces between perfect metal spheres in the presence of a perfect metal plate, analytically and numerically, and how this effect is modified with the distance between the spheres and with the plate.
Finally, the relevance of the effect presented here is discussed in Sect.~\ref{Discussion_chapII-2}.

The contents of this Chapter is based on the work published in~\cite{Rodriguez-Lopez_1}.

\section{General approach}\label{General Approach_chapII-2}
As presented in Ref.~\cite{Emig:2008ee} the Casimir energy of two bodies in the presence of a perfectly conducting sidewall can be obtained using the scattering approach by employing the method of images, which introduces fluctuating currents on the mirror bodies.
The
Casimir energy of the original system is then given by the energy of
the original and the image objects and it can be expressed as
an integral over imaginary wave number,
\begin{equation}
  \label{eq:energy}
  \mathcal{E} = \frac{\hbar c}{2\pi} \int_0^\infty d\kappa \,\log \Det{ {\mathbb M}\, {\mathbb M}_\infty^{-1}}
\end{equation}
with the matrix
\begin{equation}
  \label{eq:M-matrix}
  {\mathbb M} = \begin{pmatrix}
 {\mathbb T}^{-1} + {\mathbb U}^{I,11} &  {\mathbb U}^{12} + {\mathbb U}^{I,12} \\
{\mathbb U}^{21} + {\mathbb U}^{I,21} &  {\mathbb T}^{-1} + {\mathbb U}^{I,22} 
\end{pmatrix} \, ,
\end{equation}
which is given by the ${\mathbb T}$-matrix that relates the regular and scattered electromagnetic (EM) fields for each body, and by the $\mathbb U$-matrices that describe the interaction between the multipoles of object $\alpha$ and object $\beta$, ${\mathbb U}^{\alpha\beta}$, and between the multipoles of object $\alpha$ and the image of object $\beta$, ${\mathbb U}^{I,\alpha\beta}$.
The $\mathbb{T}$-matrices depend only on the properties of the individual bodies such as polarizability, size, and shape.
The ${\mathbb U}$-matrices depend only on the distance vector between the objects and decay exponentially with distance and wave number $\kappa$.
The matrix ${\mathbb
  M}_\infty$ accounts for the subtraction of the object's
self-energies and hence follows from ${\mathbb M}$ by taking the limit
of infinite separations, i.e., by setting all $\mathbb{U}$-matrices to zero. For
a multipole expansion the matrix elements are computed in a vector
spherical basis for the EM field with partial wave numbers $l\ge 1$,
$m=-l,\ldots, l$~\cite{Emig:2008ee}.

In the following we study the force $F = -\partial \mathcal{E}/\partial L$
between the two objects at separation $L$ and hence eliminate the
contributions to the energy that depend only on the sidewall
separation $H$, see Fig.~\ref{fig:1}. We expand the determinant of Eq.~\eqref{eq:energy} as
\begin{align}
\Det{\mathbb{M}}\Det{\mathbb{M}}_{\infty}^{-1} & = 
\Det{\mathbbm{1} +  \mathbb{T}\mathbb{U}^{I} }
\Det{\mathbbm{1} +  \mathbb{T}\mathbb{U}^{I} }\times\nonumber\\
& \times\Det{ \mathbbm{1} - ( \mathbbm{1} +  \mathbb{T}\mathbb{U}^{I} )^{-1}
  \mathbb{T} ( \mathbb{U}^{21} + \mathbb{U}^{I,21} ) 
 ( \mathbbm{1} +  \mathbb{T} \mathbb{U}^{I} )^{-1} 
  \mathbb{T} ( \mathbb{U}^{12} + \mathbb{U}^{I,12} )}\label{eq:det-M-1}.
\end{align}

The first two determinants on Eq.~\eqref{eq:det-M-1} yield together twice the
interaction energy between a single object and the sidewall, since
${\mathbb U}^{I}\equiv{\mathbb U}^{I,11}= {\mathbb U}^{I,22}$
describes the multipole coupling between one object and its image and
hence depends only on $H$. Hence, we consider only the energy
$\mathcal{E}_{\underline{\circ\circ}}$ that corresponds to the last
determinant of Eq.~\eqref{eq:det-M-1} and provides the potential
energy of the two objects in the presence of the sidewall so that
$F=-\partial \mathcal{E}_{\underline{\circ\circ}}/\partial L$. In the absence
of the sidewall, $H\to\infty$, the matrices ${\mathbb
  U}^{I,\alpha\beta}$ all vanish and $\mathcal{E}_{\underline{\circ\circ}}$
simplifies to the energy between two spheres \cite{Kardar-Geometrias-Arbitrarias}. For
an interpretation in terms of multiple scatterings, it is instructive
to use the relation $\log\Det{A} = \tr\log(A)$ and to Taylor expand the logarithm and the inverse matrices,
\begin{equation}\label{eq:energy-2}
  \mathcal{E}_{\underline{\circ\circ}} = - \frac{\hbar c}{2\pi} \int_{0}^{\infty}d\kappa
\sum_{p=1}^{\infty}\frac{1}{p} \tr\left(\begin{array}{l}
\sum_{n=0}^\infty (-1)^{n} ( \mathbb{T}\mathbb{U}^{I})^{n} 
  \mathbb{T} ( \mathbb{U}^{21} + \mathbb{U}^{I,21} ) \times\\
 \times\sum_{n'=0}^{\infty} (-1)^{n'} ( \mathbb{T}\mathbb{U}^{I})^{n'} 
  \mathbb{T} ( \mathbb{U}^{12} + \mathbb{U}^{I,12} )
\end{array}\right)^{p}.
\end{equation}
The trace acts on an alternating product of $\mathbb{T}$- and
$\mathbb{U}$-matrices which describe scattering and free propagation of EM
fluctuations, respectively. Multiple scatterings between an object and
its image (${\mathbb T} {\mathbb U}^I$) are followed by a propagation
to the other object--image pair, either to the object (${\mathbb
  U}^{21}$) or its image (${\mathbb U}^{I,21}$), between which again
multiple scatterings occur before the fluctuations are scattered back
to the initial object or its image (${\mathbb U}^{12}$ or ${\mathbb
  U}^{I,12}$) and the process repeats. This expansion is useful for small
objects or large separations.

\section{Two atoms}\label{Two atoms_chapII-2}

First, we consider the case of two identical,
ground state atoms near a wall, see Fig.~\ref{fig:1}.
The separation between the atoms is $L$ and the separation of each of them from the wall is $H$.
In dipole approximation, the retarded limit of the interaction
is described by the static electric ($\alpha_z$, $\alpha_\|$) and
magnetic ($\beta_z$, $\beta_\|$) dipole polarizabilities of the atoms
which can be different in the directions perpendicular ($z$) and parallel ($\|$) to the
wall.  The $\mathbb{T}$-matrix of the atoms is diagonal and has finite elements
only for the dipole channel (partial waves with $l=1$), given by
\begin{equation}\label{TE_atomo}
T^{E}_{10} = \frac{2}{3} \alpha_{z}\kappa^{3}, 
\hspace{1.0cm}
T^{E}_{1m} = \frac{2}{3} \alpha_{\parallel}\kappa^{3}
\end{equation}
for electric and
\begin{equation}\label{TM_atomo}
T^{M}_{10} =
\frac{2}{3} \beta_{z}\kappa^{3},
\hspace{1.0cm}
T^{M}_{1m} = \frac{2}{3}\beta_{\parallel}\kappa^{3}
\end{equation}
for magnetic polarization with $m=\pm 1$. For atoms, the
polarizability is much smaller than $L^3$, and hence it is sufficient
to compute the interaction to second order in the
polarizabilities. This amounts to neglecting all terms other than $p=1$ and $n=n'=0$ in Eq.~\eqref{eq:energy-2}. The resulting energy
$\mathcal{E}_{\underline{\circ\circ}}$ is then compared to the well-known
Casimir--Polder (CP) interaction energy between two atoms (without the wall),
\begin{equation}\label{eq:E_CP}
\mathcal{E}_{2,|}(L) = -\frac{\hbar c}{8\pi L^{7}} \!\!
\left[ 33\alpha_{\parallel}^{2} +\! 13 \alpha_{z}^{2} - \! 14 \alpha_{\parallel}\beta_{z}
 + 33\beta_{\parallel}^{2} +\! 13 \beta_{z}^{2} - \! 14 \beta_{\parallel}\alpha_{z} \right],
\end{equation}
which corresponds to the sequence $ {\mathbb T} {\mathbb U}^{21}
{\mathbb T} {\mathbb U}^{12}$ in Eq.~\eqref{eq:energy-2}. The total
interaction energy is
\begin{equation}\label{eq:E_CP_plane}
  \mathcal{E}_{\underline{\circ\circ}}(L,H) = \mathcal{E}_{2,|}(L) + \mathcal{E}_{2,\backslash}(D,L) + \mathcal{E}_3(D,L) 
\end{equation}
with $D=\sqrt{L^2+4H^2}$. The 2-body energy $\mathcal{E}_{2,\backslash}(D,L)$
comes from the sequence $ {\mathbb T} {\mathbb U}^{I,21} {\mathbb T}
{\mathbb U}^{I,12}$ in Eq.~\eqref{eq:energy-2} and hence is the usual
CP interaction between one atom and the image of the other atom (see
Fig.~\ref{fig:1}).  The change in the relative orientation of the
atoms with $\ell=L/D$ leads to the modified CP potential
\begin{equation}\label{eq:E_CP_diag}
\mathcal{E}_{2,\backslash}(D,L) = -\frac{\hbar c}{8\pi D^{7}} \!\!\left[\begin{array}{l}
\!\! 26\alpha_{\|}^{2} +\! 20 \alpha_{z}^{2} -\! 14 \ell^{2} (4\alpha_{\|}^{2}-9\alpha_{\|}\alpha_{z} +5\alpha_{z}^{2})
\!+\! 63\ell^4 (\alpha_{\|} - \alpha_{z})^{2}  \\
\!\!+ 26\beta_{\|}^{2} +\! 20 \beta_{z}^{2} -\! 14 \ell^{2} (4\beta_{\|}^{2}-9\beta_{\|}\beta_{z} +5\beta_{z}^{2})
\!+\! 63\ell^4 (\beta_{\|} - \beta_{z})^{2}  \\
\!\!- 14\!\left(\alpha_{\|} \beta_{\|}(1\!-\!\ell^{2}) +\!\ell^{2} \alpha_{\|} \beta_{z} \!\right) - 14\!\left(\beta_{\|} \alpha_{\|}(1\!-\!\ell^{2}) +\!\ell^{2} \beta_{\|} \alpha_{z} \!\right)
\end{array}\!\!\right]\!\!.
\end{equation}

\begin{figure}[htb]
\includegraphics[width=1.\linewidth]{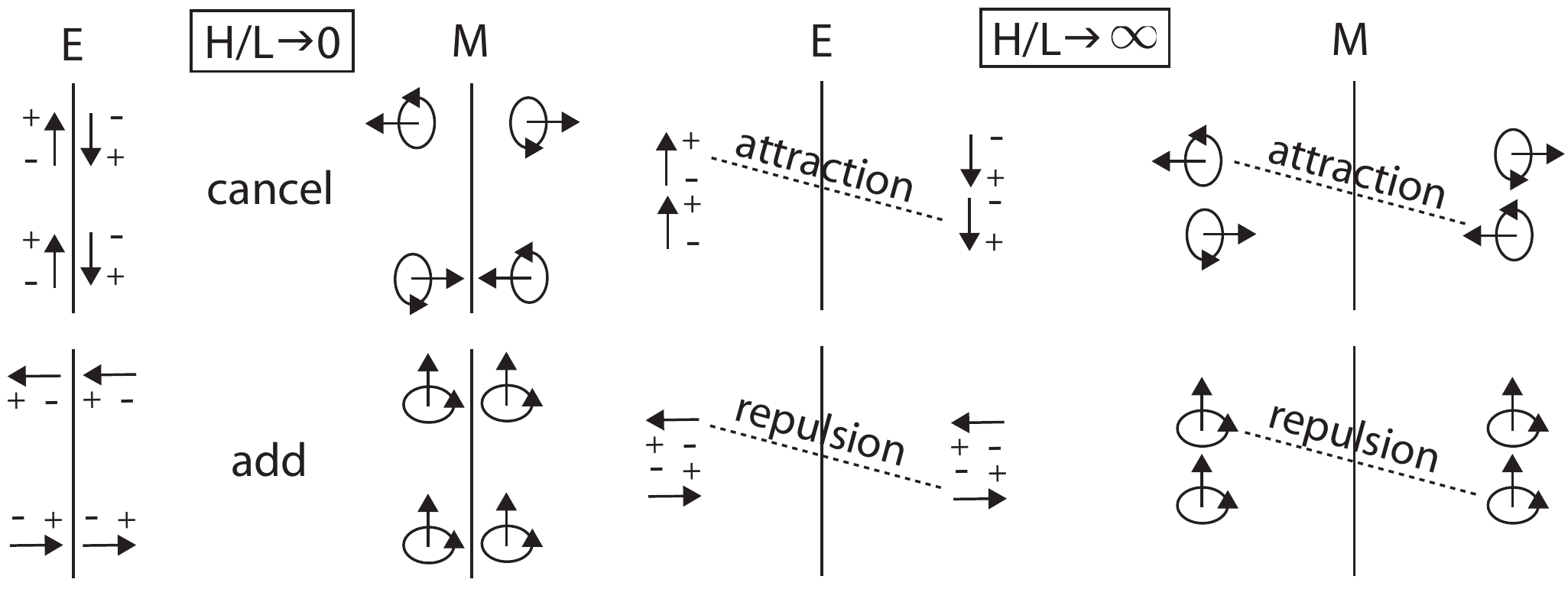}
\caption{\label{fig:fig2}Typical orientations of electric (E) and
  magnetic (M) dipoles and image dipoles for $H/L\to 0$ and
  $H/L\to\infty$.}
\end{figure}

\begin{figure}[htb]
\begin{center}
\includegraphics[width=0.48\linewidth]{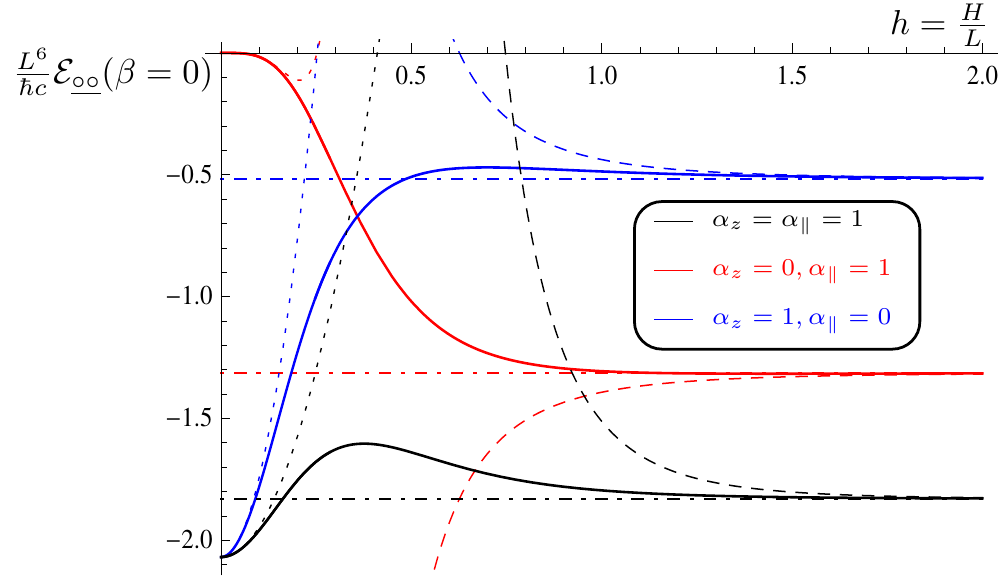}
\includegraphics[width=0.48\linewidth]{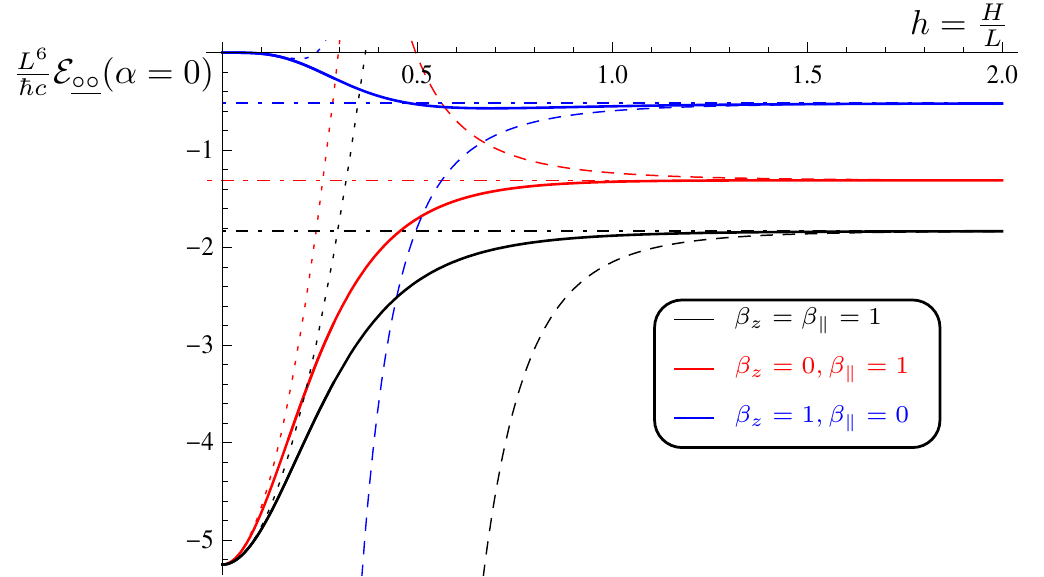}
\end{center}
  \caption{\label{Energia_atomos}Contributions to the energy between 2 atoms described by Eqs.~\eqref{TE_atomo} and \eqref{TM_atomo}, splitted in contributions of electric polarizabilities (left figure) and magnetic polarizabilities (right figure). In blue, the perpendicular to the plate ($z$) contribution, in red the parallel to the plate ($\|$) contribution and, in black, the energy for isotropic polarizabilities. The limits for $H\ll L$ (dotted curve), $H\gg L$ (dashed curve) and the limit $H\to\infty$ (dot--dashed curve) are also plotted. As the limits for $H\ll L$ and $H\gg L$ show, all figures plotted here are nonmonotonous with $H$, except the case of isotropic magnetic polarizabilities (black curve of the right plot).}
\end{figure}

The 3-body energy $\mathcal{E}_3(D,L)$ corresponds to the matrix products ${\mathbb T} {\mathbb U}^{21} {\mathbb T} {\mathbb U}^{I,12}$ and ${\mathbb T} {\mathbb U}^{I,21} {\mathbb T} {\mathbb U}^{12}$ in Eq.~\eqref{eq:energy-2} and hence describes the collective interaction between the two atoms and one image atom. 
It is given by
\begin{equation}\label{eq:E_3}
\mathcal{E}_3(D,L) =  \frac{4\hbar c}{\pi} \frac{1}{L^{3}D^{4}(\ell+1)^{5}}\left[\begin{array}{l}
\left(3\ell^{6} +15\ell^{5} + 28\ell^{4} + 20\ell^{3} + 6\ell^{2} - 5\ell - 1\right)\left(\alpha_{\|}^{2} - \beta_{\|}^{2}\right)\\ 
- \left(3\ell^{6} + 15\ell^{5} + 24\ell^{4} - 10\ell^{2} - 5\ell - 1\right) \left(\alpha_{z}^{2} - \beta_{z}^{2}\right)\\
+4\left(\ell^{4} + 5\ell^{3} + \ell^{2}\right)\left(\alpha_{z}\beta_{\|} - \alpha_{\|}\beta_{z}\right)
\end{array}\right].
\end{equation}
For isotropic electric polarizable atoms this result agrees with that
of Ref.~\cite{Power:1982cs}. It is instructive to consider the two
limits $H\ll L$ and $H\gg L$.  For $H\ll L$ one has $D\to L$ and the
2-body potentials are identical, $\mathcal{E}_{2,\backslash}(L,L)
=\mathcal{E}_{2,|}(L)$. The 3-body energy becomes
\begin{equation}\label{eq:E_3_small_H}
\mathcal{E}_{3}(L,L) = -\frac{\hbar c}{4\pi L^{7}} \!\!\left[ - 33\alpha_{\|}^{2} + 13\alpha_{z}^{2} + 14\alpha_{\|}\beta_{z} + 33\beta_{\|}^{2} - 13\beta_{z}^{2} - 14\beta_{\|}\alpha_{z}\right] \, .
\end{equation}
The total energy $\mathcal{E}_{\underline{\circ\circ}}$ is now twice the
energy of Eq.~\eqref{eq:E_CP} plus the energy of
Eq.~\eqref{eq:E_3_small_H}, and hence
\begin{equation}
\lim_{H\ll L}\mathcal{E}_{\underline{\circ\circ}}(L,H) = -\frac{\hbar c}{8\pi L^{7}} \!\!
\left[ 13 (2\alpha_{z})^{2} + 33(2\beta_{\parallel})^{2} - 14(2\beta_{\parallel})(2\alpha_{z})\right].
\end{equation}
Then, $\mathcal{E}_{\underline{\circ\circ}}$ becomes the CP potential of Eq.~\eqref{eq:E_CP} with the replacements
$\alpha_z\to 2\alpha_z$, $\alpha_\|\to 0$, $\beta_z\to 0$,
$\beta_\|\to 2\beta_\|$, as can be observed in Fig.~\ref{Energia_atomos}. The 2-body and 3-body contributions add
constructively or destructively, depending on the relative orientation
of a dipole and its image which together form a dipole of zero or twice
the original strength (see Figs.~\ref{fig:fig2} and \ref{Energia_atomos}). For $H \gg L$ 
the leading correction to the CP potential of Eq.~\eqref{eq:E_CP_plane} comes
from the 3-body energy which in this limit becomes (up to order $H^{-6}$)
\begin{equation}\label{eq:E_3_large_H}
\mathcal{E}_{3}(H,L) = \frac{\hbar c}{\pi} \!\!\left[\!\frac{\alpha_{z}^{2} - \alpha_{\|}^{2}}{4 L^{3}H^{4}} - \frac{\beta_{z}^{2} - \beta_{\|}^{2}}{4 L^{3}H^{4}} + \frac{9\alpha_{\|}^{2} - \alpha_{z}^{2} - 2\alpha_{\|}\beta_{z}}{8LH^{6}} - \frac{9\beta_{\|}^{2} - \beta_{z}^{2} - 2\beta_{\|}\alpha_{z}}{8LH^{6}}\!\right] .
\end{equation}
The signs of the polarizabilities in the leading term $\sim H^{-4}$
can be understood from the relative orientation of the dipole of one
atom and the image dipole of the other atom, see Fig.~\ref{fig:fig2}.
If these two electric dipoles are almost perpendicular to
their distance vector they contribute attractively to
the potential between the two original atoms, but if these electric 
dipoles are almost parallel to their distance vector they
yield a repulsive contribution.
Contrary to the electric behavior, if these two magnetic dipoles are almost 
perpendicular to their distance vector they contribute repulsively to
the potential between the two original atoms, but if these magnetic 
dipoles are almost parallel to their distance vector they
yield a attractive contribution.
For isotropic polarizabilities, the leading term of Eq.~\eqref{eq:E_3_large_H}
vanishes and the electric part $\sim H^{-6}$ of the 3-body
energy is always repulsive, while the magnetic contribution $\sim H^{-6}$ of the 3-body
energy is always attractive.

The above results show how the force between the two
particles varies with $H$.
If the two particles have only either $\alpha_z$
or $\beta_\|$ polarizability, their attractive force is reduced
when they approach the wall from large $H$ due to the repulsive 3-body
interaction, as shown in Fig.~\ref{Energia_atomos}.
At close proximity to the wall, the fluctuations of the dipole
and its image add up to yield a force between the particles that is 
enhanced by a factor of $4$ compared to the force for
$H\to\infty$. Corresponding arguments show that the force between
particles with either $\alpha_\|$ or $\beta_z$ polarizability is 
enhanced at large $H$ and reduced to zero for $H\to 0$. This
proves that the force between particles which both have either of the
four polarizabilities is always non-monotonic. The situation can be
different if more than one polarizability is finite, especially for
isotropic particles. In the latter case all contributions (electric,
magnetic, mixed) are enhanced for $H\to 0$ and only the electric term
is reduced at large $H$ so that only the electric part gives a
non-monotonic force (Fig.~\ref{Energia_atomos}). In general, the monotonicity property depends on
the relative strength and anisotropy of the electric and magnetic
polarizabilities.

\section{Two macroscopic metallic spheres}\label{Two macroscopic metallic spheres_chapII-2}
\subsection{Analytical results}
\begin{figure}[htb]
\begin{center}
\includegraphics[width=0.48\linewidth]{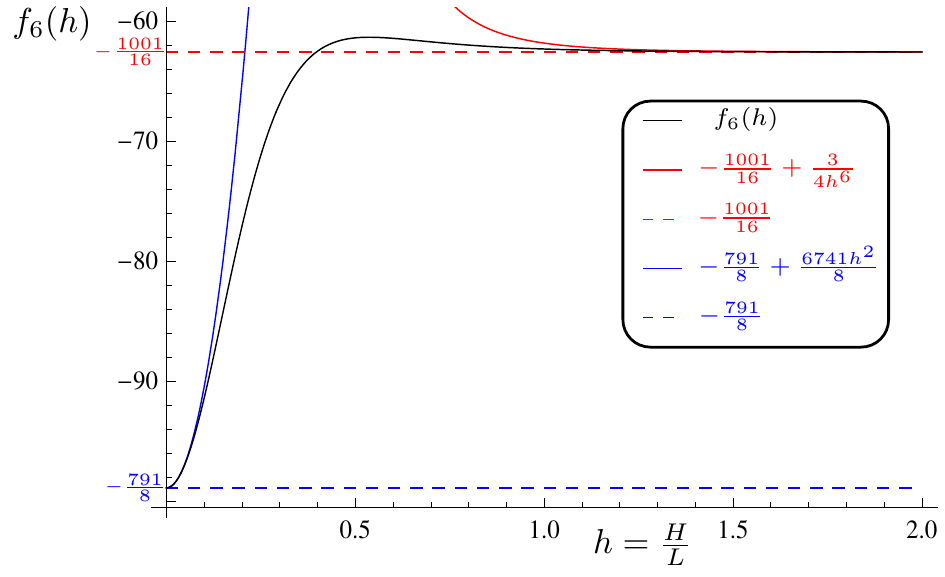}
\includegraphics[width=0.48\linewidth]{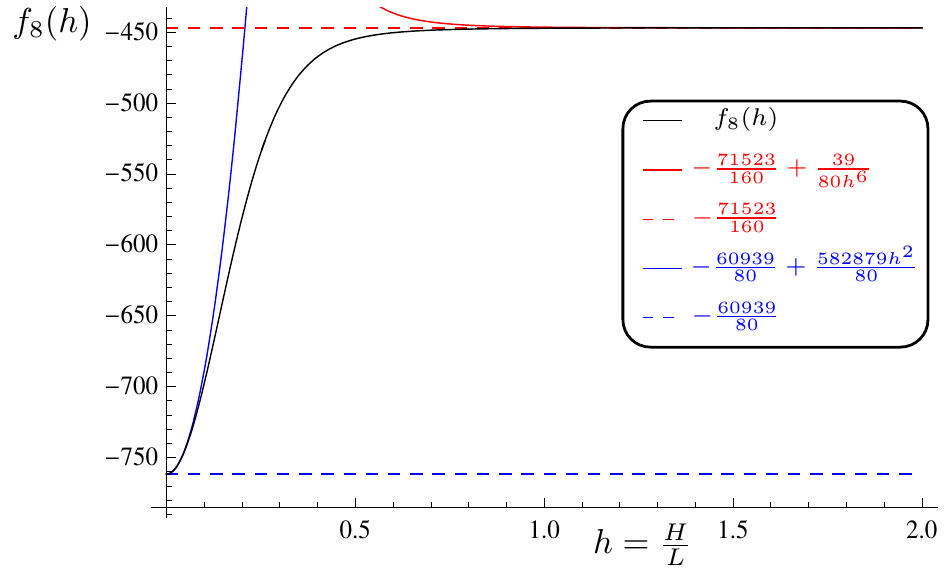}
\end{center}
  \caption{\label{fuerza_f6_f8}Functions $f_{6}(h)$ (left figure) and $f_{8}(h)$ (right figure) (Eq.~\eqref{eq:h-fcts}) as a function of $h = \frac{H}{L}$. The black curve is the exact function. The series expansion for $H\ll L$ (blue curve), $H\gg L$ (red curve) are also plotted, and the limits $h\to 0$ (dashed blue curve), and $h\to\infty$ (dashed red curve) are also represented to show the nonmonotonicity of $f_{6}(h)$ and $f_{8}(h)$.}
\end{figure}

Secondly, we study two macroscopic perfect metallic
spheres of radius $R$ for the same geometry as in the case of atoms
where the lengths $L$ and $H$ are measured now from the centers of the
spheres, see Fig.~\ref{fig:1}. The $\mathbb{T}$-matrix is diagonal and the
elements 
\begin{equation}
T^{M}_{lm} = (-1)^{l}\frac{\pi}{2}\frac{I_{l+1/2}(\kappa R)}{K_{l+1/2}(\kappa R)}
\end{equation}
\begin{equation}
T^{E}_{lm} = (-1)^{l}\frac{\pi}{2}\frac{I_{l+1/2}(\kappa R) + 2\kappa R I'_{l+1/2}(\kappa R)}{K_{l+1/2}(\kappa R) + 2\kappa R K'_{l+1/2}(\kappa R)}
\end{equation}
 are given in
terms of the modified Bessel functions $I_\nu$, $K_\nu$. First, we expand the
energy in powers of $R$ by using Eq.~\eqref{eq:energy-2} which implies
that we expand the $\mathbb{T}$-matrices for small frequencies but use the exact
expressions for the $\mathbb{U}$-matrices. For $R \ll L,\, H$ and arbitrary $H/L$
the result for the force can be written as
\begin{equation}\label{eq:force-of-L}
  F  = \frac{\hbar c}{\pi R^2} \sum_{j=6}^\infty  f_j(H/L) \left(\frac{R}{L}\right)^{j+2} \, .
\end{equation}
The functions $f_j$ can be computed exactly. We have obtained them up to $j=11$
and the first three are (with $s\equiv \sqrt{1+4h^2}$)
\begin{eqnarray}\label{eq:h-fcts}
f_6(h) & = &  -\frac{1}{16h^8}\Big[s^{-9}(18 + 312 h^2 + 2052 h^4 + 6048 h^6 +  5719 h^8) + 18 - 12 h^2 + 1001 h^8\Big]\nonumber\\
f_7(h) & = & 0\nonumber\\
f_8(h) & = & - \frac{1}{160 h^{12}}\left[\begin{array}{l}
s^{-11} (6210 + 140554 h^2 + 1315364 h^4 + 6500242 h^6   )\\
+ s^{-11}(17830560 h^8 + 25611168 h^{10} + \! 15000675 h^{12})\\
 - 6210 - 3934 h^2 + 764 h^4 - 78 h^6 + 71523 h^{12}
\end{array}\right]
\end{eqnarray}
The coefficient $f_7$ of $R^7$ vanishes since a multipole of order $l$
contributes to the $\mathbb{T}$-matrix at order $R^{2l+1}$ so that beyond the
two-dipole term $\sim R^6$ the next term comes from a dipole ($l=1$)
and a quadrupole ($l=2$), yielding $f_8$. For $H \gg L$ one has
$f_6(h) = -1001/16 +3/(4h^6)+ {\cal O}(h^{-8})$,
$f_8(h)=-71523/160+39/(80h^6)+ {\cal O}(h^{-8})$ so that the wall
induces weak repulsive corrections. For $H \ll L$,
$f_6(h)=-791/8+6741 h^2/8 +{\cal O}(h^4)$, $f_8(h)=-60939/80 + 582879
h^2/80 +{\cal O}(h^4)$ so that the force amplitude decreases  when the spheres are moved a small
distance  away from the wall. This proves the existence of a minimum in
the force amplitude as a function of $H/R$ for fixed, sufficiently
small $R/L$, see Fig.~\ref{fuerza_f6_f8}. We note that all $f_j(h)$ are finite for
$h\to \infty$ but some diverge for $h\to 0$, e.g., $f_9 \sim f_{11}
\sim h^{-3}$, making them important for small $H$.

\subsection{Numerical results}
\begin{figure}[h]
\includegraphics[width=0.4\linewidth]{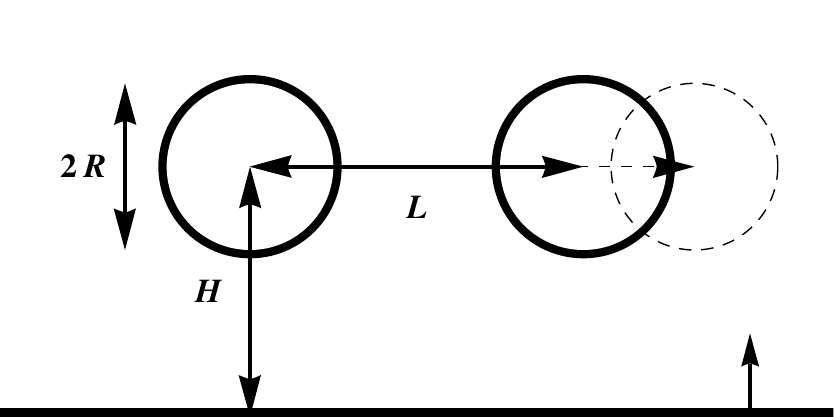}
\includegraphics[width=0.6\linewidth]{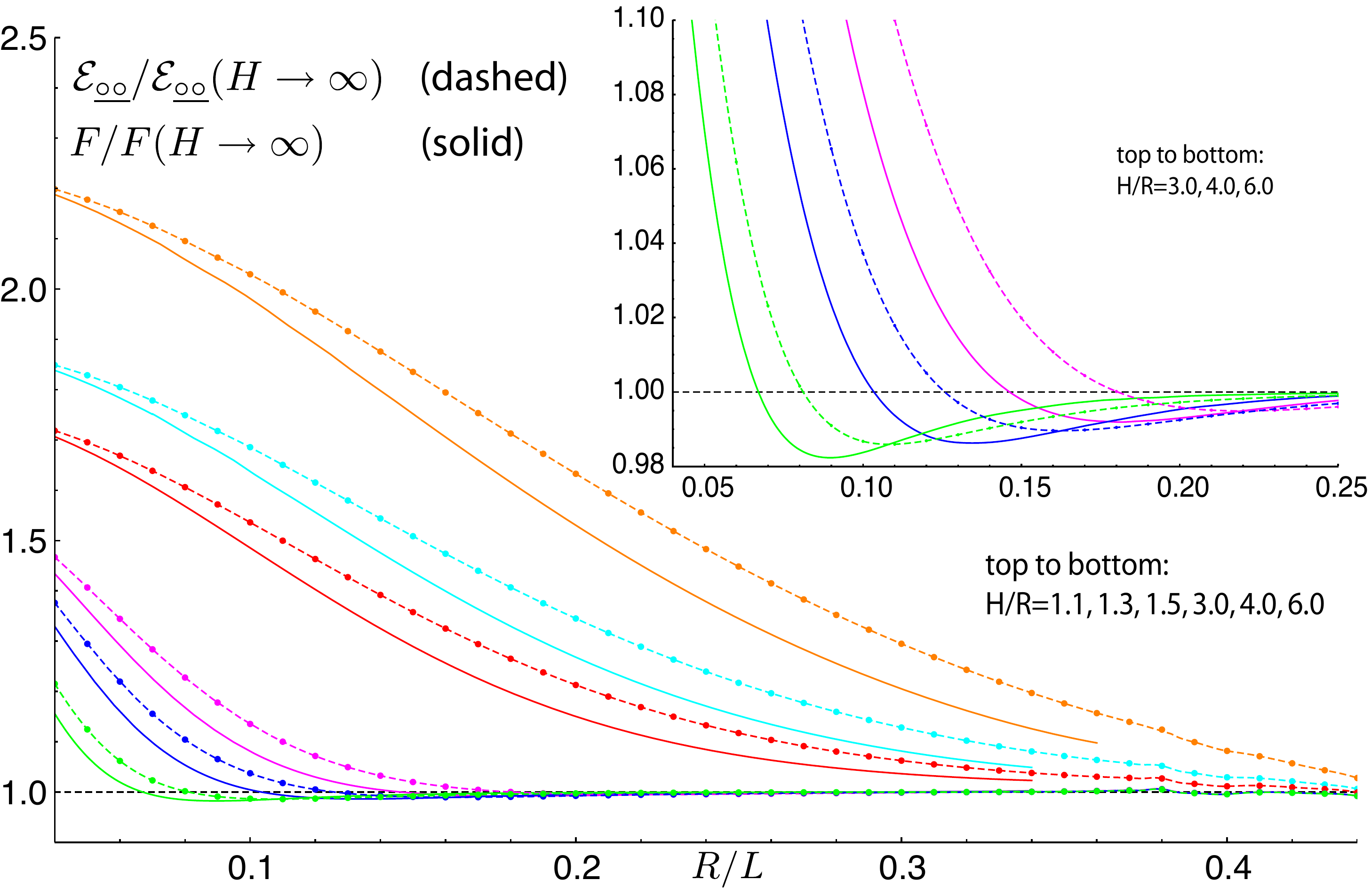}
\caption{\label{fig:3}Numerical results for the potential energy
  (dashed curves) and force (solid curves) between two spheres as
  function of $R/L$ for different sidewall separations $H/R$. Both
  force and energy are normalized to their values in the absence of
  the sidewall. Inset: Magnification of behavior for small $R/L$.}
\end{figure}

\begin{figure}[h]
\includegraphics[width=0.4\linewidth]{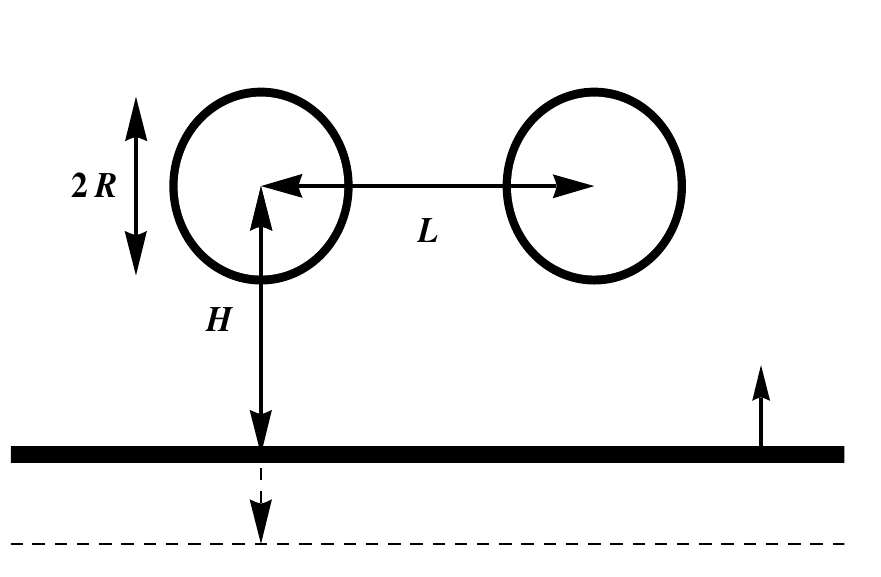}
\includegraphics[width=0.6\linewidth]{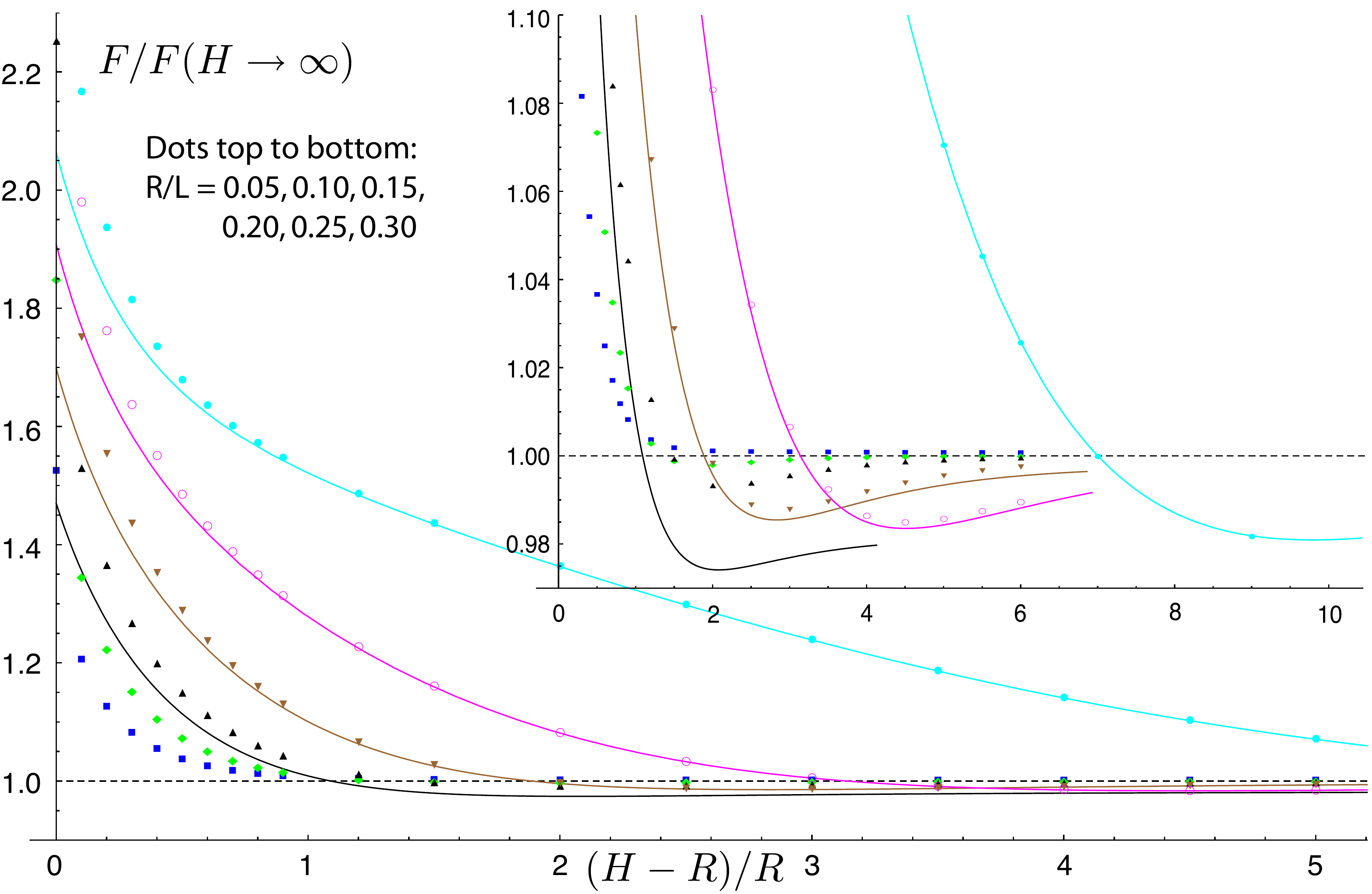}
\caption{\label{fig:4}Numerical results for the force (dots) between
  two spheres as function of the sidewall separation $H/R$ for
  different sphere separations $R/L$. Shown are also the analytical
  results of Eq.~\eqref{eq:force-of-L}, including terms up to $j=10$
  for $R/L\le 0.2$ (solid curves). Inset: Magnification of the non-monotonicity.}
\end{figure}

To obtain the interaction at smaller separations or larger radius, we
have computed the energy $\mathcal{E}_{\underline{\circ\circ}}$ and force
$F=-\partial \mathcal{E}_{\underline{\circ\circ}} /\partial L$ between the
spheres numerically.  For the energy, we have computed the last
determinant of Eq.~\eqref{eq:det-M-1} and the integral over $\kappa$
of Eq.~\eqref{eq:energy} numerically. The force is obtained by
polynomial interpolation of the data for the energy. The matrices are
truncated at a sufficiently large number of partial waves (with a
maximum truncation order $l_{max}=17$ for the smallest separation) so
that the relative accuracy of the values for
$\mathcal{E}_{\underline{\circ\circ}}$ is $\approx 10^{-3}$. The data for
$H/R=1$ are obtained by extrapolation in $l_{max}$. The results are
shown in Figs.~\ref{fig:3} and \ref{fig:4}.  In order to show the effect
of the wall, the figures display the energy and force normalized to
the results for two spheres without a wall. Fig.~\ref{fig:3} shows the
energy and force as a function of the (inverse) separation between the
spheres for different fixed wall distances.
%Energy and force show an
%increasing relative enhancement due to the wall with increasing $L$,
%with the maximal enhancement for small $H$.
The proximity of the wall generally increases the interaction energy and the force between the two spheres. The effect is more pronounced, the further the two spheres are separated.
For sufficiently large
$H/R$, the energy and force ratios are non-monotonic in $L$ and can be
slightly smaller than they would be in the absence of the wall.
Fig.~\ref{fig:4} shows the
force between the two spheres as a function of the wall distance for fixed $L$. When the spheres
approach the wall, the force first decreases slightly if $R/L \lesssim
0.3$ and then increases strongly under a further reduction of $H$. For
$R/L \gtrsim 0.3$ the force increases monotonically as the spheres approach the wall.
This agrees with the prediction of the large distance
expansion. The expansion of Eq.~\eqref{eq:force-of-L} with $j=10$
terms is also shown in Fig.~\ref{fig:4} for $R/L\le 0.2$. Its validity
is limited to large $L/R$ and not too small $H/R$; it fails completely
for $R/L>0.2$ and hence is not shown in this range.


\section{Discussion}\label{Discussion_chapII-2}
In this Chapter we have demonstrated  that nonmonotonicities of Casimir force appears not only between cylinders \cite{Rodriguez:2007a}\cite{Rahi:2008bv}\cite{Rahi:2008kb}, but also between compact objects. In particular, nonmonotonicities of Casimir force appears between two atoms and between two perfect metal spheres, in presence of a perfect metal plate.
For perfect metal spheres, there exists a critical distance from which this nonmonotinicity of the force disappears, although the variation (enhancement in this case) of the force between the spheres continues because the perfect metal plate contribution is not pairwise additive.
As shown by Eq.~\eqref{eq:det-M-1}, Casimir energy between three objects is not pairwise additive, but the effect of the inclusion of a third object in the system is not a simple enhancement of the force between objects (although in general this seems to be the case). We have shown here that, depending of the relative distance between the three objects, the effect of the non--pairwise behavior can be either an increase or a decrease of the force between objects, when it is compared with the isolated case.

The results presented here for isotropic atoms are potentially relevant to the interaction
between trapped Bose-Einstein condensates and a surface
\cite{Harber:2005a} at close surface separations. The results for
macroscopic spheres could be important for the design of
nano-mechanical devices where small components operate in close
vicinity to metallic boundaries.  Generally, the wall-induced
enhancement of the interaction can make the experimental observation
of Casimir forces between small particles more feasible. The dependence on the anisotropy of
polarizabilities applies not only to atoms but to general polarizable objects and suggests interesting
effects for  objects of non-spherical shape, e.g. spheroids.

%We acknowledge helpful discussions with N.~Graham, R.~L.~Jaffe, and M.~Kardar, and the hospitality of the Institute of Theoretical Physics, University of Cologne.  This research was supported by projects MOSAICO, UCM/PR34/07-15859 and a PFU MEC grant (PR) and by DFG through grant EM70/3 (TE).

% PSA and PSA between topological insulators
\begin{savequote}[10cm] % this sets the width of the quote
\sffamily
``Science is built upon approaches that gradually approach the truth.'' 
\qauthor{Isaac Asimov - Nightfall}
\end{savequote}

\newcommand{\ecref}[1]{Eq.~\eqref{#1}}

\chapter{Pairwise Summation Approximation of Casimir energy}\label{chb3}
\graphicspath{{02-Casimir_Multiscattering/chb3_PSA/Figuras/}} 

In this Chapter, we obtain the Pairwise Summation Approximation (PSA) of the Casimir energy from first principles in the soft dielectric and soft diamagnetic limit, this analysis let us find that the PSA is an asymptotic approximation of the Casimir energy valid for large distances between the objects. We also obtain the PSA for the electromagnetic (EM) coupling part of the Casimir energy, so we are able to complete the PSA limit for the first time for the complete electromagnetic field.

\section{Introduction}
Since 1948, when Casimir introduced the energy that got his name \cite{Casimir_Placas_Paralelas}, calculation formulas have been looked for. Many analytical and numerical methods have been proposed, such as the zeta function technique, the heat kernel method, semiclassical methods or Green function (local) methods just to mention a few of them \cite{Review_Casimir}. However, exact results have been obtained only for some simple geometries. 

The asymptotic analysis of Casimir energies has also a long history. In fact, Casimir himself and Polder in year 1948 gave the first asymptotic formula for the Casimir energy between two electrically neutral bodies in terms of their electric induced dipoles \cite{VdW-int.electrica}. Some time later, a generalization of that formula, known as the Pairwise Summation Approximation (PSA) was derived \cite{Lifshitz} for electric media. The main assumption is a linear superposition of the Casimir--Polder interactions between the induced polarizabilities of each element of volume body. Then, the PSA energy is expressed as an integral over the two object's volumes and it is proportional to the objects polarizabilities. The formula has been recently reobtained by R. Golestanian in \cite{Golestanian} and by K. A. Milton et. al. in \cite{MiltonPRL} in the soft dielectric limit.

Besides, a new asymptotic method for calculating Casimir energies in term of the induced multipoles of the interacting bodies has been proposed in \cite{Kardar-Geometrias-Arbitrarias} and \cite{Functional_Determinant_Method_2}. This formula provides a procedure for the calculation of the Casimir energy between \textit{N} arbitrary shaped compact objects \cite{Rodriguez-Lopez_1}.

In this Chapter, the asymptotic calculation of the electromagnetic Casimir energy and a presentation of a systematic asymptotic expansion procedure for the integral dipole formula on higher orders is presented. We will reobtain the classical results of Casimir and Polder \cite{VdW-int.electrica}, and of Feinberg--Sucher in \cite{VdW_int.magnetica} in the rarefied or soft dielectric and soft diamagnetic limit.

The result we obtain here is a generalization of Milton et. al. one \cite{MiltonPRL} but assuming that the bodies are soft diamagnetic as well as soft dielectric. For this purpose, we will use the multi-scattering expansion of the Casimir energy formalism given in \cite{Kardar-Geometrias-Arbitrarias}. To our knowledge, it is the first time this formalism is used to derive the PSA. We obtain that the PSA is the first order of a perturbation expansion in the difference between the electric and magnetic permeability constants of the objects respect the electric and magnetic permeability constants of the medium were the objects are placed. The interest of this result is that we obtain this pairwise summation formula for the complete electromagnetic Casimir energy which is derived as an asymptotic limit of an exact and free of divergences formula. That means that now we can establish the range of validity for that approximation. In fact, we will justify why this formula is valid in the far distance objects limit. This derivation also gives us a perturbative procedure for corrections of this approximation and  posterior expansions to more than two objects or finite temperature cases.

We will follow this plan for the Chapter:
Using the soft dielectric and soft diamagnetic approximation, in Sect. \ref{sec: 2_PSA} we will obtain the PSA of Casimir energy for the zero temperature case starting from the exact Casimir energy formula given in \cite{Kardar-Geometrias-Arbitrarias}. We will also study the far distance limit to reobtain the asymptotic Casimir energies given in \cite{VdW-int.electrica} and \cite{VdW_int.magnetica}. In Sect. \ref{sec: 3_PSA} we will obtain PSA Casimir energy formulas for any temperature, studying the high and low temperature limits. In Sect. \ref{sec: 4_PSA} we will study the PSA Casimir energy for the three bodies system and for the general $N$ bodies system. We will obtain a kind of superposition principle of Casimir energy in the PSA approximation. Finally, in Sect. \ref{sec: 5_PSA} we will study the first perturbation energy term to the PSA approximation and we will discuss about the nature of the PSA limit of the Casimir energy.

The contents of this Chapter is based on the work published in~\cite{Rodriguez-Lopez_PSA}, \cite{Placas_TI_forall_T} and \cite{Rodriguez-Lopez_5}.
\section{Diluted limit at zero temperature}\label{sec: 2_PSA}
Our objective is the calculation of the complete electromagnetic Casimir energy between two bodies in the soft dielectric limit. For this purpose, we will use the Casimir energy formula between two compact bodies at temperature $T = 0$, given in \cite{Kardar-Geometrias-Arbitrarias} as
\begin{equation}\label{Formula de Emig}
E = \frac{\hbar c}{2\pi}\int_{0}^{\infty}dk\log\Det{\mathbbm{1} - \mathbb{N}}.
\end{equation}
Where $k$ is the frequency, $\mathbbm{1}$ is the identity matrix, $\mathbb{N}$ is the matrix $\mathbb{N} = \mathbb{T}_{1}\mathbb{U}_{12}\mathbb{T}_{2}\mathbb{U}_{21}$. Here, $\mathbb{T}_{\alpha}$ is the T scattering matrix of the $\alpha$ - body under the electromagnetic field, and $\mathbb{U}_{\alpha\beta}$ is the propagation matrix of the electromagnetic field from $\alpha$ - body to $\beta$ - body. We will use the position representation instead the multipole representation used in \cite{Kardar-Geometrias-Arbitrarias}, so we can identify $\mathbb{U}_{\alpha\beta} = G_{0\alpha\beta}$, where $G_{0\alpha\beta}$ is the free dyadic Green function. Taking into account that $\log\Det{A} =\tr\log(A)$ and that $\log(1 - x) = - \sum_{p=1}^{\infty}\frac{x^{p}}{p}$, we transform \ecref{Formula de Emig} into
\begin{equation}\label{Formula de Emig en forma de traza}
E = - \frac{\hbar c}{2\pi}\sum_{p=1}^{\infty}\frac{1}{p}\int_{0}^{\infty}dk\tr\left(\mathbb{N}^{p}\right).
\end{equation}
Equation \eqref{Formula de Emig en forma de traza} is an asymptotic expansion of \ecref{Formula de Emig} in the $p_{max}<\infty$ case, which means that our calculus will be valid in the large bodies distance limit. The T operator is related with the potential $V$ (we will see what $V$ is later) by the Lippmann - Schwinger equation, that can be written as \cite{Galindo_y_Pascual}
\begin{equation}\label{Lippmann - Schwinger equation for T operator}
\mathbb{T} = \left(1 - VG_{0}\right)^{-1}V.
\end{equation}
Applying a Born expansion to \ecref{Lippmann - Schwinger equation for T operator}, we can obtain an approximation of the $T$ operator in the soft dielectric and soft diamagnetic limit as
\begin{equation}\label{Born expansion of T operator}
\mathbb{T} = \sum_{n=0}^{\infty}(VG_{0})^{n}V\simeq V.
\end{equation}
This approximation is more valid for weaker $V$, so here is where the soft dielectric and soft diamagnetic limit is applied.

Now we will study the lowest expansion order. In the lowest expansion order ($p=1$ and $\mathbb{T}=V$) we get the asymptotic approximation of the Casimir energy between two bodies as
\begin{equation}\label{Formula de Emig en soft limit}
E = - \frac{\hbar c}{2\pi}\int_{0}^{\infty}dk\tr\left(V_{1}G_{012}V_{2}G_{021}\right).
\end{equation}
We can separate the magnetic and electric part of each body potential (which is a diagonal operator in positions space, because it is local) and each field contribution in the free dyadic Green function.

To define the potential $V_{i}$, we use the generalized constitutive relations of materials with magnetoelectric coupling
\begin{align}\label{generalized_constituve_relations}
\textbf{D} & = \epsilon\textbf{E} + \alpha\textbf{H},\nonumber\\
\textbf{B} & = \beta\textbf{E} + \mu\textbf{H}.
\end{align}
In general, $\alpha_{ij} = \beta_{ij}$, because a linear magnetoelectric coupling is allowed only in materials odd under time reversal or odd under inversion $(\alpha_{ij} \to - \alpha_{ij} )$. The magnetoelectric polarizability $\alpha_{ij}$ is zero otherwise. The potential of each body can be defined as the difference of energy of the EM field because the existence of this body. Having into account that the energy of the EM is defined as
\begin{equation}
E = \frac{1}{2} \int_{\Omega}dx^{\mu}\left(\textbf{E}\cdot\textbf{D} + \textbf{H}\cdot\textbf{B}\right),
\end{equation}
where $\textbf{D} = \epsilon_{0}\textbf{E}$ and $\textbf{H} = \mu_{0}\textbf{B}$ in the vacuum, the EM energy in presence of $N$ generalized dielectrics is
\begin{equation}
S = \frac{1}{2} \int_{\Omega} dx^{\mu}\left(\epsilon_{0}\textbf{E}^{2} + \mu_{0}\textbf{H}^{2}\right) + \sum_{i=1}^{N}\Delta S_{i}.
\end{equation}
We use the generalized constitutive relations given in Eq.~\eqref{generalized_constituve_relations} to obtain the excess of energy because the existence of each dielectric as
\begin{align}
\Delta S_{i} & = \frac{1}{2}\int_{\Omega_{i}}dx^{\mu}\left(\textbf{E},\textbf{H}\right)\left(\begin{array}{c|c}
\left(\epsilon_{i} - \epsilon_{0}\right) & \alpha_{i}\\
\hline
\beta_{i} & \left(\mu_{i} - \mu_{0}\right)\end{array}\right)\left(\begin{array}{c}
\textbf{E}\\
\textbf{H}\end{array}\right),
\end{align}
then the potential is defined as
\begin{align}\label{Potencial_dielectric}
V_{i} & = \left(\begin{array}{c|c}
\tilde{\epsilon}_{i} & \alpha_{i}\\
\hline
\beta_{i} & \tilde{\mu}_{i}\end{array}\right)\chi_{i}\left(\textbf{r}\right),
\end{align}
where $\tilde{\epsilon}_{i} = \epsilon_{i} - \epsilon_{0}$, $\tilde{\mu}_{i} = \mu_{i} - \mu_{0}$ in the vacuum and $\chi_{i}\left(\textbf{r}\right)$ is the characteristic function of the $i$--body volume (1 inside the body and 0 in the rest of the space). The definition presented here let us define the potential $V_{i}$ when the medium is not the vacuum and when the objects are not isotropic (think for example in uniaxial and biaxial crystals~\cite{LL8}). Note that we represent the electromagnetic properties of the objects by their potential energy instead the complement representation by their boundary conditions \cite{BalianDuplantier-II}. It will let us to obtain the PSA as an integral over the volume of the bodies as required by the PSA. The contributions of the free dyadic Green function are obtained in the Appendix in \ecref{Acoplo EE funcion Green} - \eqref{Acoplo HH funcion Green}.

As seen in \ecref{Acoplo EE funcion Green} - \eqref{Acoplo HH funcion Green}  of the Appendix, operators are defined over three different linear spaces:
1) An $EH$ - space, whose components are the electric and the magnetic field; 2) over the space coordinates, because we are working with a vector and a pseudovector; and 3) over positions. We must solve the trace over these three spaces: $EH$ - space, vector coordinate space and position space. First we solve the trace in the $EH$ - space for the important case of non--magnetoelectric coupled objects and we obtain:
\begin{equation}
E = - \frac{\hbar c}{2\pi}\int_{0}^{\infty}dk\tr\left(\begin{array}{l}
 \phantom{+}  V_{1}^{E}G_{012}^{EE}V_{2}^{E}G_{021}^{EE} + V_{1}^{E}G_{012}^{EH}V_{2}^{H}G_{021}^{HE}\\
 + V_{1}^{H}G_{012}^{HE}V_{2}^{E}G_{021}^{EH} + V_{1}^{H}G_{012}^{HH}V_{2}^{H}G_{021}^{HH}
\end{array}\right)
\end{equation}
We identify each term with this obvious notation:
\begin{equation}\label{Energia de Milton}
E = E_{EE} + E_{EH} + E_{HE} + E_{HH}.
\end{equation}
\subsection{Purely electric and purely magnetic Casimir energy}
The purely electric case has been solved by K. A. Milton et. al. in \cite{Derivacion_caso_ee_por_Milton}. 

Using \ecref{Acoplo EE funcion Green} and \ecref{Energia de Milton}, we have to solve:
\begin{equation}
E_{EE} = - \frac{\hbar c}{2\pi}\int_{0}^{\infty}dk\tr\left(V_{1}^{E}G_{012}^{EE}V_{2}^{E}G_{021}^{EE}\right).
\end{equation}
Replacing each potential by its value and assuming isotropy, it is followed that they are proportional to the identity matrix. Then we can drop them for the coordinates trace, but not for the spatial positions trace:
\begin{equation}\label{PSA_ee_2}
E_{EE} = - \tilde{\epsilon}_{1}\tilde{\epsilon}_{2}\frac{\hbar c}{2\pi}\int_{0}^{\infty}dk\int d\textbf{r}_{1}\int d\textbf{r}_{2}\chi_{1}\chi_{2} \tr\left(G_{012}^{EE}G_{021}^{EE}\right).
\end{equation}
Using \ecref{Acoplo EE funcion Green} of the Appendix and $R = \abs{\textbf{r} - \textbf{r}'}$ we have, in matrix form:
\begin{align}
G_{0ij}^{EE}(R,k) & = - \left(3 + 3kR + k^{2}R^{2}\right)\frac{e^{-kR}}{4\pi R^{5}}\left( \begin{array}{ccc}
R_{1}^{2}  & R_{1}R_{2} & R_{1}R_{3}\\
R_{2}R_{1} & R_{2}^{2}  & R_{2}R_{3}\\
R_{3}R_{1} & R_{3}R_{2} & R_{3}^{2} 
\end{array} \right)\nonumber\\
& + \left(1 + kR + k^{2}R^{2}\right)\frac{e^{-kR}}{4\pi R^{3}}\left( \begin{array}{ccc}
1 & 0 & 0\\
0 & 1 & 0\\
0 & 0 & 1
\end{array} \right).
\end{align}
So the trace in spatial coordinates is easily solved as
\begin{equation}\label{Traza terminos de acoplo em EE}
\tr\left(G_{012}^{EE}G_{021}^{EE}\right) = \left(6 + 12 kR + 10 k^{2}R^{2} + 4 k^{3}R^{3} + 2 k^{4}R^{4}\right)\frac{e^{-2kR}}{(4\pi)^{2} R^{6}}.
\end{equation}
We can simplify our calculus using the dimensionless variable $u = kR$ to factorize spatial and frequency contributions of Eq.~\eqref{PSA_ee_2}
\begin{equation}\label{Energia Casimir EE a T = 0}
E_{EE} = - \tilde{\epsilon}_{1}\tilde{\epsilon}_{2}\frac{2\hbar c}{(4\pi)^{3}}\int d\textbf{r}_{1}\int d\textbf{r}_{2}\frac{\chi_{1}\chi_{2}}{R^{7}}\int_{0}^{\infty}due^{-2u}\left(6 + 12 u + 10 u^{2} + 4 u^{3} + 2 u^{4}\right).
\end{equation}
Finally, using $\int_{0}^{\infty}duu^{n}e^{-au} = \frac{n!}{a^{n+1}}$, the following result is obtained:
\begin{equation}
E_{EE} = -23\tilde{\epsilon}_{1}\tilde{\epsilon}_{2}\frac{\hbar c}{(4\pi)^{3}}\int_{1}\int_{2}\frac{d\textbf{r}_{1}d\textbf{r}_{2}}{\abs{\textbf{r}_{1} - \textbf{r}_{2}}^{7}}.
\end{equation}
We obtain the same result as in \cite{Derivacion_caso_ee_por_Milton} because the dyadic Green function used in \cite{Derivacion_caso_ee_por_Milton} is the pure electric part of the Green function matrix we use here.

Having into account that the pure magnetic part of the Green function matrix is equal to the pure electric part $G_{0ij}^{HH}(R,k) = G_{0ij}^{EE}(R,k)$, the calculus of the purely magnetic part of the Casimir energy is similar to the electric one. Consequently we obtain the following result for the purely magnetic part of the Casimir energy:
\begin{equation}
E_{HH} = -23\tilde{\mu}_{1}\tilde{\mu}_{2}\frac{\hbar c}{(4\pi)^{3}}\int_{1}\int_{2}\frac{d\textbf{r}_{1}d\textbf{r}_{2}}{\abs{\textbf{r}_{1} - \textbf{r}_{2}}^{7}}.
\end{equation}
To our knowledge, this is the first place where this result is obtained.

\subsection{Coupled electromagnetic Casimir energy}
Here we are going to calculate the contribution of the electromagnetic coupling part of the Casimir energy. Using \ecref{Acoplo EH funcion Green} and \ecref{Energia de Milton}, we have to solve:
\begin{equation}
E_{EH} = - \frac{\hbar c}{2\pi}\int_{0}^{\infty}dk\tr\left(V_{1}^{E}G_{012}^{EH}V_{2}^{H}G_{021}^{HE}\right).
\end{equation}
Replacing each potential by its value and assuming isotropy, it is followed that they are proportional to the identity matrix. Then we can drop them for the coordinates trace, but not for the spatial positions trace:
\begin{equation}
E_{EH} = - \tilde{\epsilon}_{1}\tilde{\mu}_{2}\frac{\hbar c}{2\pi}\int_{0}^{\infty}dk\int d\textbf{r}_{1}\int d\textbf{r}_{2}\chi_{1}\chi_{2} \tr\left(G_{012}^{EH}G_{021}^{HE}\right).
\end{equation}
Using \ecref{Acoplo EH funcion Green} of the Appendix and $R = \abs{\textbf{r} - \textbf{r}'}$ we have, in matrix form:
\begin{equation}
G_{0ij}^{EH}(R,k) = - k\frac{\partial G_{0}}{\partial R}\epsilon_{ijk}\partial_{k}R 
                  = G_{0}(R,k)\frac{k}{R}\left(k + \frac{1}{R}\right)\left( \begin{array}{ccc}
0 & R_{3} & - R_{2}\\
- R_{3} & 0 & R_{1}\\
R_{2} & - R_{1} & 0
\end{array} \right),
\end{equation}
so the trace in spatial coordinates is easily solved:
\begin{equation}\label{Traza terminos de acoplo em}
\tr\left(G_{012}^{EH}G_{021}^{HE}\right) = - 2k^{2}G_{0}^{2}(R,k)\left(k + \frac{1}{R}\right)^{2}.
\end{equation}
We can simplify our calculus using the dimensionless variable $u = kR$ to factorize spatial and frequency contributions of the formula
\begin{equation}\label{Energia Casimir acoplo em a T = 0}
E_{EH} = \tilde{\epsilon}_{1}\tilde{\mu}_{2}\frac{4\hbar c}{(4\pi)^{3}}\int d\textbf{r}_{1}\int d\textbf{r}_{2}\frac{\chi_{1}\chi_{2}}{R^{7}}\int_{0}^{\infty}due^{-2u}\left(u^{4} + 2u^{3} + u^{2}\right).
\end{equation}
Finally, using $\int_{0}^{\infty}duu^{n}e^{-au} = \frac{n!}{a^{n+1}}$, the final result is obtained as
\begin{equation}
E_{EH} = 7\tilde{\epsilon}_{1}\tilde{\mu}_{2}\frac{\hbar c}{(4\pi)^{3}}\int d\textbf{r}_{1}\int d\textbf{r}_{2}\frac{\chi_{1}\chi_{2}}{R^{7}}.
\end{equation}
There is an antisymmetry between $G^{EH}$ and $G^{HE}$ shown in \ecref{Acoplo EH funcion Green} and \ecref{Acoplo HE funcion Green} of the Appendix, that is $G_{0ij}^{EH}(R,k) = - G_{0ij}^{HE}(R,k)$. Therefore, we can obtain the coupling between the magnetic part of the first object and the electric part of the second one in a similar way.
Then, we can obtain the coupling between the magnetic part of the first object and the electric part of the second one:
\begin{equation}
E_{HE} = 7\tilde{\epsilon}_{2}\tilde{\mu}_{1}\frac{\hbar c}{(4\pi)^{3}}\int_{1}\int_{2}\frac{d\textbf{r}_{1}d\textbf{r}_{2}}{\abs{\textbf{r}_{1} - \textbf{r}_{2}}^{7}}.
\end{equation}
These two cross contributions to the Casimir energy in the soft dielectric and diamagnetic limits are the main result of this Chapter. It is interesting to note the sign change in this part of the Casimir energy with respect to the other contributions. Then, these terms can invert the typical attractive nature of the Casimir energy to repulsive in this limit for objects with very different electromagnetic nature. Finally, thanks to these new two terms, we obtain the complete electromagnetic Casimir energy between two objects in the soft dielectric and diamagnetic limit for the first time.
Then the global asymptotic electromagnetic Casimir energy is finally obtained:
\begin{equation}\label{Final Result}
E = \frac{- \hbar c}{(4\pi)^{3}} \gamma \int_{1}\int_{2}\frac{d\textbf{r}_{1}d\textbf{r}_{2}}{\abs{\textbf{r}_{1} - \textbf{r}_{2}}^{7}},
\end{equation}
where $\gamma = 23\tilde{\epsilon}_{1}\tilde{\epsilon}_{2} - 7\tilde{\epsilon}_{1}\tilde{\mu}_{2} - 7\tilde{\epsilon}_{2}\tilde{\mu}_{1} + 23\tilde{\mu}_{1}\tilde{\mu}_{2}$ is the complete multiplicative constant, taking into account all the electromagnetic effects in this limit, nor just electric effects, where $\gamma$ would be $23\tilde{\epsilon}_{1}\tilde{\epsilon}_{2}$.
\subsection{Asymptotic Casimir energy}\label{sec: 3.C}
We can also obtain from \ecref{Final Result} (always in the soft electric and magnetic limit) the first asymptotic energy order for two objects. These formulas are the first order of a multipolar expansion of the integrand in \ecref{Final Result} in the coordinate system of each object. That means that we have to assume that the distance between the objects is much greater than their characteristic lengths $R_{\alpha}$, that is $R_{\alpha}\ll R$. In this limit we can separate the problem into two scales and we can replace $\abs{\textbf{r}_{1} - \textbf{r}_{2}} = R$, where $R$ is assumed to be a constant. Then the integral can be easily solved in this limit to:
\begin{equation}
E \simeq \frac{- \hbar c}{(4\pi)^{3}}\gamma\frac{V_{1}V_{2}}{R^{7}},
\end{equation}
where $V_{\alpha}$ is the volume of the $\alpha$ - object. As it is demonstrated in the Appendix B~\ref{Apendice_limite_diluido_susceptibilidad_electrica}, in the soft limit order we can approximate the electric and magnetic polarizabilities as $\alpha^{E} = \tilde{\epsilon}\frac{V}{4\pi}$ and $\alpha^{H} = \tilde{\mu}\frac{V}{4\pi}$. In \cite{Optica-Wolf} it was derived $\alpha^{E}$ for an sphere. By using the same method with the approximation of that the effective field over the dielectric is equal to the induced field in the soft limit, we arrive at $\alpha^{E} = \tilde{\epsilon}\frac{V}{4\pi}$ for any arbitrary shaped object. Equation \eqref{Final Result} simplifies into
\begin{small} \begin{equation}
E \simeq\frac{- \hbar c}{4\pi R^{7}}\left(23\alpha_{1}^{E}\alpha_{2}^{E} + 23\alpha_{1}^{H}\alpha_{2}^{H} - 7\alpha_{1}^{E}\alpha_{2}^{H} - 7\alpha_{1}^{H}\alpha_{2}^{E}\right).
\end{equation}\end{small}
This is the soft response limit of the Feinberg and Sucher potential \cite{VdW_int.magnetica}. It coincide as well with the soft response limit of the asymptotic Casimir energy obtained in \cite{Kardar-Geometrias-Arbitrarias} for spheres. The presented scheme looks as we would use a local two point potential (where we substitute the polarizabilities of the bodies by their local susceptibilities) as in the present case with the potential given in \ecref{Final Result}, and integrate over the volume of each body to obtain the PSA of the Casimir energy. If we also study the asymptotic distance limit of the PSA, then we reobtain the Feinberg and Sucher potential in the soft response limit, now proportional to the polarizabilities of the objects instead the susceptibilities.

\section{Diluted limit at any temperature $T$}\label{sec: 3_PSA}
In this Section we will calculate the Casimir energy in the diluted limit at any finite temperature. Then we will focus on different approximations to low an high temperature limits, recovering the zero temperature case showed before and giving new formulas of the PSA for any temperature. We begin this study with the Casimir energy formula \eqref{Formula de Emig} for any temperature:
\begin{equation}\label{Formula de Emig T}
E = k_{B}T{\sum_{n = 0}^{\infty}}'\log\Det{\mathbbm{1} - \mathbb{N}(k_{n})},
\end{equation}
with Matsubara frequencies $k_{n} = 2\pi\frac{k_{B}T}{\hbar c}n = \Lambda n$. The tilde means that the $n=0$ case is weighted by a $1/2$ factor. As usual, we apply the Born approximation to the $T$ matrix scattering obtaining at first order $\mathbb{T}_{\alpha} = V_{\alpha}$ (where $\alpha$ labels the body in interaction). We apply again that $\log\Det{A} = \tr\log(A)$ and $\log(1 - x) = - \sum_{p=1}^{\infty}\frac{x^{p}}{p}$ , then we transform \ecref{Formula de Emig T} into:
\begin{equation}
E = - k_{B}T\sum_{p=1}^{\infty}\frac{1}{p}{\sum_{n = 0}^{\infty}}'\tr\left(\mathbb{N}^{p}(\Lambda n)\right).
\end{equation}
At first order in $p$ we obtain:
\begin{equation}
E = - k_{B}T{\sum_{n = 0}^{\infty}}'\tr\left(\mathbb{N}(\Lambda n)\right)
  = - k_{B}T{\sum_{n = 0}^{\infty}}'\tr\left(\mathbb{T}_{1}\mathbb{U}_{12}\mathbb{T}_{2}\mathbb{U}_{21}\right).
\end{equation}
As we are working in the positions representation space instead in the multipolar representation space, we have to represents the operators $\mathbb{T}_{\alpha}$ and $\mathbb{U}_{\alpha\beta}$ in the positions space. For that issue, we have into account that $\mathbb{U}_{\alpha\beta}$ matrices represent the field propagation between two points. Then they are represented by the free vacuum Green function of the interaction field. Using the Born approximation, we obtain the next formula for the diluted approximation of the Casimir energy:
\begin{equation}
E \simeq E_{T} = - k_{B}T{\sum_{n = 0}^{\infty}}'\tr\left(V_{1}G_{012}V_{2}G_{021}\right).
\end{equation}
As we did with \eqref{Formula de Emig en soft limit}, we can separate the magnetic and electric part of each body potential and each field contribution in the free dyadic Green function. We trace over the $EH$ space obtaining:
\begin{equation}
E_{T} = - k_{B}T{\sum_{n = 0}^{\infty}}'\tr\left(\begin{array}{l}
\phantom{+} V_{1}^{E}G_{012}^{EE}V_{2}^{E}G_{021}^{EE} + V_{1}^{E}G_{012}^{EH}V_{2}^{H}G_{021}^{HE}\\
 + V_{1}^{H}G_{012}^{HE}V_{2}^{E}G_{021}^{EH} + V_{1}^{H}G_{012}^{HH}V_{2}^{H}G_{021}^{HH}
\end{array}\right).
\end{equation}
As before, we identify each term with this obvious notation:
\begin{equation}
E_{T} = E_{EE} + E_{EH} + E_{HE} + E_{HH}.
\end{equation}
\subsection{Purely electric and purely magnetic energy}
Using \ecref{Acoplo EE funcion Green} and \ecref{Acoplo HH funcion Green} of the Appendix, we can solve the trace over the coordinates of $E_{EE}$ and, similarly, of $E_{HH}$. The matricial form of the purely electric part of the dyadic Green function \label{Acoplo EE funcion Green} and of the purely magnetic part of the dyadic Green function\label{Acoplo HH funcion Green} are:
\begin{align}
G_{0ij}^{EE}(R,k) & = - R_{i}R_{j}\left(3 + 3kR + k^{2}R^{2}\right)\frac{e^{-kR}}{4\pi R^{5}} + \delta_{ij}\left(1 + kR + k^{2}R^{2}\right)\frac{e^{-kR}}{4\pi R^{3}}\nonumber\\
& = G_{0ij}^{HH}(R,k).
\end{align}
Where $\textbf{R} = \textbf{r}_{\alpha} - \textbf{r}_{\beta}$. The trace in spatial coordinates is easily solved obtaining:
\begin{equation}\label{Caso diagonal T neq 0}
E_{EE} = - k_{B}T{\sum_{n = 0}^{\infty}}'\tilde{\epsilon}_{1}\tilde{\epsilon}_{2}\int_{1}d\textbf{r}_{1}\int_{2}d\textbf{r}_{2}\frac{e^{-2kR}}{(4\pi)^{2} R^{6}}\left[6 + 12kR + 10k^{2}R^{2} + 4k^{3}R^{3} + 2k^{4}R^{4}\right].
\end{equation}
Replacing $k$ by $\Lambda n$ we obtain:
\begin{align}\label{Caso diagonal T neq 0 2}
E_{EE} & = - \frac{k_{B}T}{(4\pi)^{2}}\tilde{\epsilon}_{1}\tilde{\epsilon}_{2}\int_{1}\int_{2}\frac{d\textbf{r}_{1}d\textbf{r}_{2}}{R^{6}}{\sum_{n = 0}^{\infty}}'e^{-2\Lambda Rn}\times\nonumber\\
& \times\left[6 + 12\Lambda Rn + 10\Lambda^{2}R^{2}n^{2} + 4\Lambda^{3}R^{3}n^{3} + 2\Lambda^{4}R^{4}n^{4}\right].
\end{align}
Having into account that $\sum_{n = 0}^{\infty}e^{-an} = \frac{1}{1 - e^{-a}}$, it is easily deduced:
\begin{equation}
\partial_{a}^{t}\sum_{n = 0}^{\infty}e^{-an} = \sum_{n = 0}^{\infty}(-n)^{t}e^{-an} = \partial_{a}^{t}(1 - e^{-a})^{-1}.
\end{equation}
Denoting $\lambda = R\Lambda$, we can carry out the sum obtaining:
\begin{eqnarray}\label{Caso diagonal T neq 0 resuelto}
E_{EE} & = & - \frac{k_{B}T}{(4\pi)^{2}}\tilde{\epsilon}_{1}\tilde{\epsilon}_{2}\int_{1}\int_{2}\frac{d\textbf{r}_{1}d\textbf{r}_{2}}{R^{6}}\frac{1}{(e^{2\lambda} - 1)^{5}}\nonumber\\
& & \times[e^{2\lambda}\left( - 9 - 12\lambda + 10\lambda^2 - 4\lambda^3 + 2\lambda^4 \right)
 + e^{4\lambda}\left(  6 + 36\lambda - 10\lambda^2 - 12\lambda^3 + 22\lambda^4 \right)\nonumber\\
& & + e^{6\lambda}\left(  6 - 36\lambda - 10\lambda^2 + 12\lambda^3 + 22\lambda^4 \right)
 + e^{8\lambda}\left(- 9 + 12\lambda + 10\lambda^2 +  4\lambda^3 +  2\lambda^4 \right)\nonumber\\
& & + 3 e^{10\lambda} + 3].
\end{eqnarray}
We obtain a similar result for the purely magnetic Casimir energy replacing $\tilde{\epsilon}_{1}\tilde{\epsilon}_{2}$ by $\tilde{\mu}_{1}\tilde{\mu}_{2}$.
\subsection{Coupled magnetic - electric Casimir energy terms}
We perform the same calculations using \eqref{Acoplo EH funcion Green} and \eqref{Acoplo HE funcion Green} of the Appendix as for the purely electric case. The trace over spatial coordinates is already done in \eqref{Traza terminos de acoplo em}, so we obtain the formula for the coupling terms of the Casimir energy just replacing the integral in \eqref{Energia Casimir acoplo em a T = 0} by a sum:
\begin{equation}\label{Energia Casimir acoplo em a T neq 0}
E_{EH} = \frac{2k_{B}T}{(4\pi)^{2}}\tilde{\epsilon}_{1}\tilde{\mu}_{2}\int_{1}\int_{2}\frac{d\textbf{r}_{1}d\textbf{r}_{2}}{R^{6}}{\sum_{n = 0}^{\infty}}'e^{-2kR}\left(R^{4}k^{4} + 2R^{3}k^{3} + R^{2}k^{2}\right).
\end{equation}
Here $k_{n} = 2\pi\frac{k_{B}T}{\hbar c}n = \Lambda n$ are the Matsubara frequencies. After solving this sum and denoting $\lambda = R\Lambda$, we obtain:
\begin{align}\label{Energia Casimir acoplo em a T neq 0 resuelto}
E_{EH} = & \frac{2k_{B}T}{(4\pi)^{2}}\tilde{\epsilon}_{1}\tilde{\mu}_{2}\int_{1}\int_{2}\frac{d\textbf{r}_{1}d\textbf{r}_{2}}{R^{6}}\frac{\lambda^2}{(e^{2\lambda} - 1)^{5}}\times\nonumber\\
 & \times\left[\begin{array}{l}
\phantom{+} e^{2\lambda}\left(  1 - 2\lambda +\lambda^2 \right) + e^{4\lambda}\left( -1 - 6\lambda + 11\lambda^2\right)\\
 + e^{6\lambda}\left( -1 + 6\lambda + 11\lambda^2\right) + e^{8\lambda}\left( 1 + 2\lambda +\lambda^2 \right)
\end{array}\right]
\end{align}
Replacing $\tilde{\epsilon}_{1}\tilde{\mu}_{2}$ by $\tilde{\mu}_{1}\tilde{\epsilon}_{2}$, we obtain the formula for $E_{HE}$. These formulas for the Casimir energy in the diluted limit are valid for any temperature, but they are too much complicate for analytical analysis at any temperature. Therefore, we study the limits at high and low temperatures.
\subsection{Low and zero temperature limit}
Here we will recover the zero temperature limit of the diluted limit. Then we will make a perturbative analysis valid for low temperatures. We can make it easily by expanding the Taylor series in $\lambda = R\Lambda$ of the integrand and studying just the first orders deleting the rest ones. By using \eqref{Caso diagonal T neq 0 resuelto} and \eqref{Energia Casimir acoplo em a T neq 0 resuelto},  we reobtain \eqref{Final Result} for the zero temperature case. We need to take the fifth order series term of the Taylor expansion of \ecref{Caso diagonal T neq 0 resuelto} and \ecref{Energia Casimir acoplo em a T neq 0 resuelto} to get the next non zero perturbation term of the Casimir energy as
\begin{equation}\label{diluted_Limit_correction_low_T_diagonal_term}
\frac{\Delta_{5}E_{EE}}{\tilde{\epsilon}_{1}\tilde{\epsilon}_{2}} = \frac{\Delta_{5}E_{HH}}{\tilde{\mu}_{1}\tilde{\mu}_{2}} = - \frac{22\pi^{3}}{945}k_{B}T\left(\frac{k_{B}T}{\hbar c}\right)^{5}\int_{1}\int_{2}\frac{d\textbf{r}_{1}d\textbf{r}_{2}}{R},
\end{equation}
\begin{equation}\label{diluted_Limit_correction_low_T_non_diagonal_term}
\frac{\Delta_{5}E_{EH}}{\tilde{\epsilon}_{1}\tilde{\mu}_{2}} = \frac{\Delta_{5}E_{HE}}{\tilde{\mu}_{1}\tilde{\epsilon}_{2}} = - \frac{2\pi^{3}}{189}k_{B}T \left(\frac{k_{B}T}{\hbar c}\right)^{5}\int_{1}\int_{2}\frac{d\textbf{r}_{1}d\textbf{r}_{2}}{R},
\end{equation}
because $\Delta_{1}E = \Delta_{2}E = \Delta_{3}E = \Delta_{4}E = 0 $, where $\Delta_{n}E$ is the n-th order correction term to the low temperature expansion.
\subsection{High temperature and classical limit}
In this Section we will obtain the high temperature limit of the Casimir energy in the diluted limit, whose first term will be the classical limit of the Casimir energy. To obtain this limit, instead of solving the sum in \ecref{Caso diagonal T neq 0} and \ecref{Energia Casimir acoplo em a T neq 0}, we will just keep the first term of the sum. Then the classical limit of the Casimir energy in the diluted limit reads the first sum term:
\begin{equation}
\frac{E_{EE}^{cl}}{\tilde{\epsilon}_{1}\tilde{\epsilon}_{2}} = \frac{E_{HH}^{cl}}{\tilde{\mu}_{1}\tilde{\mu}_{2}} = - 3 \frac{k_{B}T}{(4\pi)^{2}}\int_{1}\int_{2}\frac{d\textbf{r}_{1}d\textbf{r}_{2}}{R^{6}},
\end{equation}
\begin{equation}
\frac{E_{EH}^{cl}}{\tilde{\epsilon}_{1}\tilde{\mu}_{2}} = \frac{E_{HE}^{cl}}{\tilde{\mu}_{1}\tilde{\epsilon}_{2}} = 0.
\end{equation}
And the first perturbation to that limit is the next sum term:
\begin{align}\label{diluted_Limit_correction_high_T_diagonal_term}
\frac{\Delta_{1}E_{EE}^{cl}}{\tilde{\epsilon}_{1}\tilde{\epsilon}_{2}} &= 
\frac{\Delta_{1}E_{HH}^{cl}}{\tilde{\mu}_{1}\tilde{\mu}_{2}}\\
& = - \frac{k_{B}T}{(4\pi)^{2}}\int_{1}\int_{2}\frac{d\textbf{r}_{1}d\textbf{r}_{2}}{R^{6}}e^{-2\Lambda R}\left(6 + 12\Lambda R + 10\Lambda^{2}R^{2} + 4\Lambda^{3}R^{3} + 2\Lambda^{4}R^{4}\right),\nonumber
\end{align}
\begin{equation}\label{diluted_Limit_correction_high_T_non_diagonal_term}
\frac{\Delta_{1}E_{EH}^{cl}}{\tilde{\epsilon}_{1}\tilde{\mu}_{2}} = \frac{\Delta_{1}E_{HE}^{cl}}{\tilde{\mu}_{1}\tilde{\epsilon}_{2}} = \frac{2k_{B}T}{(4\pi)^{2}}\int_{1}\int_{2}\frac{d\textbf{r}_{1}d\textbf{r}_{2}}{R^{6}}e^{-2R\Lambda}\left(R^{4}\Lambda^{4} + 2R^{3}\Lambda^{3} + R^{2}\Lambda^{2}\right).
\end{equation}
Where $\Delta_{1}E^{cl}$ is the first correction to the classic limit of the Casimir energy $E^{cl}$. With these results we see that we loose the coupling between electric and magnetic energy terms in the classical limit.

Finally we must remark that, if we take the asymptotic distance approximation as in Sect. \ref{sec: 3.C}, we reobtain the results given in \cite{Barton} in the soft response limit.

\section{PSA for three bodies system}\label{sec: 4_PSA}
In this Section we are going to calculate the Casimir energy between three bodies in the Pairwise Summation approximation. In this asymptotic limit we will obtain that the energy of the system will be the addition of the PSA energy of each pair of objects. This linear behavior of the Casimir energy was expected in that approximation although we know that it is in general false~\cite{Rodriguez-Lopez_1}. We begin this study from the Casimir Energy formula for three objects given in \cite{Emig_Casimir_caso_escalar}\cite{Rodriguez-Lopez_1}:
\begin{equation}\label{Casimir N objects}
E = \frac{\hbar c}{2\pi}\int_{0}^{\infty}dk\log\left(\frac{\abs{\mathbb{M}}}{\phantom{_{\infty}}\abs{\mathbb{M}}_{\infty}}\right).
\end{equation}
The $M$ matrix (whose coefficients are non commutative matrices) is:
\begin{displaymath}
\mathbb{M} = \left(\begin{array}{c c c}
\phantom{-}\mathbb{T}_{1}^{-1} & - \mathbb{U}_{12}   & - \mathbb{U}_{13}\\
 - \mathbb{U}_{21}  & \phantom{-}\mathbb{T}_{2}^{-1} & - \mathbb{U}_{23}\\
 - \mathbb{U}_{31}  & - \mathbb{U}_{32}   & \phantom{-}\mathbb{T}_{3}^{-1}
\end{array}\right).
\end{displaymath}
So, using the logarithm product rule, the Casimir energy between three bodies is
\begin{equation}\label{Energia Casimir 3 cuerpos}
E_{3} = \frac{\hbar c}{2\pi}\int_{0}^{\infty}dk\log\Det{\mathbbm{1} - \mathbb{N}_{12}}
 + \frac{\hbar c}{2\pi}\int_{0}^{\infty}dk\log\Det{\mathbbm{1} - \mathbb{N}_{13}}
 + \frac{\hbar c}{2\pi}\int_{0}^{\infty}dk\log\Det{\mathbbm{1} - \mathbb{R}}.
\end{equation}
Here the $N$ matrix is $\mathbb{N}_{\alpha\beta} = \mathbb{T}_{\alpha}\mathbb{U}_{\alpha\beta}\mathbb{T}_{\beta}\mathbb{U}_{\beta\alpha}$, and:
\begin{equation}
\mathbb{R} = \left(\mathbbm{1} - \mathbb{N}_{13}\right)^{-1}\left(\mathbb{T}_{3}\mathbb{U}_{32} + \mathbb{N}_{31}\right)\left(\mathbbm{1} - \mathbb{N}_{12}\right)^{-1}\left(\mathbb{T}_{2}\mathbb{U}_{23} + \mathbb{N}_{21}\right).
\end{equation}
Expanding the inverses we obtain the double series
\begin{equation}
\mathbb{R} = \sum_{n_{1}=0}^{\infty}\left(\mathbb{N}_{13}\right)^{n_{1}}\left(\mathbb{T}_{3}\mathbb{U}_{32} + \mathbb{N}_{31}\right)
\sum_{n_{2}=0}^{\infty}\left(\mathbb{N}_{12}\right)^{n_{2}}\left(\mathbb{T}_{2}\mathbb{U}_{23} + \mathbb{N}_{21}\right).
\end{equation}
The lowest order of the Born series of the $\mathbb{R}$ matrix will come from the first order expansion series making $n_{1} = n_{2} = 0$, that is
\begin{equation}
\mathbb{R} \simeq \left(\mathbb{T}_{3}\mathbb{U}_{32} + \mathbb{N}_{31}\right)\left(\mathbb{T}_{2}\mathbb{U}_{23} + \mathbb{N}_{21}\right).
\end{equation}
In addition to that, we just consider the sum term with the minimum number of $\mathbb{T}$ matrices products, because that will be the lowest order expansion in susceptibilities of the $\mathbb{R}$ matrix. That means that we reduce the highly non linear $\mathbb{R}$ matrix to
\begin{equation}\label{sol limit R matrix}
\mathbb{R} \simeq \left(\mathbb{T}_{3}\mathbb{U}_{32}\mathbb{T}_{2}\mathbb{U}_{23}\right) = \mathbb{N}_{23}.
\end{equation}
Replacing \ecref{sol limit R matrix} in \ecref{Energia Casimir 3 cuerpos}, we obtain the next PSA of the Casimir energy between three objects:
\begin{equation}\label{3-bodies PSA}
E_{3} = \frac{\hbar c}{2\pi}\int_{0}^{\infty}dk\log\Det{\mathbbm{1} - \mathbb{N}_{12}}
 + \frac{\hbar c}{2\pi}\int_{0}^{\infty}dk\log\Det{\mathbbm{1} - \mathbb{N}_{13}}
 + \frac{\hbar c}{2\pi}\int_{0}^{\infty}dk\log\Det{\mathbbm{1} - \mathbb{N}_{23}}.
\end{equation}
Therefore we obtain that the PSA approximation of the Casimir energy between three objects is the sum of the PSA energy of each pair of objects. In other words, we rederive a kind of superposition law of energies in the diluted PSA limit as expected, because the usual presentation of the PSA approximation is the assumption that we have a superposition behavior in that asymptotic limit. Here we have justified the validity of that approximation.

If we take the asymptotic distance limit to \ecref{3-bodies PSA}, we will obtain a superposition of two bodies PSA energies in the asymptotic limit. The nonlinearity of the Casimir energy must be given by higher orders expansion terms even for systems with three bodies.
The same superposition behavior is expected for the general $N$ body case, and it will be proven in the next Section.

\subsection{PSA for general N bodies system}
In this Section we are going to generalize the PSA energy of three bodies given by \ecref{3-bodies PSA}. For this purpose we are going to use an iterative procedure which will give us the new terms to include to the PSA of the Casimir energy of $n-1$ bodies when we include another new object to our system. Let us represent the $\mathbb{M}$ matrix of \ecref{Casimir N objects} as the sum of its diagonal and its non-diagonal parts as \cite{Emig_Casimir_caso_escalar}:
\begin{equation}
\mathbb{M}_{\alpha\beta} = \delta_{\alpha\beta}\mathbb{T}_{\alpha}^{-1} + (\delta_{\alpha\beta} - 1)\mathbb{U}_{\alpha\beta},
\end{equation}
or symbolically as
\begin{equation}
\mathbb{M} = \mathbb{T}^{-1} + \mathbb{U},
\end{equation}
then it is easy to find that the inverse of $\mathbb{M}$ is the next perturbative series
\begin{equation}
\mathbb{M}^{-1} = \mathbb{T}\sum_{n=0}^{\infty}\left(-\mathbb{U}\mathbb{T}\right)^{n}.
\end{equation}
On the other hand, the $\mathbb{M}$ matrix of the $N$ objects system is related with the $\mathbb{M}$ matrix of the $N-1$ objects system by block matrices in the next way
\begin{displaymath}
\mathbb{M}_{N} = \left(\begin{array}{c c c}
\phantom{-}\mathbb{M}_{N-1\phantom{,\gamma}} & - \mathbb{U}_{N-1,\gamma}\\
 - \mathbb{U}_{\gamma,N-1}  & \phantom{-}\mathbb{T}_{N\phantom{-1,\gamma}}^{-1}
\end{array}\right).
\end{displaymath}
Where $\gamma$ index goes from 1 to $N-1$. With this result we calculate $\Det{\mathbb{M}_{N}}$ obtaining
\begin{equation}
\abs{\mathbb{M}_{N}} = \abs{\mathbb{M}_{N-1}}\abs{\mathbb{T}_{N}}^{-1}\abs{\mathbbm{1} - \mathbb{T}_{N}\mathbb{U}_{\gamma ,N-1}\mathbb{M}_{N-1}^{-1}\mathbb{U}_{N-1,\gamma}}.
\end{equation}
In the PSA approximation is valid the substitution $\mathbb{M}^{-1} \simeq \mathbb{T}$, where $\mathbb{T}$ is a diagonal matrix whose $N-1$ diagonal elements are $\mathbb{T}_{\gamma}$ with $\gamma$ index defined as before.
Then, using this approximation we can approximate the $\mathbb{M}_{N}$ determinant as:
\begin{equation}
\abs{\mathbb{M}_{N}} = \abs{\mathbb{M}_{N-1}}\abs{\mathbb{T}_{N}}^{-1}\abs{\mathbbm{1} - \mathbb{T}_{N}\mathbb{U}_{\gamma ,N-1}\mathbb{T}\mathbb{U}_{N-1,\gamma}}.
\end{equation}
Where we multiply by blocks the matrix of the last determinant obtaining
\begin{eqnarray}
\abs{\mathbb{M}_{N}} & = & \abs{\mathbb{M}_{N-1}}\abs{\mathbb{T}_{N}}^{-1}\Det{\mathbbm{1} - \left(\begin{array}{c c c c}
\mathbb{N}_{1N} & 0 & \dots & 0\\
0 & \mathbb{N}_{2N} & \dots & 0\\
\vdots & \vdots & \ddots & \vdots\\
0 & 0 & \dots & \mathbb{N}_{N-1,N}
\end{array}\right)}.\nonumber\\
& &
\end{eqnarray}
Finally, we obtain the desired result:
\begin{equation}
\abs{\mathbb{M}_{N}} = \abs{\mathbb{M}_{N-1}}\abs{\mathbb{T}_{N}}^{-1}\prod_{\gamma =1}^{N-1}\abs{\mathbbm{1} - \mathbb{N}_{\gamma N}}.
\end{equation}
The other needed matrix is
\begin{equation}
\abs{\mathbb{M}_{\infty, N}} = \abs{\mathbb{M}_{\infty ,N-1}}\abs{\mathbb{T}_{N}}^{-1}.
\end{equation}
And with the initial conditions:
\begin{equation}
\abs{\mathbb{M}_{1}} = \abs{\mathbb{M}_{\infty, 1}} = \abs{\mathbb{T}_{1}}^{-1},
\end{equation}
we obtain these determinants in a closed form as
\begin{equation}
\abs{\mathbb{M}_{N}} = \prod_{k = 1}^{N}\abs{\mathbb{T}_{k}}^{-1}\prod_{l = 2}^{N}\prod_{m = 1}^{l-1}\abs{\mathbbm{1} - \mathbb{N}_{lm}}
\end{equation}
and
\begin{equation}
\abs{\mathbb{M}_{\infty, N}} = \prod_{k = 1}^{N}\abs{\mathbb{T}_{k}}^{-1}.
\end{equation}
Taking the logarithms of \ecref{Casimir N objects}, the Casimir energy for the N bodies system is approximated as
\begin{equation}\label{N-bodies PSA}
E_{N} = \frac{\hbar c}{2\pi}\int_{0}^{\infty}dk\sum_{l = 2}^{N}\sum_{m = 1}^{l-1}\log\Det{\mathbbm{1} - \mathbb{N}_{lm}}
\end{equation}
in the PSA limit. \ecref{N-bodies PSA} is the generalization of \ecref{3-bodies PSA} for the $N$ body case, and it show us that in the PSA limit, we always obtain a superposition principle of the two body PSA Casimir energy despite the fact that Casimir energy is not a nonadittive interaction. We will obtain the non-linear effects in the next orders of the expansion.

This result is qualitatively different of the usual PSA procedure of integrating each point of a $N$ point potential over the volume of each object. The reason is simple, a $N$ point potential is proportional to $N$ polarizabilities \cite{Thiru}, but in the soft limit approximation, the lower allowed term is proportional to two polarizabilities. Therefore we should study the $(N-1)$ expansion term to obtain the first one proportional to $N$ polarizabilities.
So in a consistent calculation of PSA energies for three or more objects, the asymptotic approximation of the PSA energy in the soft material approximation is different of the $N$ point potential function.
In the next Section we will find where the contribution of the three point potential is relevant in the diluted limit.

\section{Second order expansion of diluted limit}\label{sec: 5_PSA}
In this Section we are going to study the second order expansion of the PSA in the diluted limit. This is a complicated long calculus so, instead the complete electromagnetic case, we will restrict ourselves to the purely electric case. That means that we will ignore the magnetic properties of the bodies. It could be interesting to show the complete study of that series term, but it is a long calculation to show here. However, it possesses theoretical utility, as it shows the lack of the superposition principle in the Casimir energy calculations. The next results have been done with the help of {\tt Mathematica} \cite{Mathematica}.

Again we start from the Casimir energy between two compact objects as given by \ecref{Formula de Emig en forma de traza}. We studied the first order expansion of the energy in Sect. \ref{sec: 2_PSA}. Now we are interested in the next order expansion term in polarizabilities, so we maintain the study of the fist term of \ecref{Formula de Emig en forma de traza} taking just the case $p = 1$. In addition to that, we take the second order approximation of the Born series of the $\mathbb{T}$ matrix:
\begin{equation}
\mathbb{T} = \sum_{n=0}^{\infty}(VG_{0})^{n}V\simeq V + VG_{0}V.
\end{equation}
Applying the linearity of the trace and taking our attention just to the second order expansion in polarizabilities (that is, to the third order term in polarizabilities), we obtain the following result for the second order of the diluted limit result:
\begin{equation}\label{segundo orden PSA}
E_{2} = - \frac{\hbar c}{2\pi}\int_{0}^{\infty}dk\tr\left(V_{1'}G_{01'1}V_{1}U_{12}V_{2}U_{21'}\right)
 - \frac{\hbar c}{2\pi}\int_{0}^{\infty}dk\tr\left(V_{1}U_{12'}V_{2'}G_{02'2}V_{2}U_{21}\right).
\end{equation}
We will center our study just on the first integral of \eqref{segundo orden PSA}, because the analysis of these two integrals is the same. First we make the trace over the $EH$ space, which is automatic here because we have canceled the magnetic properties of the bodies. In other words, we are not studying the coupling between electric and magnetic induced dipoles. After that we trace over the space coordinates, so we need the matricial form of the electric part of the dyadic Green function given in \ecref{Acoplo EE funcion Green} of the Appendix, where $\textbf{R} = \textbf{r} - \textbf{r}'$. We will need to define the function:
\begin{equation}
\mathcal{R}[\alpha ,\beta]_{i}(k) = R_{\alpha\beta}^2 \left(1+k R_{\alpha\beta}+k^2 R_{\alpha\beta}^2\right)
 + \left(3+3 k R_{\alpha\beta}+k^2 R_{\alpha\beta}^2\right)R_{i\alpha\beta}^{2}.
\end{equation}
Where $k$ is the frequency, $\alpha$ and $\beta$ labels the bodies coordinates and $i$ shows the vector component of $\textbf{R}_{\alpha\beta}$ used. So we can solve the trace over the space coordinates obtaining the following result for the first term of \eqref{segundo orden PSA}:
\begin{eqnarray}
E_{2a} & = & - \frac{\hbar c}{2\pi}\int_{1}d\textbf{r}_{1}\int_{2}d\textbf{r}_{2}\int_{1'}d\textbf{r}_{1'}\int_{0}^{\infty}dk
\frac{e^{-k (R_{12} + R_{21'} + R_{1'1})}}{64\pi^{3}R_{12}^{5}R_{21'}^{5}R_{1'1}^{5}}\tilde{\epsilon}_{1}\tilde{\epsilon}_{2}\tilde{\epsilon}_{1'}\nonumber\\
& & \times[\mathcal{R}[1,2]_{x}(k)\mathcal{R}[2,1']_{x}(k)\mathcal{R}[1',1]_{x}(k)\nonumber\\
& & + \mathcal{R}[1,2]_{y}(k)\mathcal{R}[2,1']_{y}(k)\mathcal{R}[1',1]_{y}(k)\nonumber\\
& & + \mathcal{R}[1,2]_{z}(k)\mathcal{R}[2,1']_{z}(k)\mathcal{R}[1',1]_{z}(k)].
\end{eqnarray}
Here the space integrations are performed just over the body volumes because it is there where dielectric and diamagnetic potentials are defined. It is possible to perform the $k$ integral, because we have an exponential multiplied by a polynomial integrated between zero and infinity. It is not shown here because it contains 216 sum terms.

For any finite temperature we also obtain a long result, but in the classical limit we can at least obtain a tractable formula. This limit consists of keeping only the $k = 0$ mode. In that case we obtain the result
\begin{eqnarray}
E_{2a}^{cl} & = & - \frac{k_{B}T}{128\pi^{3}}\int_{1}d\textbf{r}_{1}\int_{2}d\textbf{r}_{2}\int_{1'}d\textbf{r}_{1'}
\frac{\tilde{\epsilon}_{1}\tilde{\epsilon}_{2}\tilde{\epsilon}_{1'}}{R_{12}^{5}R_{21'}^{5}R_{1'1}^{5}}\nonumber\\
& & \times[\mathcal{R}[1,2]_{x}(0)\mathcal{R}[2,1']_{x}(0)\mathcal{R}[1',1]_{x}(0)\nonumber\\
& & + \mathcal{R}[1,2]_{y}(0)\mathcal{R}[2,1']_{y}(0)\mathcal{R}[1',1]_{y}(0)\nonumber\\
& & + \mathcal{R}[1,2]_{z}(0)\mathcal{R}[2,1']_{z}(0)\mathcal{R}[1',1]_{z}(0)].
\end{eqnarray}
Where $\mathcal{R}[\alpha ,\beta]_{i}(0) = \left(R_{\alpha\beta}^{2} + 3R_{\alpha\beta i}^{2}\right)$, so we can write:
\begin{eqnarray}\label{Divergent integrand example}
E_{2a}^{cl} & = & - \frac{k_{B}T}{128\pi^{3}}\int_{1}d\textbf{r}_{1}\int_{2}d\textbf{r}_{2}\int_{1'}d\textbf{r}_{1'}
\frac{\tilde{\epsilon}_{1}\tilde{\epsilon}_{2}\tilde{\epsilon}_{1'}}{R_{12}^{5}R_{21'}^{5}R_{1'1}^{5}}\nonumber\\
& & \times[\left(R_{12}^{2} + 3R_{12 x}^{2}\right)\left(R_{21'}^{2} + 3R_{21' x}^{2}\right)\left(R_{1'1}^{2} + 3R_{1'1 x}^{2}\right)\nonumber\\
& & + \left(R_{12}^{2} + 3R_{12 y}^{2}\right)\left(R_{21'}^{2} + 3R_{21' y}^{2}\right)\left(R_{1'1}^{2} + 3R_{1'1 y}^{2}\right)\nonumber\\
& & + \left(R_{12}^{2} + 3R_{12 z}^{2}\right)\left(R_{21'}^{2} + 3R_{21' z}^{2}\right)\left(R_{1'1}^{2} + 3R_{1'1 z}^{2}\right)].
\end{eqnarray}
When performing this integral, a term proportional to $R_{1'1}^{-5}$ and another one to $R_{1'1}^{-3}$ appear. These terms can be problematic because $R_{1'1} = 0$ belongs to the integration interval for all the points of the $1$ body. But these singularities can be removed with the appropriate regularization procedure.
Note that if we make $1' = 3$ in \ecref{segundo orden PSA}, this integral would also be one of the contributions of the second term expansion of a three objects system. In that case we have not got any problem in the volumes integration for this term and this expansion term looks quite similar to the local analog of the three points potential given in \cite{Thiru}. In fact, in the three (or more) bodies system that term is the first term that breaks the superposition behavior founded in \ref{sec: 4_PSA}.
Replacing the $k^{2}$ term by the operator $-\Delta$ in Eqs.~\eqref{Acoplo EE funcion Green} and \eqref{Acoplo HH funcion Green} of the Appendix,  and replacing the $k$ term by the operator $\sqrt{-\Delta}$ in Eqs.~\eqref{Acoplo HE funcion Green} and \eqref{Acoplo EH funcion Green} of the Appendix, which is a change valid on shell because $G_{0}(R,k) = \frac{e^{-kR}}{4\pi R}$, we recover the formalism of Power and Thirunamachandran~\cite{Thiru} in the soft dielectric and soft diamagnetic limit. This scheme let us take into account not only electric phenomena as in \cite{Thiru}, but also magnetic and electro-magnetic coupling effects so as an extension to finite temperature cases. Taking this argument into account, we can understand the structure of the perturbation terms of the PSA limit of the electromagnetic Casimir energy between $N$ bodies. The Casimir energy in the PSA limit has the structure of a series of infinity terms whose $n$-order term is the sum of integrals over $n+2$ bodies (which can be repeated or not) of the $(n+2)$ points local em potential. Each series term come from the expansions made in \ecref{Formula de Emig en forma de traza} and in \ecref{Born expansion of T operator} and is proportional to the product of $n+2$ permittivities. If we take the asymptotic distance limit of these integrals, we will obtain a series of $n$ points EM potential now with polarizabilities instead susceptibilities as in \cite{Thiru}, but in the soft response limit. When we use perturbations of the tree term of the Born expansion, it appears divergent terms in the integrand as seen in \ecref{Divergent integrand example}. These singularities can be removed with the appropriate regularization procedure~\cite{PhysRevA.80.012519}.

\section{PSA for objects with general magneto--electric coupling}\label{sec: 6}
In this Section, we extend the study of PSA to materials with a general magneto--electric linear coupling with the electromagnetic field, we use the potential $V_{i}$ given in Eq.~\eqref{Potencial_dielectric} to obtain the PSA Casimir energy of frequency independent dielectric with magnetoelectric couplings at zero temperature as
\begin{equation}
E = \frac{-\hbar c}{(4\pi)^{3}}\gamma_{0}\int_{1}\int_{2}\frac{d\textbf{r}_{1}d\textbf{r}_{2}}{\abs{\textbf{r}_{1} - \textbf{r}_{2}}^{7}},
\end{equation}
where
\begin{equation}
\gamma_{0} = 23\tilde{\epsilon}_{1}\tilde{\epsilon}_{2} - 7\tilde{\epsilon}_{1}\tilde{\mu}_{2} - 7\tilde{\mu}_{1}\tilde{\epsilon}_{2} + 23\tilde{\mu}_{1}\tilde{\mu}_{2} + 7\alpha_{1}\alpha_{2} + 23\alpha_{1}\beta_{2} + 23\beta_{1}\alpha_{2} + 7\beta_{1}\beta_{2}.
\end{equation}
This result extend the Feinberg and Sucher potential~\cite{VdW_int.magnetica} to objects with magnetoelectric couplings, but only in the diluted limit. In the high temperature limit, the PSA of Casimir energy is
\begin{equation}
E = \frac{- k_{B}T}{(4\pi)^{2}}\gamma_{cl}\int_{1}\int_{2}\frac{d\textbf{r}_{1}d\textbf{r}_{2}}{\abs{\textbf{r}_{1} - \textbf{r}_{2}}^{6}},
\end{equation}
where
\begin{equation}
\gamma_{cl} = 3\tilde{\epsilon}_{1}\tilde{\epsilon}_{2} + 3\tilde{\mu}_{1}\tilde{\mu}_{2} + 3\alpha_{1}\beta_{2} + 3\beta_{1}\alpha_{2}.
\end{equation}

It is also possible to obtain results valid at any fixed temperature from Eq.~\eqref{Formula de Emig T}, the results are Eq.~\eqref{Caso diagonal T neq 0 resuelto}, but with $\left(\alpha_{1}\beta_{2} + \beta_{1}\alpha_{2}\right)$ instead $\tilde{\epsilon}_{1}\tilde{\epsilon}_{2}$, and Eq.~\eqref{Energia Casimir acoplo em a T neq 0 resuelto}, but with $\left(- \alpha_{1}\alpha_{2} - \beta_{1}\beta_{2}\right)$ instead $\tilde{\epsilon}_{1}\tilde{\mu}_{2}$. Then the correction terms to high and low temperature cases are given by Eqs.~\eqref{diluted_Limit_correction_low_T_diagonal_term}, \eqref{diluted_Limit_correction_low_T_non_diagonal_term}, \eqref{diluted_Limit_correction_high_T_diagonal_term} and \eqref{diluted_Limit_correction_high_T_non_diagonal_term}, being careful with the sign for terms proportional to $\alpha_{1}\alpha_{2}$ and $\beta_{1}\beta_{2}$.

As we can see, if magnetoelectric couplings have different signs and are strong enough to compensate the usual electric and magnetic coupling of dielectrics, a repulsive Casimir energy between objects can be achieved in the diluted limit. We will study the appearance of this repulsion in a more physical system of Topological Insulators in the next Section.

\section{PSA for Topological Insulators}\label{sec: 7}
Recently, in \cite{GC10}, A. Grushin and A. Cortijo demonstrated the existence of an equilibrium distance between topological insulators with opposite topological polarizabilities sign between parallel plates. This result was extended for all temperatures in \cite{Placas_TI_forall_T}, leading to a reduction of the equilibrium distance with the temperature until the disappearance of the equilibrium distance in the high temperature regime. Regions of Casimir repulsion for all distances were also reported for some values of $w = \frac{\omega_{e}}{\omega_{R}}$ and $\abs{\theta}$ (where $w$ is the ratio between the oscillator strength $\omega_{e}$ and the resonant frequency $\omega_{R}$ of the material, while $\theta$ is the magnetoelectric topological susceptibility, their definition is given below). In this part of the article, we apply the formalism developed in the previous Section to TI. 

Following \cite{GC10} and \cite{Placas_TI_forall_T}, the electromagnetic response of three-dimensional topological insulators, which determines the reflection coefficients, is governed by the Lagrangian
\begin{equation}\label{Lagrangiano_TI}
\mathcal{L} = \mathcal{L}_{0} + \mathcal{L}_{\theta} = \textbf{E}\cdot\textbf{D} + \textbf{B}\cdot\textbf{H}.
\end{equation}
The first term of Eq.~\eqref{Lagrangiano_TI} reads $\mathcal{L}_{0} = \frac{1}{2}\left(\epsilon\textbf{E} - \mu^{-1}\textbf{B}\right)$, and it is the usual electro--magnetic term, which gives the usual constitutive relations of dielectrics without magneto--electric couplings. The second term of Eq.~\eqref{Lagrangiano_TI} is a nontrivial axionic~\cite{W87} or topological magnetoelectric term which gives rise to the topological magnetoelectric~\cite{QHZ08}~\cite{EMB09} term in the Lagrangian $\mathcal{L}_{\theta} = \frac{\alpha\theta}{2\pi}\textbf{E}\cdot\textbf{B}$, where $\alpha$ is the fine structure constant ($\alpha = \frac{e^{2}}{\hbar c}$) and $\theta$ is the magnetoelectric topological susceptibility. Therefore, the constitutive relations of TI are easily obtained from the Lagrangian of TI given in Eq.~\eqref{Lagrangiano_TI} as
\begin{align}
\textbf{D} & = \epsilon\textbf{E} + \alpha\left(\frac{\theta}{\pi}\right)\textbf{B},\nonumber\\
\textbf{H} & = \mu^{-1}\textbf{B} - \alpha\left(\frac{\theta}{\pi}\right)\textbf{E}.
\end{align}
Time-reversal symmetry indicates that $\theta = 0,\pi (\text{mod }2\pi)$ being $\theta = \pi$ the case for TI and $\theta = 0$ the case for trivial insulators.
When the boundary of the considered object is included, the action corresponding
to the Lagrangian Eq.~\eqref{Lagrangiano_TI}, $S_{0} + S_{\theta}$, is a fair description of the
TI only when a time-reversal breaking perturbation is induced on the surface to gap the surface
states~\cite{QHZ08}. In this work we consider that the time-reversal perturbation is a magnetic coating of small thickness $l$ compared with the rest scales of the problem (therefore we will neglect its contribution to Casimir effect) which gaps the surface states. In the described
situation, $\theta$ is quantized in odd integer values of $\pi$ such that
\begin{equation}
\theta = (2 n + 1)\pi,
\end{equation}
where $n\in\mathbb{Z}$, determined by the nature of the magnetic coating, but independent of the absolute value of the magnetization of the coating. Positive or negative values of $\theta$ are related to different signs of the magnetization on the surface \cite{QLZZ09}, which we consider is perpendicular to the surface of the body. Being a topological contribution, $\theta$ is defined in the bulk as a constant whenever the bulk Brillouin zone is defined \cite{EMB09}. %Surface effects are not taken into account in this model~\cite{Private_Communication}, but the model presented here can be easily generalized to cope with them. We just should add a surface contribution with the dependence on the imaginary frequency of the magnetoelectric coupling of the studied model, with a penetration length $\delta$ of the surface contribution. In the model studied here, $\delta = 0$ is assumed and therefore the surface contribution to Casimir energy is neglected.
Following the discussion in \cite{Placas_TI_forall_T}, we assume that the frequency dependent dielectric function $\epsilon(\omega)$ is described by an oscillator model of the form:
\begin{equation}
\epsilon(i\kappa) = \epsilon_{0} + \sum_{i}\frac{\omega_{e,i}^{2}}{\omega_{R,i}^{2} + \gamma_{R,i}c\kappa + c^{2}\kappa^{2}}.
\end{equation}
In this model, $\omega_{R,i}$ is the resonant frequency of the $i$th oscillator, while $\omega_{e,i}$ accounts for the oscillator strength. The damping parameter $\gamma_{R,i}$ satisfies $\gamma_{R,i} \ll \omega_{R,i}$, playing therefore a secondary role. As there are few experimental results of $\epsilon(i\kappa)$ for TI, we will assume that there is just one resonance and do not take into account the contribution of the damping constant, therefore the dielectric function considered here has the form:
\begin{equation}
\epsilon(i\kappa) = \epsilon_{0} + \frac{w^{2}}{1 + \frac{c^{2}}{\omega_{R}^{2}}\kappa^{2}},
\end{equation}
where $w = \frac{\omega_{e}}{\omega_{R}}$. In addition to that, we assume $\mu = \mu_{0}$.

In order to apply the formalism developed in the last Section, we need the fields $\textbf{D}$ and $\textbf{B}$ as a function of $\textbf{E}$ and $\textbf{H}$, then
\begin{align}
\textbf{D} & = \left(\epsilon + \mu\alpha^{2}\left(\frac{\theta}{\pi}\right)^{2}\right)\textbf{E} + \mu\alpha\left(\frac{\theta}{\pi}\right)\textbf{H},\nonumber\\
\textbf{B} & = \mu\alpha\left(\frac{\theta}{\pi}\right)\textbf{E} + \mu\textbf{H},
\end{align}
where $\alpha$ is the fine structure constant ($\alpha = \frac{e^{2}}{\hbar c}$). Note that the PSA can be widely applied to TI, because it is expected that the magnetoelectric topological susceptibility $\theta$ was small (and the magnetoelectric coupling proportional to the fine structure constant), and the interesting behavior of TI because of the Casimir effect appears also for the soft dielectric response of the TI.

The case of multiple resonances or any other more exact model of $\epsilon(i\kappa)$~\cite{GC10} and magnetoelectric coupling (which could have into account the contribution of the surface of the TI to the magnetoelectric response) can be easily performed within the formalism presented here.

\begin{figure}[h]
\begin{center}
\includegraphics[width=0.9\columnwidth]{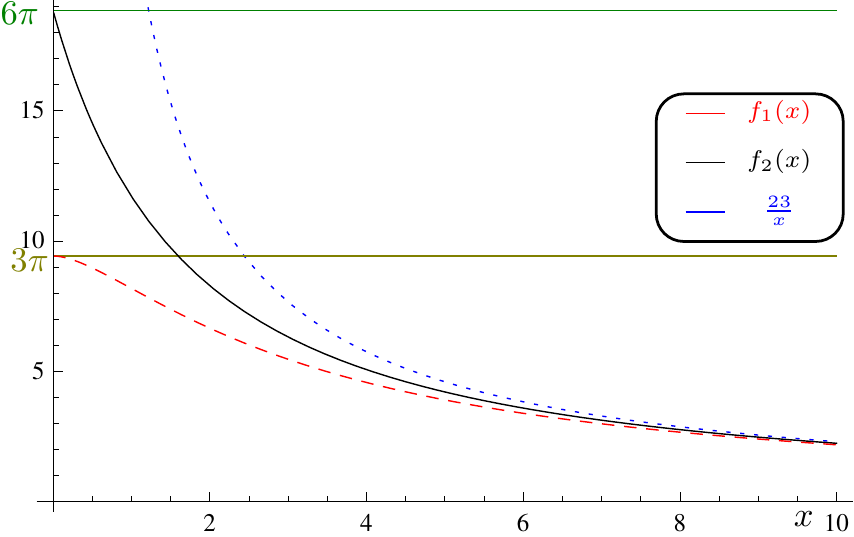}
\caption{\label{Plot_f00_f02} Representation of $f_{1}(x)$ given in Eq.~\eqref{Eq_f00} (red color) and $f_{2}(x)$ given in Eq.~\eqref{Eq_f02} (black color). The blue curve is their common limit at large $x$, the green curve is the limit of $f_{2}(x)$ at small $x$ and the yellow curve is the limit at small $x$ of $f_{1}(x)$.}
\end{center}
\end{figure}

%\section{Asymptotic Casimir energy between 2 compact TI}\label{sec: 7.1}
Because of the frequency dependence of $\epsilon(i\kappa)$ for TI, it is not possible to obtain the integrand of the PSA energy at zero temperature in a closed form in terms of simple functions. Then the PSA Casimir energy at zero temperature for topological insulators is given by the integral
\begin{equation}
E_{0}^{PSA} = - \frac{\hbar c}{(4\pi)^{3}}\int_{1}\int_{2}\frac{d\textbf{r}_{1}d\textbf{r}_{2}}{\abs{\textbf{r}_{1} - \textbf{r}_{2}}^{7}}\gamma_{0},
\end{equation}
where $\gamma_{0}$ now depends on the distance between points of the TI and on the dielectric properties of both TI as 
\begin{equation}\label{Integrando_PSAE_for_TI}
\gamma_{0} = 
w_{1}^{2}w_{2}^{2} x f_{1}(x) + 60 \bar{\alpha}_{1}\bar{\alpha}_{2} + 23\bar{\alpha}_{1}^{2} \bar{\alpha}_{2}^{2}
  + \left(\bar{\alpha}_{1}^{2}w_{2}^{2} + \bar{\alpha}_{2}^{2}w_{1}^{2}\right)x f_{2}(x),
\end{equation}
where $\bar{\alpha}_{i} = \alpha\frac{\theta_{i}}{\pi}$, $w = \frac{\omega_{e}}{\omega_{R}}$, $x = \frac{\omega_{R}}{c}\abs{\textbf{r}_{1} - \textbf{r}_{2}}$ is the adimensional distance and $f_{1}(x)$ and $f_{2}(x)$ the next functions:
\begin{equation}\label{Eq_f00}
f_{1}(x) = 
\frac{2}{\sqrt{\pi }} \left(\begin{array}{l}
6 x G_{1,3}^{3,1}\left(x^2|
\begin{tiny}
\begin{array}{c}
 0 \\
 0,\frac{1}{2},1
\end{array}
\end{tiny}
\right) + 3 G_{1,3}^{3,1}\left(x^2|
\begin{tiny}
\begin{array}{c}
 \frac{1}{2} \\
 0,\frac{1}{2},\frac{3}{2}
\end{array}
\end{tiny}
\right)+ 5 G_{1,3}^{3,1}\left(x^2|
\begin{tiny}
\begin{array}{c}
 \frac{1}{2} \\
 1,\frac{3}{2},\frac{3}{2}
\end{array}
\end{tiny}
\right) \\
+ x^{4}G_{1,3}^{3,1}\left(x^2|
\begin{tiny}
\begin{array}{c}
 -\frac{3}{2} \\
 -\frac{1}{2},0,\frac{1}{2}
\end{array}
\end{tiny}
\right) + 2 x^{3} G_{1,3}^{3,1}\left(x^2|
\begin{tiny}
\begin{array}{c}
 -1 \\
 0,0,\frac{1}{2}
\end{array}
\end{tiny}
\right)\end{array}\right),
\end{equation}

\begin{equation}\label{Eq_f02}
f_{2}(x) = \frac{2}{\sqrt{\pi }}\left(\begin{array}{l}
6 x G_{1,3}^{3,1}\left(x^{2}|
\begin{tiny}
\begin{array}{c}
 -1 \\
 0,0,\frac{1}{2}
\end{array}
\end{tiny}
\right) + 6 x G_{1,3}^{3,1}\left(x^{2}|
\begin{tiny}
\begin{array}{c}
 0 \\
 0,\frac{1}{2},1
\end{array}
\end{tiny}
\right) + 3 G_{1,3}^{3,1}\left(x^{2}|
\begin{tiny}
\begin{array}{c}
 -\frac{1}{2} \\
 0,\frac{1}{2},\frac{1}{2}
\end{array}
\end{tiny}
\right)\\
 + 5 G_{1,3}^{3,1}\left(x^{2}|
\begin{tiny}
\begin{array}{c}
 -\frac{1}{2} \\
 \frac{1}{2},1,\frac{3}{2}
\end{array}
\end{tiny}
\right) + 3 G_{1,3}^{3,1}\left(x^{2}|
\begin{tiny}
\begin{array}{c}
 \frac{1}{2} \\
 0,\frac{1}{2},\frac{3}{2}
\end{array}
\end{tiny}
\right) + 5 G_{1,3}^{3,1}\left(x^{2}|
\begin{tiny}
\begin{array}{c}
 \frac{1}{2} \\
 1,\frac{3}{2},\frac{3}{2}
\end{array}
\end{tiny}
\right)\\
 + x^{4} G_{1,3}^{3,1}\left(x^{2}|
\begin{tiny}
\begin{array}{c}
 -\frac{5}{2} \\
 -\frac{3}{2},0,\frac{1}{2}
\end{array}
\end{tiny}
\right) + x^{4}G_{1,3}^{3,1}\left(x^{2}|
\begin{tiny}
\begin{array}{c}
 -\frac{3}{2} \\
 -\frac{1}{2},0,\frac{1}{2}
\end{array}
\end{tiny}
\right)\\
 + 2 x^{3}G_{1,3}^{3,1}\left(x^{2}|
\begin{tiny}
\begin{array}{c}
 -2 \\
 -1,0,\frac{1}{2}
\end{array}
\end{tiny}
\right) + 2 x^{3}G_{1,3}^{3,1}\left(x^{2}|
\begin{tiny}
\begin{array}{c}
 -1 \\
 0,0,\frac{1}{2}
\end{array}
\end{tiny}
\right)\end{array}\right),
\end{equation}
where $G_{pq}^{mn}\left(x\left|\begin{tiny}
\begin{array}{c}
 a_1,\ldots ,a_p \\
 b_1,\ldots ,b_q
\end{array}
\end{tiny}\right.\right)$ are the Meijer $G$ functions. Eqs.~\eqref{Eq_f00} and \eqref{Eq_f02} and their limits at short and large $x$ are plotted in Fig. \ref{Plot_f00_f02}. All coefficients of $\gamma_{0}$ are positive ($xf_{1}(x)$, $xf_{2}(x)$, $23$ and $60$, because $f_{1}(x)$ and $f_{2}(x)$ are smooth positive functions in the real positive axis, as seen in Fig. \ref{Plot_f00_f02}) for any value of $\theta_{1}$ and $\theta_{2}$.

At short distances we can approximate $f_{1}(x)\approx 3\pi$ and $f_{2}(x)\approx 6\pi$, then $\gamma_{0}$ can be approximated by
\begin{equation}\label{Integrando_PSAE_for_TI_short}
\gamma_{0} = 60\bar{\alpha}_{1}\bar{\alpha}_{2} + 23 \bar{\alpha}_{1}^{2}\bar{\alpha}_{2}^{2}.
\end{equation}
If $\text{sign}(\theta_{1}) = - \text{sign}(\theta_{2})$, this energy becomes positive (i.e. repulsive) for all
\begin{equation}\label{Repulsion_condition_short_distance_limit}
\theta_{1}\theta_{2} < - \frac{60\pi^{2}}{23\alpha^{2}}.
\end{equation}
For diluted materials, this condition is beyond the positive energy condition $\abs{\theta_{1}\theta_{2}} < \frac{\pi^{2}}{\alpha^{2}}\sqrt{\epsilon_{1}(0)\epsilon_{2}(0)} < \frac{60\pi^{2}}{23\alpha^{2}}$ \cite{BHS68}, then we can consider it is always fulfilled for diluted dielectrics.

At large distances we can approximate $f_{1}(x)\approx\frac{23}{x}$ and $f_{2}(x)\approx\frac{23}{x}$, then $\gamma_{0}$ can be approximated by
\begin{equation}\label{Integrando_PSAE_for_TI_large}
\gamma_{0} = w_{1}^{2}w_{2}^{2} + \bar{\alpha}_{1}^{2}\bar{\alpha}_{2}^{2} + \bar{\alpha}_{1}^{2}w_{2}^{2} + \bar{\alpha}_{2}^{2}w_{1}^{2} + \frac{60}{23}\bar{\alpha}_{1}\bar{\alpha}_{2}.
\end{equation}
If $\text{sign}(\theta_{1}) = - \text{sign}(\theta_{2})$ and $w_{1} = w_{2} = w$, the energy becomes positive (i.e. repulsive) for all
\begin{equation}\label{Repulsion_condition_large_distance_limit}
w^{2} < \frac{23}{46}\left(- \bar{\alpha}_{1}^{2} - \bar{\alpha}_{2}^{2} + \sqrt{\bar{\alpha}_{1}^{4} + \bar{\alpha}_{2}^{4} - \frac{240}{23}\bar{\alpha}_{1}\bar{\alpha}_{2} - \frac{46}{23}\bar{\alpha}_{1}^{2}\bar{\alpha}_{2}^{2}}\right).
\end{equation}

As a consequence, there are several different regimes for the system in the quantum limit. When $\text{sign}(\theta_{1}) = \text{sign}(\theta_{2})$, the Casimir energy is enlarged because of the contribution of topological charge $\theta$, but when $\text{sign}(\theta_{1}) = - \text{sign}(\theta_{2})$, different regimes appear.

When $\text{sign}(\theta_{1}) = - \text{sign}(\theta_{2})$ and for low enough absolute values of $\theta_{1}$ and $\theta_{2}$, the condition given in Eq.~\eqref{Repulsion_condition_short_distance_limit} is fulfilled while the condition given in Eq.~\eqref{Repulsion_condition_large_distance_limit} is not, then we have a repulsive Casimir energy at short distances and an attractive Casimir energy at large distances, then there must exist an stable equilibrium distance in this case.

But if  the condition given in Eq.~\eqref{Repulsion_condition_large_distance_limit} is also fulfilled, the magnitude of the positive arguments of Eq.~\eqref{Integrando_PSAE_for_TI} is not large enough to compensate the repulsion because topological charges $\theta$, then we obtain repulsion for all distances.

\begin{figure}[h]
\begin{center}
\includegraphics[width=\columnwidth]{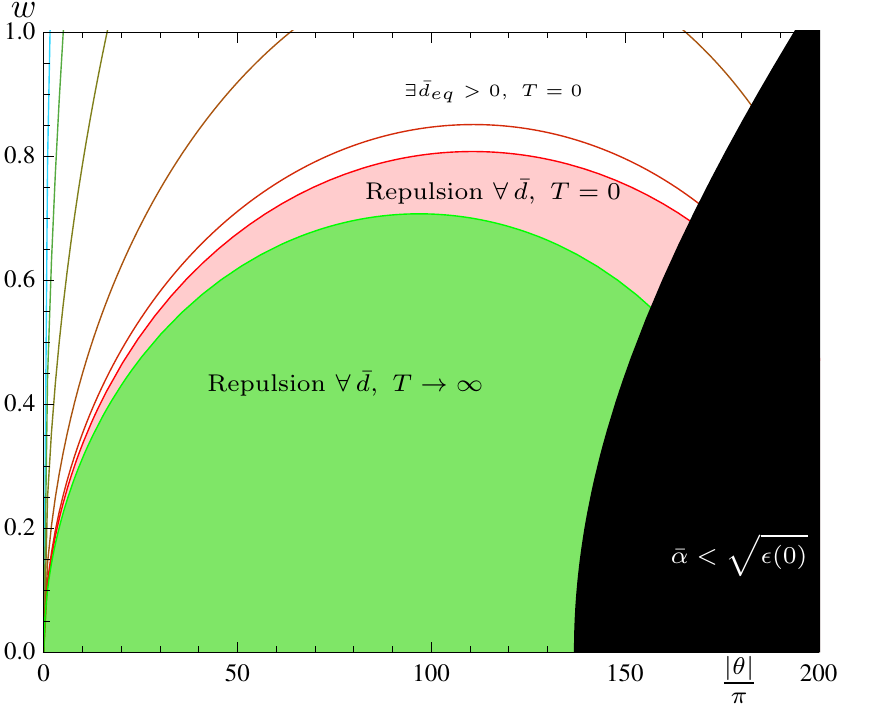}
\caption{\label{Mapa_Fases_TI}Attraction versus Repulsion in the classical limit ($T\to\infty$) and in the quantum limit ($T\to 0$) as a function of $w$ and $\abs{\theta}$. For the Classical limit, the green region shows repulsion for all distances and the rest of phase space shows attraction for all distances. For the quantum limit, the red and green regions show repulsion for all distances and the white region shows parameters for which an equilibrium distance appears. Several curves of constant adimensional equilibrium distance $\bar{d}_{eq} = \frac{\omega_{R}}{c}\abs{\textbf{r}_{1} - \textbf{r}_{2}}$ are also plotted. The black region is a forbidden region for the parameters due to the positive energy condition $\abs{\bar{\alpha}} < \sqrt{\epsilon(0)}$ \cite{BHS68}.}
\end{center}

\end{figure}

In the high temperature limit, another result is obtained. In this case the PSA gives the Casimir energy as
\begin{equation}
E_{cl}^{PSA} = - \frac{3 k_{B}T}{(4\pi)^{2}}\gamma_{cl}\int_{1}\int_{2}\frac{d\textbf{r}_{1}d\textbf{r}_{2}}{\abs{\textbf{r}_{1} - \textbf{r}_{2}}^{6}},
\end{equation}
where, contrary to the zero temperature case, $\gamma_{cl}$ does not depend on the distance between points, and it is given by
\begin{equation}
\gamma_{cl} = w_{1}^{2}w_{2}^{2} + \bar{\alpha}_{1}^{2}w_{2}^{2} + \bar{\alpha}_{2}^{2}w_{1}^{2} + \bar{\alpha}_{1}^{2}\bar{\alpha}_{2}^{2} + 2\bar{\alpha}_{1}\bar{\alpha}_{2}.
\end{equation}
Depending on the values of $w$ and $\abs{\theta}$, we obtain either repulsion for all distances or attraction for all distances. The condition to be fulfilled in order to obtain repulsion in the classical limit when $w_{1} = w_{2} = w$ is
\begin{equation}\label{Repulsion_condition_classical_limit}
w^{2} < \frac{1}{2}\left(- \bar{\alpha}_{1}^{2} - \bar{\alpha}_{2}^{2} + \sqrt{\bar{\alpha}_{1}^{4} + \bar{\alpha}_{2}^{4} - 8\bar{\alpha}_{1}\bar{\alpha}_{2} - 2\bar{\alpha}_{1}^{2}\bar{\alpha}_{2}^{2}}\right).
\end{equation}
In Fig. \ref{Mapa_Fases_TI}, we represent the behavior of the PSA energy as a function of $\theta$ and $w$ when $w_{1} = w_{2}$ and $\theta_{1} = \theta = - \theta_{2}$ in the quantum and in the classical limit, having into account that there is a forbidden region because of the positive energy condition \cite{BHS68}. In both cases we find different regions of parameters of repulsion for all distances, but in the quantum limit there is a region of existence of an equilibrium distance, while in the classical limit a region of attraction for all distances appears.

It is expected that at finite temperatures the equilibrium distance found in the quantum limit would be reduced when temperature increases, and Fig. \ref{Mapa_Fases_TI} suggests that behavior, but we have let this study as a future work.

\subsection{Casimir energy between spheres and plates}\label{sec: 7.2}
A claim would be done to the method performed here. It is a method just valid in the diluted limit. Then the results presented here are general results in the geometry of the bodies, but not in the dielectric response of materials.

In this Section we will use the PSA formulas presented in the previous Chapter to obtain the Casimir energy to specific systems. In particular, we study the sphere--plate system because of its experimental relevance and the two infinite plates system to compare the PSA results with the exact ones. Both systems were studied at zero temperature.

We were not able to obtain analytical results, so a numerical method has been implemented.

\begin{figure}[h]
\begin{center}
\includegraphics[width=0.9\columnwidth]{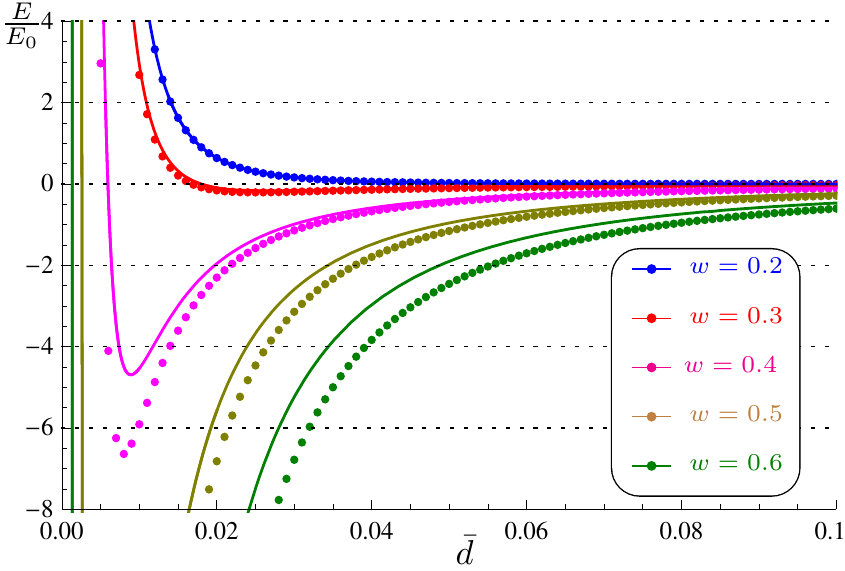}
\caption{\label{Comparacion_PSA_Multiscattering} Comparison between exact Casimir energies in units of $E_{0} = A\hbar c/(2\pi)^{2}\left(\omega_{R}/c\right)^{3}$(full curves) and PSA (points) at zero temperature for TI infinity plates with $\theta_{1} = - \theta_{2} = \pi$ and different $\epsilon(0)$ as a function of the adimensional distance $\bar{d}$~\cite{GC10}. PSA tends to overestimate the absolute value of the energy, but captures the equilibrium distance without appreciable error. The overestimation of the energy increases with $\epsilon(0)$, as expected because of the nature of the approximation.}
\end{center}
\end{figure}

The two infinite parallel TI plates have been already studied in \cite{GC10}. Our interest here is to compare the results obtained by the PSA with the exact ones. We impose $\theta_{1} = - \theta_{2} = \pi$ in order to obtain an equilibrium distance, and different $w$ from $w = 0.2$ to $w = 0.6$. The obtained results of PSA and the exact ones are compared in Fig. \ref{Comparacion_PSA_Multiscattering}, where we find that PSA tends to overestimate the magnitude of the Casimir energy the more the larger $w$. This fact reflects the nature of the approximation made in PSA, where we must assume the diluted limit is valid. On the other side, an excellent approximation of the equilibrium distance between plates is obtained, then the PSA gives a good qualitative result.

\begin{figure}[h]
\begin{center}
\includegraphics[width=0.9\columnwidth]{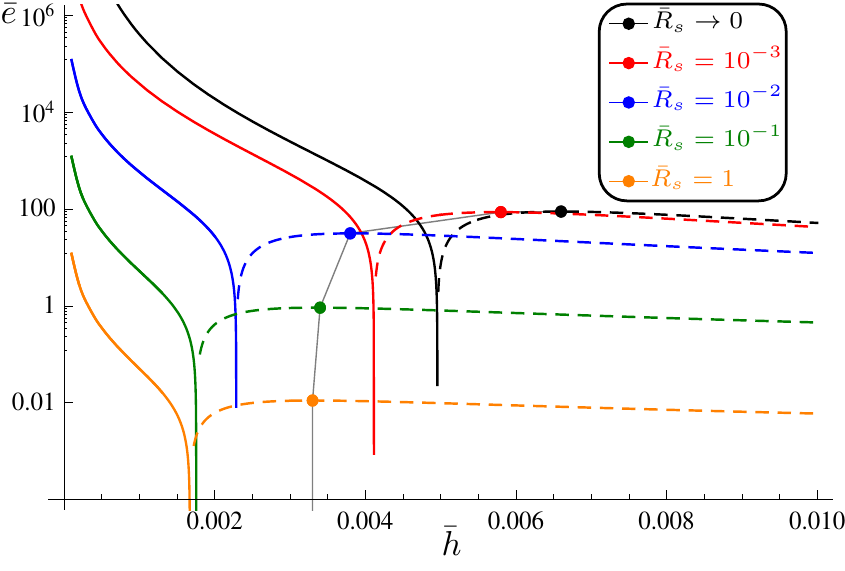}
\caption{\label{Figura_Esfera_Placa}Dimensionless PSA Casimir energies in the sphere-plate TI system per unit of volume of the sphere $\bar{e} = \frac{c^{3}}{\hbar\omega_{R}^{4}}\frac{E}{V}$ with $\theta_{1} = - \theta_{2} = \pi$ and $w = 0.45$ as a function of the dimensionless distance between the plate and the contact point of the sphere $\bar{h} = \bar{d} - \bar{R}_{s}$, where $\bar{d}$ is the dimensionless distance from the surface of the plate to the center of the sphere. Positive energies are the whole curves, while negative energies are the dashed curves. The equilibrium distance for each sphere radius is represented by a point. These points are joined by a soft line to have an idea of the dependence of the equilibrium distance with the radius of the sphere.}
\end{center}
\end{figure}

We have also used the PSA to study the sphere-plate system because of its experimental relevance in Casimir effect experiments. In this case, we also impose $\theta_{1} = - \theta_{2} = \pi$ and $w = 0.45$ for both plate and sphere in order to obtain an equilibrium distance at $T=0$ (see Fig. \ref{Mapa_Fases_TI}). We vary the dimensionless radius of the sphere $\bar{R}_{s} = \frac{\omega_{R}}{c}R_{s}$ from $\bar{R}_{s}\to 0$ until $\bar{R}_{s} = 1$ to observe its effect in the equilibrium distance. In Fig.~\ref{Figura_Esfera_Placa}, the PSA energy per unit of volume of the sphere is plotted as a function of the distance between the nearest points of the plate and the sphere. As a result, an equilibrium distance has been obtained for all the studied sphere radii. This equilibrium distance reduces when the radius increases until reaching a constant value. The appearance of an equilibrium distance is easy to understand because of the nature of the integrand in PSA calculations at zero temperature. As discussed in Eq. \eqref{Integrando_PSAE_for_TI}, from a given distance, the points of the sphere nearer to the plate tend to increase the Casimir energy (giving a repulsive contribution), while the rest of points of the sphere tend to reduce this energy (giving an attractive contribution). As at short distances Eq. \eqref{Integrando_PSAE_for_TI} diverges as $\frac{1}{d^{7}}$ and $\int_{0}^{d}\frac{dx}{x^{7}}$ diverges, it is evident that there is always a distance where repulsion compensates attraction, reaching the system an equilibrium distance.

\section{Conclusions}
In this Chapter, we have calculated the full electromagnetic Casimir energy in the diluted limit between two bodies using the formalism given in \cite{Kardar-Geometrias-Arbitrarias} reobtaining the energy given in \cite{MiltonPRL} in the pure electric case. Although this formalism fails in the perfect metal case, it is valid for soft dielectric and diamagnetic bodies and for bodies with magnetoelectric coupling. This formalism has been generalized to finite temperatures cases (giving exact results at finite temperatures) and extended for an arbitrary number of bodies. We have shown that the Casimir energy, which is a nonadittive interaction, has a superposition behavior in the first expansion of the PSA and how this superposition behavior is broken when the integration of different $N$ point potentials appears in the perturbation series to contribute in the whole energy as the next perturbation terms of the PSA.

This formalism let us reobtain known results of the literature as particular cases. For example, high and low temperature correction terms were obtained in~\cite{Barton} from a functional formalism, here we have obtained the exact result at any finite temperature. In addition to that, we obtain a generalization of the $N$ point potentials formalism of Power and Thirunamachandran~\cite{Thiru} valid for diamagnetic objects and at finite temperatures.

The extension of this formalism to anisotropic materials is straightforward, although we have not done it here, it is shown how to do it for the case where it would be needed.

We have also used the results derived here to study the Casimir energy for a two infinity TI plates, to compare the PSA with the exact results given in \cite{GC10}. As a result, we found that PSA gives better quantitative results for lower $w$, as expected, because PSA works better in the diluted limit. On the other side, we obtain an excellent approximation of the equilibrium distance between plates, therefore we obtain good qualitative results for all $\theta$ and $w$ studied. We have also used the PSA to study the Casimir energy between an infinite TI plate and a TI sphere because of the experimental relevance of this geometry.

The results presented in this article are valid for objects of arbitrary shape, but in the diluted limit. Then we have a systematic procedure to calculate Casimir energies valid for bodies with soft dielectric, soft diamagnetic and soft magnetoelectric response.

%\acknowledgements
%I acknowledge helpful discussions with R.~Brito and T.~Emig.  This research was supported by projects MOSAICO, UCM/PR34/07-15859 and a FPU MEC grant.

% Apéndice
%\appendix
\section{Appendix A: Obtention of the matricial Green function}\label{sec:Appendix}
Here we have calculated the Casimir energy using a matricial Green function for the electromagnetic fields instead of the dyadic Green function used in \cite{Kardar-Geometrias-Arbitrarias}. In this Appendix we will see that both formalisms are equivalent just using as field sources the induced polarization and magnetization vectors instead of the induced currents into the bodies. The induced currents into the bodies are \cite{Jackson}:
\begin{equation}
\textbf{j} = -ik\textbf{P} + \nabla\times\textbf{M}.
\end{equation}
We introduce this linear change of variable in the partition function, so there is not any new relevant term in the action of the problem. This action will be transformed in the next way:
\begin{equation}
S = \int d\textbf{x}\bar{\textbf{j}}\mathcal{G}_{0}\textbf{j},
\end{equation}
\begin{eqnarray}
S & = & \int d\textbf{x}k^{2}\bar{\textbf{P}}\mathcal{G}_{0}\textbf{P} + \int d\textbf{x}\nabla\times\bar{\textbf{M}}\mathcal{G}_{0}\nabla\times\textbf{M}\nonumber\\
  &   & + \int d\textbf{x}ik\bar{\textbf{P}}\mathcal{G}_{0}\nabla\times\textbf{M} - \int d\textbf{x}\nabla\times\bar{\textbf{M}}\mathcal{G}_{0}ik\textbf{P}.
\end{eqnarray}
Where
\begin{equation}
\mathcal{G}_{0} = \left[\delta_{ij} - \frac{1}{k^{2}}\nabla_{i}\nabla_{j}\right]G_{0},
\end{equation}
is the Green dyadic function. We integrate by parts each action term obtaining:
\begin{eqnarray}
S_{EE} & = & \int d\textbf{x}k^{2}\bar{\textbf{P}}\mathcal{G}_{0}\textbf{P},\nonumber\\
       & = & \int d\textbf{x}\bar{\textbf{P}}_{i}\left[k^{2}\delta_{ij} - \nabla_{i}\nabla_{j}\right]G_{0}\textbf{P}_{j},\nonumber\\
       & = & \int d\textbf{x}\bar{\textbf{P}}_{i}G_{ij}^{EE}\textbf{P}_{j},
\end{eqnarray}
\begin{eqnarray}
S_{EH} & = & \int d\textbf{x}ik\bar{\textbf{P}}\mathcal{G}_{0}\vec{\nabla}\times\textbf{M}, \nonumber\\
       & = & \int d\textbf{x}ik\bar{\textbf{P}}_{i}\left[\delta_{ij} - \frac{1}{k^{2}}\nabla_{i}\nabla_{j}\right]G_{0}\epsilon_{j\alpha\beta} \nabla^{\alpha}\textbf{M}^{\beta},\nonumber\\
& &
\end{eqnarray}
\begin{equation}
S_{EH} = \int d\textbf{x}\bar{\textbf{P}}_{i}\left[ik\epsilon_{i\alpha\beta}\nabla^{\beta}\right]G_{0}\textbf{M}^{\alpha}
  + \int d\textbf{x}\bar{\textbf{P}}_{i}\left[ik\frac{1}{k^{2}}\nabla_{i}\nabla_{j}\epsilon_{j\alpha\beta}\nabla^{\beta}\right]G_{0}\textbf{M}^{\alpha}.
\end{equation}
The second term is zero:
\begin{equation}
\epsilon_{j\alpha\beta}\nabla_{j}\textbf{M}^{\alpha}\nabla^{\beta}G_{0} = \textbf{M}\cdot\vec{\nabla}\times\vec{\nabla}G_{0} = \textbf{M}\cdot\textbf{0} = 0.
\end{equation}
So we finally get:
\begin{eqnarray}
S_{EH} & = & \int d\textbf{x}\bar{\textbf{P}}_{i}\left[-ik\epsilon_{ij\beta}\nabla^{\beta}\right]G_{0}\textbf{M}^{j},\nonumber\\
       & = & \int d\textbf{x}\bar{\textbf{P}}_{i}G_{ij}^{EH}\textbf{M}^{j}.
\end{eqnarray}
$S_{HE}$ is similar to $S_{EH}$, but with a changed sign because of the complex conjugation of the induced current into the action:
\begin{equation}
S_{HE} =  - \int d\textbf{x}\nabla\times\bar{\textbf{M}}\mathcal{G}_{0}ik\textbf{P},
\end{equation}
\begin{equation}
S_{HE} = \int d\textbf{x}\bar{\textbf{M}}_{i}G_{ij}^{HE}\textbf{P}^{j} = \int d\textbf{x}\bar{\textbf{M}}_{i}\left[- G_{ij}^{EH}\right]\textbf{P}^{j}.
\end{equation}
And the last term:
\begin{equation}
S_{HH} = \int d\textbf{x}\nabla\times\bar{\textbf{M}}\mathcal{G}_{0}\vec{\nabla}\times\textbf{M},
\end{equation}
\begin{equation}
S_{HH} = \int d\textbf{x}\epsilon^{i\alpha\beta}\nabla_{\alpha}\bar{\textbf{M}}_{\beta}\left[\delta_{ij} - \frac{1}{k^{2}}\nabla_{i}\nabla_{j}'\right]G_{0}\epsilon_{jab}\nabla'^{a}\textbf{M}^{b},
\end{equation}
\begin{equation}
S_{HH} = \int d\textbf{x}\epsilon^{i\alpha\beta}\bar{\textbf{M}}_{\alpha}\nabla_{\beta}\left[\delta_{ij} - \frac{1}{k^{2}}\nabla_{i}\nabla_{j}'\right]G_{0}\epsilon_{jab}\textbf{M}^{a}\nabla'^{b},
\end{equation}
\begin{equation}
S_{HH} = \int d\textbf{x}\bar{\textbf{M}}_{\alpha}\left(\epsilon^{i\alpha\beta}\nabla_{\beta}\delta_{ij}G_{0}\epsilon_{jab}\nabla'^{b}\right)
\textbf{M}^{a}
 - \int d\textbf{x}\bar{\textbf{M}}_{\alpha}\left(\epsilon^{i\alpha\beta}\nabla_{\beta}\frac{1}{k^{2}}\nabla_{i}\nabla'^{j}G_{0}\epsilon_{jab}\nabla'^{b}\right)\textbf{M}^{a},
\end{equation}
Using $\epsilon^{i\alpha\beta}\delta_{i}^{j}\epsilon_{jab} = \delta_{a}^{\alpha}\delta_{b}^{\beta} - \delta_{b}^{\alpha}\delta_{a}^{\beta}$ and $\nabla_{b}\nabla'^{b}G_{0} = k^{2}G_{0}$, we get the next result:

\begin{eqnarray}
\left(\epsilon^{i\alpha\beta}\nabla_{\beta}\delta_{ij}G_{0}\epsilon_{jab}\nabla'^{b}\right) & = & \left[\delta_{a}^{\alpha}\delta_{b}^{\beta} - \delta_{b}^{\alpha}\delta_{a}^{\beta}\right]\nabla_{\beta}\nabla'^{b}G_{0}, \nonumber\\
 & = & \left[\delta_{a}^{\alpha}\nabla_{b}\nabla'^{b} - \nabla_{a}\nabla'^{\alpha}\right]G_{0},\nonumber\\
 & = & \left[\delta_{a}^{\alpha}k^{2} - \nabla_{a}\nabla'^{\alpha}\right]G_{0}.
\end{eqnarray}
The other term requires even a more tedious work, but it is easy to obtain that

\begin{equation}
\frac{-1}{k^{2}}\epsilon^{i\alpha\beta}\epsilon_{jab}\nabla_{\beta}\nabla_{i}\nabla'^{j}\nabla'^{b}G_{0} = 0,
\end{equation}
because this differential operator is zero. So $S_{HH}$ is
\begin{eqnarray}
S_{HH} & = & \int d\textbf{x}\bar{\textbf{M}}_{\alpha}\left(\epsilon^{i\alpha\beta}\nabla_{\beta}\delta_{ij}G_{0}\epsilon_{jab}\nabla'^{b}\right)
\textbf{M}^{a}, \nonumber\\
 & = & \int d\textbf{x}\bar{\textbf{M}}_{\alpha}\left[\delta_{a}^{\alpha}k^{2} - \nabla_{a}\nabla'^{\alpha}\right]G_{0}\textbf{M}^{a},\nonumber\\
 & = & \int d\textbf{x}\bar{\textbf{M}}_{i}G_{ij}^{HH}\textbf{M}_{j}.
\end{eqnarray}
After a Wick rotation, we obtain the used form of the matricial Green function, which components are, using $G_{0}(R,k) = \frac{e^{-kR}}{4\pi R}$ and $R = \abs{\textbf{r} - \textbf{r}'}$:
\begin{eqnarray}
G_{0ij}^{EE}(R,k) & = & \left[k^{2}\delta_{ij} + \nabla_{i}\nabla_{j}'\right]G_{0}(R,k),\label{Acoplo EE funcion Green}\\
G_{0ij}^{EH}(R,k) & = & - k\epsilon_{ijk}\nabla_{k}G_{0}(R,k),\label{Acoplo EH funcion Green}\\
G_{0ij}^{HE}(R,k) & = & k\epsilon_{ijk}\nabla_{k}G_{0}(R,k),\label{Acoplo HE funcion Green}\\
G_{0ij}^{HH}(R,k) & = & \left[k^{2}\delta_{ij} + \nabla_{i}\nabla_{j}'\right]G_{0}(R,k)\label{Acoplo HH funcion Green}.
\end{eqnarray}
Which are the results used in \ecref{Formula de Emig en soft limit}. It is also possible to obtain the same result from fluctuation - dissipation theorem, as made in \cite{Agarwal}.

\section{Appendix B: Dielectric polarizability in the diluted limit}\label{Apendice_limite_diluido_susceptibilidad_electrica}
In this Appendix we are going to demonstrate that, in the diluted limit, the dielectric polarizability is given by
\begin{equation}\label{App_Diluted_Limit_alpha_1}
\alpha^{E} = \tilde{\epsilon}\frac{V}{4\pi}.
\end{equation}
Following \cite{Optica-Wolf}, the macroscopic induced polarization vector $\textbf{P}$ of a given dielectric is defined as
\begin{equation}\label{App_Diluted_Limit_alpha_2}
\textbf{P} = \chi_{E}\textbf{E},
\end{equation}
where $\textbf{E}$ is the external applied electric field and $\chi_{E}$ is the electric susceptibility of the dielectric, defined as $\epsilon = 1 + 4\pi\chi_{E}$. Then $\chi_{E} = \frac{\tilde{\epsilon}}{4\pi}$. On the other side, the microscopic induced polarization vector $\textbf{p}$ of a given dielectric is defined as
\begin{equation}\label{App_Diluted_Limit_alpha_3}
\textbf{p} = \alpha^{E} \textbf{E}',
\end{equation}
where $\textbf{E}'$ is the local effective electric field. This local effective electric field is defined in terms of the local electric field $\textbf{E}$ and the macroscopic polarization field $\textbf{P}$, but in the diluted limit we can assume that the contribution of $\textbf{P}$ is negligible, then, just in the diluted limit,
\begin{equation}\label{App_Diluted_Limit_alpha_4}
\textbf{E}' = \textbf{E}.
\end{equation}
Because of the definition $\textbf{P} = \frac{d\textbf{p}}{dV}$ for homogeneous bodies, we can combine Eqs.~\eqref{App_Diluted_Limit_alpha_1} and \eqref{App_Diluted_Limit_alpha_2},
\begin{equation}\label{App_Diluted_Limit_alpha_5}
\textbf{p} = \alpha^{E} \textbf{E}' = V\textbf{P} = \chi_{E}V\textbf{E}.
\end{equation}
As we are in the diluted limit, using Eq.~\eqref{App_Diluted_Limit_alpha_4}, we obtain
\begin{equation}\label{App_Diluted_Limit_alpha_6}
\alpha^{E} = \chi_{E}V = \tilde{\epsilon}\frac{V}{4\pi},
\end{equation}
which is the result we wanted to demonstrate. Exactly the same reasoning can be applied to obtain the diluted limit of the magnetic polarizability as $\alpha^{H} = \tilde{\mu}\frac{V}{4\pi}$.

% Negative entropies between spheres
\begin{savequote}[9cm] % this sets the width of the quote
\sffamily
``In this house, we obey the laws of Thermodynamics!'' 
\qauthor{Homer Simpson}
\end{savequote}

\chapter{Casimir energy and entropy in the sphere--sphere Geometry}\label{Chap: Casimir Energy and Entropy in the Sphere--Sphere Geometry}
\graphicspath{{02-Casimir_Multiscattering/chb5_Entropia_Esferas/Figuras/}}

In this Chapter, we calculate the Casimir energy and entropy for two spheres described by the perfect metal model, plasma model, and Drude model in the large separation limit. We obtain nonmonotonic behavior of the Helmholtz free energy with separation and temperature for the perfect metal and plasma models, leading to parameter ranges with negative entropy, and also nonmonotonic behavior of the entropy with temperature and the separation between the spheres. This nonmonotonic behavior has not been found for Drude model. The appearance of this anomalous behavior of the entropy is discussed as well as its thermodynamic consequences.

In 1948, Casimir predicted the attraction between perfect metal parallel plates \cite{Casimir_Placas_Paralelas} and between neutral polarizable atoms \cite{VdW-int.electrica} due to quantum fluctuations of the electromagnetic field. Some years later, Schwinger extended this formalism to dielectric plates at finite temperature \cite{Schwinger}. Recently, a multiscattering formalism of the Casimir effect for the electromagnetic field has been presented \cite{Kardar-Geometrias-Arbitrarias}\cite{RE09} (see also \cite{Multiscattering_Lambrecht} and \cite{Multiscattering_Milton}).

The Casimir effect has some peculiarities. In particular, it is a non-pairwise interaction; the Casimir thermal force (the excess of Casimir energy at a given temperature compared with the Casimir energy in the zero temperature case) between two isolating bodies is not necessarily monotonic in their separation, as seen in the sphere--plate and cylinder--plate cases \cite{Metodo_Caminos_Esfera_Placa_Escalar}. In addition, for some geometries, intervals of negative entropy appear, as in the case of two parallel plates described by the Drude model \cite{Entropias_negativas_placas_Drude} or, as recently shown, in the interaction between a perfect metal plate and sphere \cite{Canaguier-Durand_Caso_Esfera_Placa_PRA}.

In this Chapter we study the Casimir effect between two spheres in the large separation approximation using different models for the electric permeability: (1)~the perfect metal model, (2)~the plasma model, and (3)~the Drude model. As a result, we find negative entropies in certain ranges of temperature and separation between the spheres for the perfect metal model, and for low penetration length for the plasma model. In addition, we find nonmonotonic behavior of the entropy with the separation while the force is attractive for all separations, making it appear as though the natural evolution of the system tends to increase the entropy in certain ranges of temperature and separation. For long plasma length and for the Drude model we do not find this anomalous behavior of the entropy. We also discuss the thermodynamical meaning and consequences of negative entropies in Casimir effect.

The remainder of the Chapter is arranged as follows: In Sect. \ref{Presentacion del modelo de Emig del Efecto Casimir}, we describe the multiscattering model used herein to obtain the Casimir energies and entropies for the two spheres. In Sect. \ref{Perfect metal spheres}, we obtain the large separation limit of the Casimir energy between perfect metal spheres, as well as the entropy and force. We also obtain entropies at smaller separations numerically.
We study the plasma model system in Sect. \ref{Plasma Model} and the Drude model in Sect. \ref{Drude Model}. We discuss the thermodynamic consequences of these results in Sect. \ref{Sect:Thermodynamic}. Finally, we discuss the results obtained in the Conclusions.

The contents of this Chapter is based on the work published in~\cite{Rodriguez-Lopez_3}.
\section{Multiscattering formalism of the Casimir energy}\label{Presentacion del modelo de Emig del Efecto Casimir}
To calculate the Casimir energy, entropy, and forces between two spheres, we employ the multiscattering formalism for the electromagnetic field \cite{Kardar-Geometrias-Arbitrarias}\cite{RE09}. This formalism relates the Casimir interaction between objects with the scattering of the field from each object. The Casimir contribution to the Helmholtz free energy at any temperature $T$ is given by
\begin{equation}\label{Energy_T_finite8}
E = k_{B}T{\sum_{n=0}^{\infty}}'\log\Det{\mathbbm{1} - \mathbb{N}(\kappa_{n})},
\end{equation}
where $\kappa_{n} = \frac{n}{\lambda_{T}}$ are the Matsubara frequencies and $\lambda_{T} = \frac{\hbar c}{2\pi k_{B}T}$ is the thermal wavelength. The prime indicates that the zero Matsubara frequency contribution has height of $1/2$. All the information regarding the system is described by the $\mathbb{N}$ matrix. For a system of two objects, this matrix is $\mathbb{N} = \mathbb{T}_{1}\mathbb{U}_{12}\mathbb{T}_{2}\mathbb{U}_{21}$. $\mathbb{T}_{i}$ is the T scattering matrix of the $i^{th}$ object, which accounts for all the geometrical information and electromagnetic properties of the object. $\mathbb{U}_{ij}$ is the translation matrix of electromagnetic waves from object $i$ to object $j$, which accounts for all information regarding the relative positions between the objects of the system.

For a sphere of radius $R$ with electric and magnetic permeabilities $\epsilon$ and $\mu$, the $\mathbb{T}$ matrix is diagonal in $\left(\ell m P,\ell' m'P'\right)$ space, with elements given by \\ $\mathbb{T}^{PP'}_{\ell m,\ell' m'} =  - \delta_{\ell\ell'}\delta_{mm'}\delta_{PP'}T^{P}_{\ell m}$, where the $T^{P}_{\ell m}$ are defined as \cite{RE09}
\begin{equation}\label{TMM_general}
\hspace{-5pt}T^{M}_{\ell m} \hspace{-2pt}=\hspace{-2pt} \frac{i_{\ell}(\kappa R)\partial_{R}(R i_{\ell}(n \kappa R)) \hspace{-2pt}-\hspace{-2pt} \mu\partial_{R}(R i_{\ell}(\kappa R))i_{\ell}(n \kappa R)}{k_{\ell}(\kappa R)\partial_{R}(R i_{\ell}(n \kappa R)) \hspace{-2pt}-\hspace{-2pt} \mu\partial_{R}(R k_{\ell}(\kappa R))i_{\ell}(n \kappa R)}\hspace{-1pt},
\end{equation}
\begin{equation}\label{TEE_general}
\hspace{-5pt}T^{E}_{\ell m} \hspace{-2pt}=\hspace{-2pt} \frac{i_{\ell}(\kappa R)\partial_{R}(R i_{\ell}(n \kappa R)) \hspace{-2pt}-\hspace{-2pt} \epsilon\partial_{R}(R i_{\ell}(\kappa R))i_{\ell}(n \kappa R)}{k_{\ell}(\kappa R)\partial_{R}(R i_{\ell}(n \kappa R)) \hspace{-2pt}-\hspace{-2pt} \epsilon\partial_{R}(R k_{\ell}(\kappa R))i_{\ell}(n \kappa R)}\hspace{-1pt},
\end{equation}
where $i_{\ell}(x) = \sqrt{\frac{\pi}{2x}}I_{\ell + \frac{1}{2}}(x)$, $k_{\ell}(x) = \frac{2}{\pi}\sqrt{\frac{\pi}{2x}}K_{\ell + \frac{1}{2}}(x)$, and $n = \sqrt{\mu\epsilon}$. Expressions for the $\mathbb{U}_{\alpha\beta}$ matrices can be found in \cite{RE09} and \cite{Wittmann}.

\section{Perfect metal spheres}\label{Perfect metal spheres}
In the perfect metal limit, we apply $\epsilon\to\infty$ for any $\mu$ to Eqs.~\eqref{TMM_general} and \eqref{TEE_general}. In this case, the $\mathbb{T}$ matrix elements are independent of $\mu$, taking the well-known universal form
\begin{equation}
\mathbb{T}^{MM}_{\ell m,\ell' m'} = - \delta_{\ell\ell'}\delta_{mm'}\frac{\pi}{2}\frac{I_{\ell + \frac{1}{2}}(\kappa R)}{K_{\ell + \frac{1}{2}}(\kappa R)},
\end{equation}
\begin{equation}
\mathbb{T}^{EE}_{\ell m,\ell' m'} \hspace{-2pt}=\hspace{-2pt} - \delta_{\ell\ell'}\delta_{mm'}\frac{\pi}{2}\frac{\ell I_{\ell + \frac{1}{2}}(\kappa R) - \kappa R I_{\ell - \frac{1}{2}}(\kappa R)}{\ell K_{\ell+ \frac{1}{2}}(\kappa R) + \kappa R K_{\ell - \frac{1}{2}}(\kappa R)}.
\end{equation}

\subsection{Casimir energy in the large separation limit}
To obtain the large separation limit of the Casimir energy, we need the dominant part of the $\mathbb{T}$ matrix in this limit. We define the adimensional frequency $q$ by $\kappa = q/d$. The main contribution in the large separation limit comes from the lowest-order expansion term of the $\mathbb{T}$ matrix elements in $1/d$. As it is known that, at small $\kappa$, the $\mathbb{T}$ elements scale as $\kappa^{2\ell+1}$ \cite{Kardar-Geometrias-Arbitrarias}, the dominant contribution comes from the dipolar polarizabilities part of the $\mathbb{T}$ matrix, taking the form
\begin{equation}
\mathbb{T}^{MM}_{1 m,1 m'} = - \frac{1}{3}\left(\frac{qR}{d}\right)^{3},
\hspace{20pt}
\mathbb{T}^{EE}_{1 m,1 m'} =   \frac{2}{3}\left(\frac{qR}{d}\right)^{3}.
\end{equation}
By the use of the universal relationship between determinants and traces, $\log\abs{A} = \text{Tr}\log(A)$, and applying a Taylor expansion in terms of $\frac{1}{d}$ of Eq.~\eqref{Energy_T_finite8}, we obtain
\begin{equation}
E = - k_{B}T{\sum_{n = 0}^{\infty}}'\text{Tr}\left(\mathbb{N}(\lambda_{T}^{-1}n)\right).
\end{equation}
Using the translation matrices in a spherical vector multipole basis \cite{RE09} and the large separation approximation of the $\mathbb{T}$ matrix, the trace of the $\mathbb{N}$ matrix is obtained by straightforward calculus. 
Here we denote with a sub--index $T$ the results valid for all temperatures, with a sub--index $0$ the results in the quantum limit ($T\to 0$), and with a sub--index $cl$ the results in the classical limit ($\hbar\to 0$), which is equivalent to the high $T$ limit.
Carrying out the sum over Matsubara frequency, we obtain the Casimir contribution to the Helmholtz free energy as
\begin{eqnarray}\label{Energia_finite_T_metal_perfecto}
E_{T} & = & - \frac{\hbar c R^{6}}{2\pi d^{7}}\frac{z e^{5z}}{2\left(e^{2z} - 1\right)^{5}}\times\nonumber\\
& & \left(2\left(15 - 29z^{2} + 99z^{4}\right)\cosh(z) + 15\cosh(5z)\right.\nonumber\\
& & \left. + \left( - 45 + 58z^{2} + 18z^{4}\right)\cosh(3z)\right.\\
& & \left. + 24z\left( 6z^{2} - 5 + \left(5 + 3z^{2}\right)\cosh(2z)\right)\sinh(z)\right),\nonumber
\end{eqnarray}
where $z = d/\lambda_{T}$. We also define the adimensional Casimir energy as $E_{ad}(z) = \frac{2\pi d^{7}}{\hbar c R^{6}}E_{T}$. From this result, the quantum ($T\to 0$) and classical ($\hbar\to 0$) limits with their first corrections are easily obtained as
\begin{equation}\label{Low_T_corrections_Energy}
E_{0} = - \frac{143 \hbar c R^{6}}{16\pi d^{7} } - \frac{8 k_{B}T\pi^{5}R^{6}}{27 d}\left(\frac{k_{B}T}{\hbar c}\right)^{5},
\end{equation}
\begin{equation}\label{High_T_corrections_Energy}
E_{cl} =  - \frac{15 k_{B}T R^{6}}{4 d^{6}} - k_{B}T\frac{R^{6}}{2d^{6}}\left(15 + 30 z + 29 z^{2} + 18 z^{3} + 9 z^{4}\right)e^{-2z},
\end{equation}
where $z = d/\lambda_{T}$. Note that in Eq. \eqref{Low_T_corrections_Energy}, the first correction to zero temperature case is proportional to $T^{6}$, contrary to the plate--sphere case \cite{Canaguier-Durand_Caso_Esfera_Placa_PRA}\cite{Metodo_Caminos_Esfera_Placa_Escalar} and to the cylinder--plate case \cite{Metodo_Caminos_Esfera_Placa_Escalar}, where a result proportional to $T^{4}$ were obtained for both systems. The result presented here is new for perfect metal spheres, but expected, because it has been obtained also for compact objects in the diluted limit \cite{Rodriguez-Lopez_PSA}\cite{Barton}. As corrections have the same negative sign of the Casimir energy in both limits, they describe an increase of the magnitude of the Casimir energy in high and low temperature limits. It is no longer the case if we study the energy for all temperatures. In fact, for some ranges of separation and temperature, thermal photons tend to reduce the Casimir energy between the spheres, as shown in Fig. \ref{Energia_2_esferas_metalicas}. This effect is captured in the contribution of the next term to the quantum limit, this term is $\Delta E_{0} = +\frac{2288 d k_{B}T\pi^{7}R^{6}}{1575}\left(\frac{k_{B}T}{\hbar c}\right)^{7}$ and tends to reduce the magnitude of the energy. Fig. \ref{Energia_2_esferas_metalicas} is an indicator of the appearance of negative entropy in this system because of the negative slope of the energy curve.
\begin{figure}
\begin{center}
\includegraphics[width=0.9\columnwidth]{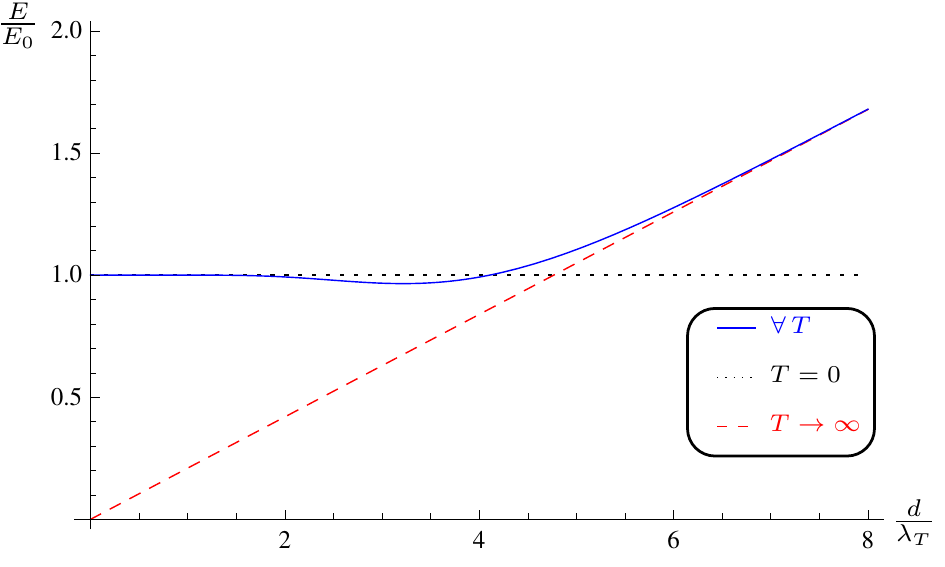}
\caption{Casimir energy between perfect metal spheres as a function of $\frac{d}{\lambda_{T}}$ compared with the energy at zero temperature. The dotted curve is the quantum limit, the dashed curve is the classical limit, and the solid curve is the asymptotic finite-temperature Casimir energy. Note that these curves are independent of the radius of the spheres but are only valid in the large separation limit.}
\label{Energia_2_esferas_metalicas}
\end{center}
\end{figure}
It is clear in Fig. \ref{Energia_2_esferas_metalicas} that for some temperatures and distances, the Casimir energy between spheres is lower than in the zero temperature case. But it is less evident the validity Eq. \eqref{Low_T_corrections_Energy}. In fact, there is a tiny increase of $E/E_{0}$ at low $\frac{d}{\lambda_{T}}$ until reaching a local maximum at $\frac{d}{\lambda_{T}}\approx 1.0388$ of $E/E_{0}\approx 1 + 10^{-4}$. It is not visible in Fig. \ref{Energia_2_esferas_metalicas} because the difference of scales. The local minimum is reached at $\frac{d}{\lambda_{T}}\approx 3.21733$, where $E/E_{0}\approx 1 - 3.46\times 10^{-2}$. This minimum is reached, in the important cases of room temperature ($T = 300K$) and at the boiling point of $\text{N}_{\text{2}}$ ($T = 77K$) at $d_{T=300} = 3.89\mu m$ and $d_{T=77} = 15.21\mu m$ respectively.

\subsection{Casimir entropy in the large separation limit}
\begin{figure}%[H]
\begin{center}
\includegraphics[width=0.9\columnwidth]{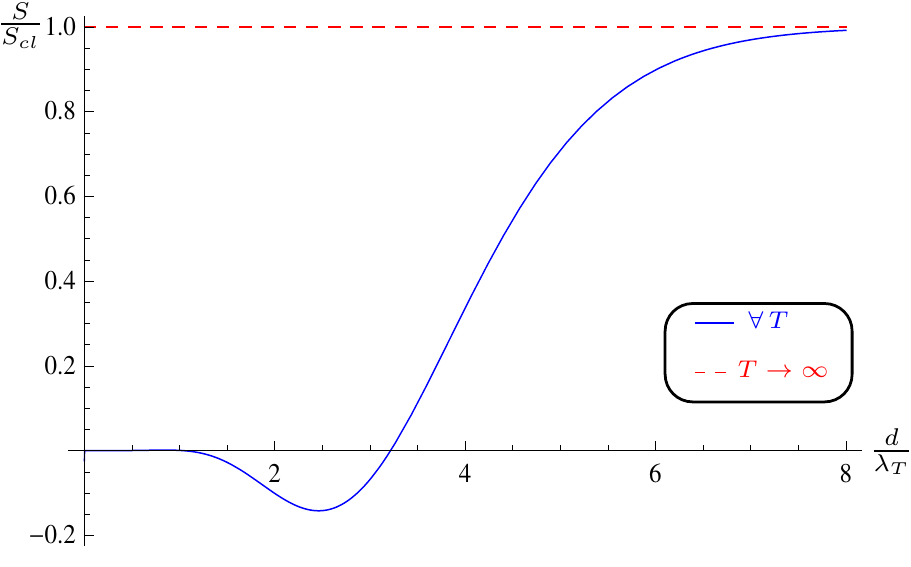}
\caption{Casimir entropy between perfect metal spheres as a function of $\frac{d}{\lambda_{T}}$ compared with the classical limit. The dashed curve is the classical limit, and the solid curve is the asymptotic finite-temperature Casimir entropy. Note the regions of negative entropy and negative slope of the entropy with the parameter $\frac{d}{\lambda_{T}}$}
\label{Entropia_2_esferas_metalicas}
\end{center}
\end{figure}

In the canonical ensemble, the entropy is defined as $S = - \partial_{T}E$, where with an abuse of the notation we denote the Helmholtz free energy by $E$. In the large separation limit, the Helmholtz free energy depends on the adimensional variable $z = \frac{d}{\lambda_{T}} = 2\pi\frac{dk_{B}T}{\hbar c}$, so we can write the entropy as
\begin{equation}
S = - \frac{\partial z}{\partial T}\frac{\partial E}{\partial z} = - 2\pi\frac{dk_{B}}{\hbar c}\frac{\partial E}{\partial z},
\end{equation}
and define the adimensional entropy as $S_{ad}(z) = \frac{d^{6}}{k_{B}R^{6}}S = - \partial_{z}E_{ad}(z)$. From this result, the quantum ($T\to 0$) and classical ($\hbar\to 0$) limits are easily obtained as
\begin{equation}\label{Low_T_Entropy}
S_{0} = 0 + \frac{16  k_{B} \pi^{5}R^{6} }{9 d }\left(\frac{k_{B}T}{\hbar c}\right)^{5},
\end{equation}
\begin{equation}\label{High_T_Entropy}
S_{cl} = \frac{15 k_{B} R^6}{4 d^6} + k_{B}\frac{R^{6}}{2d^{6}}\left(15 + 30z + 27z^{2} + 14z^{3} + 9z^{4} - 18z^{5}\right)e^{-2z},
\end{equation}
where $z = \frac{d}{\lambda_{T}}$. So, the entropy is a growing function with temperature in both limits, but this is not the case for all temperatures (the next low temperature expansion term of Eq. \eqref{Low_T_Entropy} is $\Delta S_{0} = -\frac{18304\pi^{7}}{1575}k_{B}d R^{6}\left(\frac{k_{B}T}{\hbar c}\right)^{7}$, which already indicates that the entropy could change its growing behavior with the temperature), as we can observe in Fig. \ref{Entropia_2_esferas_metalicas}, where a region of negative entropy and another region of negative slope of the entropy are observed.

Because of the limit at low temperature of the entropy, we know that the entropy is positive for low $T$ (not seen in Fig. \ref{Entropia_2_esferas_metalicas} because it is small compared with $S/S_{\text{cl}}$, but it can be observed in Fig. \ref{LogLogPlot_Entropia_2_esferas_metalicas_forall_distances}), so there are three points where $S=0$, including the origin. These two new points where the entropy nulls correspond to the local maximum and minimum observed in the Casimir energy in Fig. \ref{Energia_2_esferas_metalicas}.

Negative entropy of the Casimir effect has already been obtained between Drude parallel plates in \cite{Entropias_negativas_placas_Drude} and in \cite{Ingold}, and more recently between a perfect metal plate and sphere in \cite{Canaguier-Durand_Caso_Esfera_Placa_PRA} and between a Drude sphere and plate in \cite{Zandi_Emig_Placa_Esfera_varios_modelos} and \cite{Bordag_Entropia_Esfera_Placa}. This is the first time, to the best of our knowledge, that negative entropy appears between spheres because of the Casimir effect.

These results are only valid when the separation between the spheres is large compared with their radius, regardless of the radius of each one. Therefore, it is possible that the interval of negative entropy would disappear when the separation between the spheres reduces, regardless of the temperature of the system. A numerical exploration of entropies at smaller separation has been performed to verify this issue.

\subsection{Numerical study at smaller separations}
As noted in the previous Subsection, asymptotic results are no longer valid when the separation between the spheres becomes comparable to their radius. For this reason, a numerical study of entropy was performed for these cases. We computed Eq.~\eqref{Energy_T_finite8} numerically for all temperatures from $T=0$ until reaching the classical limit for fixed ratio between the radius $R$ and separation, $r = \frac{R}{d}$.

For spheres, the $\mathbb{T}$ matrices are diagonal but infinite matrices, so a cutoff in $(\ell,\ell')$ space is needed to obtain finite-dimensional matrices. In addition, another cutoff in Matsubara frequency is needed to obtain a finite series.

The proposed method is an asymptotic approximation, at small separations more and more modes are needed to obtain convergent results. This means that there exists a minimum separation between the spheres below which we are not able to take into account enough multipoles to obtain good results. We use multipoles up to $\ell \leq 15$, which means that we are restricted to $r_{\text{max}} = \frac{R}{d_{\text{min}}} \leq 0.45$, when contact is reached at $r_{\text{contact}} = \frac{R}{d_{\text{contact}}} = \frac{R}{2R} = 0.5$, where the energy diverges.

In the small separation limit, the proximity force approximation (PFA) is known to be a good approximation to the Casimir energy \cite{Kardar-Geometrias-Arbitrarias}. It is also known that perfect metal plates do not experience negative entropies, so we do not expect to observe negative entropies between spheres in the small separation limit.

\begin{figure}%[H]
\begin{center}
\includegraphics[width=0.9\columnwidth]{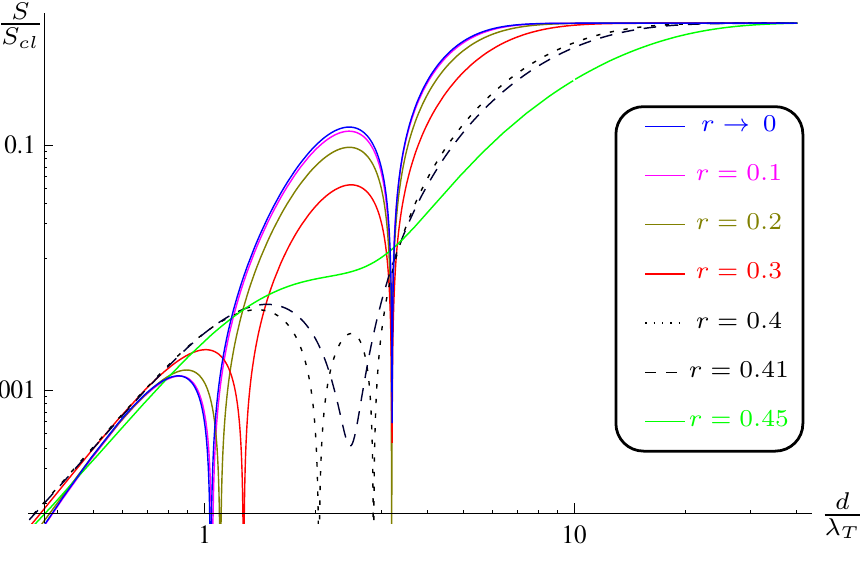}
\caption{Log--log plot of the absolute value of the entropy divided by its classical limit for perfect metal spheres at constant $r$ as a function of $\frac{d}{\lambda_{T}}$. Starting from the asymptotic solid curve ($r\to 0$), we increase $r$ in steps of $0.1$ up to $r = 0.4$ (dotted curve). Dashed curve is the case $r = 0.41$, where the interval of negative entropy has disappeared. The curve for $r=0.45$ is also shown}
\label{LogLogPlot_Entropia_2_esferas_metalicas_forall_distances}
\end{center}
\end{figure}

In Fig. \ref{LogLogPlot_Entropia_2_esferas_metalicas_forall_distances}, the entropy of the system of two perfect metal spheres is plotted as a function of $z = \frac{d}{\lambda_{T}}$ for constant $r = \frac{R}{d}$. The large separation result is shown too. We choose a log--log representation of the absolute value of the entropy divided by its corresponding classical limit. Therefore, zeros are observed as log divergences, and we can also observe the cases of negative entropy. Starting in the large separation regime, we observe an interval of negative entropy. As we increase $r$ (reducing the separation between the spheres), the region of negative entropy tends to reduce until it disappears between $r = 0.40$ and $r = 0.41$. Power-law decay of the entropy \eqref{Low_T_Entropy} at low temperatures is observed as a linear decay of the curve at low $z$, and constant behavior in the high-temperature limit \eqref{High_T_Entropy} is also reached in the simulation.

\subsection{Casimir force in the large separation limit}
\begin{figure}%[H]
\begin{center}
\includegraphics[width=0.9\columnwidth]{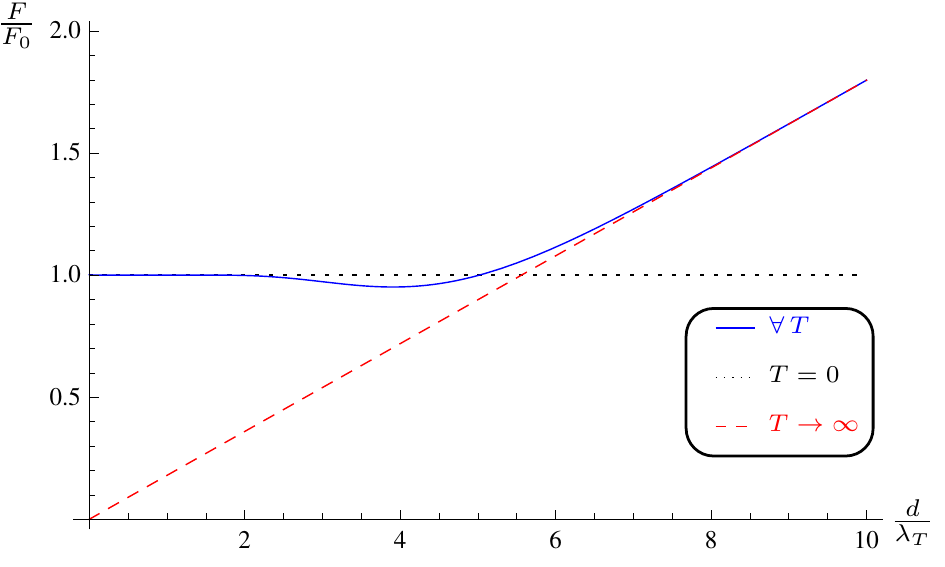}
\caption{Asymptotic force divided by its zero-temperature limit as a function of $\frac{d}{\lambda_{T}}$. The dotted curve is the zero-temperature result. The dashed curve is the classical limit, and the solid curve is the result at finite temperatures. The nonmonotonic behavior of the force with temperature results from the negative slope of the solid curve at constant separation}
\label{Fuerza_2_esferas_metalicas}
\end{center}
\end{figure}

In the previous Subsection we demonstrated that intervals of negative entropy appear due to the Casimir effect between perfect metal spheres. In these cases, for any given temperature, a minimum of entropy exists for a given, nonzero separation. Naively, this would imply a violation of the third law of Thermodynamics if the Casimir force is not zero at the minimum of the entropy. However, this is not the case, as will be explained in Sect. \ref{Sect:Thermodynamic}. In addition to that, we find that the force is always attractive, independent of the increase or decrease of the entropy, which also would naively imply a violation of the second law, because the system can be enforced to perform a process in which the entropy is reduced instead increased. However, this is not the case, as will be also explained in Sect. \ref{Sect:Thermodynamic}. However, the appearance of negative entropy does have an effect on the force. The asymptotic Casimir force $F = -\partial_{d}E$ can be written in terms of the adimensional Casimir energy as
\begin{equation}\label{Fuerza_adimensional}
F_{ad}(z) = \frac{2\pi}{\hbar c}\frac{d^{8}}{R^{6}}F = 7E_{ad}(z) - z\partial_{z}E_{ad}(z),
\end{equation}
where $z = \frac{d}{\lambda_{T}}$. In Fig. \ref{Fuerza_2_esferas_metalicas}, the adimensional asymptotic force between the perfect metal spheres compared with the zero-temperature force is plotted as a function of $\frac{d}{\lambda_{T}}$. Here, for constant temperature, we observe nonmonotonic behavior of the force with the adimensional parameter $\frac{d}{\lambda_{T}}$.

Nonmonotonic behavior of the adimensional Casimir force implies nonmonotonic force behavior with temperature, but not with separation, because of the extra dependence of the force on the separation in Eq.~\eqref{Fuerza_adimensional}. In fact, it is easy to verify that the force behaves monotonically with separation, and the nonmonotonicity of the entropy with separation implies nonmonotonic behavior of the force with temperature, because
\begin{equation}\label{Relacion_Fuerza_Entropia}
\frac{\partial F}{\partial T} = - \frac{\partial^{2} E}{\partial T\partial d} = \frac{\partial S}{\partial d},
\end{equation}
so the appearance of negative slopes of the entropy with separation implies nonmonotonicity of the Casimir force with temperature, despite the attractive force for all separations and temperatures.

%As observed in Fig. \ref{Fuerza_2_esferas_metalicas}, force between spheres is always attractive, then asymptotically for any given temperature, exist a range of distances where the system tends to increase the entropy because Casimir effect.

As observed in Fig. \ref{Fuerza_2_esferas_metalicas}, the force between the spheres is always attractive, but it is not monotonic with temperature; asymptotically, for any given temperature, there exists a range of separations for which the force decreases with temperature. This is the first time, to the best of our knowledge, that nonmonotonic behavior of the Casimir force with temperature has been described between compact objects. Nonmonotonicities of the force between a plate and cylinder and between a plate and sphere were already obtained in \cite{Metodo_Caminos_Esfera_Placa_Escalar}, but in that case the nonmonotonicity already appears for the scalar field. Nonmonotonicity does not appear between spheres for the scalar field; this is a characteristic effect of the electromagnetic field, because cross-polarization terms of the Casimir energy are essential for this nonmonotonicity to appear.

\section{Plasma model}\label{Plasma Model}
In this Section, we assume that the electric susceptibility of both spheres is described by the plasma model, i.e.,
\begin{equation}
\epsilon(ic\kappa) = 1 + \frac{(2\pi)^{2}}{\lambda_{P}^{2}\kappa^{2}}
\end{equation}
and $\mu = 1$. To obtain the large separation limit of the Casimir energy, we need the dominant part of the $\mathbb{T}$ matrix in this limit. The main contribution comes from the dipolar polarizabilities part of $\mathbb{T}$ matrix, taking the form \cite{Canaguier-Durand_Caso_Esfera_Placa_PRA}\cite{Zandi_Emig_Placa_Esfera_varios_modelos}
\begin{equation}
\mathbb{T}^{MM}_{1 m,1 m'} = - \frac{R\lambda_{P}^{2}q^{3}}{12\pi^{2}d^{3}} \left(3 + y^{2} - 3y\coth\left(y\right)  \right),
\end{equation}
\begin{equation}
\mathbb{T}^{EE}_{1 m,1 m'} =   \frac{2}{3}\left(\frac{qR}{d}\right)^{3},
\end{equation}
with $y = 2\pi\frac{R}{\lambda_{P}}$. The coefficients of the magnetic polarizability depend on the plasma frequency, tending to the perfect metal result as $\lambda_{P}\to 0$ and to zero in the transparent limit $\lambda_{P}\to\infty$ \cite{Zandi_Emig_Placa_Esfera_varios_modelos}.

\subsection{Casimir energy in the large separation limit}
Once we have the asymptotic $\mathbb{T}$ matrix, the Casimir energy can be obtained by a straightforward but long calculation. It is possible to obtain analytical results for finite temperatures, but they are too long to show here. The zero-temperature Casimir energy is
\begin{eqnarray}
E_{0} & = & - \frac{\hbar c R^{6} }{16\pi d^{7}y^{4}}\times\\
& & \left( 207 + 222 y^{2} + 143 y^{4} - 414y\coth(y) + 207 y^{2}\coth(y)^{2} - 222 y^{3}\coth(y)\right),\nonumber
\end{eqnarray}
and the high-temperature limit is given by
\begin{equation}
E_{cl} \hspace{-2pt}=\hspace{-2pt} - k_{B}T\frac{3 R^{6}}{d^{6}}\left(1 \hspace{-2pt}+\hspace{-2pt} \frac{1}{4y^{4}}\left(3 \hspace{-2pt}+\hspace{-2pt} y^{2} \hspace{-2pt}-\hspace{-2pt} 3y\coth(y)\right)^{2}\right),
\end{equation}
with $y = 2\pi\frac{R}{\lambda_{P}}$. The perfect metal limit is reached for $\lambda_{P}\to 0$, as expected. When the plasma wavelength $\lambda_{P} \ll R$, only the electric sector of the $\mathbb{T}$ matrix contributes, so in this case the energy in the zero- and high-temperature limits is given by
\begin{equation}
E_{0} = -\frac{23 \hbar c R^{6} }{4\pi d^{7}},
\hspace{20pt}
E_{cl} = - k_{B}T\frac{3 R^{6}}{d^{6}}.
\end{equation}

\subsection{Casimir entropy in the large separation limit}
\begin{figure}[h]
\begin{center}
\includegraphics[width=0.9\columnwidth]{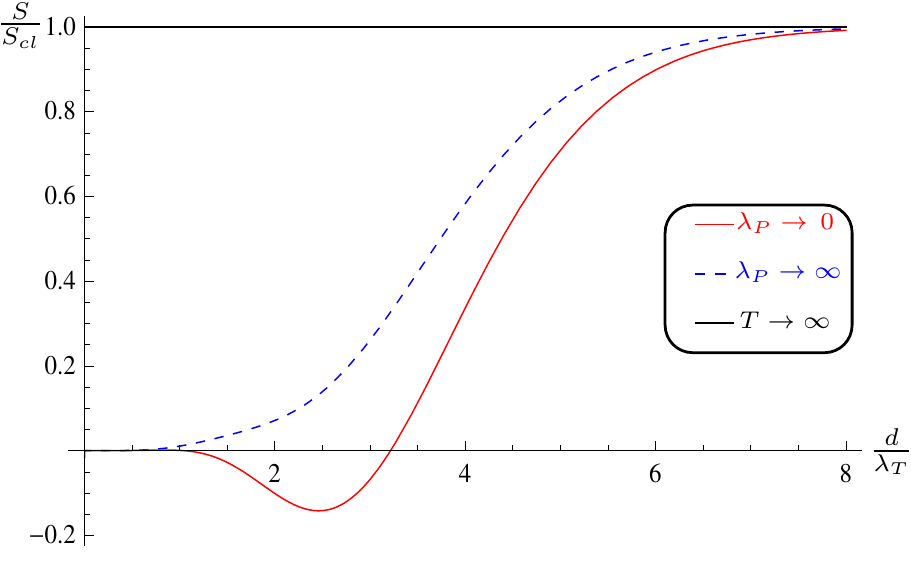}
\caption{Entropy between plasma spheres as a function of $\frac{d}{\lambda_{T}}$ divided by the classical limit for the two limit cases. The solid curve is the entropy in the perfect metal limit $\lambda_{R} \ll R$, where a region of negative entropy and nonmonotonicities of entropy with separation and temperature are present; the dashed curve is the entropy in the transparent limit $\lambda_{R} \gg R$, where the usual monotonic behavior of entropy is shown}
\label{Entropia_2_esferas_Plasma_Casos_Limite}
\end{center}
\end{figure}

\begin{figure}[h]
\begin{center}
\includegraphics[width=0.9\columnwidth]{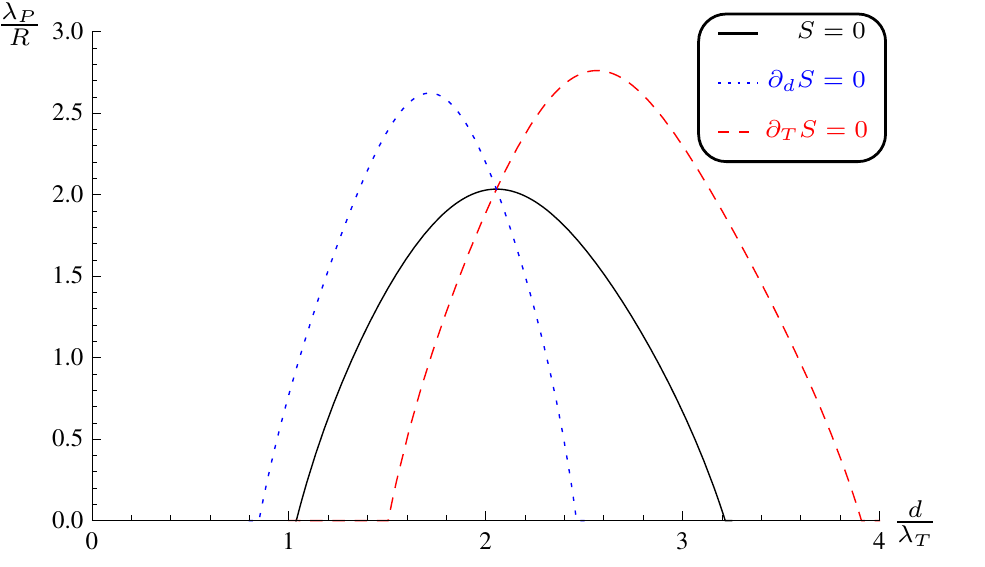}
\caption{Representation, in the $\left(\frac{d}{\lambda_{T}},\frac{\lambda_{P}}{R}\right)$ space, of zeros of the entropy (dotted curve), the temperature derivative of the entropy (dashed curve), and the separation derivative of the entropy (dotted curve)}
\label{Ceros_Entropia_2_esferas_Plasma}
\end{center}
\end{figure}

The entropy of the system is obtained in the same way as for the perfect metal case, but for the plasma model we have two different regimes. When $\lambda_{P}\ll R$, we reach the perfect metal limit, obtaining negative entropy and nonmonotonic behavior of the entropy with separation and temperature as before. However, when $\lambda_{P}\gg R$, the spheres are transparent to the magnetic field and the problem reduces to a scalar problem. In this case, the entropy is always positive. In Fig. \ref{Entropia_2_esferas_Plasma_Casos_Limite}, the adimensional asymptotic entropy, compared with its classical limit, is plotted as a function of $\frac{d}{\lambda_{T}}$ for these two limits. In Fig. \ref{Ceros_Entropia_2_esferas_Plasma}, we present the points $(\frac{d}{\lambda_{T}},\frac{\lambda_{P}}{R})$ corresponding to zeros of the entropy (continuous curve), its temperature derivative (dashed curve), or its separation derivative (dotted curve).
As discussed in the case of perfect metal plates, the anomalous behavior of the entropy found here would naively imply a violation of second and third laws of Thermodynamics, because the Casimir force is always attractive, irrespective of the slope of the entropy. However, this is not the case, as we will discuss in Sect. \ref{Sect:Thermodynamic}.
As noted above, in the perfect metal limit we obtain a region of negative entropy, while for low $\lambda_{P}$ this region disappears. The point where negative entropies appear depends on the magnetic susceptibility $\mu$. For $\mu = 1$, negative entropies appear around $\lambda_{P} \approx 2 R$.

\section{Drude model}\label{Drude Model}
In this Section we assume that the electric susceptibility of both spheres is described by the Drude model, i.e.,
\begin{equation}
\epsilon(ic\kappa) = 1 + \frac{4 \pi^{2}}{\lambda_{P}^{2}\kappa^{2} + \frac{\pi c\kappa}{\sigma}}
\end{equation}
and $\mu = 1$. To obtain the large separation limit of the Casimir energy, we need the dominant part of the $\mathbb{T}$ matrix in this limit. The main contribution comes from the dipolar polarizabilities part of the $\mathbb{T}$ matrix, taking the form \cite{Canaguier-Durand_Caso_Esfera_Placa_PRA}\cite{Zandi_Emig_Placa_Esfera_varios_modelos}
\begin{equation}
\hspace{-4pt}\mathbb{T}^{MM}_{1 m,1 m'} \hspace{-2pt}=\hspace{-2pt} - \frac{4\pi}{45}\frac{R\sigma}{c}\left(\frac{qR}{d}\right)^{4}\hspace{-2pt},
\hspace{2cm}
\mathbb{T}^{EE}_{1 m,1 m'} \hspace{-2pt}=\hspace{-2pt}   \frac{2}{3}\left(\frac{qR}{d}\right)^{3}\hspace{-2pt}.
\end{equation}
Now, the response of the material changes dramatically from the case of the plasma model. In principle, one would expect to obtain the $\mathbb{T}$ matrix for the plasma model as $\sigma\to\infty$, but it is obvious that this is not the case. Not only do we not recover the plasma model in this limit, but also the $\mathbb{T}$ matrix diverges. The reason is that the plasma model is some kind of singular limit of Drude model, with a qualitative different behavior. Now the dominant contribution is given by the electric sector of the $\mathbb{T}$ matrix, whereas the magnetic part does not contribute for asymptotic energies.

\subsection{Casimir energy and entropy in the large separation limit}
\begin{figure}
\begin{center}
\includegraphics[width=0.9\columnwidth]{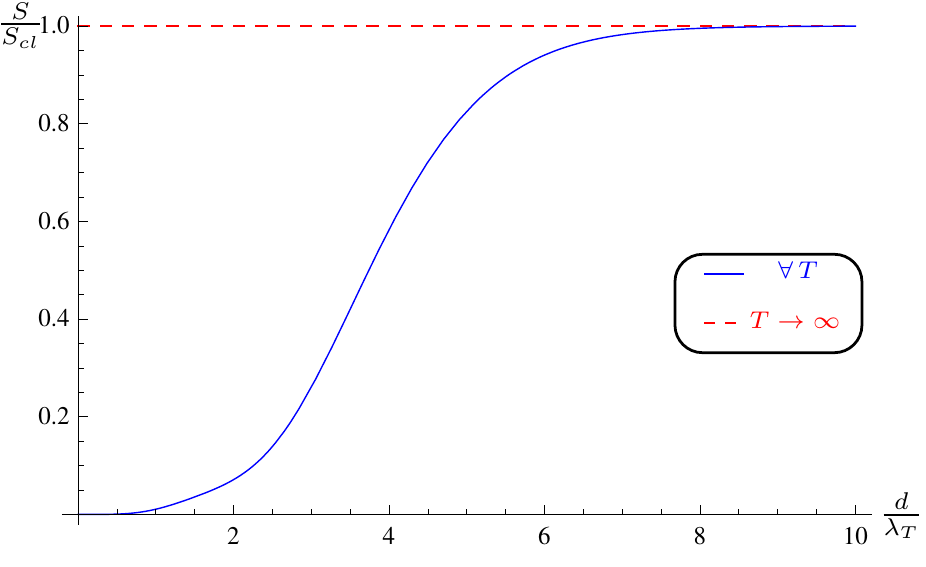}
\caption{Entropy as a function of the adimensional parameter $\frac{d}{\lambda_{T}}$ divided by its classical limit for the large separation limit of the system of Drude spheres. Entropy is positive and monotonic in temperature and separation}
\label{Entropia_2_esferas_Drude}
\end{center}
\end{figure}

Once we have the asymptotic $\mathbb{T}$ matrix, obtaining the Casimir energy is a straightforward calculation. Carrying out the sum over Matsubara frequency, we obtain the Casimir contribution to the Helmholtz free energy as
\begin{align}
E_{T} & = -\frac{2 k_{B}T R^{6}e^{5 z}}{d^{6}\left(e^{2 z} - 1\right)^5}\times\nonumber\\
& \left(\left(6-10 z^2+22 z^4\right) \cosh(z) + \left(-36z+12z^3\right) \sinh(z) \right.\nonumber\\
& + \left(12 z+4z^3\right) \sinh(3 z) + 3 \cosh(5 z) + \left.\left(-9 + 10z^{2} + 2z^{4}\right) \cosh(3 z) \right),
\end{align}
where $z = d/\lambda_{T}$. The zero- and high-temperature Casimir energy limits are given by
\begin{equation}
E_{0} = -\frac{23 \hbar c R^{6}}{4\pi d^{7}},
\hspace{2cm}
E_{cl} = - k_{B}T\frac{3 R^{6}}{d^{6}}.
\end{equation}
These asymptotic Casimir energies are universal because they do not depend on any material property. The lack of magnetic polarizability contributions is characteristic of the Drude model, leading to an effective scalar Dirichlet problem.
The entropy is obtained as in the perfect metal case. As only the electric polarization contributes, we expect the usual behavior of the entropy as a monotonic function of $\frac{d}{\lambda_{T}}$, as shown in Fig. \ref{Entropia_2_esferas_Drude}. The zero- and high-temperature Casimir entropy limits are given by
\begin{equation}
S_{0} = \frac{704 k_{B}R^{6}}{315 d}\left(\frac{\pi k_{B}T}{\hbar c}\right)^{5} - \frac{11776 k_{B}d R^{6}}{1575 }\left(\frac{\pi k_{B}T}{\hbar c}\right)^{7},
\end{equation}
\begin{equation}
S_{cl} = 3k_{B}\frac{R^{6}}{d^{6}}.
\end{equation}
In this case, the correction term to the quantum limit tends to reduce the entropy, but it is not strong enough to change the sign of it, as seen in Fig. \ref{Entropia_2_esferas_Drude}.

\section{Thermodynamical consequences}\label{Sect:Thermodynamic}
In this Section we discuss the thermodynamical consequences of the obtained results. In this article we have obtained and compared the large separation limit of the Casimir energy and entropy for two spheres using three different electric susceptibility models. For perfect metal spheres, at any nonzero fixed temperature, we observe an interval of separations for which entropy is negative, while at zero temperature the entropy is always zero.

In addition, the Casimir force is attractive for all separations and temperatures. So, we could naively think that we have possible violations of the third and second laws of Thermodynamics, due to the existence of processes where the entropy of the system tends to increase and to the negative entropy intervals at finite temperature and distances, respectively.

According to the Krein formula \cite{Wirzba}, we know that the Helmholtz free energy of the electromagnetic field has three independent additive contributions (see Eq. \eqref{Definicion_DoS4} of the Appendix): one is from the thermal bath, being proportional to the volume of the space \cite{SPECTRA_FINITE_SYSTEMS}. Another is the sum of contributions of objects immersed in the bath considered as isolated objects, each contribution being also a function of the volume and surface of each object \cite{SPECTRA_FINITE_SYSTEMS}. The third is of geometrical nature, which we could call the Casimir part of the Helmholtz free energy and that we actually calculate in Eq. \eqref{Energy_T_finite8}. This term depends on the electromagnetic nature of the objects in the system, but it has a geometric nature, because it depends on the relative separations and orientations between the objects, being zero iff the relative separations between all the considered objects become infinite.

Considering the whole system, the nonmonotonicity of the Casimir entropy with temperature is compensated by the contribution of the vacuum, because one scales with the global volume~\cite{SPECTRA_FINITE_SYSTEMS} (Eq. \eqref{Definicion_DoS_Weyl}) and the other with the separation between the spheres (Eq. \eqref{High_T_Entropy}). This is not the case for the nonmonotonicity with separation. At constant temperature, the entropies of the thermal bath and of each object remain constant, but the Casimir force is always attractive while the entropy can increase or decrease with separation. Therefore, the potential violation of the second law of Thermodynamics still requires an explanation.

The Krein formula also states that internal sources of entropy inside the spheres cannot appear in order to compensate these regions of anomalous behavior of the entropy, unless these internal sources are independent of the separation between the spheres.

The second law of Thermodynamics states that global entropy must increase for any process, but only in closed systems. As we are working in the canonical ensemble, we are implicitly assuming that there exists an external reservoir which keeps our system at constant temperature, so the system is not isolated and entropy can increase or decrease without violation of the second law. In the canonical ensemble, the condition equivalent to the second law is that the global Helmholtz free energy must decrease for any process, and this is true for the studied system. Therefore, the appearance of nonmonotonic entropy behavior in the canonical ensemble just implies nonmonotonic behavior of the force with temperature, as seen in Eq. \eqref{Relacion_Fuerza_Entropia}.

\section{Conclusions}
In this Chapter, we have obtained and compared the large separation limit of the Casimir energy and entropy for two spheres using three different electric susceptibility models. For perfect metal spheres, at any nonzero fixed temperature, we observe an interval of separations for which entropy is negative, while at zero temperature the entropy is always zero.

We can trace the origin of this anomalous effect to the functional form of the Casimir energy in this large separation regime. It consists of four components, two of which are of attractive nature, equivalent to the sum of two scalar problems, one for the electric polarization and the other for the magnetic polarization. The other two components come from the cross-coupling between the electric and magnetic polarizations of the two spheres, and tend to reduce the Casimir energy between the spheres (Eq. 12 of \cite{Kardar-Geometrias-Arbitrarias} and Eq. 6 of \cite{Rodriguez-Lopez_PSA}). These cross-coupling polarization terms are responsible for the impossibility of factorization of the electromagnetic Casimir energy into two equivalent scalar problems in general, and appear because the translation matrix $\mathbb{U}_{\alpha\beta}$ is not diagonal in polarization space. In this article we show that these cross terms not only reduce the Casimir energy, but also for some separations and temperatures, their contribution to the entropy is greater than the contributions of direct coupling between the electric and magnetic polarizabilities, resulting in an interval of negative entropy and nonmonotonic behavior of the entropy with separation and temperature.

In addition, the Casimir force is attractive for all separations and temperatures. So, we could naively think that we have possible violations of the third and second laws of Thermodynamics, due to the existence of processes where the entropy of the system tends to increase and to the negative entropy intervals at finite temperature and distances, respectively. 

In Sect. \ref{Sect:Thermodynamic}, having into account the complete thermodynamical system, that the system is described by the canonical ensemble, and with the help of the Krein formula and Weyl formula, we have demonstrated that there are not violations of second and third laws respectively. Therefore, the appearance of nonmonotonic entropy behavior in the canonical ensemble just implies nonmonotonic behavior of the force with temperature, as seen in Eq. \eqref{Relacion_Fuerza_Entropia}.

The interval of negative entropy for the spheres appears because of the cross-coupling between the polarizations of the electromagnetic field, which leads us to conclude that it is a characteristic of the electromagnetic field and does not have an analog in the Casimir effect due to scalar fields.

Applying the PFA to this problem, we obtain that, for perfect metal spheres near contact, the entropy is always positive, so we performed a numerical study of the entropy between perfect metal spheres of equal radius at smaller separations. The results showed that there exists a minimum separation between the spheres for which the negative entropies disappear. In addition, the region of negative slope of the entropy with separation disappears for another smaller separation.

We have also obtained the energy and entropy for the plasma model, for which similar results are obtained for plasma penetration length $\lambda_{P} \lesssim 2R$, while for $\lambda_{P} > 2R$, the negative entropies disappear, and for another greater $\lambda_{P}$, the nonmonotonic behavior of the entropy with separation also disappears.

In the transparent limit ($\lambda_{P} \gg R$), the magnetic polarization does not contribute to the asymptotic Casimir effect, so the problem reduces to a scalar field problem and we find that entropy is a positive monotonic function.

When we study the asymptotic Casimir effect with Drude model spheres, the system is qualitatively different. The magnetic polarization does not contribute to the asymptotic Casimir effect because it depends on $d^{-4}$ instead of the $d^{-3}$ dependence of the electric polarization term. Then also for Drude spheres the asymptotic limit reduces to a scalar field problem, resulting in a positive monotonic entropy.

Nonmonotonicities of the Casimir force between objects are not unusual. As the Casimir effect is not pairwise additive, the interaction between two objects is affected by the presence of a third, leading to nonmonotonicities of the Casimir force between cylinders \cite{Fuerza_cilindros_en_presencia_de_plato}\cite{Fuerza_cilindros_en_presencia_de_plato2}, or between spheres \cite{Rodriguez-Lopez_1} in the presence of a plate. The nonmonotonicity presented in this article has a different nature, as it does not come from interactions with a third object but rather from correlations with temperature, as seen in Eq. \eqref{Relacion_Fuerza_Entropia}.

In addition, in the assumptions of the multiscattering formalism it is implicitly assumed that the objects are stationary, at fixed positions in space, so all results shown here are valid for quasistatic processes. Therefore, we must carefully consider when this quasistatic assumption does not apply, because the stationary system could abandon equilibrium, requiring more careful study \cite{Intravaia_non_eq_and_dynamic_Casimir_effects}\cite{Dedkov_Kyasov}.

\section{Appendix A: Krein formula from multiscattering formalism}
In this Appendix, we will derive the Krein formula of the electromagnetic field \cite{Wirzba} from the multiscattering formula \cite{Kardar-Geometrias-Arbitrarias}.

Krein formula factorizes the density of states of the electromagnetic field in three parts \cite{Wirzba}. Here we present a derivation of such formula from the Multiscattering formalism of the Casimir effect \cite{Kardar-Geometrias-Arbitrarias}\cite{RE09}.

The partition function ($\mathcal{Z} = {\sum_{n=0}^{\infty}}'\mathcal{Z}_{n}$) of the EM field in the presence of $N$ general dielectrics of arbitrary geometry is obtained, before regularization of the Casimir energy, as~\cite{RE09}
\begin{equation}\label{Z_Casimir_Completa}
\mathcal{Z}_{n} = \frac{1}{\Det{S_{n}}\Det{\mathcal{M}}},
\end{equation}
where $\Det{S_{n}}^{-1} = \Det{\Delta + \kappa_{n}^{2}}^{-1}$ is the $\text{n}^{\underline{\text{th}}}$ Matsubara frequency contribution to the partition function of the thermal bath, and $\mathcal{M}$ is the next $N\times N$ block matrix
\begin{equation}
\mathcal{M}_{\alpha\beta} = \delta_{\alpha\beta}\mathbb{T}_{\alpha}^{-1} + (\delta_{\alpha\beta} - 1)\mathbb{U}_{\alpha\beta}.
\end{equation}
If we multiply and divide Eq. \eqref{Z_Casimir_Completa} by $\Det{\mathcal{M}_{\infty}}$ (being $\mathcal{M}_{\infty}$ the same $\mathcal{M}$ matrix as above, but with each object at an infinite distance from each other), the Helmholtz free energy of the whole system can be written as ($\beta\mathcal{F} = \log(\mathcal{Z})$)
\begin{equation}\label{E_Helmholtz_global}
\beta\mathcal{F} = - {\sum_{n=0}^{\infty}}'\left[\log\Det{S_{n}} + \log\Det{\mathcal{M}_{\infty}} + \log\left(\frac{\Det{\mathcal{M}}}{\Det{\mathcal{M}_{\infty}}}\right)\right],
\end{equation}
where 
\begin{itemize}
\item $\log\Det{S_{n}}$ is the contribution of the thermal bath in absence of immersed objects, which leads to the Planck formula of the blackbody spectrum. 
\item $\log\Det{\mathcal{M}_{\infty}} = \sum_{\alpha = 1}^{n}\log\Det{\mathbb{T}_{\alpha}}$ is the contribution to the Helmholtz free energy of each dielectric object immersed in the bath considered as an isolated object.
\item $\log\left(\frac{\Det{\mathcal{M}}}{\Det{\mathcal{M}_{\infty}}}\right) = \log\Det{\mathbbm{1} - \mathbb{N}}$ is the Casimir part of the Helmholtz free energy, which depends on the electromagnetic nature of the objects in the system, but it has a geometric nature, because it depends on the relative separations and orientations between the objects, being zero iff the relative separations between all the considered objects become infinite. This term is the Casimir energy that we actually calculate in Eq. \eqref{Energy_T_finite8}.
\end{itemize}

The energy of a quantum system is related with the Density of States $\rho(\omega)$ by~\cite{Wirzba}
\begin{equation}\label{Definicion_DoS}
E = \int_{0}^{\infty}d\omega\frac{\hbar}{2}\omega\rho(\omega).
\end{equation}
Using Eq. \eqref{E_Helmholtz_global}, the same energy in the zero temperature case can be written as
\begin{equation}\label{E_Helmholtz_global1}
E = \frac{\hbar c}{2\pi}\int_{0}^{\infty}d\kappa\left[\log\Det{S_{\kappa}} - \sum_{\alpha = 1}^{n}\log\Det{\mathbb{T}_{\alpha}} + \log\Det{\mathbbm{1} - \mathbb{N}}\right].
\end{equation}
Because the dispersion relation of the massless EM field $\omega = c\kappa$, and integrating by parts, we obtain
%\begin{equation}\label{E_Helmholtz_global2}E = \int_{0}^{\infty}d\omega\frac{\hbar}{2}\omega\frac{1}{\pi}\left[- \partial_{\omega}\log\Det{S_{0}} + \sum_{\alpha = 1}^{n}\partial_{\omega}\log\Det{\mathbb{T}_{\alpha}} - \partial_{\omega}\log\Det{\mathbbm{1} - \mathbb{N}}\right].\end{equation}
\begin{align}\label{E_Helmholtz_global2}
E & = \int_{0}^{\infty}d\omega\frac{\hbar}{2}\omega\frac{1}{\pi}\times\\
& \times\left[- \partial_{\omega}\log\Det{S_{\omega}} + \sum_{\alpha = 1}^{n}\partial_{\omega}\log\Det{\mathbb{T}_{\alpha}} - \partial_{\omega}\log\Det{\mathbbm{1} - \mathbb{N}}\right].\nonumber
\end{align}
Then, because Eqs. \eqref{Definicion_DoS} and \eqref{E_Helmholtz_global2}, we can relate the Density of States with scattering properties of the system as
\begin{equation}\label{Definicion_DoS2}
\rho(\omega) = \frac{1}{\pi}\left[- \partial_{\omega}\log\Det{S_{\omega}} + \sum_{\alpha = 1}^{n}\partial_{\omega}\log\Det{\mathbb{T}_{\alpha}} - \partial_{\omega}\log\Det{\mathbbm{1} - \mathbb{N}}\right],
\end{equation}
by the use of the universal relationship between trace and determinant $\log\Det{A} = \tr\log(A)$, this result can be written as
\begin{equation}\label{Definicion_DoS3}
\rho(\omega) = \frac{1}{\pi}\left[- \tr\left(S_{\omega}^{-1}\partial_{\omega}S_{\omega}\right) + \sum_{\alpha = 1}^{n}\tr\left(\mathbb{T}_{\alpha}^{-1}\partial_{\omega}\mathbb{T}_{\alpha}\right) + \tr\left(\left(\mathbbm{1} - \mathbb{N}\right)^{-1}\partial_{\omega}\mathbb{N}\right)\right].
\end{equation}
As a conclusion, we can factorize the Density of States as the sum of three independent terms
\begin{equation}\label{Definicion_DoS4}
\rho(\omega) = \rho_{0}(\omega) + \sum_{\alpha = 1}^{n}\rho_{\alpha}(\omega) + \rho_{C}(\omega),
\end{equation}
where 
\begin{itemize}
\item $\rho_{0}(\omega) = - \frac{1}{\pi}\partial_{\omega}\log\Det{S_{\omega}}$ is the Density of States of the free EM field.
\item $\rho_{\alpha}(\omega) = \frac{1}{\pi}\partial_{\omega}\log\Det{\mathbb{T}_{\alpha}}$ is the change of the vacuum Density of States of the EM field because each $\alpha^{\text{\underline{th}}}$ dielectric object considered as a isolated object in the medium.
\item $\rho_{C}(\omega) = - \frac{1}{\pi}\partial_{\omega}\log\Det{\mathbbm{1} - \mathbb{N}}$ is the change of the Density of States related with the relative positions of the $N$ dielectric objects, which is zero when all the objects are infinitely apart from each other. $\rho_{C}(\omega)$ is the source of the Casimir effect.
\end{itemize}
Eq. \eqref{Definicion_DoS4} is the Krein formula \cite{Wirzba}, derived from the Multiscattering formalism of the Casimir effect for the EM field. In \cite{Wirzba}, Eq. \eqref{Energy_T_finite8} at zero temperature is obtained from the Krein formula removing the divergent contributions. Here we have followed the opposite way, a derivation of the density of states of the a thermal bath with intrusions and it factorization in Eq. \eqref{Definicion_DoS4}.

Terms $\rho_{0}(\omega)$ and $\rho_{\alpha}(\omega)$ diverges with $\omega$ because main contribution to Weyl formula for 3D volumes of the EM field~\cite{SPECTRA_FINITE_SYSTEMS}\cite{Libro_Tejero}, 
\begin{equation}\label{Definicion_DoS_Weyl}
\rho_{V}(\omega) = \frac{V}{\pi^{2}c^{3}}\omega^{2}\theta(\omega),
\end{equation}
but $\rho_{C}(\omega)$ converges if our system consists on $N$ compact objects, because in this case, $\mathbb{N}$ can be demonstrated to be a trace class operator for all $\omega$, then the determinant is well defined~\cite{Casimir_forces_in_a_T-operator_approach}.

% Casimir energy between non-parallel cylinders
\begin{savequote}[10cm] % this sets the width of the quote
\sffamily
``I think that it is a relatively good approximation to truth -- which is much too complicated to allow anything but approximations.'' 
\qauthor{John von Neumann}
\end{savequote}

\chapter{Casimir energy between non parallel cylinders}\label{Chap: Casimir energy between non parallel cylinders}
\graphicspath{{ch1/}} 
\graphicspath{{02-Casimir_Multiscattering/chb6_Cilindros_No_Paralelos/Figuras/}}

Casimir energies between cylinders have been studied from long time ago \cite{Barash-and-Kyasov}, but always between parallel cylinders by the help of the axial symmetry of the system \cite{Rahi:2008bv} \cite{RE09}. However, the case of non-parallel cylinders has not been studied until recently in \cite{non-parallel-cylinders-London-energies} and \cite{PSA-non-parallel-cylinders-salt-suspenssions}, where they studied asymptotic results. In addition to that, in \cite{PhysRevLett.103.040401} the authors performed a numerical method and studied the case of two perpendicular elongated objects of length $L$, when they incremented the length $L$ of the objects, fixed at a given distance, the energy finally saturates to a finite result independent of the length.

In this Chapter, we present the formalism to study the Casimir energy in systems with non parallel cylinders (see Fig.~\ref{Cylinders_Plot}) with the multiscattering formalism~\cite{Kardar-Geometrias-Arbitrarias}\cite{RE09}. This formalism will let us perform the numerical evaluation of the energy for all distances, not only asymptotic results. In particular, we study two non parallel cylinders for the scalar case for cylinders subject to Dirichlet or Neumann boundary conditions and the electromagnetic case for perfect metals. This study can be easily extended to dielectric cylinders.

Our goal is the calculation of Casimir energy between two non parallel cylinders, extending the study of Casimir energies to new geometries. Parallel cylinders case is easily reobtained as a limiting case of our formula.

The interest of the proposed problem is to expand the study of Casimir energies for systems where non parallel cylinders appear. Here we remind that in the microscopic world, a fiber could be actually considered as an infinite cylinder. For example, Casimir energy could have a contribution in the stability orientation of cylinders, as in the case of carbon nanotubes.

We follow this plan for the Chapter:
In Sect. \ref{sec: 2} we present the system under study. In Sect. \ref{sec: 3}, in order to use the multiscattering formalism, we obtain the translation matrices between cylindrical multipole basis when their axial axis are tilted an angle $\gamma$ and their origin are separated by a distance $d$, for scalar and electromagnetic cases. In Sect. \ref{sec: 4} we obtain the asymptotic far distance approximation of electromagnetic and scalar Casimir energies at zero and high temperature limits. In Sect. \ref{sec: 6} we present some numerical evaluation of Casimir energies between perpendicular and half--perpendicular perfect metals cylinders at zero temperature. The mathematical details of the derivation of translation matrices are left to Appendix \ref{App. A}. In Appendix \ref{App. B} we show how to cover the limit of parallel cylinders from our formalism. Finally, in Appendix \ref{sec: Appendix.C} we obtain the Proximity Force Approximation of the studied system, at the zero and high temperature limits.

The content of this Chapter is based on the work published in~\cite{Rodriguez-Lopez_4}.
\setcounter{equation}{0}
\section{Geometry of the system under study}\label{sec: 2}
Recently, a multiscattering formalism of Casimir energies has been developed \cite{Kardar-Geometrias-Arbitrarias} \cite{RE09}. It relates the Casimir interaction between objects with the scattering of the field from each object. The Casimir energy for a two objects system at any temperature is 
\begin{equation}\label{Energy_T_finite}
E_{T} = k_{B}T{\sum_{n=0}^{\infty}}'\log\abs{\mathbbm{1} - \mathbb{N}(\kappa_{n})},
\end{equation}
with $\kappa_{n} = 2\pi\frac{k_{B}T}{\hbar c}n$ as the Matsubara frequencies. The tilde means that the zero contribution has a height of $1/2$. The system is described by the $\mathbb{N}$ matrix. For a system of two objects, this matrix is $\mathbb{N} = \mathbb{T}_{1}\mathbb{U}_{12}\mathbb{T}_{2}\mathbb{U}_{21}$. $\mathbb{T}_{i}$ is the T scattering matrix of $i^{\text{\underline{th}}}$ object, which accounts for all the geometrical information and electromagnetic properties of each object. $\mathbb{U}_{ij}$ are the translation matrices of electromagnetic waves from object $i$ to object $j$, which account with all the information of the relative positions between the objects of our system.

From Eq.~\eqref{Energy_T_finite}, the quantum ($T\to 0$) and classical ($\hbar\to 0$) limits are easily obtained as
\begin{equation}\label{Energy_zero_T}
E_{0} = \frac{\hbar c}{2\pi}\int_{0}^{\infty}dk\log\abs{\mathbbm{1} - \mathbb{N}(\kappa)}
\end{equation}
and
\begin{equation}\label{Energy_high_T}
E_{cl} = \frac{k_{B}T}{2}\log\abs{\mathbbm{1} - \mathbb{N}(0)}
\end{equation}
respectively. Here we denote with a sub--index $T$ the results valid for all temperatures, with a sub--index $0$ the results in the quantum limit ($T\to 0$), and with a sub--index $cl$ the results in the classical limit ($\hbar\to 0$), which is equivalent to the high $T$ limit.

In the case of cylinders, $\mathbb{T}$ matrices are generally known \cite{RE09}, as well as $\mathbb{U}$ matrices for parallel cylinders \cite{RE09}. For this particular case, Casimir energy scales with the infinite length $L$ of the cylinders. It is a natural question if this scaling with the length remains for non parallel cylinders. On one hand, for the non parallel case, the scale of cylinders is still the length $L$, but on the other hand, for the parallel case, all points of the cylinders were equally relevant in order to contribute to the Casimir energy and it is no longer true for the non parallel case, where necessarily there is a principal contribution of the nearest points and there are points infinitely far away.

The Proximity Force Approximation (PFA) applied to the electromagnetic (EM) Casimir energy leads to the next results for perfect metal parallel cylinders in the zero \cite{Rahi:2008kb} and high temperature limits:
\begin{equation}\label{PFA_Energy_T_0_parallel}
E_{0,\parallel}^{PFA} = - \frac{\pi^{3}}{1920}\hbar c L\sqrt{\frac{R}{(d - 2R)^{5}}},
\end{equation}
\begin{equation}\label{PFA_Energy_T_cl_parallel}
E_{cl,\parallel}^{PFA} = - \frac{\zeta(3)}{16}k_{B}T L\sqrt{\frac{R}{(d - 2R)^{3}}},
\end{equation}
where $R$ is the radius of the cylinders and $d$ is the distance between their centers. These energies scales with the length of the cylinders $L$. If we calculate the PFA for the non parallel case, we obtain (see Appendix \ref{sec: Appendix.C})
\begin{equation}\label{PFA_Energy_T_0_non_parallel}
E_{0,\gamma}^{PFA} = - \frac{\pi^{3}}{720}\frac{\hbar c}{\abs{\sin(\gamma)}}\frac{R}{(d - 2R)^{2}},
\end{equation}
\begin{equation}\label{PFA_Energy_T_cl_non_parallel}
E_{cl,\gamma}^{PFA} = - \frac{\zeta(3)}{4}\frac{k_{B}T}{\abs{\sin(\gamma)}}\frac{R}{(d - 2R)}.
\end{equation}
For the non-parallel case, the energy scales with $1/\abs{\sin(\gamma)}$ instead with $L$. This result suggests that the parallel and the non-parallel problems have different scales. However, PFA is an approximation valid for short distances between the bodies. At large distances, the energy can be qualitatively different, as it is the case of prolate spheroids \cite{Spheroids_Emig}.

For non-parallel cylinders, Eq.~\eqref{Energy_T_finite} remains valid, as well as $\mathbb{T}$ matrices. But now $\mathbb{U}$ matrices depend not only on the relative distance $d$ between cylinders, but also on their relative orientation. Then, in addition to the translation contribution, we have to consider the rotation around the translation axis an angle $\theta$. For non crossed cylinders this can be demonstrated to be the more general transformation.

\begin{figure}[h]
\begin{center}
\includegraphics[width=0.9\columnwidth]{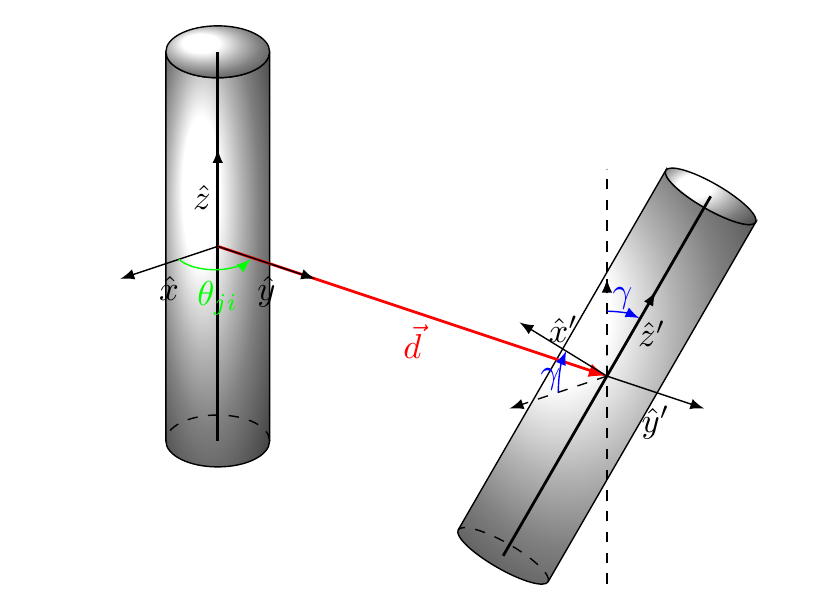}
\caption{\label{Cylinders_Plot}System of non--parallel cylinders studied in this Chapter for the particular case $\theta_{ji} = \frac{\pi}{2}$. The coordinate basis are also represented.}
\end{center}
\end{figure}

In particular, we will study the case of a translation over a $\textbf{d}$ vector of the $(x,y)$-plane, which will also be the rotation axis (see Fig. \ref{Cylinders_Plot}), then the change of spatial coordinates is
\begin{equation}\label{Cordenadas espaciales a prima}
\textbf{x}' = \mathbb{R}_{\hat{\textbf{d}},\gamma}\textbf{x} + \textbf{d},
\end{equation}
with
\begin{equation}\label{Translation_Vector}
\textbf{d} = \left(\begin{array}{c}
d\cos(\theta_{ji})\\
d\sin(\theta_{ji})\\
0
\end{array}\right)
\end{equation}
as the translation vector over the $(x,y)$-plane, and
\begin{align}\label{Rotation_Matrix}
& \mathbb{R}_{\hat{\textbf{d}},\gamma} = \\
& \left(\begin{array}{c|c|c}
\cos(\gamma) + \cos^{2}(\theta_{ji})(1 - \cos(\gamma)) & \sin(\theta_{ji})\cos(\theta_{ji})(1 - \cos(\gamma)) & \sin(\theta_{ji})\sin(\gamma)\\
\hline
\sin(\theta_{ji})\cos(\theta)(1 - \cos(\gamma)) & \cos(\gamma) + \sin^{2}(\theta_{ji})(1 - \cos(\gamma)) & - \cos(\theta_{ji})\sin(\gamma)\\
\hline
- \sin(\theta_{ji})\sin(\gamma) & - \cos(\theta_{ji})\sin(\gamma) & \cos(\gamma)
\end{array} \right)\nonumber
\end{align}
as the rotation matrix over the $\hat{\textbf{d}}$ axis an angle $\gamma$.
We also will use the transformation in the momentum space
\begin{equation}\label{Cordenadas momentos a prima}
\left(\begin{array}{c}
k_{x}'\\
k_{y}'\\
k_{z}'
\end{array}\right) = \mathbb{R}_{\hat{\textbf{d}},\gamma}\left(\begin{array}{c}
k_{x}\\
k_{y}\\
k_{z}
\end{array}\right)
\end{equation}
The derivation of the translation matrices will be given in the next Section.

\section{Translation matrices between non-parallel cylindrical multipole basis}\label{sec: 3}
In this Section, we derive the transformation matrices $\mathbb{U}$ between scalar and vectorial cylindrical multipoles of non parallel basis. In particular, we need to transform the outgoing cylindrical wave functions with imaginary frequency of a $B'$ basis in terms of regular cylindrical wave functions with the same imaginary frequency of another basis $B$ translated by a distance $d$ and whose axes are tilted against each other by an angle $\gamma$ , as shown in Eq.~\eqref{Cordenadas espaciales a prima}.

We perform this analysis for scalar and vectorial cylindrical waves.

\subsection{Scalar waves}
We consider regular and outgoing cylindrical wave functions, which on the imaginary frequency axis are given by 
\begin{equation}\label{Energy_T_0_scalar_reg}
\phi_{n,k_{z}}^{reg}(\textbf{x}) = I_{n}(\rho p)e^{in\theta}e^{ik_{z}z},
\end{equation}
\begin{equation}\label{Energy_T_0_scalar_out}
\phi_{n,k_{z}}^{out}(\textbf{x}) = K_{n}(\rho p)e^{in\theta}e^{ik_{z}z},
\end{equation}
with $p = \sqrt{k_{x}^{2} + k_{y}^{2}}$ and $\rho = \sqrt{x^{2} + y^{2}}$. The outgoing waves at position $\textbf{x}'$ can be expanded in terms of regular waves at $\textbf{x}$ by the linear relation
\begin{equation}\label{U_matrix_definition_crossed_cylinders}
\phi^{out}_{n',k'_{z}}(\textbf{x}') = \sum_{n\in\mathbb{Z}}\int_{-\infty}^{\infty}\frac{dk_{z}}{\frac{2\pi}{L}}\mathbb{U}_{n'k'_{z},nk_{z}}(d,\gamma)\phi_{n,k_{z}}^{reg}(\textbf{x}),
\end{equation}
which defines the translation matrix $\mathbb{U}_{n'k'_{z},nk_{z}}(d,\gamma)$. The basis have a normal distance $d$ and their axes are tilted against each other by an angle $\gamma$ as shown in Eq.~\eqref{Cordenadas espaciales a prima}. The mathematical details of the derivation of this translation matrix are given in the Appendix A \ref{App. A}, and the translation matrix defined in Eq.~\eqref{U_matrix_definition_crossed_cylinders} is obtained as
\begin{equation}\label{U matriz cilindros girados theta escalar bueno}
\mathbb{U}_{n'k'_{z},nk_{z}}(d,\gamma) \hspace{-1pt}=\hspace{-1pt} \frac{2\pi}{L}\frac{(-1)^{n}}{\abs{\sin(\gamma)}}e^{i(n' - n)\theta_{ji}}\hspace{-1pt}\left(\xi' + \sqrt{{\xi'}^{2} + 1}\right)^{n'}\hspace{-5pt}\left(\xi + \sqrt{{\xi\phantom{'}}^{2} + 1}\right)^{-n}\hspace{-3pt}\frac{e^{- d\sqrt{{k'}_{\perp}^{2} + {p'}^{2}} }}{2\sqrt{{k'}_{\perp}^{2} + {p'}^{2}}},
\end{equation}
where $\xi = \frac{k_{\perp}}{p}$, $\xi' = \frac{k'_{\perp}}{p'}$, $k_{\perp} = \frac{\cos(\gamma)k_{z} - k'_{z}}{\sin(\gamma)}$ and $k'_{\perp} = \frac{k_{z} - \cos(\gamma)k'_{z}}{\sin(\gamma)}$.

For the calculation of Casimir energies, we need the translation matrices $\mathbb{U}_{ij}$ and $\mathbb{U}_{ji}$, which are the translation matrices from object $i$ to object $j$. For this particular problem, $\mathbb{U}_{12} = \mathbb{U}_{n'k'_{z},nk_{z}}(d, \theta_{21},\gamma)$, and $\mathbb{U}_{21} = \mathbb{U}_{n'k'_{z},nk_{z}}(d, \pi + \theta_{21}, \gamma)$. Notice that the sign of $\gamma$ is not modified. Therefore, using Eq.~\eqref{U matriz cilindros girados theta escalar bueno}, it is easy to verify that
\begin{equation}\label{U_scalar_matrix_to_minus_hat_x}
\mathbb{U}_{n'k'_{z},nk_{z}}(d, \pi + \theta_{21}, \gamma) = (-1)^{n' - n}\mathbb{U}_{n'k'_{z},nk_{z}}(d, \theta_{21},\gamma).
\end{equation}

\subsection{Electromagnetic waves}
For the EM field the translation matrix couples the polarizations. However, it can expected that, for large $d$, the EM energy is given by the Dirichlet (TM) part only due to the logarithmic dependence on d for TM modes compared to the power-law decay of TE modes. However, the TM modes are coupled now by the translation matrix of vector waves, resulting in an attenuation of the asymptotic energy. The vector waves are defined in terms of the scalar cylinder waves as
\begin{equation}
\textbf{M}_{n,k_{z}}^{out} = \frac{1}{p}\nabla\times\left(\phi_{n,k_{z}}^{out}\hat{\textbf{z}}\right),
\hspace{2cm}
\textbf{N}_{n,k_{z}}^{out} = \frac{1}{\kappa p}\nabla\times\left(\phi_{n,k_{z}}^{out}\hat{\textbf{z}}\right).
\end{equation}
and equivalently for regular vector waves. To derive the transformation matrices, we first multiply both sides of Eq.~\eqref{U_matrix_definition_crossed_cylinders} by $\hat{\textbf{z}}'$ and apply $\frac{1}{p'}\nabla'\times$ to both sides. Then the left hand side gives the components of $\textbf{M}_{n,k_{z}}^{out}$ in the coordinate frame $\left(x',y',z'\right)$. To obtain the components in the frame $\left(x,y,z\right)$ we multiply both sides by the inverse rotation matrix $\mathbb{R}_{\hat{\textbf{d}},\gamma}^{-1}$ where $\mathbb{R}_{\hat{\textbf{d}},\gamma}$ is defined by $\textbf{x}' = \mathbb{R}_{\hat{\textbf{d}},\gamma}\textbf{x} + \textbf{d}$ (see Eq.~\eqref{Cordenadas espaciales a prima}). This yields, using $\nabla' = \mathbb{R}_{\hat{\textbf{d}},\gamma}\nabla$,
\begin{align}\label{Before_substitution}
\textbf{M}_{n',k'_{z}}^{out,\,\gamma}(\textbf{x}') & = \mathbb{R}_{\hat{\textbf{d}},\gamma}^{-1}{\textbf{M}'}_{n',k'_{z}}^{out,\,\gamma}(\textbf{x}')\nonumber\\
 & = \frac{1}{p'}\sum_{n\in\mathbb{Z}}\int_{-\infty}^{\infty}\frac{dk_{z}}{2\pi}\mathbb{U}_{n'k'_{z},nk_{z}}(d,\theta)\nabla\times\left[\phi^{reg}_{n,k_{z}}(\textbf{x})\left(\begin{array}{l}
\phantom{+}\sin(\theta)\sin(\gamma)\hat{\textbf{x}}\\
 - \cos(\theta)\sin(\gamma)\hat{\textbf{y}}\\
 + \cos(\gamma)\hat{\textbf{z}}\end{array}\right)\right],
\end{align}
where we have used the definition of $\hat{\textbf{z}}'$ in the initial coordinate basis given in Eq.~\eqref{Cordenadas espaciales a prima}. Eq \eqref{Before_substitution} shows we need to express $\nabla\times\left(\phi^{reg}_{n,k_{z}}(\textbf{x})\hat{\textbf{x}}\right)$ and $\nabla\times\left(\phi^{reg}_{n,k_{z}}(\textbf{x})\hat{\textbf{y}}\right)$ in terms of $\textbf{M}_{n,k_{z}}^{reg}(\textbf{x})$ and $\textbf{N}_{n,k_{z}}^{reg}(\textbf{x})$. One can show (by expressing $\nabla$, $\hat{\textbf{x}}$ and $\hat{\textbf{y}}$ in cylinder coordinates) that
\begin{equation}
\nabla\times\left(\psi_{n,k_{z}}^{reg}(\textbf{x})\hat{\textbf{x}}\right) = 
\frac{-i\,k_{z}}{2}\left(\textbf{M}_{n-1,k_{z}}^{reg}(\textbf{x}) + \textbf{M}_{n+1,k_{z}}^{reg}(\textbf{x})\right) + 
\frac{i\,\kappa}{2}\left(\textbf{N}_{n-1,k_{z}}^{reg}(\textbf{x}) - \textbf{N}_{n+1,k_{z}}^{reg}(\textbf{x})\right),
\end{equation}
\begin{equation}
\nabla\times\left(\psi_{n,k_{z}}^{reg}(\textbf{x})\hat{\textbf{y}}\right) = 
\frac{k_{z}}{2}\left(\textbf{M}_{n-1,k_{z}}^{reg}(\textbf{x}) - \textbf{M}_{n+1,k_{z}}^{reg}(\textbf{x})\right) - 
\frac{\kappa}{2}\left(\textbf{N}_{n-1,k_{z}}^{reg}(\textbf{x}) + \textbf{N}_{n+1,k_{z}}^{reg}(\textbf{x})\right).
\end{equation}
After applying this result into Eq.~\eqref{Before_substitution}, we need to shift the index $n$ in order to express the right hand side in terms of vector waves with the same index $n$. This can be done by using
\begin{equation}\label{Subida_bajada_indice_matriz_U}
\mathbb{U}_{n'k'_{z},n\pm 1\,k_{z}}(d,\gamma) = - e^{\mp i\theta_{ji}}\left(\xi + \sqrt{1 + \xi^{2}}\right)^{\mp 1}\mathbb{U}_{n'k'_{z},nk_{z}}(d,\gamma).
\end{equation}
This yields after some elementary algebra to the transformation formula for vector waves,
\begin{equation}\label{U_em_matrix_crossed_cylinders}
\left(\begin{array}{c}
\textbf{M}_{n',k'_{z}}^{\gamma}\\
\textbf{N}_{n',k'_{z}}^{\gamma}
\end{array}\right)^{out}\hspace{-10pt}(\textbf{x}') = \sum_{n\in\mathbb{Z}}\int_{-\infty}^{\infty}\frac{dk_{z}}{\frac{2\pi}{L}}
\mathbb{U}_{n'k'_{z},nk_{z}}^{EM}(d,\gamma)\left(\begin{array}{c}
\textbf{M}_{n,k_{z}}\\
\textbf{N}_{n,k_{z}}
\end{array}\right)^{reg}\hspace{-10pt}(\textbf{x}),
\end{equation}
where we have defined the EM translation matrix $\mathbb{U}_{n'k'_{z},nk_{z}}^{EM}(d,\theta)$ in terms of the translation matrix for the scalar field given in Eq.~\eqref{U matriz cilindros girados theta escalar bueno} as
\begin{equation}\label{U_em_matrix_crossed_cylinders2}
\mathbb{U}_{n'k'_{z},nk_{z}}^{EM}(d,\gamma) = \mathbb{U}_{n'k'_{z},nk_{z}}(d,\gamma)\frac{p}{p'}\left(\begin{array}{c|c}
\cos(\gamma) + \sin(\gamma)\frac{k_{z}}{p}\xi & \sin(\gamma)\frac{\kappa}{p}\sqrt{1 + \xi^{2}}\\
\hline
- \sin(\gamma)\frac{\kappa}{p}\sqrt{1 + \xi^{2}} & \cos(\gamma) + \sin(\gamma)\frac{k_{z}}{p}\xi
\end{array}\right).
\end{equation}

Here we have used $\textbf{N} = \frac{1}{\kappa}\nabla\times\textbf{M}$ and $\frac{1}{\kappa}\nabla\times\textbf{N} = - \textbf{M}$ to obtain the transformation formula for $\textbf{N}$.

The transformation formula for the inverse coordinate transformation is given by Eq.~\eqref{U_em_matrix_crossed_cylinders} with $\mathbb{U}_{n'k'_{z},nk_{z}}$ replaced by $(-1)^{n+n'}\mathbb{U}_{n'k'_{z},nk_{z}}$, see Eq.~\eqref{U_scalar_matrix_to_minus_hat_x}, then
\begin{equation}\label{U_EM_matrix_to_minus_hat_x}
\mathbb{U}_{n'k'_{z},nk_{z}}^{EM}(d, \pi + \theta_{ji}, \gamma) = (-1)^{n' - n}\mathbb{U}_{n'k'_{z},nk_{z}}^{EM}(d, \theta_{ji},\gamma).
\end{equation}

\section{Asymptotic Casimir energy}\label{sec: 4}
In this Section we obtain the approximation at large distances of electromagnetic and scalar Casimir energies at zero and high temperature limits. The approximation performed in this Section is valid for small cylinder radius or for large distance between the cylinders.

To perform this limit we first apply the change to adimensional frequencies $q_{z} = d\,k_{z}$, and $q = d\,\kappa$ for the zero temperature case. As a consequence, $\mathbb{U}$ matrices depend on adimensional variables, while the $\mathbb{T}_{i}$ matrix of each cylinder scales with the ratio between the radius of this cylinder and the distance between their centers $r_{i} = \frac{R_{i}}{d}$. In this Chapter, we assume that both cylinders have the same radius $R$, but the obtained results can be easily extended to the most general case without difficulty.

After that, we apply the relation $\log\abs{A} = \text{Tr}\log(A)$ and perform the Taylor expansion to $\log(\mathbbm{1} - \mathbb{N}) = - \sum_{p=1}^{\infty}\frac{\text{Tr}\left(\mathbb{N}^{p}\right)}{p}$ to Eqs.~\eqref{Energy_zero_T} and \eqref{Energy_high_T} in the adimensional variable $r = \frac{R}{d}$, then it is an expansion on large distances, but also on small radius. Then the integrand is now $\log(\mathbbm{1} - \mathbb{N}) \approx - \text{Tr}\left(\mathbb{N}\right)$, but with the $\mathbb{T}$ matrices in $\mathbb{N}$ substituted by the tree level of their Taylor expansion over $r$.

\subsection{Scalar field}
For Dirichlet and Neumann boundary conditions, the scattering amplitudes of a cylinder of radius R are given by the known
expressions,
\begin{equation}
\mathbb{T}_{n'k'_{z},nk_{z}}^{D} = - \frac{2\pi}{L}\frac{I_{n}(pR)}{K_{n}(pR)}\delta_{n,n'}\delta(k_{z} - k'_{z}).
\end{equation}
\begin{equation}
\mathbb{T}_{n'k'_{z},nk_{z}}^{N} = - \frac{2\pi}{L}\frac{I'_{n}(pR)}{K'_{n}(pR)}\delta_{n,n'}\delta(k_{z} - k'_{z}).
\end{equation}
To obtain the asymptotic energy for Dirichlet boundary conditions cylinders at large distances $d$ (small radius $R$), we need to consider only the terms with $n = n' = 0$ and $p_{\text{max}} = 1$ of $\mathbb{N}$ matrix of Eqs.~\eqref{Energy_zero_T} and \eqref{Energy_high_T}. Then the main contribution at large distance of the $\mathbb{T}$ matrix with argument $z$ is $\frac{I_{0}(z)}{K_{0}(z)} = - \left(\gamma + \log\left(\frac{z}{2}\right)\right)^{-1} + \mathcal{O}\left[z^{2}\right]\approx - \log^{-1}(z)$, with the Euler constant $\gamma$. For $z = r\sqrt{q^{2} + q_{z}^{2}}$, we have $\frac{I_{0}(z)}{K_{0}(z)}\approx - \log^{-1}(r)$ up to logarithmic corrections. Then the Asymptotic Casimir energy at zero temperature between non parallel Dirichlet cylinders is
\begin{equation}\label{Scalar_Dirichlet_SRAEnergy_zero_T}
E_{0} = - \frac{\hbar c}{8 d \abs{\sin(\gamma)}\log^{2}\left(\frac{R}{d}\right)}.
\end{equation}
For parallel cylinders of length $L\gg (d, R)$, one has \cite{RE09}
\begin{equation}
E_{0}^{\parallel} = - \frac{\hbar c L}{8\pi d^{2}\log^{2}\left(\frac{R}{d}\right)},
\end{equation}
so that one could identify $\abs{\sin(\gamma)} \leftrightarrow \pi\frac{d}{L}$.

In order to study the classical limit we just need to consider the zero Matsubara frequency contribution to obtain the energy, but a simple scaling analysis shows that $\mathbb{N}_{0\,k_{z},0\,k_{z}}(\kappa = 0)\approx\frac{1}{\abs{k_{z}}}$ for $k_{z}\to\,0$ with logarithmic corrections. Hence the trace of $\mathbb{N}$ is not well defined, i.e., $\mathbb{N}_{0\,k_{z},0\,k_{z}}(\kappa = 0)$ is not a trace class operator, so that $\abs{\mathbbm{1} - \mathbb{N}_{0\,k_{z},0\,k_{z}}(\kappa = 0)}$ is not well defined. Kenneth and Klich showed that $\mathbb{N}$ is a trace class operator for compact objects \cite{Kenneth_and_Klich}. While parallel cylinders are also non-compact objects, they constitute effectively a 2D problem with two compact discs. It appears that two tilted infinitely long cylinders provide the first example of non-compact objects whose geometry cannot be reduced to a lower dimensional one.

However, we expect the force to be well defined, which means that the operator $\partial_{d}\mathbb{N}_{0\,k_{z},0\,k_{z}}(\kappa = 0)$ should be a trace class operator. This is indeed the case, as the following calculation shows. We have that the force in the high $T$ limit is
\begin{equation}
F_{cl} = - \frac{k_{B}T\pi}{4 d\abs{\sin(\gamma)}\log^{2}\left(\frac{R}{d}\right)},
\end{equation}
and once we have the force, it is a straightforward exercise to obtain the energy as $E_{cl} = \int_{d}^{\infty}dx\,F_{cl}(x)$, then it is
\begin{equation}\label{Scalar_Dirichlet_SRAEnergy_high_T}
E_{cl} = \frac{k_{B}T\pi}{4\abs{\sin(\gamma)}\log\left(\frac{R}{d}\right)}.
\end{equation}
Note that the sign of the energy is negative because the power of the logarithm has been reduced from two to one and $R\ll d$ in this limit. 
This is the only case when we find a non trace class operator for scalar fields. When frequency increases or the boundary conditions of one of the cylinders is not of the Dirichlet type, all  $\mathbb{N}$ operators are trace class operators, then they give finite contributions to finite energies.

For example, for Neumann boundary conditions we have to consider the terms with $p_{\text{max}} = 1$, $\abs{n}\leq 1$ and $\abs{n'}\leq 1$ since $\frac{I'_{n}(z)}{K'_{n}(z)} = - \frac{z^{2}}{2} + \mathcal{O}\left[z^{4}\right]$ for $\abs{n}\leq 1$. With this expansion we get to lowest order in $\frac{R}{d}$ the energy
\begin{equation}
E_{0} = - \frac{\hbar c R^{4}}{320 d^{5}\abs{\sin(\gamma)}}(167 + \cos(2\gamma)),
\end{equation}
and the classical limit for this case is directly the finite result
\begin{equation}
E_{cl} = - \frac{3 k_{B}T\pi R^{4}}{1024 d^{4}\abs{\sin(\gamma)}}(98 + \cos(2\gamma)).
\end{equation}

\subsection{Electromagnetic Casimir energy}
For perfect metals, the scattering amplitudes of a cylinder of radius $R$ are given by expressions
\begin{equation}
\mathbb{T}_{n'k'_{z},nk_{z}}^{EE} = - \frac{2\pi}{L}\frac{I_{n}(pR)}{K_{n}(pR)}\delta_{n,n'}\delta(k_{z} - k'_{z}),
\end{equation}
\begin{equation}
\mathbb{T}_{n'k'_{z},nk_{z}}^{MM} = - \frac{2\pi}{L}\frac{I'_{n}(pR)}{K'_{n}(pR)}\delta_{n,n'}\delta(k_{z} - k'_{z}),
\end{equation}
where non diagonal terms are zero. We calculate the energies with the expansions of Eqs.~\eqref{Energy_zero_T} and \eqref{Energy_high_T}. The leading part of the energy in the far distance approximation is given by the term that is quadratic in the T-matrix for the TM waves. The polarization mixing in the translation matrices given in Eq.~\eqref{U_em_matrix_crossed_cylinders} plays now a role, because the part of the polarization which is transformed from TM to TE does not contribute to the lowest order of Casimir energy, so the asymptotic energy must be lower than in the scalar case, and equal to the scalar case just in these cases when polarizations do not mix. It is for parallel cylinders, as already known, and for the classical limit, because mixing terms are proportional to $\sin(\gamma)$ and $\kappa$. Then the asymptotic Casimir energy for perfect metal cylinders can be written as
\begin{equation}\label{SRA_em_Energy_nonparallel_cylinders}
E_{0} = - \frac{\hbar c }{8d\abs{\sin(\gamma)}\log^{2}\left(\frac{R}{d}\right)}\Omega(\gamma),
\end{equation}
where the function $\Omega(\gamma)$ is defined as

\begin{equation}\label{Ecuacion_Omega_de_gamma}
\Omega(\gamma) = \frac{1}{2\pi}\int_{0}^{\frac{\pi}{2}}\hspace{-4mm}d\varphi\sin(\varphi)\int_{0}^{2\pi}\hspace{-4mm}d\phi\frac{\left(\cos(\gamma)\sin(\varphi) + \sin(\gamma)\cos(\varphi)\sin(\phi)\right)^{2}}{\left(\sin(\gamma)\cos(\varphi) + \cos(\gamma)\sin(\varphi)\sin(\phi)\right)^{2} + \sin^{2}(\varphi)\cos^{2}(\phi)}.
\end{equation}

\begin{figure}[h] %[h] para here [b] para bottom [t] para top, pero sólo son sugerencias,  [H] ES IMPERATIVO
\begin{center}
\includegraphics[width=0.9\columnwidth]{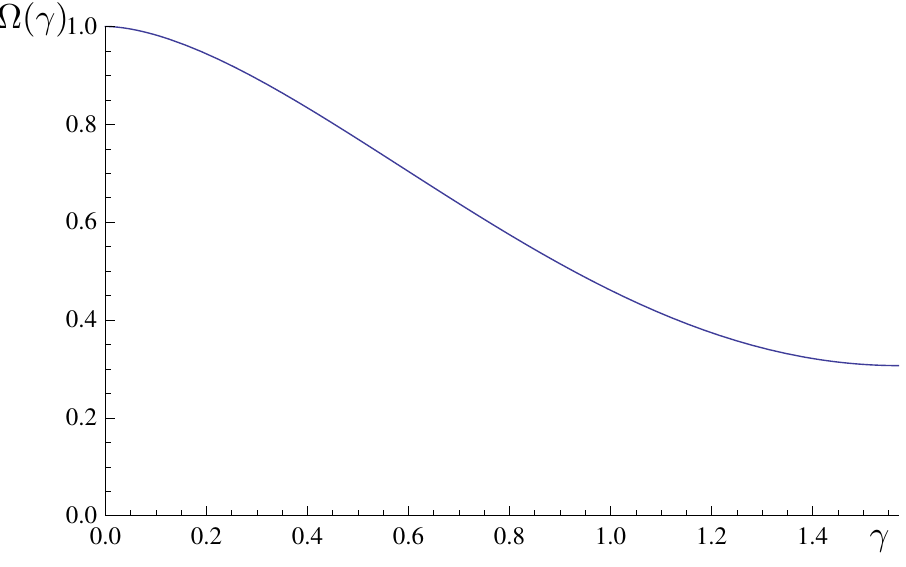}
\caption{\label{Omega(gamma)}Function $\Omega(\gamma)$ given in Eq.~\eqref{Ecuacion_Omega_de_gamma}, as a function of the tilted angle $\gamma$. }% The greater difference is seen for the parallel case.}
\end{center}
\end{figure}

We have $\Omega(0) = 1$ so that for $\gamma\to\,0$ the Dirichlet result given in Eq.~\eqref{Scalar_Dirichlet_SRAEnergy_zero_T} is recovered as expected. For orthogonal cylinders one has $\Omega(\frac{\pi}{2}) = 1 - \log(2)$. We have not been able to perform the integral definition of $\Omega(\gamma)$. Anyway, a Fourier series expansion can be performed numerically giving us a good enough solution. Because the symmetry of the system, the series just contains cosines of even terms. then $\Omega(\gamma)\approx\sum_{n=0}^{\infty}\Omega_{2n}\cos(2n\gamma)$. We find numerically that $\Omega_{0} = 0.6137$, $\Omega_{2} = 0.3333$, $\Omega_{4} = 0.0333$ and $\Omega_{6} = 0.0096$. The function $\Omega(\gamma)$ is plotted in the Fig.~\ref{Omega(gamma)}.

% The maximun $\Omega(0) = 1$ and the minimum $\Omega(\frac{\pi}{2}) = 1 - \log(2)$ are plotted just by eye orientation.}

As in the classical limit the polarization mixing disappears, we have actually two scalar problems. One for TM modes, equivalent to a scalar field between cylinders subject to Neumann boundary conditions because of the perfect diamagneticity, which leads to a well defined trace class operator, and another one for TE modes, equivalent to another scalar field between cylinders subject to Dirichlet boundary conditions because of the perfect reflexibility. This part of the system is the dominant one at large distances and presents the same problem in the classical limit as its scalar analog.

When we try to calculate the classical limit of the energy just considering the zero Matsubara frequency contribution, we find again the same problem as in the scalar field with Dirichlet boundary conditions. A simple scaling analysis shows that $\mathbb{N}_{0\,k_{z},0\,k_{z}}(\kappa = 0)\approx\frac{1}{\abs{k_{z}}}$ for $k_{z}\to\,0$ with logarithmic corrections also for the leading order of classical limit of the energy for perfect metal cylinders. Also in this case $\mathbb{N}_{0\,k_{z},0\,k_{z}}(\kappa = 0)$ is not a trace class operator, so that $\abs{\mathbbm{1} - \mathbb{N}_{0\,k_{z},0\,k_{z}}(\kappa = 0)}$ is not well defined \cite{Kenneth_and_Klich}. It is another more physical case when non compact objects lead to a non trace class $\mathbb{N}$ operator.

However, as in the scalar case, we are able to obtain the classical limit of the force, because $\partial_{d}\mathbb{N}_{0\,k_{z},0\,k_{z}}(\kappa = 0)$ is a trace class operator as shown before. Then the far distance approximation of the classical limit of the force is
\begin{equation}
F_{cl} = - \frac{k_{B}T\pi}{4 d\abs{\sin(\gamma)}\log^{2}\left(\frac{R}{d}\right)},
\end{equation}
and the corresponding energy is obtained with the integral $E_{cl} = \int_{d}^{\infty}dx\,F_{cl}(x)$ as
\begin{equation}
E_{cl} = \frac{k_{B}T\pi}{4\abs{\sin(\gamma)}\log\left(\frac{R}{d}\right)}.
\end{equation}
Then we obtain the same result as in Eq.~\eqref{Scalar_Dirichlet_SRAEnergy_high_T} of the scalar case, as expected. Once again, this anomalous behavior of the high temperature limit just can appear when the leading contribution of cylinder $\mathbb{T}$ matrices of both cylinders presents a log-divergence when the frequency $\kappa$ goes to zero. As in the scalar case, if just one cylinder presents this divergence, $\mathbb{N}$ matrix is still of trace class. In particular, for two dielectric cylinders (or even for the case of a metallic cylinders and a dielectric cylinder) this anomalous behavior is not present.

\section{Numerical results}\label{sec: 6}
Following the analytical asymptotic results of Casimir energies, we perform a numerical evaluation of Casimir energies between non parallel cylinders in order to study intermediate regimes and the convergence to asymptotic results. We center our study in the zero temperature case, then we perform a numerical implementation of Eq.~\eqref{Energy_zero_T}. Finite temperatures could be performed on the same way, just being careful with the zero Matsubara frequency contribution for two perfect metals cylinders. This procedure is simple, but with some subtleties. If we observe the expression of the $\mathbb{N}$ matrix, we see that there is an internal integration loop over $k_{\perp}$ in order to obtain the matrix in $(n k_{z},n'k'_{z})$ labels. After that we have to perform the logarithm of a linear operator over one set of discrete variables $(n,n')$ and over another set of continuous variables $(k_{z},k'_{z})$, because now the $\mathbb{N}$ matrix is not diagonal in $k_{z}$. Obviously, the number of elements of $\mathbb{N}$ must be finite, which means we will use cylindrical multipoles with label $n\in[-n_{\text{max}},n_{\text{max}}]$. The effect of this cutoff is the transformation of the exact formula given by Eq.~\eqref{Energy_zero_T} into an asymptotic result, valid at a given error until a characteristic distance between the cylinders which depends on the $n_{\text{max}}$ used as a numerical cutoff to obtain a numerical tractable method. In addition to that, we have to calculate the determinant of an operator defined over the continuous variables $(k_{z},k'_{z})$. One possibility would be the approximation of the determinant by a series of traces of powers of $\mathbb{N}$, but this procedure introduces another new asymptotic approximation to the problem. The other possibility is to apply a discretization to the $(k_{z},k'_{z})$ space and evaluate the determinant as usual with enough discretization points in order to minimize the discretization error. We expect that this procedure would be some kind of trapezium rule of integration for determinants of continuous operators. It is important to note that the evaluation of each $\mathbb{N}$ matrix is now several orders of magnitude harder than the parallel cylinders case. The final reason is the symmetry breaking of the system, which not only produces that the $\mathbb{N}$ matrix is not diagonal in $k_{z}$, making the evaluation of the determinant harder, it also means that an internal integration over $k_{\perp}$ must be performed numerically, because it depends on the $\mathbb{T}$ matrix of one of the cylinders, while in the parallel case this integration is divided into two integrals independent of the $\mathbb{T}$ matrix and which can been performed analytically (see Appendix B \ref{App. B}). As a consequence, we just have obtained numerical results for perpendicular cylinders $\gamma = \frac{\pi}{2}$ and for $\gamma = \frac{\pi}{4}$ for all distances $r = \frac{R}{d}\leq 0.45$. Then we are far of the range of validity of PFA. We have used cylindrical multipoles just until $\abs{n}\leq 3$. It is enough to observe corrections to the small radius approximation and a soft convergence to PFA results.

\begin{figure}[ht]
\begin{center}
\includegraphics[width=0.9\columnwidth]{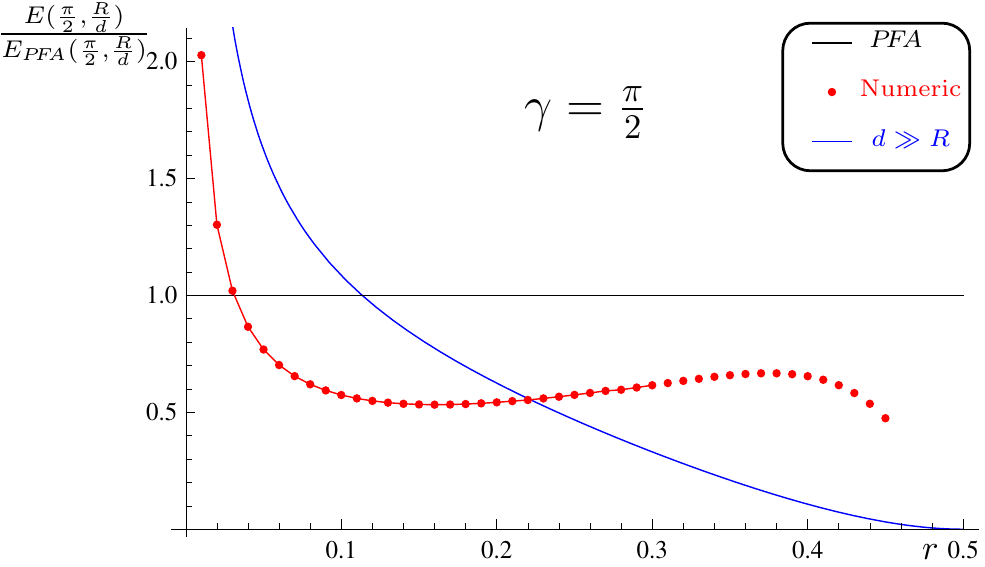}
\caption{\label{Fig 1}Energies compared with PFA as a function of $r = \frac{R}{d}$ for $\gamma = \frac{\pi}{2}$. The black curve is the PFA energy itself. The blue curve is the asymptotic distance energy and the red points are the numerical Casimir energies. The red curve joins the points $r\le 0.3$, where our numerical method converges properly to the correct result. A convergence to the PFA result is expected when $r\to\frac{1}{2}$.}
\end{center}
\end{figure}

\begin{figure}[ht]
\begin{center}
\includegraphics[width=0.9\columnwidth]{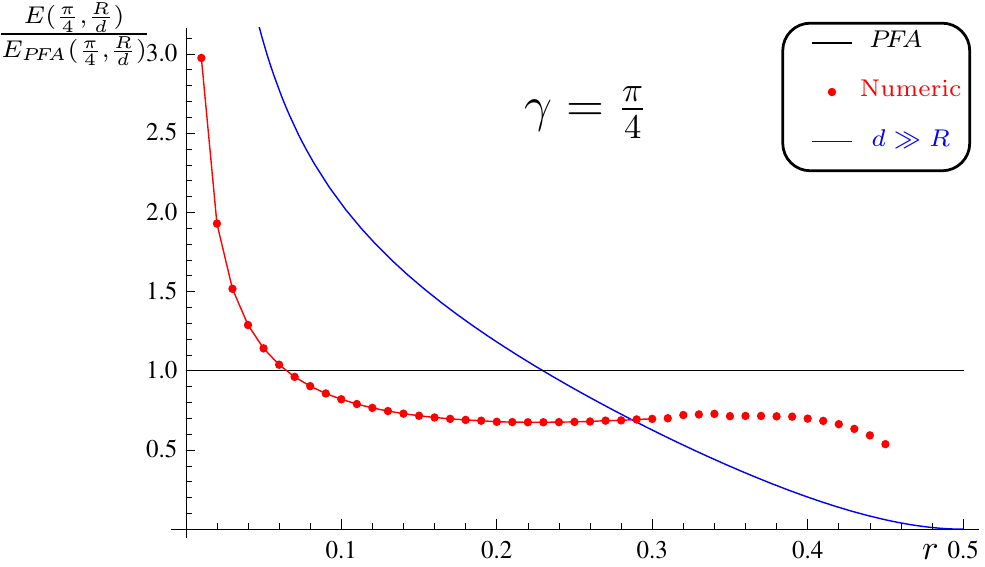}
\caption{\label{Fig 2}Energies compared with PFA as a function of $r = \frac{R}{d}$ for $\gamma = \frac{\pi}{4}$. The black curve is the PFA energy itself. The blue curve is the asymptotic distance energy and the red points are the numerical Casimir energies. The red curve joins the points $r\le 0.3$, where our numerical method converges properly to the correct result. A convergence to the PFA result is expected when $r\to\frac{1}{2}$.}
\end{center}
\end{figure}

In Fig. \ref{Fig 1} and in Fig. \ref{Fig 2} we show the energies scaled with the PFA energy for cylinders with $\gamma = \frac{\pi}{2}$ and with $\gamma = \frac{\pi}{4}$ respectively. In the two figures we observe common characteristics with the parallel cylinders case. It looks like the energies (red curves) do not converge to their asymptotic result (blue curves) at large distances, i.e. small $r$, but it is not the case. As shown in Eq.~\eqref{SRA_em_Energy_nonparallel_cylinders}, the asymptotic distance Casimir energy $E_{ass}$ decays with $\log^{-2}(r)$, then the convergence is reached at extremely small distances and it is not observable in out plots. At intermediate distances, higher order cylindrical contributions get to produce relevant contributions to Casimir energy, and the energy became to be an appreciable fraction of the PFA result (black curve), until the approximations of the numerical method become relevant and we loose the convergence to PFA at shorter distances. The use of cylindrical multipoles with $\abs{n}\leq n_{\text{max}}$ is valid for a given error until a characteristic distance between the cylinders which depends on the $n_{\text{max}}$ used as a numerical cutoff to obtain a numerical tractable method. We have checked numerically that the results obtained at $r = \frac{R}{d}\leq 0.3$ are well described with $n_{\text{max}} = 3$. To study even shorter distances and the convergence to PFA result, we should include more multipoles into our study.

\begin{figure}[h]
\begin{center}
\includegraphics[width=0.9\columnwidth]{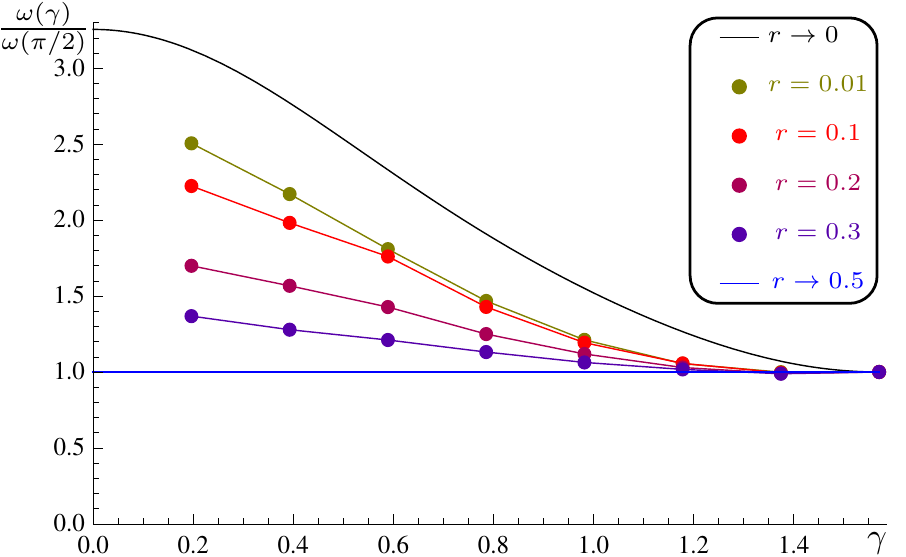}
\caption{\label{Fig 3}Ratio $\frac{\omega(r,\gamma)}{\omega(r,\frac{\pi}{2})}$, where $\omega(r,\gamma) = E(r,\gamma)\abs{\sin(\gamma)}$ for angles $\gamma\in\left[\frac{\pi}{16},\frac{\pi}{2}\right]$. The asymptotic distance result and the PFA are also plotted. Curves are smoothing from the asymptotic result to the constant value of the PFA result when the distance is reduced, as expected.}
\end{center}
\end{figure}

We are also interested in the dependence of the energy with the tilted angle $\gamma$ between the cylinders at a given distance. As seen in Eq.~\eqref{SRA_em_Energy_nonparallel_cylinders} and in Eq.~\eqref{PFA_em_Energy_nonparallel_cylinders_zero_T}, the asymptotic distance approximation and PFA energies are inversely proportional to $\abs{\sin(\gamma)}$, but they have different angular dependence. To study the angular dependence of the energy at intermediate distances, we represent in Fig. \ref{Fig 3} how is modified the product $\omega(r,\gamma) = E(r,\gamma)\abs{\sin(\gamma)}$ for all angles compared with the same function at $\gamma = \frac{\pi}{2}$, so we eliminate the dependence with the distance and we are able to study the relative dependence of the energy with the angle. We can observe that, at large distances, the dependence of $\omega(r,\gamma)$ with $\gamma$ is the one of the asymptotic results, but when we reduce the distance between the cylinders, $\omega(r,\gamma)$ becomes more and more independent of the angle, showing that it tends to converge to the constant value of the PFA result.

\section{Conclusions}
In this Chapter we have expanded the study of Casimir energy to non parallel cylinders, for the scalar field and for the electromagnetic field.

In order to study this system, we have obtained the translation matrices of outgoing cylindrical multipoles in terms of regular cylindrical multipoles of a translated non parallel basis in Sect. \ref{sec: 3} for scalar and vectorial cases.

We have studied the asymptotic far distance regime in Sect. \ref{sec: 4} and near contact regime in Appendix \ref{sec: Appendix.C}, at the zero and high temperature regimes.

We have found that the zero Matsubara frequency case leads to a non trace class $\mathbb{N}$ operator for the scalar field when both cylinders impose Dirichlet boundary conditions and for the electromagnetic field when both cylinders are perfect metals. The reason is that, contrary to the case of parallel cylinders, non parallel cylinders system is not reducible to an effectively 2D system of compact objects, then the theorem of Kenneth and Klich \cite{Kenneth_and_Klich} is not applicable to this system, and $\mathbb{N}$ matrix could be a trace-class operator or not. Anyway we can recover an energy towards the integration of the force, because $\partial_{d}\mathbb{N}(\kappa=0)$ is a trace-class operator. We should remark that the zero Matsubara frequency contributes to Casimir energy at any non zero temperature, then this result is relevant not only in the high temperature limit.

We have studied numerically the zero temperature Casimir energy between perfect metal cylinders for all distances and angles, showing how asymptotic large and short limits are connected at intermediate distances, and the variation of the angular dependence at a constant distance.

We have to remark that all the study made for perfect metal cylinders could be easily repeated for dielectrics. In this case at large distances the energy decays with a power law behavior, the zero Matsubara frequency leads to a trace class operator and the energy is generally smaller than in the perfect metal case.

The study of Casimir energy between cylinders as a function of their relative angle $\gamma$ could be experimentally relevant. We could use the different scales of the energy when cylinders are parallel, when $E_{\parallel}\propto\frac{L}{d^{2}}$ instead $E_{\gamma}\propto\frac{1}{d\abs{\sin(\gamma)}}$ for the non parallel case. It turns from an irrelevant value for non parallel cylinders to a much greater value proportional to their length $L$, which could be experimentally studied. In addition to that, Casimir effect could be relevant in nematic order of elongated fibers, because it flavors the parallel configuration of them.

%\acknowledgements
%We acknowledge helpful discussions with R.~Brito.  P.R.-L.’s research is supported  by projects MOSAICO, UCM/PR34/07-15859, MODELICO (Comunidad de Madrid) and a FPU MEC grant.

% Apéndice
%\renewcommand{\thesection}{\arabic{section}}
%\appendix
\section{Appendix A: Derivation of translation matrices for scalar cylindrical multipoles between non-parallel system of coordinates}\label{App. A}
In this Appendix, we provide the mathematical details of the derivation the representation of Eq.~\eqref{U matriz cilindros girados theta escalar bueno}. It is the representation of $\phi^{out}_{n',k'_{z}}(\textbf{x}')$, outgoing scalar cylindrical multipoles of a given basis in terms of $\phi_{n,k_{z}}^{reg}(\textbf{x})$, regular cylindrical multipoles of another basis translated a distance $d$ and rotated around the axis defined by the translation vector an angle $\theta$, as shown by the coordinate transformation given in Eq.~\eqref{Cordenadas espaciales a prima}. Then the translation matrix $\mathbb{U}_{n'k'_{z},nk_{z}}(d,\gamma)$ is defined by the linear relation \eqref{U_matrix_definition_crossed_cylinders}. We start from the 2D Fourier transform of the outgoing waves,
\begin{equation}\label{Outgoing_2D_Fourier_Transform}
K_{n}(\rho p)e^{in\theta} = \int_{-\infty}^{\infty}\frac{dk_{x}}{2\pi}\int_{-\infty}^{\infty}\frac{dk_{y}}{2\pi}2\pi(-i)^{n}\left(\frac{k_{x} + ik_{y}}{p}\right)^{n}\frac{e^{i(k_{x}x + k_{y}y)}}{k_{x}^{2} + k_{y}^{2} + p^{2}}.
\end{equation}
First of all, we apply a change of variables from $(k_{x},k_{y})$ to $(k_{\|},k_{y\perp})$ defined as
\begin{equation}\label{Cambio1}
\left(\begin{array}{c}
k_{\|}\\
k_{\perp}
\end{array}\right) = \left(\begin{array}{c|c}
\phantom{-} \cos(\theta_{ji}) & \sin(\theta_{ji})\\
\hline
 - \sin(\theta_{ji}) & \cos(\theta_{ji})
\end{array} \right)\left(\begin{array}{c}
k_{x}\\
k_{y}
\end{array}\right),
\end{equation}
and the same change of variables to the conjugate variables in the position space as
\begin{equation}\label{Cambio2}
\left(\begin{array}{c}
r_{\|}\\
r_{\perp}
\end{array}\right) = \left(\begin{array}{c|c}
\phantom{-} \cos(\theta_{ji}) & \sin(\theta_{ji})\\
\hline
 - \sin(\theta_{ji}) & \cos(\theta_{ji})
\end{array} \right)\left(\begin{array}{c}
x\\
y
\end{array}\right),
\end{equation}
therefore, the integral of Eq.~\eqref{Outgoing_2D_Fourier_Transform} can be written in terms of the new integration variables as
\begin{equation}\label{Outgoing_2D_Fourier_Transform_v2}
K_{n}(\rho p)e^{in\theta} = \int_{-\infty}^{\infty}\frac{dk_{\|}}{2\pi}\int_{-\infty}^{\infty}\frac{dk_{\perp}}{2\pi}2\pi(-i)^{n}\left(\frac{k_{\|} + ik_{\perp}}{p}\right)^{n}e^{in\theta_{ji}}\frac{e^{i(k_{\|}r_{\|} + k_{\perp}r_{\perp})}}{k_{\|}^{2} + k_{\perp}^{2} + p^{2}}.
\end{equation}

The integration over $k_{\|}$ can be easily performed using the residue theorem. The integrand has the poles $k_{\|} = \pm i\sqrt{p^{2} + k_{\perp}^{2}}$. Hence we obtain
\begin{equation}
K_{n}(\rho p)e^{in\theta} = e^{in\theta_{ji}}\int_{-\infty}^{\infty}dk_{\perp}\left(\frac{k_{\perp} \pm \sqrt{k_{\perp}^{2} + p^{2}}}{p}\right)^{n}\frac{e^{ik_{\perp}r_{\perp} \mp r_{\|}\sqrt{k_{\perp}^{2} + p^{2}} }}{2\sqrt{k_{\perp}^{2} + p^{2}}}
\end{equation}
where the minus (plus) sign applies to $r_{\|} > 0$ ($r_{\|} < 0$). After multiplying by $e^{ik_{z}z}$, we obtain
\begin{equation}
K_{n}(\rho p)e^{in\theta}e^{ik_{z}z} = e^{in\theta_{ji}}\int_{-\infty}^{\infty}dk_{\perp}\left(\frac{k_{\perp} \pm \sqrt{k_{\perp}^{2} + p^{2}}}{p}\right)^{n}\frac{e^{i(k_{\perp}r_{\perp} + k_{z}z) \mp r_{\|}\sqrt{k_{\perp}^{2} + p^{2}} }}{2\sqrt{k_{\perp}^{2} + p^{2}}}
\end{equation}
Writing the same formula in the prime coordinate system:
\begin{align}
K_{n'}(\rho' p')e^{in'\theta'}e^{ik'_{z}z'} & = e^{in'\theta_{ji}}\int_{-\infty}^{\infty}dk'_{\perp}\left(\frac{k'_{\perp} \pm \sqrt{{k'}_{\perp}^{2} + {p'}^{2}}}{p'}\right)^{n'}\frac{e^{i(k'_{\perp}r'_{\perp} + k'_{z}z') \mp r'_{\|}\sqrt{{k'}_{\perp}^{2} + {p'}^{2}} }}{2\sqrt{{k'}_{\perp}^{2} + {p'}^{2}}},
\end{align}
and using $p = \sqrt{\kappa^{2} + k_{z}^{2}}$, $p' = \sqrt{\kappa^{2} + {k'}_{z}^{2}}$, it is easy to obtain $k_{z}^{2} + k_{\perp}^{2} = {k_{z}'}^{2} + {k_{\perp}'}^{2}$, $k_{\perp}r_{\perp} + k_{z}z = k'_{\perp}r'_{\perp} + k'_{z}z'$ and $k_{\|} = {k'}_{\|} = \sqrt{{k'}_{\perp}^{2} + {p'}^{2}}$ by the use of Eq.~\eqref{Cordenadas momentos a prima}. Then, applying the change of coordinates given in Eq.~\eqref{Cordenadas espaciales a prima}, we have the outgoing cylindrical functions of one basis in terms of outgoing plane waves of the translated and rotated basis as
\begin{align}\label{Formula_cambiada_de_coordenadas}
K_{n'}(\rho' p')e^{in'\theta'}e^{ik'_{z}z'} & = e^{in'\theta_{ji}}\int_{-\infty}^{\infty}dk'_{\perp}\left(\frac{k'_{\perp} + \sqrt{{k'}_{\perp}^{2} + {p'}^{2}}}{p'}\right)^{n'}\frac{e^{i(k_{\perp}r_{\perp} + k_{z}z + k_{\|}r_{\|}) - d\sqrt{{k'}_{\perp}^{2} + {p'}^{2}} }}{2\sqrt{{k'}_{\perp}^{2} + {p'}^{2}}},
\end{align}
Thanks to the relationship $(xk_{x} + yk_{y}) = (r_{\|}k_{\|} + r_{\perp}k_{\perp})$, we can use the expansion of 2D plane waves in cylindrical waves as,
\begin{equation}
e^{i(r_{\|}k_{\|} + r_{\perp}k_{\perp})} = e^{i(xk_{x} + yk_{y})} = \sum_{n\in\mathbb{Z}}i^{n}J_{n}\left(\rho\sqrt{k_{x}^{2} + k_{y}^{2}}\right)e^{in\theta}e^{-in\cos^{-1}\left(\frac{k_{x}}{\sqrt{k_{x}^{2} + k_{y}^{2}}}\right)},
\end{equation}
where $\rho = \sqrt{x^{2} + y^{2}}$. Obviously we have $\cos^{-1}\left(\frac{k_{x}}{\sqrt{k_{x}^{2} + k_{y}^{2}}}\right) = \sin^{-1}\left(\frac{k_{y}}{\sqrt{k_{x}^{2} + k_{y}^{2}}}\right)$. Applying the relation $p = i\sqrt{k_{x}^{2} + k_{y}^{2}}$, we transform the equality to
\begin{equation}
e^{i(r_{\|}k_{\|} + r_{\perp}k_{\perp})} = \sum_{n\in\mathbb{Z}}i^{n}J_{n}\left(i\rho p\right)e^{in\theta}e^{-in\sin^{-1}\left(\frac{k_{y}}{ip}\right)},
\end{equation}
which can be written as
\begin{equation}
e^{i(r_{\|}k_{\|} + r_{\perp}k_{\perp})} = \sum_{n\in\mathbb{Z}}(-1)^{n}I_{n}\left(\rho p\right)e^{in\theta}\left(\frac{k_{y} + \sqrt{k_{y}^{2} + p^{2}}}{p}\right)^{-n}.
\end{equation}
Using the relation $p = i\sqrt{k_{x}^{2} + k_{y}^{2}}$, it is easy to obtain that $\sqrt{k_{y}^{2} + p^{2}} = ik_{x}$. Therefore, we can write $(k_{y} + \sqrt{k_{y}^{2} + p^{2}}) = (k_{y} - ik_{x}) = e^{i\theta_{ji}}(k_{\perp} - ik_{\|}) = e^{i\theta_{ji}}(k_{\perp} + \sqrt{p^{2} + k_{\perp}^{2}})$ and the series acquire an additional phase proportional to $\theta_{ji}$
\begin{equation}
e^{i(r_{\|}k_{\|} + r_{\perp}k_{\perp})} = \sum_{n\in\mathbb{Z}}(-1)^{n}I_{n}\left(\rho p\right)e^{in\theta}e^{-in\theta_{ji}}\left(\frac{k_{\perp} + \sqrt{k_{\perp}^{2} + p^{2}}}{p}\right)^{-n}.
\end{equation}
By the insertion of this sum in the integrand, we are able to bring the result as a sum of modes and a integration over $k_{\perp}'$ frequencies as
\begin{align}\label{U_matrix_before_change_from_ky_to_kz}
K_{n'}(\rho' p')e^{in'\theta'}e^{ik'_{z}z'} &= \sum_{n\in\mathbb{Z}}\int_{-\infty}^{\infty}dk'_{\perp}(-1)^{n}\left(\frac{k'_{\perp} + \sqrt{{k'}_{\perp}^{2} + {p'}^{2}}}{p'}\right)^{n'}\left(\frac{k_{\perp} + \sqrt{k_{\perp}^{2} + p^{2}}}{p}\right)^{-n}\times\nonumber\\
&\times\frac{e^{- d\sqrt{{k'}_{\perp}^{2} + {p'}^{2}} }}{2\sqrt{{k'}_{\perp}^{2} + {p'}^{2}}}e^{i(n' - n)\theta_{ji}}I_{n}\left(\rho p\right)e^{in\theta}e^{ik_{z}z}.
\end{align}
In order to bring the result into the form of Eq.~\eqref{U_matrix_definition_crossed_cylinders} we change the variable of integration from $k_{\perp}$ to $k_{z}$ with the help of the relation $k_{z} = \cos(\gamma)k'_{z} + \sin(\gamma)k'_{\perp}$, then $dk'_{\perp} = \frac{dk_{z}}{\abs{\sin(\gamma)}}$, and the expansion of outgoing cylindrical function in terms of regular waves of a non-parallel cylindrical basis is
\begin{align}
K_{n'}(\rho' p')e^{in'\theta'}e^{ik'_{z}z'} &= \sum_{n\in\mathbb{Z}}\frac{L}{2\pi}\int_{-\infty}^{\infty}dk_{z}\frac{2\pi}{L}\frac{(-1)^{n}}{\abs{\sin(\gamma)}}e^{i(n' - n)\theta_{ji}}\left(\frac{k'_{\perp} + \sqrt{{k'}_{\perp}^{2} + {p'}^{2}}}{p'}\right)^{n'}\times\nonumber\\
& \times\left(\frac{k_{\perp} + \sqrt{k_{\perp}^{2} + p^{2}}}{p}\right)^{-n}\frac{e^{- d\sqrt{{k'}_{\perp}^{2} + {p'}^{2}} }}{2\sqrt{{k'}_{\perp}^{2} + {p'}^{2}}}I_{n}\left(\rho p\right)e^{in\theta}e^{ik_{z}z},
\end{align}

where
\begin{equation}
k_{\perp} = \frac{\cos(\gamma)k_{z} - k'_{z}}{\sin(\gamma)},
\hspace{40pt}
k'_{\perp} = \frac{k_{z} - \cos(\gamma)k'_{z}}{\sin(\gamma)}.
\end{equation}
Then the translation matrix defined in Eq.~\eqref{U_matrix_definition_crossed_cylinders} is:
\begin{align}\label{U matriz cilindros girados theta escalar bueno2}
& \mathbb{U}_{n'k'_{z},nk_{z}}(d,\gamma) = \nonumber\\
& \frac{2\pi}{L}\frac{(-1)^{n}}{\abs{\sin(\gamma)}}e^{i(n' - n)\theta_{ji}}\left(\xi' + \sqrt{{\xi'}^{2} + 1}\right)^{n'}\left(\xi + \sqrt{{\xi\phantom{'}}^{2} + 1}\right)^{-n}\frac{e^{- d\sqrt{{k'}_{\perp}^{2} + {p'}^{2}} }}{2\sqrt{{k'}_{\perp}^{2} + {p'}^{2}}},
\end{align}
with $\xi = \frac{k_{\perp}}{p}$ and $\xi' = \frac{k'_{\perp}}{p'}$, which is the result given in Eq.~\eqref{U matriz cilindros girados theta escalar bueno}. It is interesting to note that an additional translation over the $\hat{\textbf{z}}$ axis, this is to say, for cases with $d_{z}\neq 0$, the generalization of the translation matrices showed here is straightforward and results in just an additional phase factor $e^{-ik_{z}d_{z}}$ (which appears in the coordinate transformation of $z'$ showed in Eq.~\eqref{Formula_cambiada_de_coordenadas}) in the expression of $\mathbb{U}_{n'k'_{z},nk_{z}}(d,\gamma)$ given in Eq.~\eqref{U matriz cilindros girados theta escalar bueno2}.

\section{Appendix B: Translation matrices for scalar cy\-lin\-dri\-cal multipoles between parallel system of coordinates}\label{App. B}
In this Appendix, we reobtain the translation matrix between parallel cylindrical multipoles as a limiting case when $\gamma\to\,0$ of Eq.~\eqref{U matriz cilindros girados theta escalar bueno} for scalar and Eq.~\eqref{U_em_matrix_crossed_cylinders} for vectorial multipoles.

In the parallel case, the cross polarization matrix which appears in Eq.~\eqref{U_em_matrix_crossed_cylinders} is reduced to the identity matrix. Then for the parallel case polarizations do not couple. As a consequence,  Eq.~\eqref{U_em_matrix_crossed_cylinders} is reduced to two scalar decoupled translation problems, one for each independent polarization. So it is enough to obtain the translation matrix for scalar multipoles in order to obtain the result for the vectorial case. For the parallel case we have $k_{\perp} = k_{\perp}'$ and $k_{z} = k_{z}'$. As a consequence, $\frac{p}{p'} = 1$ and $\xi' = \xi$. Then we can follow the derivation of the scalar translation matrix until Eq.~\eqref{U_matrix_before_change_from_ky_to_kz}, where for the parallel case we have
\begin{align}
& \mathbb{U}_{n'k'_{z},nk_{z}}(d,\gamma = 0) = \nonumber\\
& (-1)^{n'}e^{i(n' - n)\theta_{ji}}\int_{-\infty}^{\infty}dk_{\perp}\frac{2\pi}{L}\delta(k_{z} - k'_{z})\left(\xi + \sqrt{ 1 + \xi^{2} }\right)^{n - n'}\frac{e^{-d\sqrt{\kappa^{2} + k_{y}^{2} + k_{z}^{2}}}}{2\sqrt{\kappa^{2} + k_{y}^{2} + k_{z}^{2}}}.
\end{align}
Instead of applying the change of variable from $k_{\perp}$ to $k_{z}$, which would be singular in the parallel case, we carry out the integration over $k_{\perp}$, resulting in the known translation matrix between parallel cylindrical scalar multipoles and for each polarization of vector multipoles as
\begin{equation}
\mathbb{U}_{n'k'_{z},nk_{z}}(d,\gamma = 0) = (-1)^{n'}e^{i(n' - n)\theta_{ji}}K_{n - n'}(d\sqrt{\kappa^{2} + k_{z}^{2}})\frac{2\pi}{L}\delta(k_{z} - k'_{z}).
\end{equation}
Then we recover the parallel translation matrices \cite{RE09} as a particular case.

\section{Appendix C: Proximity Force Approximation of Casimir energy}\label{sec: Appendix.C}
Small Radius Approximation is valid for large distance. In the short distance regime Proximity Force Approximation (PFA) is the valid one. We just perform the calculus for perfect metal cylinders. The scalar case is just a half of the perfect metal result.

\begin{figure}[h]
\begin{center}
\includegraphics[width=0.9\columnwidth]{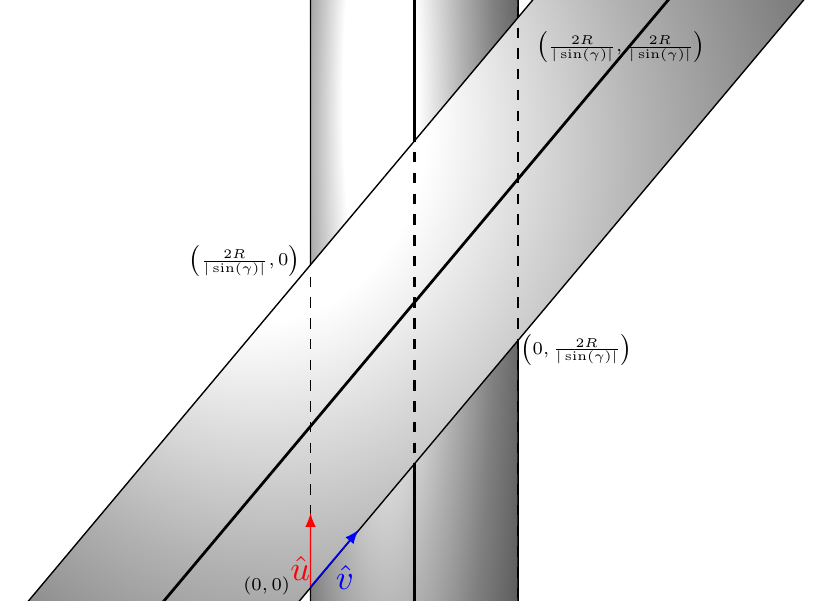}
\caption{\label{Cilindros_PFA}Region of integration for PFA. The unit vectors $\hat{u}$ and $\hat{v}$ are shown too.}
\end{center}
\end{figure}

The area across which the two cylinders overlap, viewed along the axis that is perpendicular to the two cylinder axes and that intersects the axes in their crossing point, forms a parallelogram of edge length $2R/\abs{\sin(\gamma)}$. Let us denote the coordinates along the edges of this parallelogram as $u$ and $v$ (see Fig.~\ref{Cilindros_PFA}). Then the local distance $h(u,v)$ between the two cylinder surfaces, measured normal to the plane that is spanned by the cylinder axes, is given by the function
\begin{equation}
h(u,v) = d - \sqrt{R^{2} - \left(u\abs{\sin(\gamma)} - R\right)^{2}} - \sqrt{R^{2} - \left(v\abs{\sin(\gamma)} - R\right)^{2}}.
\end{equation}
where $d$ is the distance between the cylinder axes. Taking into account that a surface element of the parallelogram is given by $\sin(\gamma)dudv$, the PFA energy can be written as
\begin{equation}
E_{0}^{PFA} = - \frac{\hbar c\pi^{2}}{720}\int_{0}^{\frac{2R}{\abs{\sin(\gamma)}}}\int_{0}^{\frac{2R}{\abs{\sin(\gamma)}}}\frac{\abs{\sin(\gamma)}dudv}{h^{3}(u,v)}.
\end{equation}
This expression has the advantage that we can expand the square roots for small $l/R$, where $l = d - 2R$ is the surface-to-surface distance, which leads to
\begin{align}\label{Exact_PFA}
E_{0}^{PFA} & = - \frac{\hbar c\pi^{2}}{720}\frac{1}{\abs{\sin(\gamma)}}\frac{R}{l^{2}}\int_{-\sqrt{R/l}}^{\sqrt{R/l}}\int_{-\sqrt{R/l}}^{\sqrt{R/l}}dsdt\frac{1}{\left[\frac{l}{R} + 2 - \sqrt{1 - \frac{l}{R}s^{2}} - \sqrt{1 - \frac{l}{R}t^{2}}\right]^{3}}
\end{align}
The energy can be written after introducing new integration variables as
\begin{equation}
E_{0}^{PFA} = - \frac{\hbar c\pi^{2}}{720}\frac{1}{\abs{\sin(\gamma)}}\frac{R}{l^{2}}\int_{-\sqrt{R/l}}^{\sqrt{R/l}}\int_{-\sqrt{R/l}}^{\sqrt{R/l}}\frac{dsdt}{\left[1 + \frac{1}{2}\left(s^{2} + t^{2}\right)\right]^{3}}
\end{equation}
In this expression we can extend the integration limits to infinity to obtain the limiting behavior for small $l/R$. This yields
\begin{align}\label{PFA_em_Energy_nonparallel_cylinders_zero_T}
\lim_{\frac{l}{R}\to\,0}E_{0}^{PFA} & = - \frac{\hbar c\pi^{2}}{720}\frac{1}{\abs{\sin(\gamma)}}\frac{R}{l^{2}}\int_{0}^{2\pi}d\varphi\int_{0}^{\infty}\frac{\rho d\rho}{\left(1 + \frac{\rho^{2}}{2}\right)^{3}} = - \frac{\hbar c\pi^{3}}{720}\frac{1}{\abs{\sin(\gamma)}}\frac{R}{l^{2}}
\end{align}
For small $l/R$ this approximation deviates from the exact integral of Eq.~\eqref{Exact_PFA} by less than $1\%$ for $l/R < 0.01$. It is instructive to compare this PFA energy to the one for two spheres of radius $R$ and surface-to-surface distance $l$ which is $E_{0}^{PFA} =  - \frac{\hbar c\pi^{3}}{1440}\frac{R}{l^{2}}$~\cite{Kardar-Geometrias-Arbitrarias} and hence reduced by a factor of two compared to the case of perpendicular cylinders ($\gamma = \frac{\pi}{2}$).

With exactly the same procedure, it is possible to obtain the PFA energy for the classical limit, performing the same integration, but now over the classical limit of the energy instead of the zero temperature case. Then the classical limit of the PFA energy is
\begin{equation}
E_{cl}^{PFA} = - k_{B}T\frac{\zeta(3)}{8\pi}\int_{0}^{\frac{2R}{\abs{\sin(\gamma)}}}\int_{0}^{\frac{2R}{\abs{\sin(\gamma)}}}\frac{\abs{\sin(\gamma)}dudv}{h^{2}(u,v)}.
\end{equation}
Performing the same transformations and approximation as in the quantum case, we find that in the limiting behavior for small $l/R$, the PFA energy is
\begin{align}
\lim_{\frac{l}{R}\to\,0}E_{cl}^{PFA} & = - k_{B}T\frac{\zeta(3)}{8\pi}\frac{1}{\abs{\sin(\gamma)}}\frac{R}{l}\int_{0}^{2\pi}d\varphi\int_{0}^{\infty}\frac{\rho d\rho}{\left(1 + \frac{\rho^{2}}{2}\right)^{2}} = - k_{B}T\frac{\zeta(3)}{4}\frac{1}{\abs{\sin(\gamma)}}\frac{R}{l}
\end{align}

%%%%% Conclusion
\begin{savequote}[10cm] % this sets the width of the quote
\sffamily
``So Long, and Thanks for All the Fish.'' 
\qauthor{Douglas Adams}
\end{savequote}

\chapter{Conclusion}

This Thesis is focused on the study of Casimir effect as a consequence of fluctuations of fields and its peculiarities and properties.

We have obtained results in several categories.

\begin{itemize}
\item On the conceptual side, we have proposed a dynamical formalism of Casimir effect, which let us define the Casimir effect as \textit{the response of a fluctuanting medium to the breakdown of the translation symmetry because of the presence of intrusions in that medium.} Such formalism let us focus on the Casimir effect as an universal feature of steady states. We have defined the stochastic variable ''Stochastic Casimir force over an $\alpha$ body'' whose mean value is the Casimir force itself.

\item The formalism presented here generalizes the Casimir effect to non--equilibrium steady states systems, and it is demonstrated that the equilibrium Casimir effect is a particular but important case. In fact, we obtained the explicit equivalence between the formalism presented here, the Stress--Tensor formalism and the partition function formalism of equilibrium Casimir effect. Even the important case of Casimir effect generated from quantum--thermal fluctuations of the EM field is covered as a particular case with this formalism, via the Parisi--Wu formalism.

\item As the Casimir force is a stochastic variable, it must follows some probability distribution function, then it has interest the study of its moments, in particular, we defined the variance of the Casimir force, which let us study the fluctuations of these Fluctuation--Induced forces. In fact, we have obtained the variance of the Casimir effect between two perfect metal pistons of arbitrary cross section, at any given temperature at the first time to our knowledge.

\item On the mathematical side, we have obtained the two point correlation function of steady states generated by a kind of generalized Langevin equation when the temporal evolution operator is generalized to a polynomial of any degree. We obtained the Casimir force between two parallel Dirichlet plates in such steady states.

\item On the applied side, we used the recent developed version of the multiscattering formalism to study the physical peculiarities of the EM Casimir effect: Nonmonotonicities of Casimir forces between two objects because of the presence of a third one, or because of the appearance of intervals of negative entropies in the system; the obtention of the PSA from first principles for the complete EM field, the generalization of the $N$ points formalism of Power and Thirunamachandran, the Casimir energy between non--parallel cylinders, and the Casimir effect between Topological Insulators with their anomalous behavior.
\end{itemize}

\section{Specific outcomes}

\begin{itemize}
\item Presentation of the dynamical formalism of Casimir effect.  The Casimir effect is \textit{the response of a fluctuanting medium to the breakdown of the translation symmetry because of the presence of intrusions in that medium.}
\begin{enumerate}
\item Definition of the Stochastical variable ''Stochastic Casimir force over the object $\alpha$''
\item Generalization of Casimir effect to non--equilibrium steady states via the dynamical formalism.
\item Reobtention of the equilibrium Casimir effect as a particular case of the dynamical formalism.
\item Explicit equivalence of the Stress--Tensor formalism, the partition function formalism and the dynamical formalism of the Casimir effect in the equilibrium.
\item Reobtention of the originally proposed EM Casimir effect as a consequence of the quantum fluctuations of the EM field via Parisi--Wu formalism.
\item Definition of the variance of the Casimir force, and result for perfect metal pistons of arbitrary section.
\item Generalization of the Langevin equation to temporal derivatives greater than one, and obtention of Casimir force in such steady states.
\end{enumerate}

\item Derivation of the Pairwise Summation Approximation (PSA) for the EM Casimir effect from first principles from the multiscattering formalism of Casimir effect, including next order perturbation terms.
\begin{enumerate}
\item New results of PSA for dielectrics with electric response, magnetic response and general magnetoelectric coupled responses. We include indications on how to proceed when the electromagnetic response of the objects are not homogeneous. Results valid for any given temperature. Recovery of the superposition principle in the PSA, and breaking of it with the first correction term to PSA.
\item Reobtention of the $N$ points potential formalism of Power and Thirunamachandran as the large distance limit of PSA, and extension of such formalism to generalized (linear) dielectrics at any given temperature.
\item Use of PSA for the complete EM field for the study of the Casimir effect between Topological Insulators. Evaluation of the adequacy of the approximation as a good qualitative and quantitative approximation. Criteria for obtaining the appearance of the different regimes of behaviors of TI depending on the magnitude and sign of the magnetoelectric polarizabilities.
\end{enumerate}

\item Nonmonotonicities of the Casimir force between two objects.
\begin{enumerate}
\item Nonmonotonicities of the Casimir force between two objects because of the presence of a third one.
\item Study of the entropy of a system of two metallic spheres originated because of the Casimir effect. Appearance of intervals of negative entropy at an interval of given temperatures and distances for Perfect metal model and plasma model with lower penetration length of a critical one. No appearance of this range for all other cases of the plasma model and for the Drude model.
\item Thermodynamic consequences of the appearance of negative entropies of Casimir effect. The principles of Thermodynamics continue to be satisfied, of course, but the Casimir force is not monotonous with temperature provided that the entropy is non--monotonic with distance.
\end{enumerate}

%\item New geometries of Casimir effect.
\item Casimir energy between non--parallel cylinders.
\begin{enumerate}
\item The energy scales with the cosecant of the angle between their axis, nor with their length $L$.
\item New translation matrices between translated non--parallel cylindrical coordinate systems.
\item First counterexample of the theorem of Kenneth and Klich~\cite{Kenneth_and_Klich} in the high $T$ limit for Dirichlet cylinders and for perfect metal cylinders, because the condition of compactness of the intruders is not fulfilled.
%\item Casimir energy between non--parallel cylinders.
%\item Casimir energy between a cylinder and a sphere. With this last case, we complete all the possible cases of systems with spheres, cylinders and plates.
\end{enumerate}
\end{itemize}

\backmatter

\chapter{List of publications}
\begin{itemize}
\item[$\bullet$] \emph{Three-body Casimir effects and nonmonotonic forces}. Pablo Rodriguez-Lopez, Sahand Jamal Rahi, and Thorsten Emig. Physical Review A \textbf{80}, 022519 (2009)~\cite{Rodriguez-Lopez_1}. It corresponds to Chapter~\ref{chb2} of this Thesis.
\item[$\bullet$] \emph{Pairwise summation approximation of Casimir energy from first principles}. Pablo Rodriguez-Lopez. Physical Review E \textbf{80}, 061128 (2009)~\cite{Rodriguez-Lopez_PSA}. It corresponds to Chapter~\ref{chb3} of this Thesis.
\item[$\bullet$] \emph{Dynamical approach to the Casimir effect}. Pablo Rodriguez-Lopez, Ricardo Brito, and Rodrigo Soto. Physical Review E \textbf{83}, 031102 (2011)~\cite{PhysRevE.83.031102}. It corresponds to Chapter~\ref{Dynamical approach to the Casimir effect} of this Thesis.
\item[$\bullet$] \emph{Effect of finite temperature and uniaxial anisotropy on the Casimir effect with three-dimensional topological insulators}. Adolfo G. Grushin, Pablo Rodriguez-Lopez and Alberto Cortijo. Physical Review B \textbf{84}, 045119 (2011)~\cite{Placas_TI_forall_T}. It corresponds to Chapter~\ref{chb3} of this Thesis.
\item[$\bullet$] \emph{Casimir Energy and Entropy in the Sphere–Sphere Geometry}. Pablo Rodriguez-Lopez. Physical Review B \textbf{84}, 075431 (2011)~\cite{Rodriguez-Lopez_3}. It corresponds to Chapter~\ref{Chap: Casimir Energy and Entropy in the Sphere--Sphere Geometry} of this Thesis.
\item[$\bullet$] \emph{Casimir repulsion between topological insulators in the diluted regime}. Pablo Rodriguez-Lopez. Physical Review B \textbf{84}, 165409 (2011)~\cite{Rodriguez-Lopez_5}. It corresponds to Chapter~\ref{chb3} of this Thesis.
\item[$\bullet$] \emph{Stochastic Quantization and Casimir Forces}. Pablo Rodriguez-Lopez, Ricardo Brito, and Rodrigo Soto. Send to Referee process~\cite{Rodriguez-Lopez_2}. It corresponds to Chapter~\ref{Chap: Stochastic Quantization and Casimir forces.} of this Thesis.
\item[$\bullet$] \emph{Casimir energy between non parallel cylinders}. Pablo Rodriguez-Lopez, and Thorsten Emig. In preparation~\cite{Rodriguez-Lopez_4}. It corresponds to Chapter~\ref{Chap: Casimir energy between non parallel cylinders} of this Thesis.
\end{itemize}

%\input{erratum}

%%%%% List of symbols
% your thesis may not need this, so comment out or delete the following line
%\input{listofsymbols}

%%%%% Bibliography, in BibTeX format (the references.bib file)
\bibliography{references}

\end{document}